\documentclass[prb,twocolumn,showpacs,amsmath,amssymb,superscriptaddress]{revtex4}

\usepackage{graphicx}
\usepackage{color}
\usepackage{bbm}

\date{\today}                  

\begin{document}

\title{Surface charge theorem and topological constraints for edge states:
An analytical study of one-dimensional nearest-neighbor tight-binding models}

\author{Mikhail Pletyukhov}
\affiliation{Institut f\"ur Theorie der Statistischen Physik, RWTH Aachen, 
52056 Aachen, Germany and JARA - Fundamentals of Future Information Technology}
\author{Dante M. Kennes}
\affiliation{Institut f\"ur Theorie der Statistischen Physik, RWTH Aachen, 
52056 Aachen, Germany and JARA - Fundamentals of Future Information Technology}
\author{Jelena Klinovaja}
\affiliation{Department of Physics, University of Basel, Klingelbergstrasse 82, 
CH-4056 Basel, Switzerland}
\author{Daniel Loss}
\affiliation{Department of Physics, University of Basel, Klingelbergstrasse 82, 
CH-4056 Basel, Switzerland}
\author{Herbert Schoeller}
\email[Email: ]{schoeller@physik.rwth-aachen.de}
\affiliation{Institut f\"ur Theorie der Statistischen Physik, RWTH Aachen, 
52056 Aachen, Germany and JARA - Fundamentals of Future Information Technology}

\begin{abstract}

For a wide class of noninteracting tight-binding models in one dimension  
we present an analytical solution for all scattering and edge
states on a half-infinite system. Without assuming any symmetry constraints we 
consider models with nearest-neighbor hoppings and one orbital per site but arbitrary size of the unit
cell and generic modulations of on-site potentials and hoppings.
The solutions are parametrized by determinants which can be straightforwardly 
calculated from recursion relations. We show that this representation allows for an 
elegant analytic continuation to complex quasimomentum consistent with previous treatments
for continuum models. Two important analytical results are obtained based on the explicit 
knowledge of all eigenstates. (1) An explicit proof of the surface charge theorem is presented 
including a unique relationship between the boundary charge $Q_B^{(\alpha)}$ of a single band $\alpha$ and the
bulk polarization in terms of the Zak-Berry phase. In particular, the Zak-Berry phase
is determined within a special gauge of the Bloch states such that no unknown integer is left. 
This establishes a precise form 
of a bulk-boundary correspondence relating the {\it boundary charge of a single band} to bulk
properties. (2) We derive a topological constraint for the phase-dependence of the edge state
energies, where the phase variable describes a continuous shift of the lattice 
towards the boundary. The topological constraint is shown to be equivalent to the quantization
of a topological index $I=\Delta Q_B-\bar{\rho}\in \{-1,0\}$ introduced in an accompanying 
letter [arXiv:1911.06890]. Here $\Delta Q_B$ is the change of the boundary charge $Q_B$ for a given
chemical potential in the insulating regime when the lattice is shifted by one site towards the 
boundary, and $\bar{\rho}$ is the average charge per site (both in units of the elementary charge $e=1$). 
This establishes an interesting link between universal properties of the boundary charge and edge state 
physics discussed within the field of topological insulators. In accordance with
previous results for continuum systems, we also establish the localization of the boundary charge
and determine the explicit form of the density given by an exponential decay and a pre-exponential 
function following a power-law with generic exponent $-1/2$ at large distances.

\end{abstract}


\maketitle

\section{Introduction} 
\label{sec:introduction}

Triggered by the discovery of the quantum Hall effect \cite{klitzing_dorda_pepper_prl_80,thouless_etal_prl_82},
the study of insulating materials has received considerable interest
in the last decade due to the development of the field of topological insulators (TIs)
with interesting edge states appearing in the gap which might be useful for
quantum information processing [\onlinecite{volkov_pankratov_JETP_85}-\onlinecite{hsieh_etal_nature_08}], 
see Refs.~[\onlinecite{hasan_kane_RMP_10}-\onlinecite{asboth_book_16}] for reviews and textbooks. 
Since edge states describe interesting physical phenomena happening at the boundary of 
a system, this field is ultimately related to the study of the density at the boundary for an insulating
system, where not only the edge states but also the scattering states have a nontrivial effect. That the
boundary charge $Q_B$ (in units of the elementary charge $e=1$) for a given chemical potential in the
insulating regime does not only consist of the number $Q_E$ of occupied edge states and can appear in
quantized fractionalized units is well known and has a long history \cite{charge_fractionalization}. 
For systems with {\it local} \cite{com_1} inversion symmetry $Q_B$ is generically quantized 
in half-integer units  and the field of topological crystalline insulators 
[\onlinecite{hughes_etal_prb_83}-\onlinecite{lau_etal_prb_16}] has been put forward recently,
extending the standard classification schemes of TIs
[\onlinecite{schnyder_etal_prb_08}-\onlinecite{diez_etal_njp_15}]. An important step towards a
generic discussion of $Q_B$ has been undertaken within the so-called modern theory of polarization
(MTP) [\onlinecite{kingsmith_vanderbilt_prb_93}-\onlinecite{miert_ortix_prb_17}], see
Ref.~[\onlinecite{vanderbilt_book_2018}] for a recent textbook review over the field.
In summary, the MTP is based on two fundamental ingredients put forward in 
Refs.~[\onlinecite{kingsmith_vanderbilt_prb_93,vanderbilt_kingsmith_prb_93}]: 
(1) the establishment of a unique definition of the bulk polarization of an insulating crystal 
in terms of the Zak-Berry phase \cite{zak} and (2) the proof of the {\it surface charge theorem} 
which provides a relation between $Q_B$ and the bulk polarization. Most importantly, 
this relation is not restricted to any symmetry constraints and holds for generic models in all dimensions.
It provides an interesting variant of the common bulk-boundary correspondence discussed within
the field of TIs, where a relation between bulk topological invariants (like Chern and winding 
numbers) for infinite systems and the number of zero-energy edge states for systems with a boundary
is requested [\onlinecite{hatsugai_prl_93}-\onlinecite{bulk_boundary_correspondence}]. 
In analogy, the MTP provides a first step to 
set up such a relation between the bulk Zak-Berry phase and $Q_B$ (instead of $Q_E$). However, a limitation of
the MTP is the fact that the Zak-Berry phase is not a gauge-invariant quantity. There is still a
freedom to add any phase factor $e^{i\varphi_{\vec{k}}}=e^{i\varphi_{\vec{k}+\vec{G}}}$ to the Bloch waves 
which depends on the quasimomentum vector $\vec{k}$ and is periodic when $\vec{k}$ is shifted by
a vector $\vec{G}$ of the reciprocal lattice. The winding
of this phase factor gives rise to an arbitrary integer for the Zak-Berry phase. Moreover, also the
proof of the surface charge theorem in terms of counting Wannier function centers in a certain volume
does neither control the number of edge states nor the precise number of exponentially localized Wannier 
states close to the boundary (with wave functions differing significantly from those of the infinite 
system) in terms of bulk quantities.
Both are hard to determine analytically for generic systems and often have to be calculated numerically.
As a consequence, an important topological integer is left unknown in the 
surface charge theorem and it is even not clear how to relate this unknown integer to $Q_E$. 
This makes it very difficult to bridge the field of TIs discussing edge states 
with the MTP concentrating on $Q_B$. 

Another important topic discussed in the literature is to identify {\it universal} properties of $Q_B$.
For systems with local inversion symmetry it is known that the Zak-Berry phase is quantized in half-integer 
units of $2\pi$ \cite{zak} such that, due to the surface charge theorem, $Q_B$ is also quantized in 
half-integer units. The same quantization appears at half-filling for systems with local chiral symmetry 
like, e.g., for the Su-Schrieffer-Heeger (SSH) model \cite{SSH}. Away from symmetry constraints
a finite one-dimensional (1D) tight-binding model with nearest neighbor hopping and a harmonic on-site 
potential with a wave-length commensurable with the lattice has been studied numerically 
\cite{park_etal_prb_16}. In the insulating regime a very stable and almost linear behaviour of 
$Q_B(\varphi)$ (up to discrete jumps arising from edge states crossing the chemical potential) as 
function of a phase variable $\varphi$ controlling the offset of the potential was revealed. Later on 
this work was extended to other fillings \cite{thakurathi_etal_prb_18}, the stability against 
random disorder was demonstrated, and it was shown that the slope is universal and can be related
to the quantized Hall conductance. Obviously this slope can not be explained by edge state physics
and is triggered by scattering states. An intuitive and very simple physical argument has been put forward
\cite{park_etal_prb_16,thakurathi_etal_prb_18} in terms of charge conservation 
which, for a half-infinite system, can be 
formulated as follows: if $a$ denotes the lattice spacing, the unit cell consists of $Z$ sites for a given 
commensurate wave-length $\lambda = Za$ of the potential. This leads to an average charge
$\bar{\rho}={\nu\over Z}$ per site when $\nu$ bands are filled. Shifting the potential 
continuously by one site towards the boundary via a 
phase change by ${2\pi\over Z}$ one expects in an adiabatic picture that on average the 
charge $\bar{\rho}$ will be shifted into the boundary leading to an increase of the boundary 
charge $Q_B$ by exactly the same amount. This is fundamentally related to the fact that $Q_B$
is defined via a macroscopic average on scales much larger than $Za$, analog to the definition of the
macroscopic charge density in classical electrodynamics 
(see, e.g., Chapter 4.5.1 in Ref.~[\onlinecite{vanderbilt_book_2018}]). As a result, for large $Z$ or,
equivalently, in the large wave-length limit of the potential, one expects that $Q_B(\varphi)$
will be almost a linear function with a universal slope ${Z\bar{\rho}\over 2\pi}={\nu\over 2\pi}$
on average. However, similar to the surface charge theorem described above, this physical argument
misses an unknown integer. Since $Q_B(\varphi)=Q_B(\varphi+2\pi)$ must be periodic
when the lattice is shifted by one unit cell, the charge increase by $\nu$ due to the average linear
slope must be compensated by the net difference of edge states moving above and those moving 
below the chemical potential during the phase interval $2\pi$. It is known from the
integer quantum Hall effect (IQHE) that this number is given by the sum $C_\nu=\sum_{\alpha=1}^\nu C^{(\alpha)}$
of the Chern numbers of the occupied bands, which, due to the Diophantine equation 
\cite{dana_jpc_85,kohmoto_prb_89_jpsj_92,hatsugai_prb_93}, is given by $C_\nu=\nu-s_\nu Z$, where
$s_\nu$ is another integer topological index characteristic for gap $\nu$. To fulfil charge conservation
it is therefore possible that the average linear slope can in general  
take all values ${\nu-s_\nu Z\over 2\pi}$. Furthermore, since the slope appears only on average and 
large deviations can appear for higher lying gaps and for cases when the phase-dependence of the 
model parameters is quite strong, a precise definition of universality is required. 

The fact that an important topological integer is missing within the MTP and within the discussion
of universal properties of $Q_B(\varphi)$ was the motivation of the accompanying letter \cite{paper_prl}
to fix the appropriate gauge for the Zak-Berry phase and to identify a novel topological index to 
characterize universal properties of $Q_B(\varphi)$. In a first step, 1D nearest-neighbor 
tight-binding models with one orbital per site on a half-infinite system were studied, with arbitrary 
size $Z$ of the unit cell and generic on-site potentials $v_j$ and hoppings $t_j$
within a unit cell, where $j=1,\dots,Z$, see Fig.~\ref{fig:model_RL} for illustration. 
The $\varphi$-dependence of the parameters was chosen 
such that the phase-shift $\Delta\varphi={2\pi\over Z}$ describes a shift of the lattice by one site 
towards the boundary, i.e., $v_{j+1}(\varphi)=v_j(\varphi+{2\pi\over Z})$ and 
$t_{j+1}(\varphi)=t_j(\varphi+{2\pi\over Z})$. For constant hopping and in the 
continuum limit ($Z\rightarrow\infty$, $a\rightarrow 0$, such that the 
length $Za$ of a unit cell stays constant), one obtains at low filling the whole class of 1D solid state 
systems with generic periodic potentials    
$V(x)=V(x+Za)$. For the special case of cosine modulations with respect to $\varphi$, the
whole class of generalized Aubry-Andr\'e-Harper models \cite{AAH} is covered, discussed extensively in the
context of topological insulators, the IQHE, photonic crystals, and
cold atom systems. It was argued in Ref.~[\onlinecite{paper_prl}] that, based on the detailed 
knowledge of the eigenstates described in the present manuscript, a gauge of the Bloch states 
can be found such that the boundary 
charge $Q_B^{(\alpha)}$ of a single band $\alpha$ can be related in a unique way to the
Zak-Berry phase fixing the unknown constant of the surface charge theorem
for a {\it single} band. It was also shown that this gauge is related in a very natural way to 
the phase factors of the partial Bloch waves defining the scattering eigenstates of the half-infinite 
system. Concerning the universal properties of the phase-dependence of $Q_B^{(\alpha)}(\varphi)$, it turned out 
that the representation in terms of the Zak-Berry phase is very helpful to show that
the change of $Q_B^{(\alpha)}$ under a shift of the lattice by one site towards the boundary can be written as 
$\Delta Q_B^{(\alpha)}(\varphi)=I_\alpha(\varphi)+{1\over Z}$. Here, $I_\alpha$ is a gauge invariant topological 
index quantized in integer units, which was shown to be identical to the winding number $w_\alpha=-I_\alpha$ 
of a fundamental phase $\theta_k^{(\alpha)}$ given by the phase difference of the Bloch wave function 
between the sites right and left to the position of the boundary defining the half-infinite system. 
Interestingly, concerning the boundary charge $Q_B$ at fixed chemical potential (which includes the 
sum $\sum_{\alpha=1}^\nu Q_B^{(\alpha)}$ of the occupied bands together with the number $Q_E$ of occupied 
edge states) physical arguments were presented that the quantized topological index 
$I(\varphi)=\Delta Q_B(\varphi) - \bar{\rho} = \sum_{\alpha=1}^\nu I_\alpha(\varphi) + \Delta Q_E(\varphi)$ 
can only take the two possible values $I\in\{-1,0\}$. Here, the value $I=0$ is associated with charge 
conservation of particles, which is the argument described above 
leading to $\Delta Q_B=\bar{\rho}$.\cite{park_etal_prb_16} The other value $I=-1$ 
leading to $\Delta Q_B=\bar{\rho}-1$ is 
associated in an analog way with charge conservation of holes. We note that the duality between particles
and holes is based only on the Pauli principle and needs no further symmetry constraint. Since charge conservation 
of particles and holes is fulfilled for any model it is quite remarkable that these simple physical 
ingredients are sufficient to describe an important universal feature of $Q_B(\varphi)$. 
Without explicit proof it was argued in Ref.~[\onlinecite{paper_prl}] that other values 
of $I$ are not possible since it is always possible to choose the $\varphi$-dependence of the model parameters 
in such a way that no edge states cross the chemical potential during the shift of the lattice by one site
(keeping the model parameters fixed before and after the shift). Moreover, it was shown in 
Ref.~[\onlinecite{paper_prl}] that the quantization of $I$ together with the periodicity
$Q_B(\varphi)=Q_B(\varphi+2\pi)$ determines the generic form of $Q_B(\varphi)$ such that
the average linear slope is given in accordance with the Diophantine equation by ${\nu-s_\nu Z\over 2\pi}$,
where $s_\nu$ is a topological index characterizing a topological constraint for the phase-dependence
of the edge state energies. This constraint was stated as 
\begin{align}
\label{eq:topological_constraint}
\Delta F(\varphi) = s_\nu + I(\varphi) \in \{s_\nu-1,s_\nu\}\,,
\end{align}
where $\Delta F(\varphi)$ is the difference of the number of edge states moving below and those moving
above the chemical potential during the shift of the lattice by one site. This means that the topological
index $s_\nu$ appears at two different places in the $\varphi$-dependence of $Q_B$, in the average linear slope as
well as in the number of discrete jumps appearing when edge states cross the chemical potential. For
the calculation of the topological index $I(\varphi)$ the influence of these two dependencies cancel 
each other such that its quantization can be explained by charge conservation of particles and holes 
alone without involving any edge state physics.

The purpose of this work is to present the exact solution for all eigenstates of the
considered class of half-infinite tight-binding models and to provide a rigorous proof of all statements of the
accompanying letter regarding the unique formulation of the surface charge theorem for a single band 
and the derivation of the topological constraint (\ref{eq:topological_constraint}). 
The solutions for the scattering and edge states are presented in terms of sub-determinants of the 
matrix $h_k-\epsilon$, where $h_k$ is the Hamiltonian in quasimomentum space and $\epsilon$ is the energy
of the eigenstate, and we show how to calculate the sub-determinants efficiently 
from a set of recursion relations. To the best of our knowledge, this representation 
has not been stated before and provides a very efficient analytical and numerical tool to obtain all eigenstates 
and physical observables like the density and the boundary charge. We note that in contrast to many other 
approaches trying to find effective analytical or numerical solutions for tight-binding models on finite
systems,\cite{diagonalization_finite_tight_binding} half-infinite systems have the advantage
that the quasimomentum is continuous and the thermodynamic limit has already been carried out
such that the two ends of the system can no longer effect each other. In such systems, following 
general arguments put forward in a recent article \cite{floquet_paper} the solutions of the infinite 
system for complex quasimomentum form generically a basis to construct all scattering and edge states 
for a half-infinite 1D system (up to special bifurcation points where additional solutions with pre-exponential 
power-laws have to be taken into account). Therefore, the use of complex quasimomentum is helpful for any
system to calculate all scattering and edge states of a half-infinite system (even beyond the considered class
of models in this work). However, we note that Ref.~[\onlinecite{floquet_paper}] just describes a general 
scheme for this construction and does not present an explicit representation of the Bloch states in 
terms of the quasimomentum (which requires the explicit solution of the Schr\"odinger equation).  

Concerning the definition of the Zak-Berry phase we will elaborate on two different ways which can be chosen
for the form of the Bloch wave in tight-binding models with periodic modulation of the
parameters. The choice used in Ref.~[\onlinecite{paper_prl}] is given by
\begin{align}
\label{eq:bloch_1}
\psi^{(\alpha)}_{k,\text{bulk}}(n,j)={1\over\sqrt{2\pi}}\chi_k^{(\alpha)}(j) e^{ikn}\,,
\end{align}
where $n$ is the unit cell index and the $Z$-dimensional vector $\chi_k=\chi_{k+2\pi}$ is periodic in $k$ 
and describes the form of the Bloch wave function within the unit cell (we have set the lattice spacing 
$a=1$ such that $-\pi \le k < \pi $). The other possible choice used standardly in solid state physics is 
\begin{align}
\label{eq:bloch_2}
\psi^{(\alpha)}_{k,\text{bulk}}(m)={1\over\sqrt{2\pi}}\bar{\chi}_k^{(\alpha)}(j) e^{i{k\over Za}ma}\,,
\end{align}
where $m=Z(n-1)+j$ is the lattice site index and $\bar{\chi}_k^{(\alpha)}(j)=\chi_k^{(\alpha)}(j) e^{ik{Z-j\over Z}}$
is no longer periodic in $k$. Here, ${k\over Za}$ plays the role of the quasimomentum and $ma$ is the
position in real space. Defining the Zak-Berry phase $\bar{\gamma}_\alpha$ of band $\alpha$ with 
respect to $\bar{\chi}_k^{(\alpha)}$ (as it is standardly done within the MTP) we will prove the central result
\begin{align}
\label{eq:surface_charge_theorem}
Q_B^{(\alpha)} = -{\bar{\gamma}_\alpha \over 2\pi} - {Z-1\over 2Z}\,,
\end{align}
provided that the gauge is chosen such that the last component $\chi_k^{(\alpha)}(Z)=\bar{\chi}_k^{(\alpha)}(Z)$
is real. This gauge is ultimately related to the boundary condition for the half-infinite system since
the scattering states get the form $\psi_k^{(\alpha)}(n,j)\sim \chi_k^{(\alpha)}(j)e^{ikn}-{\rm h.c.}$ in this gauge
(here, $n=1, j=1$ denotes the first site of a half-infinite system with a left boundary or, alternatively, 
$n=0, j=Z-1$ is the last site of a half-infinite system with a right boundary, see Fig.~\ref{fig:model_RL}).
Eq.~(\ref{eq:surface_charge_theorem}) is the unique formulation of the surface charge theorem for a 
single band where the second term on the right hand side describes the polarization of the ions. We note
that a similar unique formulation of the surface charge theorem for the boundary charge 
$Q_B=\sum_{\alpha=1}^\nu Q_B^{(\alpha)} + Q_E$ at fixed chemical potential is not 
possible since it is not known how to relate the number $Q_E$ of edge states to a bulk quantity in the
absence of any symmetry constraints. Only the total number $Q_E^{\text{tot}}$ of edge states (independent
of whether they are occupied or not) is given by the quantized sum $\sum_{\alpha=1}^Z \gamma_\alpha$ 
over all Zak-Berry phases defined with respect to $\chi_k^{(\alpha)}$ in the particular gauge described
above.

Concerning the rigorous proof of the topological constraint (\ref{eq:topological_constraint}) we will
use two alternative ways. One is based on the explicit solution of the edge state wave functions in terms 
of the sub-determinants of $h_k-\epsilon$. This provides concrete conditions how the energy 
$\epsilon(\varphi)$ of the edge states can cross the chemical potential as function of the phase. 
We visualize this by a
convenient diagrammatic language where $\Delta F$ can be identified with an effective topological charge
which, due to certain diagrammatic rules can only take two possible values $s_\nu-1$ or $s_\nu$.
Based on the topological constraint we also derive useful rules how the phase-dependence of the 
edge state energies can look like in general. We complement this by a second proof based on the
analytic continuation of Bloch states to complex quasimomentum, which is straightforward by using the 
explicit solution for the eigenstates. In accordance with previous approaches for half-infinite
continuum systems \cite{rehr_kohn_prb_74} we find that the Bloch states have a pole in the complex 
plane corresponding to the edge state solutions. In addition branch cuts appear separating different 
bands on the real axis, see also Ref.~[\onlinecite{kohn_pr_59}] for continuum systems. We demonstrate that,
as function of $\varphi$, the position of the edge state poles oscillate around the branch cuts, in 
accordance with similar findings using analytic continuations based on transfer matrices within the
IQHE \cite{hatsugai_prb_93,hatsugai_prl_93,hatsugai_jphys}. We will show
that these oscillations are essential to prove that the phase-dependence of the model parameters
within a phase-interval of size ${2\pi\over Z}$ can be always chosen such that no edge state
crosses the chemical potential in a certain gap. This input was used in the accompanying letter 
\cite{paper_prl} to justify that edge states are not the driving force standing behind the topological 
constraint but are instead followers adjusting to the phase-dependence of the model parameters in 
such a way that charge conservation for particles and holes is fulfilled. 

Besides the explicit solution for all eigenstates of the considered class of tight-binding models and 
the rigorous proof of the central results (\ref{eq:topological_constraint}) and (\ref{eq:surface_charge_theorem}),
we elaborate on a number of further issues in this work. Besides the standard case where the 
wavelength $\lambda$ of the modulations is identical to the length $Za$ of the unit cell, we will
also discuss the case $\lambda=Za/p$ with a rational number $Z/p$. This is a standard choice within
the discussion of the IQHE and is also covered by our analysis for the
boundary charge. We find that the topological constraint (\ref{eq:topological_constraint}) and
the universal form of the phase-dependence of the boundary charge is not changed but, in accordance
with the IQHE [\onlinecite{dana_jpc_85}-\onlinecite{hatsugai_prb_93}], we find that the Chern number is given by 
the Diophantine equation $pC_\nu=\nu-s_\nu Z$.  

Another result concerns a useful universal relationship 
between the boundary charges $Q_B^{R,(\alpha)}$ and $Q_B^{L,(\alpha)}$ of a single band for a half-infinite 
system with a left/right boundary, respectively. Here, the system with a left boundary is starting with
site $j=1$ of the unit cell, whereas the one with a right boundary ends with $j=Z-1$ 
(see Fig.~\ref{fig:model_RL}). We note that this special relation between the left and the right boundary
corresponds to the situation where, in each gap and for each value of the phase $\varphi$, there is 
always exactly one edge state either belonging to the left or to the right boundary, see also 
Refs.~[\onlinecite{thouless_etal_prl_82,hatsugai_prb_93}]. In this case we obtain
\begin{align}
\label{eq:left_right_relation}
Q_B^{R,(\alpha)} + Q_B^{L,(\alpha)} = -{Z-1\over Z} \,,
\end{align}
which, in terms of the corresponding Zak-Berry phases, is equivalent to 
$\bar{\gamma}_\alpha^R + \bar{\gamma}_\alpha^L =0$ according to (\ref{eq:surface_charge_theorem}). 

Besides the boundary charge we discuss also the density in the insulating regime and analyse its localization 
which is essential for a unique definition of the boundary charge via a macroscopic average. In accordance 
with similar results for continuum systems \cite{kallin_halperin_prb_84} we find that the
density of an insulator falls off exponentially fast from the boundary to the bulk value of the 
infinite system. As in Ref.~[\onlinecite{kallin_halperin_prb_84}] we show that each edge state
leaves a fingerprint in the scattering state density which has exactly the same form as the edge state  
but with opposite sign, see also Ref.~[\onlinecite{rehr_kohn_prb_74}]. 
Technically this arises from a pole contribution of the Friedel density in complex quasimomentum space, 
where the Friedel density is that part which leads to the well-known $2k_F$-oscillations for 
impurities in metallic systems \cite{friedel_58}. The remaining part of the density follows from a branch cut 
contribution which leads to an exponentially decreasing contribution with localization length 
proportional to the product of a typical velocity and the inverse gap between the valence and 
conduction band. In addition to previous results for continuum systems, we show that a 
pre-exponential function occurs falling off 
generically with $1/\sqrt{n}$ at large distances, where $n$ labels the unit cells. This result looks 
very similar to a corresponding power-law found for the one-particle density matrix of an infinite system,
see Ref.~[\onlinecite{he_vanderbilt_prl_01}]. 

We note that the discussion of the phase-dependence of the bulk polarization and the pumped charge is
quite common in the literature, see e.g. 
Refs.~[\onlinecite{thouless_prb_83,hatsugai_prb_93,hatsugai_fukui_prb_16}] and
other works on generalized twisted boundary conditions 
\cite{qi_wu_zhang_prb_06,delplace_ullmo_montambaux_prb_11}. However, to the best of our knowledge, a unique
relation between the boundary charge and the bulk polarization has never been stated, and 
{\it universal} properties of the phase-dependence have been reported only with respect to
a complete cycle when the phase is changed by $2\pi$. The latter is related to the Chern number or
to the number of windings of the edge state pole around the branch cuts mentioned above
\cite{hatsugai_prb_93,hatsugai_prl_93,hatsugai_jphys}. In contrast,
the quantized winding number $w_\alpha(\varphi) = - I_\alpha(\varphi) = -\Delta Q_B^{(\alpha)}(\varphi) + {1\over Z}$, 
introduced in the accompanying letter \cite{paper_prl}, describes universal features for any given value
of the phase $\varphi\in[0,2\pi]$ and
has a direct relation to the boundary charge. Therefore, in this work, we will also present a detailed 
comparison of this winding number to the Chern number and the Zak-Berry phase and we will show explicitly that 
it contains more information compared to the topological invariants discussed so far in the literature. 

The work is organized as follows. After introducing the model in Section~\ref{sec:model} 
we present in Sections~\ref{sec:Bloch_states} and \ref{sec:dispersion} the solution of the
Bloch states and the energy dispersion for the infinite system, together with the analytic 
continuation to complex quasimomentum. The solutions for the scattering and edge states for the half-infinite
system are provided in Sections~\ref{sec:bulk} and \ref{sec:edge}. Based on the explicit conditions
how to determine the edge states we present in Section~\ref{sec:constraints_edge} 
a rigorous derivation of the topological constraint (\ref{eq:topological_constraint}) 
for the phase-dependence of the edge state energies and present rules for its visualization.
The definition of the boundary charge and the derivation of the unique relation (\ref{eq:left_right_relation})
between the boundary charges of half-infinite systems with a left or right boundary is
described in Section~\ref{sec:definition}. The particle-hole duality is reviewed in 
Section~\ref{sec:ph_duality}. Section~\ref{sec:localization} is devoted to the calculation of the
density for a half-infinite insulator, where we demonstrate the localization of the boundary charge
and derive the power-law for the pre-exponential function. 
In Section~\ref{sec:zak} we present the unique formulation of the surface charge theorem 
(\ref{eq:surface_charge_theorem}) and discuss the Zak-Berry phases in the two representations 
(\ref{eq:bloch_1}) and (\ref{eq:bloch_2}) of the Bloch wave. Furthermore, we derive the universal
relation between the change of the boundary charge of a single band under a shift of the
lattice by an arbitrary number of sites and the winding number associated with the phase difference
of the Bloch wave function between the corresponding sites. Section~\ref{sec:winding_zak_chern}
presents the comparison of this winding number to the Zak-Berry phase and the Chern number.
The physical picture underlying the universal properties of the boundary charge is the topic of 
Section~\ref{sec:invariant_physics} which reviews the derivation proposed in 
Ref.~[\onlinecite{paper_prl}]. In addition, via the analytic continuation of Bloch states, 
we will discuss in this section why the phase-dependence of the model parameters can always be chosen such that
no edge state crosses the chemical potential in a certain gap. The rigorous derivation for the 
possible quantization values of the topological invariant of a single band and for the invariant in the 
presence of a fixed chemical potential, together with the relation to the topological constraint
(\ref{eq:topological_constraint}), is presented in Sections~\ref{sec:invariant_single_band}
and \ref{sec:invariant_total}, respectively. These Sections contain also the derivation of the consequences 
for the phase-dependence of the boundary charge and the case where the wave-length of the modulations
is given by a rational number $Z/p$ in units of the lattice spacing $a$. 
We close with a summary and outlook in Section~\ref{sec:summary}.

Throughout this work we use units $\hbar=e=a=1$.

\section{Spectral properties}
\label{sec:spectral}

In this section we introduce the class of tight-binding models under consideration and 
present the analytical solution for all Bloch eigenstates of the infinite system and all
scattering and edge states of a half-infinite system with a left or a right boundary. 
The exact result for all eigenstates will form the basis for a rigorous proof of the 
unique formulation of the surface charge theorem (\ref{eq:surface_charge_theorem}) in
Section~\ref{sec:zak}.
In addition, we present the analytic continuation of the energy dispersion and the 
Bloch eigenstates to complex quasimomentum, analog to corresponding representations for continuum systems
\cite{kohn_pr_59,rehr_kohn_prb_74}. Of central importance is the determination of the conditions for 
the appearance of edge states and the analysis of their phase-dependence when the lattice is 
shifted by a continuous phase variable $\varphi$. We develop a graphical representation to 
determine the precise topological constraints (\ref{eq:topological_constraint}) for the edge states 
in Section~\ref{sec:constraints_edge} which will turn out to be the basis for the rigorous proof of the
quantization values of the invariant $I(\varphi)$ characterizing universal properties of the boundary 
charge as introduced in Ref.~[\onlinecite{paper_prl}], see Section~\ref{sec:universal}.

\subsection{The model}
\label{sec:model}

\begin{figure*}
\centering
 \includegraphics[width=\textwidth]{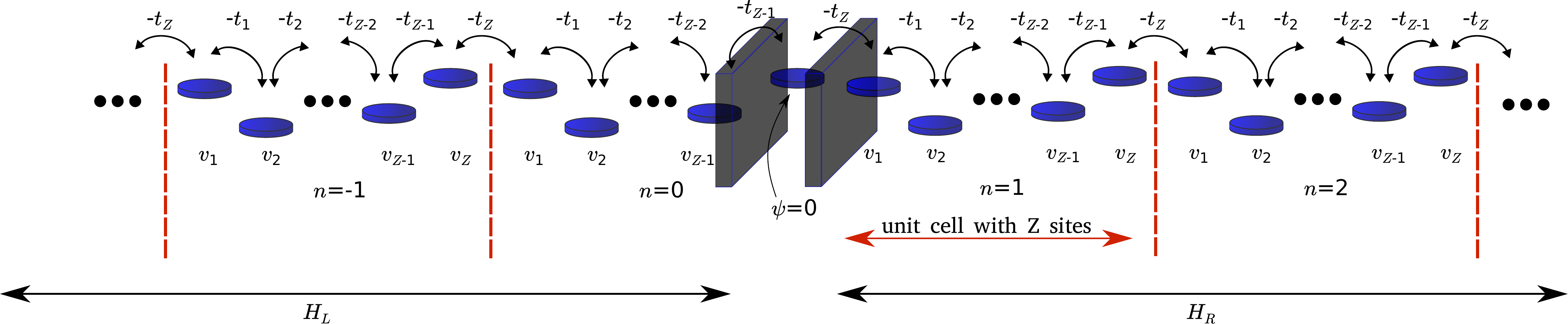}
 \caption{Sketch of the two half-infinite tight-binding models under consideration. The system
   is either extending to the right or to the left side, described by the Hamiltonians $H_R$ and
   $H_L$, respectively. The unit cells are labelled by $n=0,\pm 1,\pm 2,\dots$, the sites within a unit cell
   by $j=1,2,\dots,Z$. The absolute position of a site is labelled by $m=Z(n-1)+j$. 
   The right/left bar indicates the boundary of $H_{R/L}$. 
   The site indicated between the two bars at $n=0,j=Z$ or $m=0$ is an artificial site where the eigenstate 
   has to fulfil the boundary condition $\psi(n=0,j=Z)=\psi(m=0)=0$.
 }
\label{fig:model_RL}
\end{figure*}
We consider a generic nearest-neighbor tight-binding model with one orbital per site on a half-infinite
system, extending either to the right or to the left side, described by Hamiltonians $H_{R/L}$, respectively, 
see Fig.~\ref{fig:model_RL}(a,b) for a sketch of the system. 
The unit cells are labelled by $n=0,\pm 1,\pm 2,\dots$. Each unit cell consists of $Z$ sites labelled by
$j=1,\dots,Z$. The absolute position of a site is labelled by the index $m = Z(n-1)+j \equiv (n,j)$.
We take generic on-site potentials $\hat{v}_m\equiv \hat{v}_{nj} \equiv v_j$ and hoppings 
$\hat{t}_m \equiv \hat{t}_{nj} \equiv t_j$ that depend only on the site index $j$ within the unit cell. 
The Hamiltonians $H_{R/L}$ in the $1$-particle subspace are given by
\begin{align}
\label{eq:tb_half_R}
H_{R}&=\sum_{m\ge 1}\left\{\hat{v}_m |m\rangle\langle m| 
- \left[\hat{t}_m |m\rangle\langle m+1| + \text{h.c.}\right]\right\}\,,\\
\label{eq:tb_half_L}
H_{L}&=\sum_{m\le -1} \left\{\hat{v}_m |m\rangle\langle m| 
- \left[\hat{t}_m |m\rangle\langle m-1| + \text{h.c.}\right]\right\}\,.
\end{align}
The difference between $H_R$ and $H_L$ is the way the unit cell is cut off at the boundary. 
Whereas for $H_R$ the complete unit cell $n=1$ is included, for $H_L$ the site $j=Z$ of the first unit 
cell $n=0$ is not included. With this choice the boundary condition for both Hamiltonians becomes the
same at $m=0$ or at $n=0$ and $j=Z$, see below. 

To guarantee $H_{R/L}=H_{R/L}^\dagger$ the on-site potentials $v_j=v_j^*$ must be real. 
The hoppings $t_j=t_j^*>0$ are chosen real and positive since
possible phases can be gauged away by a unitary transformation which does not influence the
the density, see Appendix \ref{app:hoppings}. As a
consequence the Hamiltonian is real and is invariant under time
reversal transformation $T=K$ with $T^2=1$, where $K$ denotes the operator of complex conjugation
in the real-space basis of the lattice sites. We emphasize that this is {\it not} a symmetry 
constraint imposed on top of the considered tight-binding models but can always be achieved after
a unitary transformation. Furthermore, this symmetry is not relevant for inducing any topological
properties in 1D. According to the classification scheme of topological insulators
[\onlinecite{schnyder_etal_prb_08}-\onlinecite{diez_etal_njp_15}], only chiral symmetry
or particle-hole symmetry with $C^2=1$ can induce nontrivial topology in 1D 
systems. For $T^2=1$ these two symmetries are equivalent and are only fulfilled if the potentials $v_j$ are all
the same. Also inversion symmetry is a special case, where $v_j=v_{Z-j+1}$ and $t_j=t_{Z-j}$ (with
$t_0\equiv t_Z$) has to be fulfilled. 

For convenience the average of all $v_j$ is defined as zero energy
and we define by $t$ the average over all $t_j$
\begin{align}
\label{eq:vt_average}
{1\over Z}\sum_{j=1}^Z v_j = 0 \quad,\quad {1\over Z}\sum_{j=1}^Z t_j=t \,.
\end{align}

All $v_j=v_j(\varphi)$ and $t_j=t_j(\varphi)$ are taken as function of a phase variable $0\le \varphi<2\pi$,
which shifts the lattice continuously towards the boundary for $H_R$ and away from the boundary for $H_L$, 
such that a phase change by ${2\pi\over Z}$ corresponds to a shift by one lattice site, i.e.,
\begin{align}
\label{eq:shift}
v_{j+1}(\varphi)=v_j\left(\varphi+{2\pi\over Z}\right)\quad,\quad
t_{j+1}(\varphi)=t_j\left(\varphi+{2\pi\over Z}\right)\,.
\end{align}
Generically, this is achieved by using the form 
\begin{align}
\label{eq:v_form}
v_j(\varphi) &= V F_v\left({2\pi\over Z} j + \varphi\right)\,,\\
\label{eq:t_form}
t_j(\varphi) &= t + \delta t \,F_t\left({2\pi\over Z} j + \varphi\right)\,,
\end{align}
where $F_v(\varphi)=F_v(\varphi+2\pi)$ and $F_t(\varphi)=F_t(\varphi+2\pi)$ are some real and periodic 
functions of order of $O(1)$. In Appendix \ref{app:phase_dependence} we describe different ways
how we have chosen generic and random forms for the two functions $F_v$ and $F_t$ used in many figures.

For the case where the wavelength of the modulations is given by some rational and non-integer number 
$\lambda=Z/p$, with $p$ being some positive integer, the phase-dependence is chosen as
\begin{align}
\label{eq:v_form_p}
v_j(\varphi) &= V F_v\left({2\pi p\over Z} j + \varphi\right)\,,\\
\label{eq:t_form_p}
t_j(\varphi) &= t + \delta t \,F_t\left({2\pi p\over Z} j + \varphi\right)\,.
\end{align}
In this form the parameters are again periodic under a phase
change by $2\pi$ but a shift of the lattice by one site happens on the different scale $2\pi p/Z$ such
that after a phase change by $2\pi p$ the system has undergone all possible ways of how to define the
boundary. We can map this parametrization on the form of Eqs.~(\ref{eq:v_form_p}) and
(\ref{eq:t_form_p}) by the rescaling $\varphi'\equiv\varphi/p$, $F^\prime_v(\varphi')=F_v(\varphi)$ and
$F^\prime_t(\varphi')=F_t(\varphi)$ such that the parameters as function of $\varphi'$ get the
same form as above 
\begin{align}
\label{eq:v_form_p_prime}
v_j(\varphi') &= V F^\prime_v\left({2\pi \over Z} j + \varphi'\right)\,,\\
\label{eq:t_form_p_prime}
t_j(\varphi) &= t + \delta t \,F^\prime_t\left({2\pi \over Z} j + \varphi'\right)\,.
\end{align}
Therefore this case is also covered by our general ansatz and will not be treated separately.

The eigenfunctions of the half-infinite tight-binding models (\ref{eq:tb_half_R}) and (\ref{eq:tb_half_L}) 
will be constructed in terms of the Bloch eigenstates of the infinite system 
defined by the bulk Hamiltonian written in compact form as
\begin{align}
\nonumber
H_{\text{bulk}} &=\sum_{n=-\infty}^\infty \left\{|n\rangle\langle n| \otimes h(0) +
|n+1\rangle\langle n| \otimes h(1)\right. \\
\label{eq:tb_bulk}
&\hspace{3cm}
\left.+|n\rangle\langle n+1| \otimes h(-1) \right\}\,,
\end{align}
where we sum over all unit cells $n=-\infty, \dots, \infty$.
Here, $h(0)$ and $h(\pm 1)$ are $Z\times Z$-matrices in unit cell space describing the Hamiltonian
within a unit cell or the hopping from unit cell $n\rightarrow n+1$ or $n+1\rightarrow n$, respectively.
They are given by (zero matrix elements are not shown) 
\begin{align}
\nonumber
h(0)&=\sum_{j=1}^Z v_j|j\rangle\langle j| - \sum_{j=1}^{Z-1} (t_j|j+1\rangle\langle j| + \text{h.c.})\\
\label{eq:h0}&\equiv
 \left(\begin{array}{ccccc} 
v_1 & -t_1 &  &  &  \\
-t_1 & v_2 & -t_2 &   &  \\ 
 & -t_2 & \ddots & \ddots  &  \\ 
 &  & \ddots & \ddots  & -t_{Z-1} \\ 
 & & & -t_{Z-1} & v_Z \\ 
\end{array}\right)
\end{align}
and
\begin{align}
\label{eq:h1}
h(1)= -t_Z |1\rangle\langle Z| = \left(\begin{array}{cccc} 
 &  &  & -t_Z  \\
 &  &  &   \\ 
 &  &  &   \\ 
 &  &  &   \\ 
\end{array}\right)=\left[h(-1)\right]^\dagger\,.
\end{align}

The bulk Hamiltonian is translationally invariant and can be diagonalized 
by the Fourier transform 
\begin{align}
\label{eq:k_states}
|k\rangle = {1\over \sqrt{2\pi}}\sum_{n=-\infty}^\infty e^{ikn} |n\rangle,
\end{align}
where $-\pi\le k < \pi$ is the quasimomentum. We get
\begin{align}
\label{eq:tb_bulk_k}
H_{\text{bulk}}=\int_{-\pi}^\pi dk |k\rangle\langle k| \otimes h_k \,,
\end{align}
with the Bloch Hamiltonian
\begin{align}
\label{eq:h_k}
h_k &= h(0) + e^{-ik} h(1) + e^{ik} h(-1) \\
\label{eq:h_k_matrix}
&= \left(\begin{array}{ccccc} 
v_1 & -t_1 &  &  & -t_Z e^{-ik} \\
-t_1 & v_2 & -t_2 &   &  \\ 
 & -t_2 & \ddots & \ddots  &  \\ 
 &  & \ddots & \ddots  & -t_{Z-1} \\ 
-t_Z e^{ik} & & & -t_{Z-1} & v_Z \\ 
\end{array}\right).
\end{align}
For later convenience we write $h_k$ in the form
\begin{align}
\label{eq:h_k_A}
h_k= \left(\begin{array}{cc} 
A & b_k \\
b_{-k}^T & v_Z \\ 
\end{array}\right)\,,
\end{align}
where we have defined
\begin{align}
\label{eq:A}
A\,=\, \left(\begin{array}{ccccc} 
v_1 & -t_1 &  &  &  \\
-t_1 & v_2 & -t_2 &   &  \\ 
 & -t_2 & \ddots & \ddots  &  \\ 
 &  & \ddots & \ddots  & -t_{Z-1} \\ 
 & & & -t_{Z-1} & v_{Z-1} \\ 
\end{array}\right)
\end{align}
and the $(Z-1)$-dimensional column vector
\begin{align}
\label{eq:b_k}
b_k= \left(\begin{array}{c} 
-t_Z e^{-ik} \\ 0 \\ \vdots \\ 0 \\ - t_{Z-1} \\ 
\end{array}\right)\,.
\end{align}

Once an eigenstate $|\psi\rangle$ for the Hamiltonian $H_{\text{bulk}}$ of the infinite system has been found 
\begin{align}
\label{eq:eigenstate_infinite}
H_{\text{bulk}}|\psi\rangle = \epsilon |\psi\rangle\,,
\end{align}
we get also an eigenstate of the half-infinite systems $H_{R/L}$ if the boundary condition 
\begin{align}
\label{eq:boundary_RL}
\psi(n=0,j=Z) = \psi(m=0) = 0
\end{align}
and the asymptotic condition
\begin{align}
\label{eq:asymptotic_R}
\lim_{n\rightarrow\infty} \psi(n,j) \sim e^{ikn}, \quad \text{Im}(k)\ge 0  
\end{align}
for $H_R$ and 
\begin{align}
\label{eq:asymptotic_L}
\lim_{n\rightarrow -\infty} \psi(n,j) \sim e^{ikn}, \quad \text{Im}(k)\le 0 
\end{align}
for $H_L$ are fulfilled. As a consequence, we will see in Section~\ref{sec:edge} that, for given phase $\varphi$, 
we find always exactly one edge state, either for $H_R$ or $H_L$, consistent with the study in 
Ref.~[\onlinecite{hatsugai_prb_93}] for the TKNN model \cite{thouless_etal_prl_82}.

\subsection{Bloch eigenstates for the infinite system}
\label{sec:Bloch_states}

We first determine the eigenstates of the infinite system 
\begin{align}
\label{eq:bulk_eigenstates}
H_{\text{bulk}}|\psi^{(\alpha)}_{k,\text{bulk}}\rangle = \epsilon^{(\alpha)}_k |\psi^{(\alpha)}_{k,\text{bulk}}\rangle \,,
\end{align}
where $\psi^{(\alpha)}_{k,\text{bulk}}$ is the Bloch eigenstate of band $\alpha=1,\dots,Z$. We take the Bloch form
\begin{align}
\label{eq:bloch_form}
\psi^{(\alpha)}_{k,\text{bulk}}(n,j)={1\over\sqrt{2\pi}}\chi_k^{(\alpha)}(j) e^{ikn}\,,
\end{align}
where $\chi_k^{(\alpha)}=\chi_{k+2\pi}^{(\alpha)}$ are the normalized Bloch states described by
$Z$-dimensional column vectors, which are chosen periodic in $k$ (for other representations of the Bloch wave
function see the discussion in Section~\ref{sec:zak}). We will also choose the gauge such that 
$\chi_k^{(\alpha)}(Z)$ is real (which will be relevant for the boundary condition for a half-infinite system,
see Sections~\ref{sec:bulk} and \ref{sec:edge}). They are eigenstates of the Bloch Hamiltonian $h_k$
\begin{align}
\label{eq:bloch_eigenstate}
h_k \chi_k^{(\alpha)} = \epsilon^{(\alpha)}_k \chi_k^{(\alpha)}\,.
\end{align}
We note that $h_k$ is diagonalizable for all complex $k$ except at special
branching points of the dispersion where a pair of two eigenstates merge together, 
see Section~\ref{sec:dispersion}.
Since $(h_k)^T=h_{-k}$, we find that $(\chi^{(\alpha)}_{-k})^T$ 
are the left eigenvectors of $h_k$ for any complex $k$
\begin{align}
\label{eq:left_eigenstates}
(\chi^{(\alpha)}_{-k})^T h_k = \epsilon^{(\alpha)}_{-k} (\chi^{(\alpha)}_{-k})^T\,,
\end{align}
Since the eigenvalues of left and right eigenvectors are the same, we get 
\begin{align}
\label{eq:epsilon_minus_k}
\epsilon_k^{(\alpha)}=\epsilon_{-k}^{(\alpha)}
\end{align}
for any complex $k$. The orthogonality and completeness relation can be written as
\begin{align}
\label{eq:vons}
(\chi_{-k}^{(\alpha)})^T \chi_k^{(\alpha')} = \delta_{\alpha\alpha'}\quad,\quad 
\sum_{\alpha=1}^Z \chi_k^{(\alpha)} (\chi_{-k}^{(\alpha)})^T = \mathbbm{1}\,.
\end{align}
Furthermore, since $(h_k)^*=h_{-k^*}$ and $h_k=h_{k+2\pi}$ for any complex
$k$, and using the fact that the spectrum is non-degenerate (see below), we get
for any complex $k$
\begin{align}
\label{eq:epsilon_property}
& \epsilon_k^{(\alpha)}=\left(\epsilon_{-k^*}^{(\alpha)}\right)^*\,,\\
\label{eq:epsilon_periodicity}
&\epsilon_k^{(\alpha)}=\epsilon_{k+2\pi}^{(\alpha)}\,,
\end{align}
and we can choose the gauge such that
\begin{align}
\label{eq:bloch_vector_property}
\chi_k^{(\alpha)}(j) &= \left(\chi_{-k^*}^{(\alpha)}(j)\right)^*\,,\\
\label{eq:bloch_vector_periodicity}
\chi_k^{(\alpha)}(j) &= \chi_{k+2\pi}^{(\alpha)}(j)\,. 
\end{align}

\begin{figure*}
\centering
\includegraphics[width= 0.95\columnwidth]{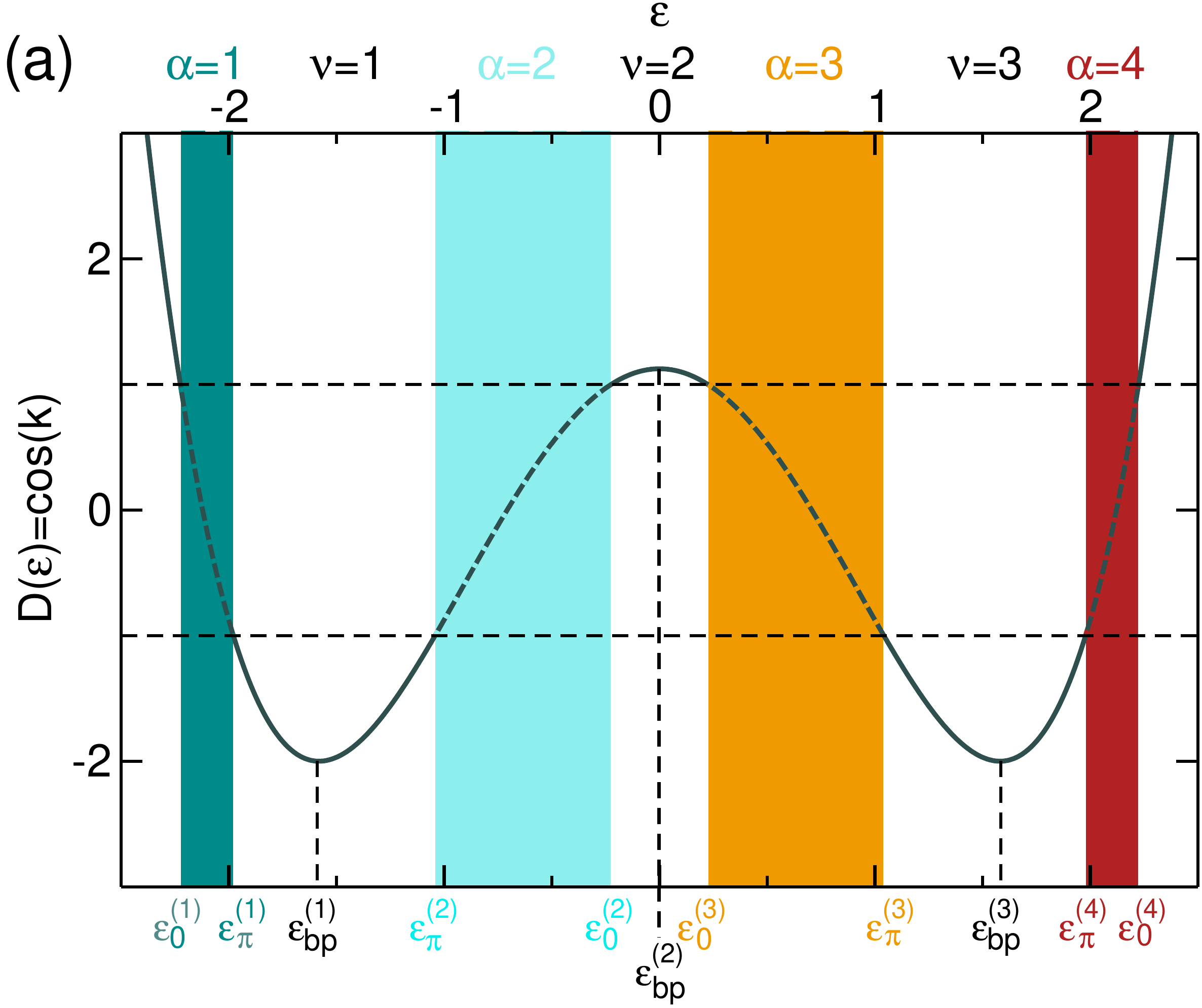} 
\hfill
\includegraphics[width= 1.\columnwidth]{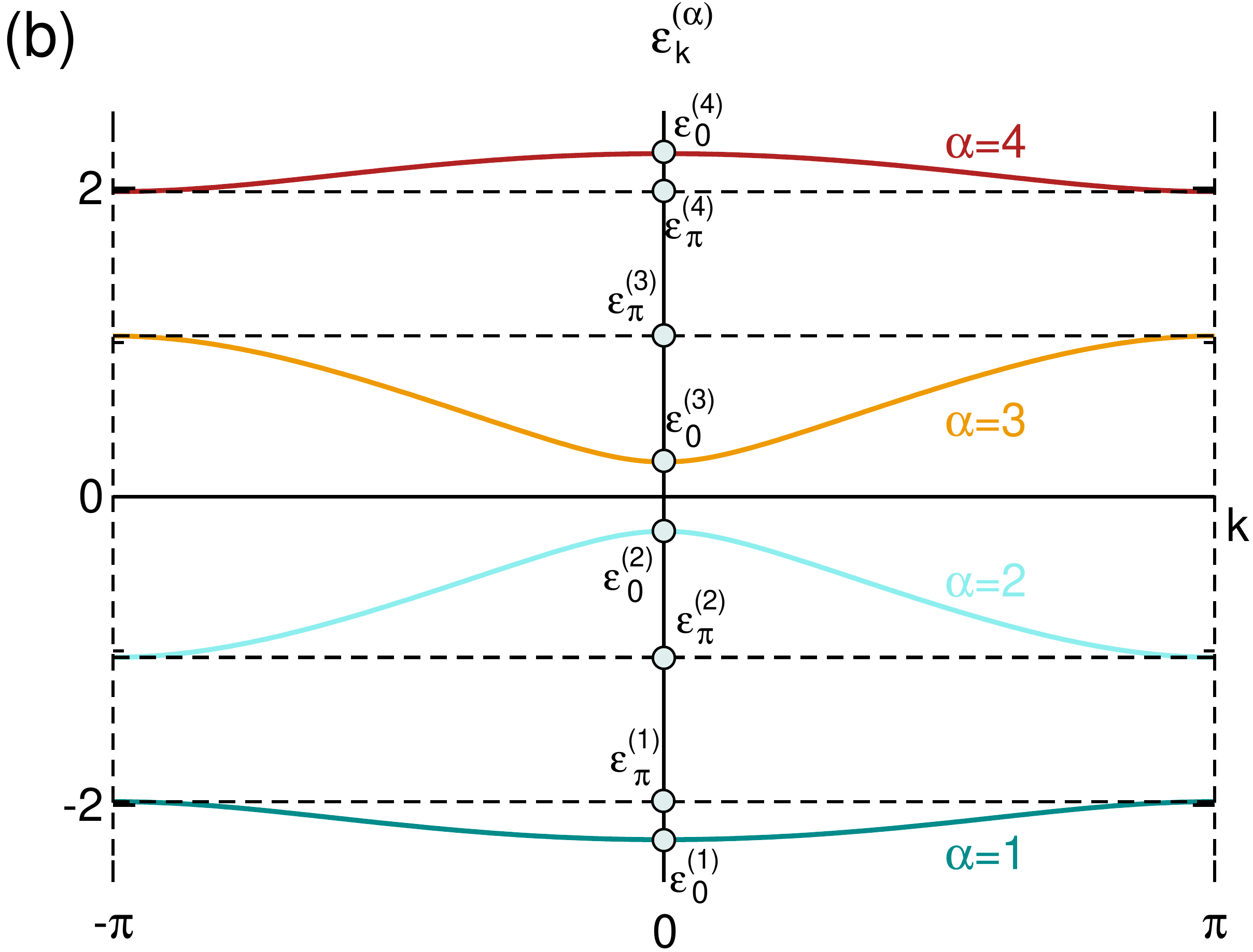}
 \caption{(a) The function $D(\epsilon)$ for $v_j=-\cos(2\pi j/Z)$ and $t_j=1$. The bands are
   formed in the regions where $D(\epsilon_k^{(\alpha)})=\cos(k)\le 1$, with $Z-1$ gaps $\nu=1,\dots,Z-1$ in
   between. The band edges are given by $\epsilon_0^{(\alpha)}$ and $\epsilon_\pi^{(\alpha)}$,
   where $\epsilon_0^{(\alpha)}$ is the band top/bottom for even/odd $\alpha$. $\epsilon_{\text{bp}}^{(\nu)}$ 
   denote the energies where ${d\over d\epsilon}D(\epsilon_{\text{bp}}^{(\nu)})=0$ and 
   $|D(\epsilon_{\text{bp}}^{(\nu)})|=|\cos(k_{\text{bp}}^{(\nu)})|>1$. 
   They correspond to complex $k$-values $k_{\text{bp}}^{(\nu)}=\pm i\kappa_{\text{bp}}^{(\nu)}$ and 
   $k_{\text{bp}}^{(\nu)}=\pm(\pi+i\kappa_{\text{bp}}^{(\nu)})$ for even and odd $\nu$, respectively, 
   with $\kappa_{\text{bp}}^{(\nu)}\ge 0$, see Eq.~(\ref{eq:k_bp}). (b) The band structure for 
   the same parameters as function of the quasimomentum $k$.
 }
\label{fig:band}
\end{figure*}

We note that the normalization and orthogonality defined in terms of the left and right 
eigenvectors is very essential to find a convenient analytic continuation to complex quasimomentum. 
To construct the exact eigenstates for any complex $k$ we use the ansatz
\begin{align}
\label{eq:chi_form}
\chi_k^{(\alpha)} = {1\over\sqrt{N_k^{(\alpha)}}}
\left(\begin{array}{c} 
a^{(\alpha)}_k \\ s(\epsilon_k^{(\alpha)}) \\ 
\end{array}\right)\,,
\end{align}
where $a_k^{(\alpha)}$ is a $(Z-1)$-dimensional column vector, and 
\begin{align}
\label{eq:N_k}
N_k^{(\alpha)} = N_{-k}^{(\alpha)} = (a_{-k}^{(\alpha)})^T a^{(\alpha)}_k + \left[s(\epsilon_k^{(\alpha)})\right]^2
\end{align}
guarantees normalization. The last component $\sim s(\epsilon_k^{(\alpha)})$ plays a special role and 
determines the gauge of the Bloch state needed for a precise definition of the Zak-Berry phase to 
obtain a unique formulation of the surface charge theorem, see Eq.~(\ref{eq:surface_charge_theorem}. 
We choose a gauge such that $\chi_k^{(\alpha)}(Z)$ is real for real $k$. Furthermore, for any complex $k$,
we will use a representation such that the normalization $N^{(\alpha)}_k$ and all components $a^{(\alpha)}_k(j)$ 
are {\it analytic} functions in the complex plane up to branch cuts arising from the dispersion 
relation $\epsilon^{(\alpha)}_k$ which occurs as a parameter in all quantities. We show that such
a representation is possible since the Hamiltonian $h_k$ is an analytic function of $k$. As we
will see this representation has many advantages and we propose it to be useful 
for the analysis of generic models even going beyond the single-channel case analysed in the present
work. 

Inserting (\ref{eq:chi_form}) into the eigenvalue problem (\ref{eq:bloch_eigenstate})
and using the form (\ref{eq:h_k_A}) of $h_k$, we obtain the two equations
\begin{align}
\label{eq:a_k_equation}
(A - \epsilon_k^{(\alpha)}) a_k^{(\alpha)} &= - s(\epsilon_k^{(\alpha)}) b_k \\
\label{eq:b_k_equation}
b_{-k}^T a_k^{(\alpha)} &= - s(\epsilon_k^{(\alpha)})\bar{v}_Z(\epsilon_k^{(\alpha)})\,,
\end{align}
where we defined
\begin{align}
\label{eq:bar_v}
\bar{v}_j(\epsilon) = v_j - \epsilon \,.
\end{align}
The first equation can be solved explicitly by
\begin{align}
\label{eq:a_k_solution}
a_k^{(\alpha)} = - B(\epsilon_k^{(\alpha)}) b_k \quad,\quad 
B(\epsilon) = {s(\epsilon)\over A - \epsilon}\,.
\end{align}
The matrix $B(\epsilon)$ is a well-defined matrix even for 
$\text{det}(A - \epsilon)=0$ if we take 
\begin{align}
\label{eq:s}
s(\epsilon) = \text{det}(A-\epsilon) \,.
\end{align}

For this choice, the matrix elements $B_{jj'}(\epsilon)$ are given
by $(-1)^{j+j'}$ times the subdeterminant of $A-\epsilon$ where the row $j'$ and the column $j$ are omitted.
Examining these subdeterminants based on the definition (\ref{eq:A}) of the $A$-matrix one finds
for $1\le j\le j'\le Z-1$
\begin{align}
\nonumber
B_{jj'}(\epsilon) &= B_{j'j}(\epsilon) \\
\label{eq:B_jj'}
&=(t_j t_{j+1}\cdots t_{j'-1}) d_{1,j-1}(\epsilon) d_{j'+1,Z-1}(\epsilon) \,,
\end{align}
where $t_j t_{j+1}\cdots t_{j'-1}\equiv 1$ for $j=j'$, and the determinants $d_{ij}(\epsilon)$ are
defined for $i\le j$ by 
\begin{align}
\label{eq:d_ij}
d_{ij}(\epsilon) = \text{det}(A^{(ij)}-\epsilon) \,,
\end{align}
with
\begin{align}
\label{eq:Aij}
A^{(ij)}\,=\, \left(\begin{array}{ccccc} 
v_i & -t_i &  &  &  \\
-t_i & v_{i+1} & -t_{i+1} &   &  \\ 
 & -t_{i+1} & \ddots & \ddots  &  \\ 
 &  & \ddots & \ddots  & -t_{j-1} \\ 
 & & & -t_{j-1} & v_j \\ 
\end{array}\right)\,.
\end{align}
By convention, we define $d_{ij}=1$ for $i>j$. Using this result in Eq.~(\ref{eq:a_k_solution}), one
obtains with Eq.~(\ref{eq:b_k}) the explicit solution for the components $j=1,\dots,Z-1$ of the Bloch state
\begin{align}
\label{eq:a_k_explicit}
a_k^{(\alpha)} = f(\epsilon_k^{(\alpha)}) e^{-ik} + g(\epsilon_k^{(\alpha)}) \,,
\end{align}
where $f(\epsilon)$ and $g(\epsilon)$ are $(Z-1)$-dimensional column vectors with components
\begin{align}
\label{eq:f}
f_j(\epsilon) &= t_1\dots t_{j-1} t_Z d_{j+1,Z-1}(\epsilon)\,,\\
\label{eq:g}
g_j(\epsilon) &= {\bar{t}^Z\over t_1\dots t_{j-1}}{1\over t_Z} d_{1,j-1}(\epsilon)\,.
\end{align}
Here, $\bar{t}$ is defined by the geometric mean
\begin{align}
\label{eq:tZ}
\bar{t} \equiv (t_1 t_2 \cdots t_Z)^{1/Z} \,.
\end{align}

The determinants $d_{ij}$ can be calculated from the recursion relations (we omit the argument $\epsilon$)
\begin{align}
\label{eq:det_recursion_1}
d_{ij} &= \bar{v}_i d_{i+1,j} - t_i^2 d_{i+2,j}\\
\label{eq:det_recursion_2}
& = \bar{v}_j d_{i,j-1} - t_{j-1}^2 d_{i,j-2}\,,
\end{align}
together with $d_{ii}=\bar{v}_i$. The two determinants $d_{2,Z-1}$ and $d_{1,Z-2}$ will play a 
special role for the determination of the edge states, see Section~\ref{sec:edge}. 
For later convenience, we define
\begin{align}
\label{eq:tilde_d_2_Z-1}
\tilde{d}_{2,Z-1}(\epsilon) &= {t_Z^2\over \bar{t}^Z} d_{2,Z-1}(\epsilon)\,,\\
\label{eq:tilde_d_1_Z-2}
\tilde{d}_{1,Z-2}(\epsilon) &= {t_{Z-1}^2\over \bar{t}^Z}d_{1,Z-2}(\epsilon)\,.
\end{align}
Further useful properties of the determinants are listed and
proven in Appendix~\ref{app:identities}. This Appendix contains also very helpful identities
for the derivatives of the determinants and the $B$-matrix. In particular we find the relation
\begin{align}
\label{eq:fg_relation}
f(\epsilon)^T {d\over d\epsilon} g(\epsilon) = g(\epsilon)^T {d\over d\epsilon} f(\epsilon) \,,
\end{align}
which will be needed in Section~\ref{sec:zak} to show the relation between the Friedel charge
and the Zak-Berry phase. 

Furthermore, we show in Appendix~\ref{app:identities} that one can write all 
components $a_k^{(\alpha)}(j)$, with $j=2,\dots,Z-1$, in terms of $a_k^{(\alpha)}(1)$ via  
\begin{align}
\nonumber
t_1\cdots t_{j-1} a_k^{(\alpha)}(j) &= d_{1,j-1}(\epsilon_k^{(\alpha)}) a_k^{(\alpha)}(1)\\
\label{eq:aj_a1}
& - d_{2,j-1}(\epsilon_k^{(\alpha)}) s(\epsilon_k^{(\alpha)}) t_Z e^{-ik}\,,
\end{align}
where $a_k^{(\alpha)}(1)$ follows from (\ref{eq:a_k_explicit}) as
\begin{align}
\label{eq:a_1_edge_tilde_d}
a^{(\alpha)}_k(1) = {\bar{t}^Z\over t_Z} \left\{\tilde{d}_{2,Z-1}(\epsilon_k^{(\alpha)})e^{-ik} + 1\right\}\,.
\end{align}
This component will play a central role to define the quantized invariant in 
Section~\ref{sec:invariant_single_band}. Since we work in a gauge where $\chi^{(\alpha)}_k(Z)$ is 
real for real $k$, we note that its phase corresponds to the gauge invariant phase-difference 
of the Bloch state $\chi_k^{(\alpha)}(j)$ between the first and last site of a unit cell.

\begin{figure*}
\centering
\begin{minipage}{0.53\textwidth}
 \includegraphics[width= 1.\columnwidth]{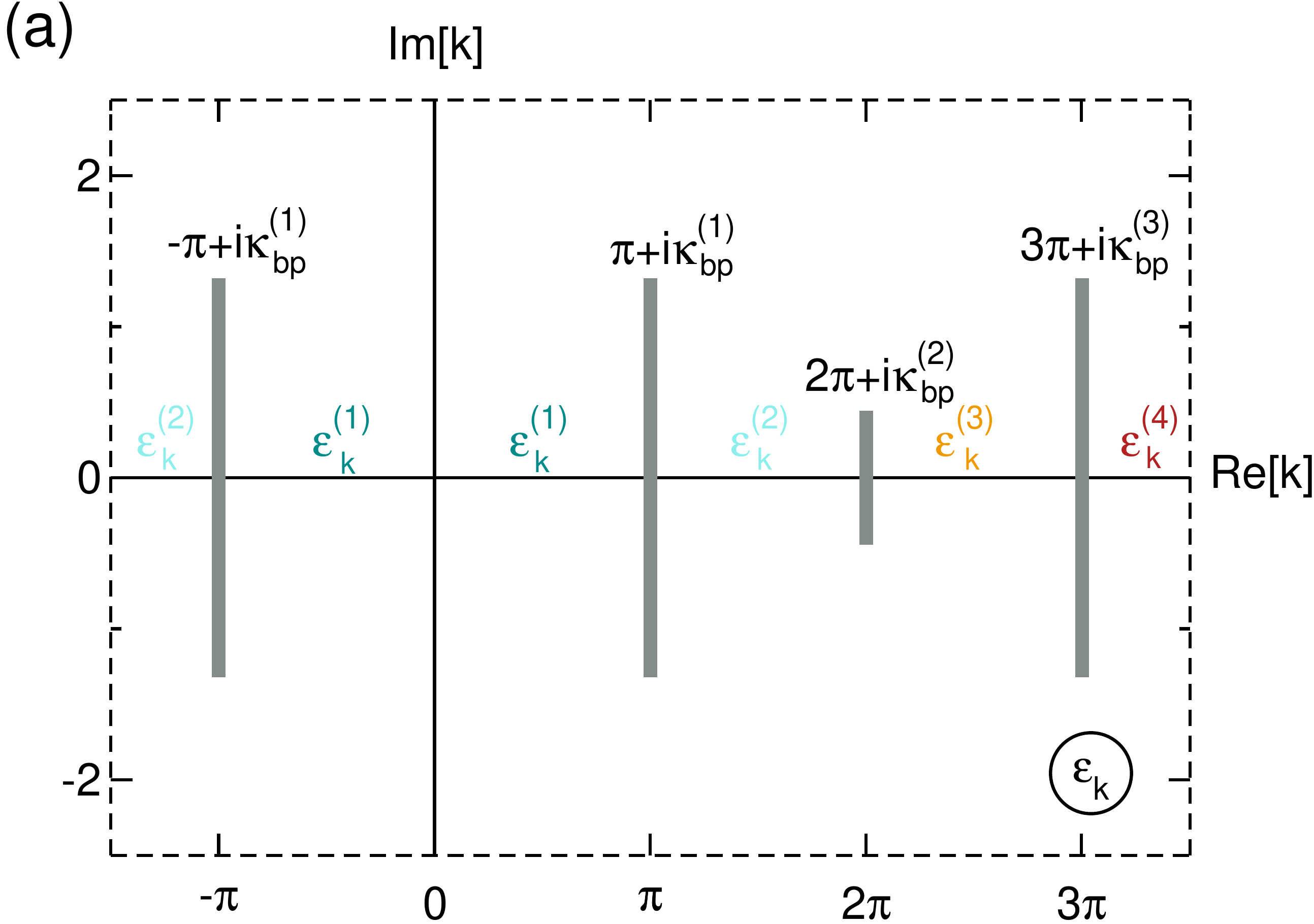}
\end{minipage}
\hfill
\begin{minipage}{0.44\textwidth}
 \includegraphics[width= 1.\columnwidth]{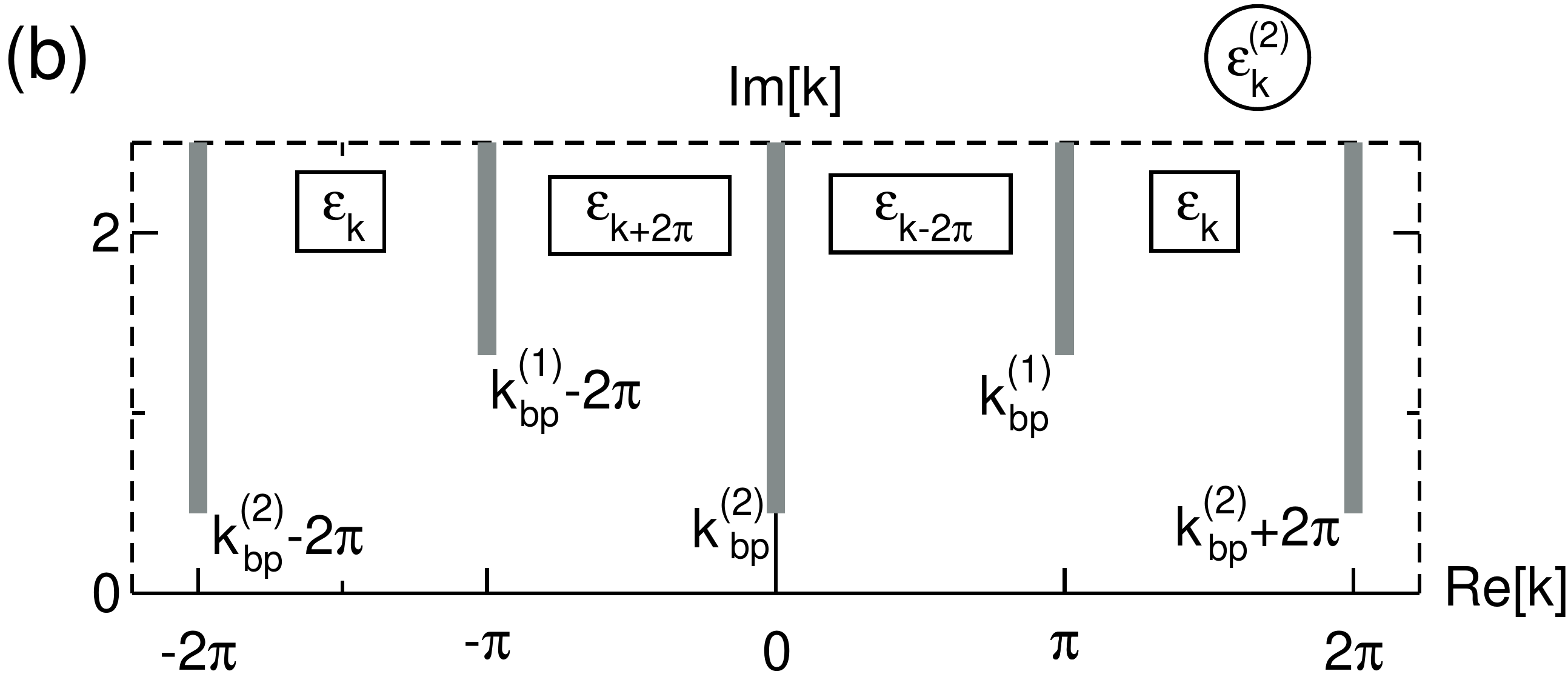}
 \includegraphics[width= 1.\columnwidth]{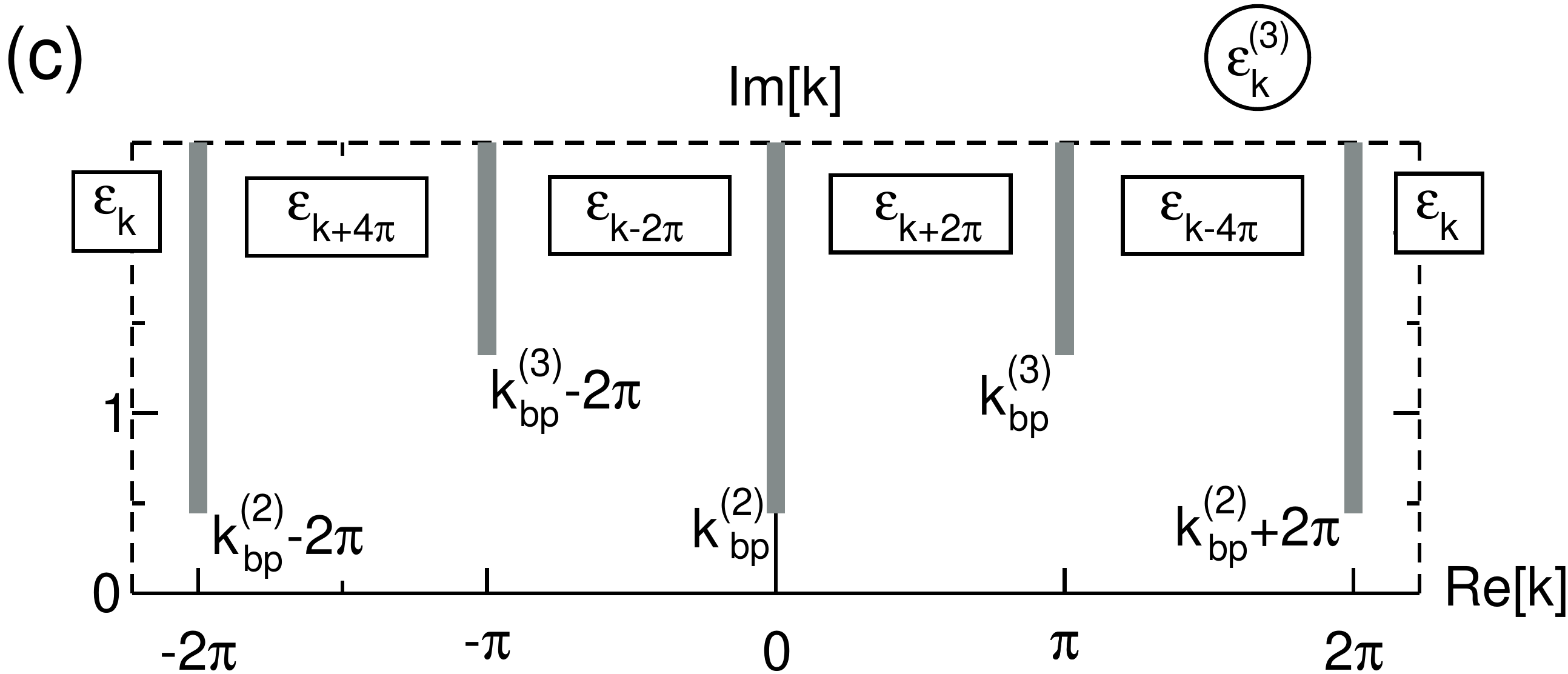}
\end{minipage}
 \caption{Two choices for the analytic continuation of the dispersion for the same parameters as
   in Fig.~\ref{fig:band}. (a) The analytic continuation defining $\epsilon_k$, with $\epsilon_k=\epsilon_k^{(\alpha)}$
   for $(\alpha-1)\pi < |k| < \alpha\pi$ and $\epsilon_k=\epsilon_k^{(Z)}$ for $|k|>Z\pi$ on the real axis. 
   Branch cuts occur at $\pm\nu\pi+i\kappa$, with $|\kappa|<\kappa_{\text{bp}}^{(\nu)}$ and  
   $\nu=1,\dots,Z-1$. (b) and (c) show the analytic continuation of
   $\epsilon_k^{(\alpha)}$ for $\alpha=2,3$, with $\epsilon_k^{(\alpha)}=\epsilon_{k+2\pi}^{(\alpha)}$ taken on the 
   whole real axis. Branch cuts are located at $k_{\text{bc}}^{(\nu)}+i\kappa$ and 
   $(k_{\text{bc}}^{(\nu)})^*-i\kappa$ (not shown), with $|\kappa|>0$ and $\nu=1,2$ (for $\alpha=2$) and
   $\nu=2,3$ (for $\alpha=3$), together with corresponding ones shifted by multiples of $2\pi$. 
   Between the branch cuts we have indicated in the boxes the relation to the analytic continuation chosen in (a).
   As one can see the values of the dispersion left and right to the branch cut starting at 
   $k_{\text{bp}}^{(2)}$ are interchanged for $\alpha=2,3$.}
 \label{fig:bc}
\end{figure*}
We note that Eq.~(\ref{eq:b_k_equation}) is not an independent equation and is automatically
fulfilled when $\epsilon_k^{(\alpha)}$ is an eigenvalue of $h_k$, i.e., when
$\text{det}(h_k-\epsilon_k^{(\alpha)})=0$. This follows from the relation
\begin{align}
\label{eq:det(hk-eps)}
\text{det}(h_k-\epsilon) = - b_{-k}^T B(\epsilon) b_k + s(\epsilon) \bar{v}_Z(\epsilon)\,,
\end{align}
together with (\ref{eq:a_k_solution}). 

Eqs.~(\ref{eq:N_k}), (\ref{eq:s}), (\ref{eq:a_k_explicit}), (\ref{eq:f}), 
and (\ref{eq:g}) show that all quantities $N^{(\alpha)}_k$, $a^{(\alpha)}_k(j)$, and $s(\epsilon^{(\alpha)}_k)$
can be written as analytic functions ${\cal{F}}(\epsilon,k)$ in $\epsilon$ and $k$, 
with $\epsilon\equiv\epsilon^{(\alpha)}_k$. Therefore, the analytic continuation of the eigenstates follows 
straightforwardly from the analytic continuation of the dispersion $\epsilon^{(\alpha)}_k$ which will 
be discussed in the next subsection. Only at the special points $N_k^{(\alpha)}=0$ additional poles
appear for the Bloch state. As we will see in Section~\ref{sec:edge} they are related to edge states, 
see also Ref.~[\onlinecite{rehr_kohn_prb_74}] with similar results for continuum models.

\subsection{Energy dispersion}
\label{sec:dispersion}

To obtain the dispersion $\epsilon_k^{(\alpha)}$, we rewrite the condition $\text{det}(h_k-\epsilon)=0$ by
using (\ref{eq:det(hk-eps)}) and inserting the form (\ref{eq:b_k}) of $b_k$ together with
the matrix elements (\ref{eq:B_jj'}) of the $B$-matrix. Using the recursion relation
$d_{1Z}=\bar{v}_Z d_{1,Z-1}-t_{Z-1}^2 d_{1,Z-2}$, one finds after a straightforward calculation
the condition
\begin{align}
\label{eq:dispersion}
\cos(k) &= D(\epsilon)\,,\\
\label{eq:D_function}
D(\epsilon)&\equiv
{1\over 2 \bar{t}^Z}\left\{d_{1Z}(\epsilon) - t_Z^2 d_{2,Z-1}(\epsilon)\right\}\,.
\end{align}
In Appendix~\ref{app:identities} we prove many helpful representations for the function
$D(\epsilon)$ and its derivatives. $D(\epsilon)$ is a polynomial of degree $Z$ in 
$\epsilon$ with real coefficients and the asymptotic behaviour
\begin{align}
\label{eq:D_asymp}
\lim_{|\epsilon|\rightarrow\infty} D(\epsilon) = {(-\epsilon)^Z\over 2 \bar{t}^Z}\,.
\end{align}
Therefore, for any given complex $k$, Eq.~(\ref{eq:dispersion}) has $Z$ solutions 
$\epsilon=\epsilon_k^{(\alpha)}$, with $\alpha=1,\dots,Z$, which fulfil the properties 
(\ref{eq:epsilon_minus_k}), (\ref{eq:epsilon_periodicity}) and (\ref{eq:epsilon_property}).
For real values $-\pi\le k<\pi$, the Hamiltonian $h_k$ is hermitian and has always $Z$ real eigenvalues
$\epsilon_k^{(\alpha)}$, with $\alpha=1,\dots,Z$, corresponding to the different bands which we label 
from bottom to top. In this case, the graphical solution of Eq.~(\ref{eq:dispersion}) is shown 
in Fig.~\ref{fig:band}(a).
From Eq.~(\ref{eq:D_asymp}) we get $D(\epsilon)>0$ for $\epsilon\rightarrow -\infty$.
Therefore, the bottom of the first band is always at $k=0$. Since for each $k$ there are $Z$
real solutions of $D(\epsilon)=\cos(k)$, the function $D(\epsilon)$ must be monotonous in each segment
where $|\cos(k)|\le 1$. As a consequence, we obtain always $Z-1$ gaps labelled by $\nu=1,\dots,Z-1$ 
and the band dispersion $\epsilon_k^{(\alpha)}$ for $\alpha$ even/odd is a monotonously 
decreasing/increasing function for $0<k<\pi$ with
band region defined by $\epsilon_{\pi/0}^{(\alpha)}<\epsilon_k^{(\alpha)}<\epsilon_{0/\pi}^{(\alpha)}$, 
see Fig.~\ref{fig:band}(b). Occasionally, two adjacent bands might touch at $k=0$ or $k=\pm\pi$ 
leading to gap closings, but the classification in $Z$ bands remains. 

The analytic continuation of the dispersion can be obtained from Eq.~(\ref{eq:dispersion}), 
analog to Ref.~[\onlinecite{kohn_pr_59}]. This is achieved by starting from
some dispersion $\epsilon_k$ on the real axis (see below for two convenient choices) and
solving a differential equation for ${d\epsilon_k\over dk}$ along an arbitrary path in the
complex plane. Inserting $\epsilon=\epsilon_k$ in Eq.~(\ref{eq:dispersion}) and taking
the derivative with respect to $k$ we get
\begin{align}
\label{eq:eps_der_1}
{d\epsilon_k\over dk} = -{\sin(k)\over D'(\epsilon_k)} \,.
\end{align}
It follows that the branching points $k_{\text{bp}}^{(\nu)}$ are given by the condition 
\begin{align}
\label{eq:bp_condition_1}
\cos(k_{\text{bp}}^{(\nu)}) &= D(\epsilon_{\text{bp}}^{(\nu)})  \,,\\
\label{eq:bp_condition_2}
D'(\epsilon_{\text{bp}}^{(\nu)}) &\equiv {d D\over d\epsilon}(\epsilon_{\text{bp}}^{(\nu)}) = 0\,,
\end{align}
such that ${d\epsilon_k\over dk}\rightarrow\pm\infty$ diverges for $k\rightarrow k_{\text{bp}}^{(\nu)}$.
Since $D'(\epsilon)$ is a polynomial of degree $Z-1$, Eq.~(\ref{eq:bp_condition_2}) has $Z-1$ 
solutions $\epsilon_{\text{bp}}^{(\nu)}$, with $\nu=1,\dots,Z-1$. As shown in Fig.~\ref{fig:band}(a), 
each $\epsilon_{\text{bp}}^{(\nu)}$ corresponds to a certain gap and has a real value between the top of 
band $\nu$ and the bottom of band $\nu+1$, with 
$D(\epsilon_{\text{bp}}^{(\nu)})=\cos(k_{\text{bp}}^{(\nu)})\ge 1$ ($\le -1$) for $\nu$ even (odd). 
By convention we define $k_{\text{bp}}^{(\nu)}$ as
\begin{align}
\label{eq:k_bp}
k^{(\nu)}_{\text{bp}} = i\kappa^{(\nu)}_{\text{bp}} + 
\begin{cases} 0 \quad \text{for} \,\,\nu\,\,\text{even} \\
\pi \quad \text{for} \,\,\nu\,\,\text{odd} \end{cases}\,,
\end{align}
with $\kappa_{\text{bp}}^{(\nu)}\ge 0$. We note that the branching points appear always 
$4$-fold as $\pm k_{\text{bp}}^{(\nu)}$ and $\pm (k_{\text{bp}}^{(\nu)})^*$. 
Furthermore, each $k_{\text{bp}}^{(\nu)}$ is defined $\text{mod}(2\pi)$ but if all of these 
replicas appear depends on the specific initial condition on the real axis which will be discussed
in the following. 

Expanding Eq.~(\ref{eq:dispersion}) around the branching point $k=k_{\text{bp}}^{(\nu)}$ and using 
$\sin(k_{\text{bp}}^{(\nu)})=i(-1)^\nu \sinh(\kappa_{\text{bp}}^{(\nu)})$ from (\ref{eq:k_bp}) and
$D''(\epsilon_{\text{bp}}^{(\nu)})=-(-1)^\nu |D''(\epsilon_{\text{bp}}^{(\nu)})|$ from
Fig.~\ref{fig:band}(a) we find (for $\alpha=\nu$ or $\alpha=\nu+1$)
\begin{align}
\label{eq:expansion_bp}
\epsilon_k^{(\alpha)} = \epsilon_{\text{bp}}^{(\nu)} +
\left(2{\sinh(\kappa_{\text{bp}}^{(\nu)})\over |D''(\epsilon_{\text{bp}}^{(\nu)})|}\right)^{1/2} 
\sqrt{i(k-k_{\text{bp}}^{(\nu)})}\,.
\end{align}
Depending on how the branch cut of the square root is chosen one obtains different ways to define
the analytic continuation. Taking the following initial condition on the real axis to solve
the differential equation (\ref{eq:eps_der_1}) 
\begin{align}
\label{eq:analytic_continuation_1}
\epsilon_k &= \epsilon_k^{(\alpha)} \quad\text{for}\quad
(\alpha-1)\pi < |k| < \alpha \pi \,,\\
\label{eq:analytic_continuation_1_Z}
\epsilon_k &= \epsilon_k^{(Z)} \quad\text{for}\quad
|k|>Z\pi \,,
\end{align}
one obtains branch cuts connecting $\pm\pi+i\kappa_{\text{bp}}^{(\nu)}$ with 
$\pm\pi-i\kappa_{\text{bp}}^{(\nu)}$, separating the band dispersions $\epsilon_k^{(\nu)}$ and
$\epsilon_k^{(\nu+1)}$ on the real axis, see Fig.~\ref{fig:bc}(a). This defines a common function
$\epsilon_k$ in the complex plane where the different band dispersion are connected analytically.
By convention, if the index $(\alpha)$ is not written in the following, $\epsilon_k$ denotes this function in the 
complex plane. Obviously, in this representation the periodicity condition (\ref{eq:epsilon_periodicity}) 
is no longer fulfilled but the properties (\ref{eq:epsilon_minus_k}) and (\ref{eq:epsilon_property}) 
remain valid
\begin{align}
\label{eq:epsilon_k_common_property}
\epsilon_k=\epsilon_{-k}=\left(\epsilon_{-k^*}\right)^*\ne\epsilon_{k+2\pi}  \,.
\end{align}
This choice for the analytic continuation has the advantage that all band dispersions are included but it is not 
very convenient to use it for calculating integrals $\int_{-\pi}^\pi dk$ by closing the integration
contour in the upper half since the integrals around the branch cuts are hard to evaluate. 

An alternative way is to define an analytic continuation
for each band $\epsilon_k^{(\alpha)}$ separately by using the initial condition
\begin{align}
\label{eq:analytic_continuation_2}
\epsilon_k = \epsilon_k^{(\alpha)} \quad\text{for}\quad -\infty < k < \infty
\end{align}
on the real axis, with $\epsilon_k^{(\alpha)}=\epsilon_{k+2\pi}^{(\alpha)}$. With this choice 
for the analytic continuation of $\epsilon_k^{(\alpha)}$ all properties stated in 
(\ref{eq:epsilon_minus_k}), (\ref{eq:epsilon_property}), and (\ref{eq:epsilon_periodicity}) 
remain valid for any $k$ in the complex plane. For each given $\alpha$,
two branching points at $k_{\text{bp}}^{(\alpha-1)}$ and $k_{\text{bp}}^{(\alpha)}$ appear 
($\text{mod}(2\pi)$, for $\alpha=1,Z$ only one branching point is present). Choosing the branch cut of
the square root in Eq.~(\ref{eq:expansion_bp}) on the negative real axis, the branch cuts
of $\epsilon_k^{(\alpha)}$ are pointing into the direction of the positive (negative) imaginary 
axis if $\text{Im}(k)>0$ ($\text{Im}(k)<0$), see Figs.~\ref{fig:bc}(b,c) for $\alpha=2,3$.

Since the initial conditions for the two choices of the analytic continuation of $\epsilon_k$
and $\epsilon_k^{(\alpha)}$ are the same on the real axis for $(\alpha-1)\pi < |k| < \alpha\pi$ and
for $|k|>Z\pi$, we obtain 
\begin{align}
\label{eq:relation_alpha}
\epsilon_k &= \epsilon_k^{(\alpha)} \quad\text{for}\quad
(\alpha-1)\pi < |\text{Re}(k)| < \alpha \pi \,,\\
\label{eq:relation_Z}
\epsilon_k &= \epsilon_k^{(Z)} \quad\text{for}\quad
|\text{Re}(k)|>Z\pi \,,
\end{align}
Using in addition $\epsilon_k^{(\alpha)}=\epsilon_{k+2\pi}^{(\alpha)}=\epsilon_{-k}^{(\alpha)}$ and 
$\epsilon_k=\epsilon_{-k}$, we can relate the analytic continuation of $\epsilon_k^{(\alpha)}$ to
the one of $\epsilon_k$, see Fig.~\ref{fig:bc}(b,c). This shows that the values left and right 
to the common branch cuts of $\epsilon_k^{(\alpha)}$
and $\epsilon_k^{(\alpha+1)}$ starting at 
$k_{\text{bp}}^{(\alpha)}$ are interchanged for the bands $\alpha$ and
$\alpha+1$. In Section~\ref{sec:localization} we will use this result to show that
the branch cut contributions to the Friedel density cancel for adjacent bands. 

We note that the differential equation (\ref{eq:eps_der_1}) determining the analytic
continuation can also be written in an alternative way by taking the derivative 
${d\over dk}$ of $(h_k-\epsilon_k)\chi_k=0$. Together with the form (\ref{eq:h_k}) of $h_k$ we find 
\begin{align}
\nonumber
&-{d\epsilon_k\over dk}\chi_k 
+ (i t_Z e^{-ik} |1\rangle\langle Z| - i t_Z e^{ik} |Z\rangle\langle 1| ) \chi_k \\
\nonumber
& + (h_k - \epsilon_k){d\over dk}\chi_k = 0 \,.
\end{align}
Multiplying from the left with $\chi_{-k}^T$ and using (\ref{eq:vons}) together with
the form (\ref{eq:chi_form}) of $\chi_k$ we get
\begin{align}
\nonumber
{d\epsilon_k\over dk} = i t_Z {s(\epsilon_k)\over N_k} 
\left(e^{-ik} a_{-k}(1) - e^{ik} a_k(1)\right)\,.
\end{align}
Finally, inserting Eq.~(\ref{eq:a_1_edge_tilde_d}) for $a_k(1)$ we obtain
\begin{align}
\label{eq:eps_der_2}
{d\epsilon_k\over dk} = 2 \bar{t}^Z {s(\epsilon_k)\over N_k}\sin(k)\,.
\end{align}
Comparing with (\ref{eq:eps_der_1}) we find the useful relation
\begin{align}
\label{eq:sND_relation}
N_k = -2\bar{t}^Z s(\epsilon_k) D'(\epsilon_k) \,.
\end{align}
For band $\alpha$ and $-\pi<k<\pi$, this gives a relation for the sign of $s(\epsilon_k^{(\alpha)})$ 
since $N_k^{(\alpha)}>0$ and the sign of $D'(\epsilon_k^{(\alpha)})$ is given by $(-1)^\alpha$ 
(see Fig.~\ref{fig:band}), providing
\begin{align}
\label{eq:sign_s}
\text{sign}(s(\epsilon_k^{(\alpha)})) = -(-1)^\alpha \quad \text{for} \quad -\pi < k < \pi \,.
\end{align}
This result will be used in Section~\ref{sec:edge} to prove that each gap hosts exactly one edge state.

The condition $N_k=0$ defines the points where $\sum_{j=1}^Z \left[\chi_k(j)\right]^2$ has a 
singularity. From Eq.~(\ref{eq:sND_relation}) it
follows that $s(\epsilon_k)=0$ or $D'(\epsilon_k)=0$ has to be fulfilled at such a point. The latter
condition corresponds to the branching points of $\epsilon_k$. The condition $s(\epsilon_k)=0$ 
corresponds to the pole positions of edge states, as will be discussed in Section~\ref{sec:edge}.
We summarize
\begin{align}
\nonumber
s(\epsilon_k)=0\,&,\, D'(\epsilon_k)\ne 0 \\ 
\label{eq:edge_poles} 
& \Leftrightarrow \text{edge pole of}\,\sum_{j=1}^Z \left[\chi_k(j)\right]^2\,,\\
\nonumber
s(\epsilon_k)\ne 0\,&,\, D'(\epsilon_k)=0 \\
\label{eq:branching_pole}
& \Leftrightarrow \text{branching point of}\,\sum_{j=1}^Z \left[\chi_k(j)\right]^2\,.
\end{align}
We note that, due to (\ref{eq:sND_relation}) and (\ref{eq:expansion_bp}), the second case 
leads to $N(\epsilon_k)\sim \epsilon_k-\epsilon_{\text{bp}}^{(\nu)} \sim \sqrt{i(k-k_{\text{bp}}^{(\nu)})}$
close to the branching point. Therefore, the factor $1/N(\epsilon_k)$ is integrable and does not have a 
pole but a branching point at this position. In contrast, the case $s(\epsilon_k)=D'(\epsilon_k)=0$ 
are special points where the edge pole and the branching point of 
$\sum_{j=1}^Z \left[\chi_k(j)\right]^2$ merge together to a branching pole,
where $N(\epsilon_k)\sim k-k_{\text{bp}}^{(\nu)})$, see Eq.~(\ref{eq:N_k_expansion_k_e_bp}) and
Appendix~\ref{app:N_k_expansion}.

\subsection{Scattering states for half-infinite system} 
\label{sec:bulk}

Once we have found the Bloch eigenstates $\psi_{k,\text{bulk}}^{(\alpha)}(n,j)$ for $H_{\text{bulk}}$ of 
the infinite system via (\ref{eq:bulk_eigenstates}), we can find a scattering eigenstate for $H_{R/L}$ of 
the half-infinite system at the same energy $\epsilon_k^{(\alpha)}$, given by
\begin{align}
\label{eq:eigenstate_H_LR}
& H_{R/L}|\psi_k^{(\alpha)}\rangle = \epsilon_k^{(\alpha)}|\psi_k^{(\alpha)}\rangle \,,\\
\label{eq:psi_H_LR}
&\psi_k^{(\alpha)}(n,j) = {1\over \sqrt{2\pi}} 
\left\{\chi_k^{(\alpha)}(j) e^{ikn} - \chi_{-k}^{(\alpha)}(j) e^{-ikn}\right\}\,.
\end{align}
The boundary condition (\ref{eq:boundary_RL}) is fulfilled since
$\chi_k^{(\alpha)}(Z)=\chi_{-k}^{(\alpha)}(Z)=s(\epsilon_k^{(\alpha)})/\sqrt{N_k}$ is real, which is
the gauge we have used in Section~\ref{sec:Bloch_states}. The $k$-values for the eigenstates of the
half-infinite system are restricted to $0<k<\pi$.

We note that the orthogonality and completeness relation of the eigenstates of the infinite system
\begin{align}
\label{eq:bulk_orthogonality}
\sum_{n=-\infty}^\infty \sum_{j=1}^Z \psi_{k,\text{bulk}}^{(\alpha)}(n,j)^*\psi_{k',\text{bulk}}^{(\alpha')}(n,j)
&= \delta_{\alpha\alpha'}\delta(k-k')\,,\\
\label{eq:bulk_completeness}
\sum_{\alpha=1}^Z \int_{-\pi}^\pi dk\, \psi_{k,\text{bulk}}^{(\alpha)}(n,j) \psi_{k,\text{bulk}}^{(\alpha)}(n',j')
&= \delta_{nn'}\delta_{jj'}
\end{align}
implies only the orthogonality relation for the eigenstates of the half-infinite system
\begin{align}
\label{eq:RL_orthogonality}
\sum_{n=-\infty}^\infty \sum_{j=1}^Z \psi_{k}^{(\alpha)}(n,j)^*\psi_{k'}^{(\alpha')}(n,j)
= \delta_{\alpha\alpha'}\delta(k-k')\,,
\end{align}
since the edge states will also contribute to the completeness relation, see Section~\ref{sec:edge}.

The result (\ref{eq:psi_H_LR}) can also be viewed as a consequence of scattering
theory: the wave function consists of a superposition of an incoming and an outgoing wave. The 
boundary condition (\ref{eq:boundary_RL}) is very simple here since we have considered only one orbital 
per site. For the multi-channel case, the eigenstates of $H_{R/L}$ will also 
involve exponentially decaying parts.

\subsection{Edge states for half-infinite system}
\label{sec:edge}

In this section we construct explicitly all edge states for a half-infinite system and discuss the relation 
of various determinants to the complex quasimomentum corresponding to the edge states. These relations
will turn out to be essential to derive the topological constraints for the phase-dependence of the 
edge state energies in the subsequent Section~\ref{sec:constraints_edge}.

To find edge states for $H_{R/L}$, 
\begin{align}
\label{eq:edge_eq}
H_{R/L} |\psi_{k_{\text{e}}}^{\text{e}}\rangle &= \epsilon_{k_{\text{e}}} |\psi^{\text{e}}_{k_\text{e}}\rangle\,,\\
\label{eq:edge_state}
\psi^{\text{e}}_{k_\text{e}}(n,j) &= \chi^{\text{e}}_{k_\text{e}}(j) e^{ik_{\text{e}}n}\,,
\end{align} 
we look for solutions of Bloch states with
$\text{Im}(k_{\text{e}})\gtrless 0$ and $s(\epsilon_{k_{\text{e}}})=\chi^{\text{e}}_{k_\text{e}}(Z)=0$ 
to fulfil the boundary and 
asymptotic conditions (\ref{eq:boundary_RL}), (\ref{eq:asymptotic_R}) and (\ref{eq:asymptotic_L}).
In contrast to $\chi_{k_{\text{e}}}$, we parametrize them with a different normalization factor
\begin{align}
\label{eq:chi_edge}
\chi^{\text{e}}_{k_{\text{e}}} = {1\over\sqrt{N^{\text{e}}_{k_{\text{e}}}}}
\left(\begin{array}{c} 
a_{k_{\text{e}}} \\ 0 \\ 
\end{array}\right)\,,
\end{align}
but with the same vector $a_{k_{\text{e}}}$. Below we will find that $e^{-ik_{\text{e}}}$ is real
[see Eq.~(\ref{eq:k_edge_values})] and, therefore, it follows from (\ref{eq:a_k_explicit}) that also the 
vector $a_{k_{\text{e}}}$ and the edge state wave function $\psi^{\text{e}}_{k_{\text{e}}}(n,j)$ are real. 
The normalization $N^{\text{e}}_{k_{\text{e}}}$ is defined 
such that the edge state wave function (\ref{eq:edge_state}) is normalized,
which means $\sum_{n=1}^\infty\sum_{j=1}^Z [\psi^{\text{e}}_{k_{\text{e}}}(n,j)]^2=1$ for an edge state of $H_R$ and
$\sum_{n=-\infty}^0\sum_{j=1}^Z [\psi^{\text{e}}_{k_{\text{e}}}(n,j)]^2$ for an edge state of $H_L$, leading to
\begin{align}
\label{eq:edge_normalization}
N^{\text{e}}_{k_{\text{e}}} = \text{sign}(\text{Im}(k_{\text{e}}))
{a_{k_{\text{e}}}^T a_{k_{\text{e}}} \over e^{-2ik_{\text{e}}}-1}\,.
\end{align}
In contrast, $N_{k_{\text{e}}}$ can not be the correct normalization since we get $N_{k_{\text{e}}}=0$ from 
$s(\epsilon_{k_{\text{e}}})=0$ and (\ref{eq:sND_relation}), such that $\sum_{j=1}^Z \left[\chi_k(j)\right]^2$ has
a pole at $k=k_{\text{e}}$. Therefore, the normalization factor 
$N_k$ of $\chi_k$ is {\it not} analytically continued to the normalization factor $N^{\text{e}}_{k_{\text{e}}}$
of $\chi^{\text{e}}_{k_{\text{e}}}$. The reason is that the Hamiltonian $h_k$ is non-hermitian for complex $k$ and has
different right and left eigenstates given by $\chi_k$ and $\chi_{-k}$, respectively. These states
get a completely different analytic continuation. In particular, at the edge pole, 
$\chi_{k_{\text{e}}}\rightarrow\infty$ and $\chi_{-k_{\text{e}}}\rightarrow 0$, 
such that the orthogonality and completeness relation (\ref{eq:vons}) remain valid. 
In Appendix~\ref{app:N_k_expansion} we will show that an expansion of $N_k$ around 
the edge pole $k_{\text{e}}$ gives the result
\begin{align}
\label{eq:N_k_expansion}
N_k = \text{sign}(\text{Im}(k_{\text{e}})) \,i N^{\text{e}}_{k_{\text{e}}}(k-k_{\text{e}}) + O(k-k_{\text{e}})^2\,,
\end{align}
which will be used in Section~\ref{sec:localization} to prove that the edge state density is 
cancelled by the pole contribution of the Friedel density of that band which belongs to 
this edge state (see below for the definition of this correspondence). 
This result applies as long as the edge pole is isolated. If it agrees with a 
branching point $k_{\text{e}}=k^{(\nu)}_{\text{bp}}$ of a certain gap $\nu$, we show in 
Appendix~\ref{app:N_k_expansion} that (\ref{eq:N_k_expansion}) gets an 
additional factor $2$ and the corrections are of $O(k-k_{\text{e}})^{3/2}$
\begin{align}
\label{eq:N_k_expansion_k_e_bp}
N_k = \text{sign}(\text{Im}(k_{\text{e}})) \,2 i N^{\text{e}}_{k_{\text{e}}}(k-k_{\text{e}}) 
+ O(k-k_{\text{e}})^{3/2}\,.
\end{align}
This will be needed to show in Section~\ref{sec:localization} that the pole contribution of 
the Friedel density of band $\alpha=\nu$ or band $\alpha=\nu+1$ cancels only half of 
the edge state density.

\begin{figure*}
\centering
\includegraphics[width= 1.\columnwidth]{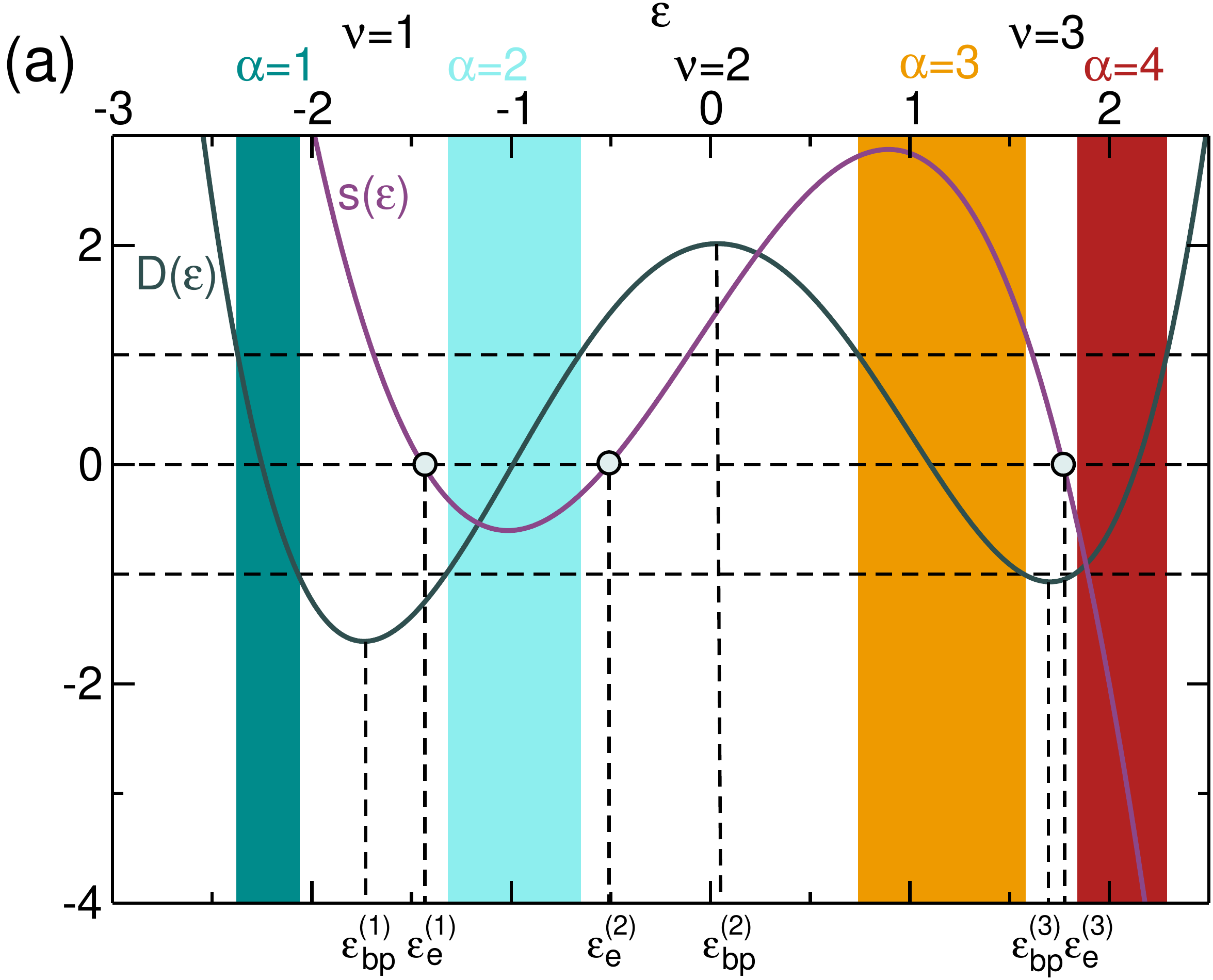} 
\includegraphics[width= 1.\columnwidth]{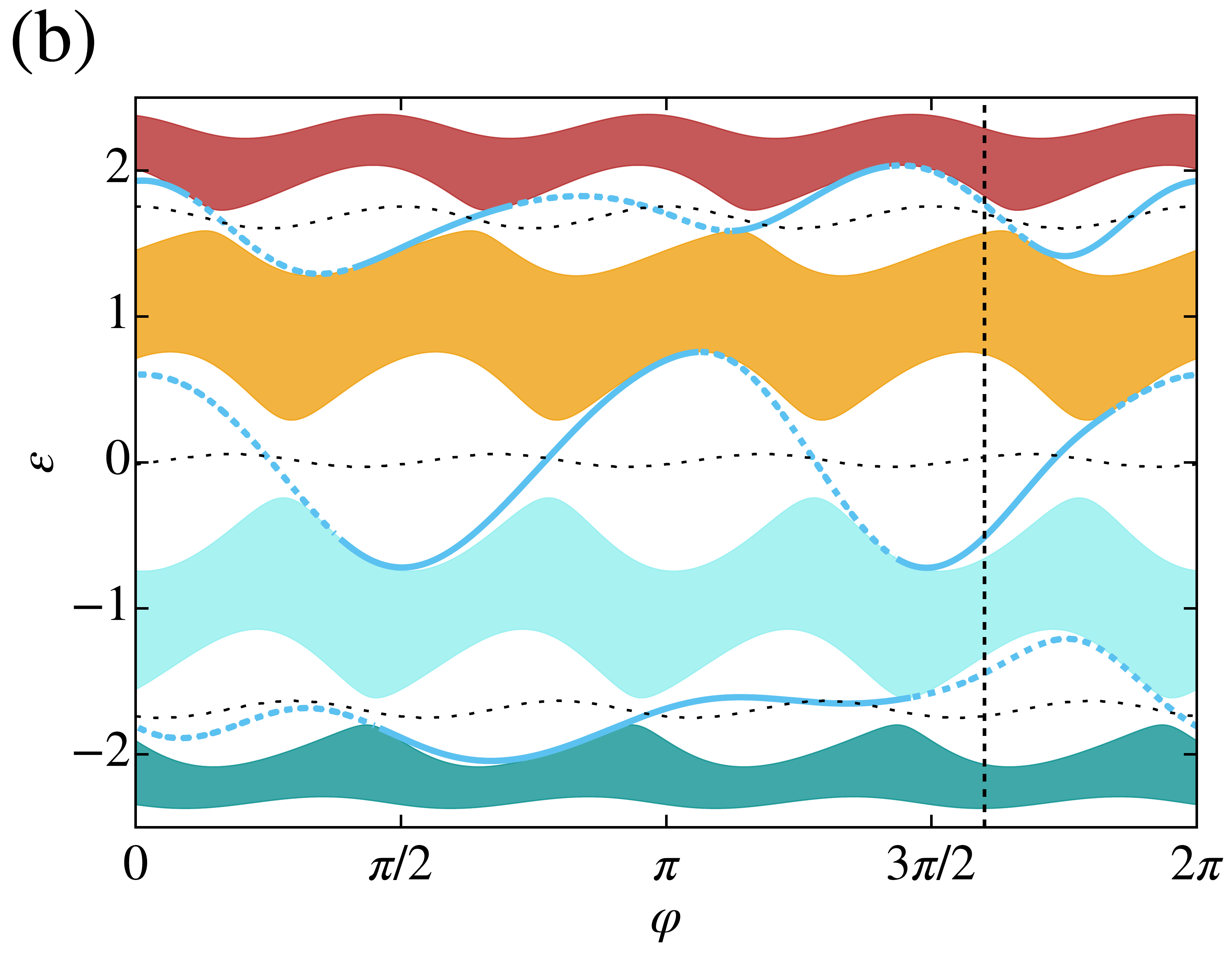}
 \caption{(a) Determination of the edge state energies $\epsilon_{\text{e}}^{(\nu)}$ from the 
   condition $s(\epsilon_{\text{e}}^{(\nu)})=0$ for $Z=4$, $V=0.5$, $t=1.1$, $\delta t=0.1$, 
   $\varphi=1.6\pi$, and three random Fourier coefficients for the real functions 
   $F_v$ and $F_t$ in Eqs.~(\ref{eq:v_form}) and (\ref{eq:t_form}) according to
   the form (\ref{eq:F_random1}), see Supplemental Material for the precise parameters \cite{SM}. 
   As explained in the 
   main text the position of the edge state energies is always located in the gaps, 
   each gap hosting exactly one edge state. The sign of $s(\epsilon)$ is
   given by $(-1)^{\alpha+1}$ in the energy regions of band $\alpha$, see Eq.~(\ref{eq:sign_s}). 
   Depending on whether $\epsilon_{\text{e}}^{(\nu)}\lessgtr\epsilon_{\text{bp}}^{(\nu)}$, 
   the edge states result from the analytic continuation of band $\nu$ or $\nu+1$. 
   (b) The band structure as function of the phase variable 
   $\varphi$ for the same parameters as in (a). The band structure is periodic under a 
   change of the phase variable by $2\pi/Z$. The edge state energies $\epsilon_{\text{e}}^{(\nu)}(\varphi)$
   for $\nu=1,2,3$ are shown by blue solid/dashed lines corresponding to edge 
   states of $H_R$/$H_L$ with $\text{Im}(k_{\text{e}}^{(\nu)})\gtrless 0$.
   As can be seen each gap hosts exactly 
   one edge state and the edge states change the boundary when they touch the bands. 
   Approximately in the middle of each gap we show by black dashed lines the phase-dependence of the 
   energies $\epsilon_{\text{bp}}^{(\nu)}(\varphi)$ at the branching points for $\nu=1,2,3$. 
   For $\epsilon_{\text{e}}^{(\nu)} \lessgtr \epsilon_{\text{bp}}^{(\nu)}$ the edge states belong
   to the analytical continuation of band $\nu$ or $\nu+1$, respectively. In (b) we have
   indicated by the vertical dashed line the phase value $\varphi=1.6\pi$ used in
   (a) and where the analytic structure and the position of the edge poles is shown in Fig.~\ref{fig:bc_edge}.
}
\label{fig:edge}
\end{figure*}
Since $s(\epsilon_{k_{\text{e}}})=0$ we get from (\ref{eq:a_k_equation}) 
\begin{align}
\label{eq:A_zero}
A a_{k_{\text{e}}} &= \epsilon_{k_{\text{e}}} a_{k_{\text{e}}}\,.
\end{align}
This eigenvalue problem for the $(Z-1)$-dimensional and hermitian matrix $A$ has exactly
$Z-1$ solutions with real eigenvalues $\epsilon_{k_{\text{e}}}$. Therefore, 
we find $Z-1$ edge states, each of them either corresponding to $H_R$ or $H_L$ depending on the
sign of $\text{Im}(k_{\text{e}})$. This sign will be determined below, see
Eqs.~(\ref{eq:edge_LR_0}) and (\ref{eq:edge_LR_pi}). The energies of these $Z-1$ edge states 
are distributed among the $Z-1$ gaps $\nu=1,\dots,Z-1$ of the bulk spectrum, each gap hosting exactly 
one edge state. This follows from (\ref{eq:sign_s}) since this equation implies 
that the sign of $s(\epsilon_k^{(\alpha)})$ 
is alternating with the band index $\alpha$ such that in each gap there must be at least one solution with 
$s(\epsilon_{k_{\text{e}}})=\text{det}(A-\epsilon_{k_{\text{e}}})=0$, see Fig.~\ref{fig:edge}(a) for 
illustration; also cf. Appendix C of Ref.~[\onlinecite{hatsugai_jphys}] for an alternative proof. Therefore, we label
the edge states by the same index $\nu$ and denote them by 
$\chi_{\text{e}}^{(\nu)}\equiv\chi^{\text{e}}_{k^{(\nu)}_{\text{e}}}$ with energies
$\epsilon^{(\nu)}_{\text{e}}\equiv\epsilon_{k^{(\nu)}_{\text{e}}}$, where $k^{(\nu)}_{\text{e}}$
denotes the complex quasimomentum of the edge state. From Fig.~\ref{fig:edge}(a) we get 
analog to (\ref{eq:k_bp})
\begin{align}
\label{eq:k_edge_values}
k^{(\nu)}_{\text{e}} = i\kappa^{(\nu)}_{\text{e}} + 
\begin{cases} 0 \quad \text{for} \,\,\nu\,\,\text{even} \\
\pi \quad \text{for} \,\,\nu\,\,\text{odd} \end{cases}\,.
\end{align}
\begin{figure*}
\centering
\includegraphics[width= 1.\columnwidth]{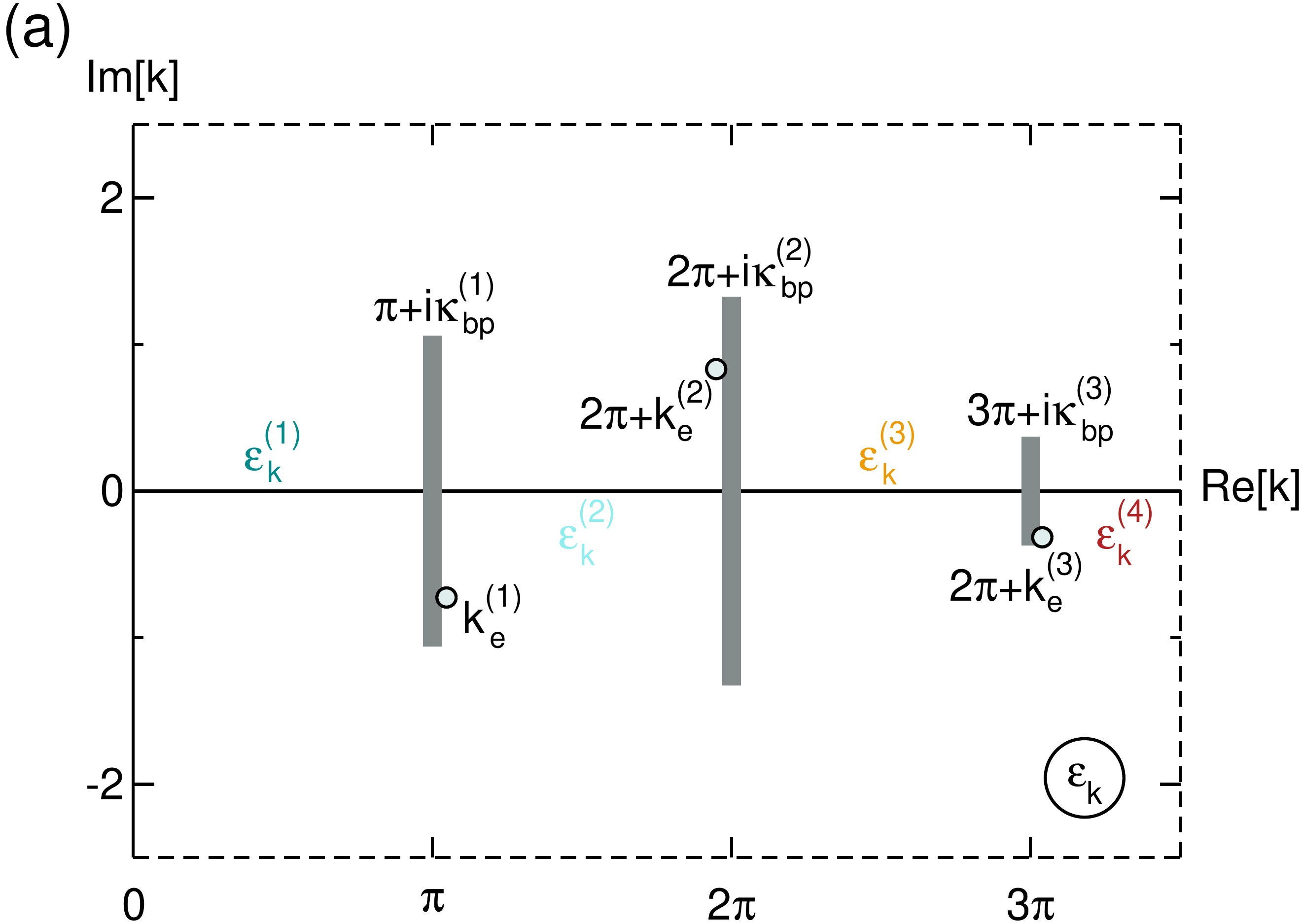} 
\includegraphics[width= 1.\columnwidth]{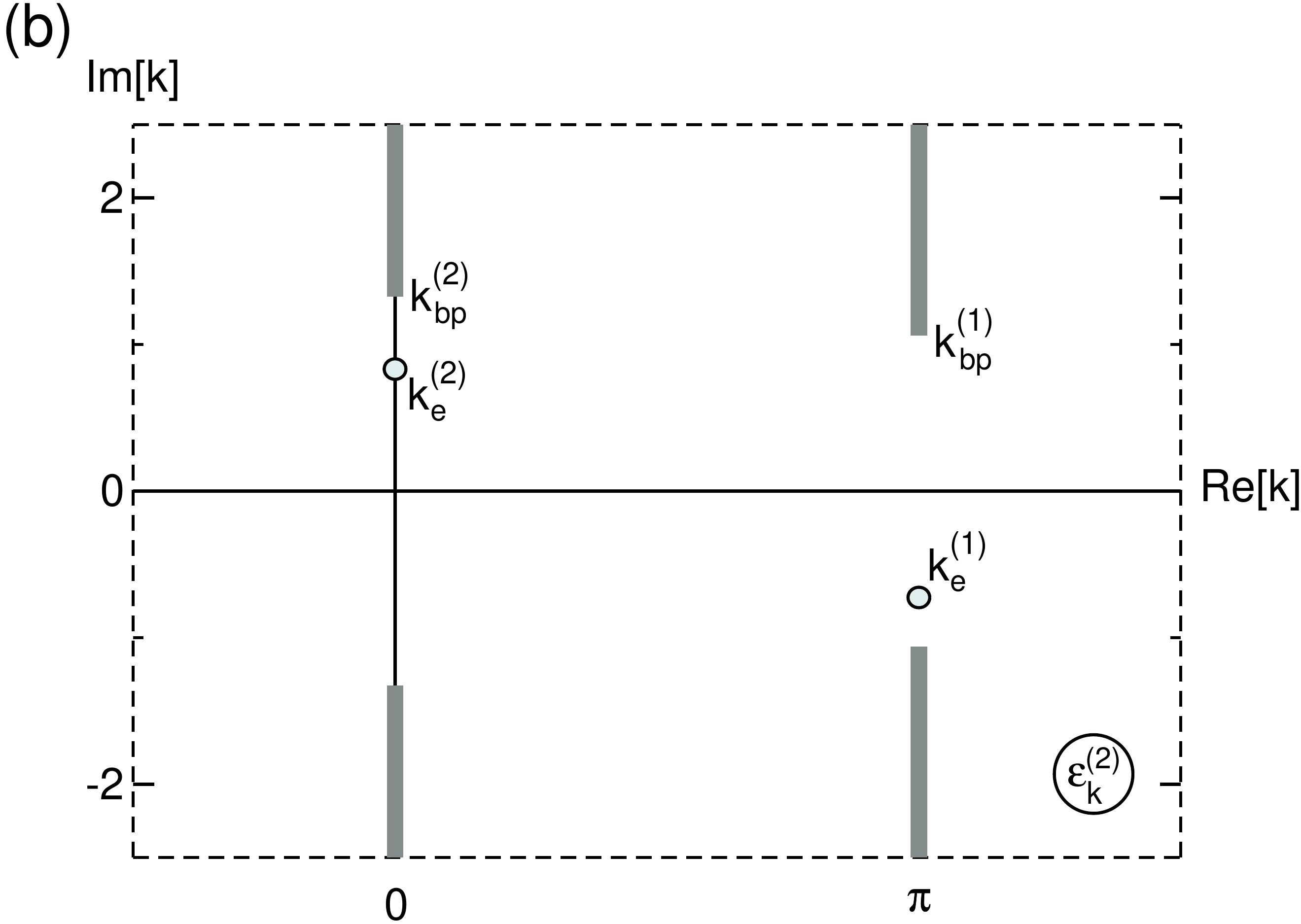}
\caption{The analytic continuation of (a) $[\chi_k(j)]^2$ and 
  (b) $[\chi^{(2)}_k(j)]^2$ analog to Fig.~\ref{fig:bc}(a,b) but for the
  parameters of Fig.~\ref{fig:edge} at the particular phase $\varphi=1.6\pi$ indicated in 
  Fig.~\ref{fig:edge}(b) by a vertical dashed line. In addition we have indicated 
  the edge pole positions $k_\text{e}^{(\nu)}$, with $\nu=1,2,3$. For $\text{Im}(k_\text{e}^{(\nu)})\gtrless 0$, 
  they correspond to edge states of either $H_R$ or $H_L$, respectively. In (a) an edge pole lying left (right) 
  to the branch cut located at $\nu\pi$ belongs to the analytic continuation of band $\nu$ ($\nu+1$).
  The edge poles move around the branch cuts as function of $\varphi$, see snapshots and videos 
  provided in the Supplemental Material \cite{SM}.}
\label{fig:bc_edge}
\end{figure*}
In Fig.~\ref{fig:edge}(b) we show an example of the phase-dependence of the band structure 
together with the edge state energies $\epsilon_{\text{e}}^{(\nu)}$ and the energies $\epsilon_{\text{bc}}^{(\nu)}$
at the branching points, with $\nu=1,2,3$. The band structure is periodic under a phase change
by $2\pi/Z$ since a shift of the whole lattice by one site does not change the bulk spectrum. 
In each gap one edge state is present which changes from $H_R$ to $H_L$ at the 
point where it touches one of the bands (where $\text{Im}(k^{(\nu)}_{\text{e}})=0$), indicated by
solid and dashed lines of the edge states in Fig.~\ref{fig:edge}(b), respectively. 
We note that the edge states of $H_R$ or $H_L$ can be shifted by
$\pm 2\pi/Z$ as function of $\varphi$ if one shifts the boundaries of $H_R$ or $H_L$ shown in 
Fig.~\ref{fig:model_RL} by one site. In Fig.~\ref{fig:bc_edge}(a,b) we show the analytic 
continuation corresponding to the band structure of Fig.~\ref{fig:edge}(b)
for a particular value of the phase. We indicate the positions of the
three edge poles $k_{\text{e}}^{(\nu)}$ for $\nu=1,2,3$ in the complex plane. 
We note that the values for $k^{(\nu)}_\text{e}$ are sitting on top of the branch cuts shown
in Fig.~\ref{fig:bc_edge}(a). Therefore, they should be shifted slightly to the left/right of the
branch cuts, depending on whether $\epsilon_{\text{e}}^{(\nu)}\lessgtr\epsilon_{\text{bp}}^{(\nu)}$ or,
equivalently, whether the edge state belongs to the analytic continuation
of band $\nu$ or band $\nu+1$. When the phase variable $\varphi$ changes, the edge poles 
move around the branch cuts in Fig.~\ref{fig:bc_edge}(b) 
as shown in snapshots and videos, available in the Supplemental Material
\cite{SM}. As can be seen from the dispersion relation $\epsilon_{\text{e}}^{(\nu)}(\varphi)$ of the
edge states in Fig.~\ref{fig:edge}, the edge pole encircles the branch cut between band $\nu$ and $\nu+1$
by an integer number when the phase has changed by $2\pi$. In Section~\ref{sec:invariant_physics} we
will explain that $\nu+nZ$ edge states of $H_R$ connect band $\nu$ and $\nu+1$, with $n=0,\pm 1,\dots$,
running either all upwards or downwards for $\nu+nZ\gtrless 0$, respectively. 
This means that the edge pole of gap $\nu$ runs 
$\nu+nZ$ times around the branch cut, either clockwise or counter-clockwise for $\nu+nZ\gtrless 0$.   

In the following we omit the index $\nu$ for simplicity, i.e., use $k_{\text{e}}\equiv k_{\text{e}}^{(\nu)}$. 
From (\ref{eq:b_k_equation}) we get 
\begin{align}
\label{eq:b_zero}
b_{-k_{\text{e}}}^T a_{k_{\text{e}}} &= 0 \,,
\end{align}
and we calculate $a_{k_{\text{e}}}$ via (\ref{eq:a_k_solution}) in terms of the well-defined matrix 
$B(\epsilon_{k_\text{e}})$
\begin{align}
\label{eq:a_k_edge}
a_{k_\text{e}} = - B(\epsilon_{k_\text{e}}) b_{k_\text{e}}\,.
\end{align}
Using (\ref{eq:aj_a1}) and (\ref{eq:a_1_edge_tilde_d}) this leads to the explicit solution
\begin{align}
\label{eq:aj_edge}
t_1\cdots t_{j-1} a_{k_\text{e}}(j) &= d_{1,j-1}(\epsilon_{k_\text{e}}) a_{k_\text{e}}(1)\,,\\
\label{eq:a1_edge}
a_{k_\text{e}}(1) &= {\bar{t}^Z\over t_Z} \left\{\tilde{d}_{2,Z-1}(\epsilon_{k_\text{e}})e^{-ik_\text{e}} + 1\right\}\,.
\end{align}

Using the form (\ref{eq:b_k}) of $b_{k_\text{e}}$, we get from (\ref{eq:b_zero}) the relation
$t_{Z-1}a_{k_\text{e}}(Z-1)=-t_Z e^{ik_\text{e}} a_{k_\text{e}}(1)$. From (\ref{eq:aj_edge}) we get for $j=Z-1$ the
result $t_{Z-1}a_{k_\text{e}}(Z-1)=-{t_Z\over \bar{t}^Z}d_{1Z}(\epsilon_{k_\text{e}})a_{k_\text{e}}(1)$. 
Since $a_{k_\text{e}}\ne 0$, this gives the relation
\begin{align}
\label{eq:d_1Z_edge}
d_{1Z}(\epsilon_{k_\text{e}}) = \bar{t}^Z e^{ik_\text{e}}
\end{align}
for all edge states. 

Furthermore, from $s(\epsilon_{k_\text{e}})=d_{1,Z-1}(\epsilon_{k_\text{e}})=0$ and the properties
(\ref{eq:det_recursion_2}) and (\ref{eq:det_property}), we get 
$(t_1\cdots t_{Z-2})^2 = d_{1,Z-2}(\epsilon_{k_\text{e}}) d_{2,Z-1}(\epsilon_{k_\text{e}})$ and 
$d_{1Z}(\epsilon_{k_\text{e}})=-t_{Z-1}^2 d_{1,Z-2}(\epsilon_{k_\text{e}})$, leading with (\ref{eq:d_1Z_edge}) to
\begin{align}
\label{eq:tilde_d_2_Z-1_edge}
\tilde{d}_{2,Z-1}(\epsilon_{k_\text{e}}) &= - e^{-ik_\text{e}} \\
\label{eq:tilde_d_1_Z-2_edge}
\tilde{d}_{1,Z-2}(\epsilon_{k_\text{e}}) &= - e^{ik_\text{e}} \\
\label{eq:d_1_Z-1_edge}
d_{1,Z-1}(\epsilon_{k_\text{e}}) &= 0 \,,
\end{align}
where $\tilde{d}_{1,Z-2}$ and $\tilde{d}_{2,Z-1}$ have been defined in (\ref{eq:tilde_d_2_Z-1}) and
(\ref{eq:tilde_d_1_Z-2}), respectively. 

From (\ref{eq:a1_edge}) and (\ref{eq:tilde_d_2_Z-1_edge}) we conclude that
\begin{align}
\label{eq:a_minus_k_edge}
a_{-k_\text{e}} = 0 \,,
\end{align}
which is consistent with $N_{k_\text{e}}=0$, see (\ref{eq:N_k}).

Using the result (\ref{eq:tilde_d_2_Z-1_edge}) for $\tilde{d}_{2,Z-1}(\epsilon_{k_\text{e}})$ we can determine
the sign of $\text{Im}(k_\text{e})$ and decide whether the edge state is an eigenstate of $H_R$ or $H_L$
\begin{align}
\label{eq:edge_LR_0}
\underline{k_\text{e}=i\kappa_\text{e}}:
\quad \tilde{d}_{2,Z-1}(\epsilon_{k_\text{e}})\lessgtr -1 \, &\Leftrightarrow \, 
\text{Im}(k_\text{e}) \gtrless 0\\
\label{eq:edge_LR_pi}
\underline{k_\text{e}=\pi + i\kappa_\text{e}}:
\quad\tilde{d}_{2,Z-1}(\epsilon_{k_\text{e}})\gtrless 1 \, &\Leftrightarrow \,
\text{Im}(k_\text{e}) \gtrless 0\,.
\end{align}
As a consequence, when the determinant $\tilde{d}_{2,Z-1}(\epsilon_{k_\text{e}(\varphi)})$ runs through the
points $\pm 1$ as function of the phase variable $\varphi$ which determines the position $k_\text{e}(\varphi)$ 
of the edge pole, an edge state is changing from $H_R$ to $H_L$ or vice
versa. The phase-dependence of the edge state energies will be discussed in all detail in the next section.

\subsection{Topological constraints for edge states}
\label{sec:constraints_edge}

In this section we will derive the topological constraints for the phase-dependence of the
edge state energies proving rigorously the central result Eq.~(\ref{eq:topological_constraint}). 
We will develop a simple diagrammatic representation to visualize the constraints and derive some rules how
the phase-dependence of the edge state energies looks like. 

To study the energy of the edge states as function of the phase $\varphi$,
we fix some gap $\nu=1,\dots,Z-1$ (not written in the following) and define the phase-dependence of the
edge state energy via 
\begin{align}
\epsilon_{\text{e}}(\varphi) = \epsilon_{k_{\text{e}}(\varphi)}(\varphi)\,,
\end{align}
where $k_e(\varphi)$ denotes the phase-dependence of the complex quasimomentum determining the
edge state and $\epsilon_k(\varphi)$ denotes the phase-dependence of the dispersion relation
via the parameters $v_j(\varphi)$ and $t_j(\varphi)$ defining the microscopic model.
We first note the important property that the dispersion can never be the 
same at $\varphi$ and $\varphi+{2\pi\over Z}$ when the edge states at both values belong either 
to $H_R$ or to $H_L$
\begin{align}
\nonumber
&\text{Im}[k_{\text{e}}(\varphi)] = \text{Im}\left[k_{\text{e}}\left(\varphi+{2\pi\over Z}\right)\right] 
\quad\Rightarrow \\
\label{eq:edge_dispersion_shift}
&\hspace{2cm}
\Rightarrow\quad\epsilon_{\text{e}}(\varphi)\ne \epsilon_{\text{e}}\left(\varphi+{2\pi\over Z}\right)\,.
\end{align}
To show this we use (\ref{eq:shift}) and find
\begin{align}
\label{eq:det_relation_edge}
\tilde{d}_{2,Z-1}(\epsilon,\varphi) = \tilde{d}_{1,Z-2}\left(\epsilon,\varphi+{2\pi\over Z}\right)\,,
\end{align}
where the dependence on $\varphi$ again indicates the one from the parameters $v_j(\varphi)$ and $t_j(\varphi)$. 
Assuming $\epsilon_{\text{e}}(\varphi)=\epsilon_{\text{e}}(\varphi+{2\pi\over Z})$
we get from (\ref{eq:tilde_d_2_Z-1_edge}), (\ref{eq:tilde_d_1_Z-2_edge}) and (\ref{eq:det_relation_edge}) 
\begin{align}
\nonumber
- e^{-ik_{\text{e}}(\varphi)} &= \tilde{d}_{2,Z-1}(\epsilon_{\text{e}}(\varphi),\varphi)\\
\nonumber
&= \tilde{d}_{1,Z-2}(\epsilon_{\text{e}}(\varphi),\varphi+{2\pi\over Z})\\
\nonumber
&= \tilde{d}_{1,Z-2}(\epsilon_{\text{e}}(\varphi+{2\pi\over Z}),\varphi+{2\pi\over Z})\\
\nonumber
&= - e^{ik_{\text{e}}(\varphi+{2\pi\over Z})}\,.
\end{align}
This is only possible for $\text{Im}[k_{\text{e}}(\varphi)]\ne\text{Im}[k_{\text{e}}(\varphi+{2\pi\over Z})]$
which proves (\ref{eq:edge_dispersion_shift}). 

\begin{figure}
  \centering
  \includegraphics[width=\columnwidth]{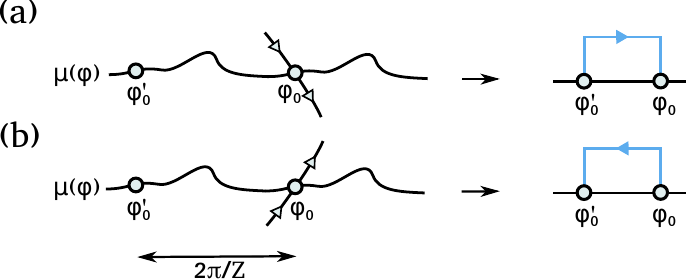}  
  \caption{edge states of $H_R$ crossing the chemical potential 
    $\mu(\varphi)=\mu(\varphi+{2\pi\over Z})$ at $\varphi=\varphi_0$ either from (a) above or 
    (b) below. At $\varphi_0^\prime=\varphi_0-{2\pi\over Z}$ it is not allowed that an edge state
    of $H_R$ crosses the chemical potential. To the right we show the way how we visualize the
    two different possibilities by contractions (in blue color). 
  }
  \label{fig:contraction}
\end{figure}
\begin{figure}
  \centering
  \includegraphics[width=0.8\columnwidth]{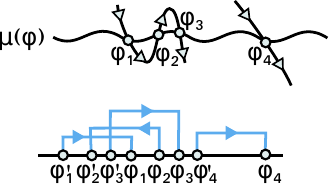}  
  \caption{Visualization of edge states of $H_R$ crossing $\mu(\varphi)$ at $\varphi=\varphi_i$,
    with $i=1,2,3,4$, in terms of sequences of contractions. 
    We defined $\varphi_i^\prime=\varphi_i-{2\pi\over Z}$ 
    such that each contraction has a fixed length ${2\pi\over Z}$. 
  }
  \label{fig:configuration}
\end{figure}
From (\ref{eq:edge_dispersion_shift}) we can deduce a first important property how edge states of
$H_R$ (analog for $H_L$) can cross any phase-dependent chemical potential 
\begin{align}
\label{eq:chemical_potential}
\mu(\varphi) = \mu\left(\varphi+{2\pi\over Z}\right)
\end{align}
chosen somewhere in gap $\nu$ and periodic with period ${2\pi\over Z}$. When an edge state of $H_R$ fulfils
$\epsilon_e(\varphi_0)=\mu(\varphi_0)$, it is not possible that an edge state of $H_R$ can cross the
chemical potential at $\varphi\pm{2\pi\over Z}$. Considering only edge states of $H_R$ we show in
Figs.~\ref{fig:contraction}(a,b) how we visualize the two possibilities of an edge state crossing the
chemical potential at $\varphi=\varphi_0$ either from above or below via contractions. The phase 
$\varphi_0$ of the right vertex of a contraction denotes the position of the edge state crossing 
$\mu(\varphi_0)$. Each contraction has a fixed length ${2\pi\over Z}$ such that the left vertex 
has phase $\varphi_0^\prime=\varphi_0-{2\pi\over Z}$. In this picture the property 
(\ref{eq:edge_dispersion_shift}) is equivalent to the fact that two vertices can never lie on top 
of each other. Each way the edge states of $H_R$ can cross the chemical potential can then be 
visualized by a sequence of contractions, see Fig.~\ref{fig:configuration}. This visualization 
will turn out to be very convenient to formulate and prove many of the following topological constraints.

If a certain topological configuration of contractions occurs for some chosen $\mu(\varphi)$ it
can not change when choosing a different chemical potential since two vertices are never allowed 
to coincide. An exception are cases when the chemical potential is moved through a local minimum
or maximum of the phase-dependence of the edge state energy. 
In this case a pair of two contractions with different directions 
fall on top of each other and are eliminated (or created). However, as explained below in all detail, 
such pairs do not change any of the topological constraints discussed in the following.
Therefore, without loss of generality, we choose for $\mu(\varphi)$ the top of the lower band $\alpha=\nu$
\begin{align}
\label{eq:mu_top}
\mu(\varphi) = \epsilon^{(\nu)}_{k_0}(\varphi)\,,
\end{align}
where $k_0=0$ for $\nu$ even and $k_0=\pi$ for $\nu$ odd. For this choice the edge states of $H_R$
can just enter or leave the band, their connection below $\mu(\varphi)$ is just meant as a guide for
the eye to formulate certain topological constraints derived in the following. 

In Appendix~\ref{app:constraints_edge} we will prove from the rules (\ref{eq:tilde_d_2_Z-1_edge})
and (\ref{eq:tilde_d_1_Z-2_edge}) for the occurrence of edge states the following 
topological condition for the allowed edge state configurations of $H_R$
\begin{align}
\nonumber
&\text{\underline{Topological constraint:}}\\
\label{eq:constraint}
&\text{Outgoing and incoming vertices must alternate.}
\end{align}

Together with the fact that the contractions have a fixed length ${2\pi\over Z}$ and have to be
ordered on an interval of size $2\pi$ with periodic boundary conditions, we can construct all
possible edge state configurations. None of them can be excluded in principle and it depends on the
model under consideration which of them appear. E.g., the configuration shown in Fig.~\ref{fig:configuration}
is obviously an allowed one consistent with (\ref{eq:constraint}).
\begin{figure}
  \centering
  \includegraphics[width=\columnwidth]{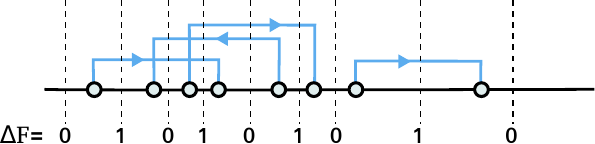}  
  \caption{Graphical rule how to determine the virtual topological charge $\Delta F(\varphi)$ 
    at some phase $\varphi$ lying between two vertices. Drawing a vertical cut at this position
    (dashed line) one has to take the number of right-going minus the number of left-going contraction
    lines through this vertical cut. 
  }
  \label{fig:topological_charge}
\end{figure}
The topological constraint can also be formulated in terms of a virtual topological charge 
$\Delta F(\varphi)$ defined by 
\begin{align}
\label{eq:topological_charge}
\Delta F(\varphi) &= F(\varphi + {2\pi\over Z}) - F(\varphi)\,,\\
F(\varphi) &= \sum_{\sigma=\pm} \sum_{i=1}^{M_\sigma}\sigma \theta(\varphi-\varphi_{i\sigma})\,,
\end{align}
where $\varphi_{i\sigma}$ are the phase values where edge states of $H_R$ enter/leave the band
and $M_\pm$ is the total number of entering/leaving edge states. 
Using the following form of the topological charge
\begin{align}
\label{eq:topological_charge_1}
\Delta F(\varphi) = \sum_{\sigma=\pm} \sum_{i=1}^{M_\sigma} 
\sigma \theta(\varphi^\prime_{i\sigma} < \varphi < \varphi_{i\sigma})\,,
\end{align}
one finds that $\Delta F(\varphi)$ can be read off for $\varphi$ lying between two adjacent vertices of 
a certain configuration by making a virtual vertical cut at $\varphi$ and taking the number of right-going 
minus the number of left-going contraction lines crossing the vertical cut, see 
Fig.~\ref{fig:topological_charge} for an example.
The topological constraint (\ref{eq:constraint}) can then alternatively be formulated as a 
constraint for the topological charge
\begin{align}
\label{eq:Delta_F_constraint}
\Delta F(\varphi) \in \{s-1,s\}\,,
\end{align}
where $s$ is a phase-independent integer characteristic of each configuration. 
E.g., in Fig.~\ref{fig:topological_charge}
we get $s=1$. We note that by reversing all direction of the contractions we change the
sign of the topological charge and get again an allowed configuration. Due to 
(\ref{eq:Delta_F_constraint}) this operation changes the parameter $s$ to $s'=-s+1$
\begin{align}
\label{eq:Delta_F_minus}
-\Delta F(\varphi) \in \{s'-1,s'\}\quad,\quad s'=-s+1\quad.
\end{align}

We note at this point that, following Ref.~[\onlinecite{paper_prl}], we will derive in 
Section~\ref{sec:universal} the topological constraint (\ref{eq:Delta_F_constraint})
by a completely different route in terms of physically intuitive arguments using charge conservation
and particle-hole duality only without involving any edge state physics. There, we will see that the 
difference $I=\Delta F - s$ is a topological invariant which can be related to the physical observable of 
the boundary charge $Q_B$ via
\begin{align}
\label{eq:IFs_QB_relation}
I(\varphi) = \Delta F(\varphi) - s = \Delta Q_B (\varphi) - \bar{\rho} \in \{-1,0\}\,,
\end{align}
where $\Delta Q_B(\varphi)=Q_B(\varphi+{2\pi\over Z})-Q_B(\varphi)$ is the difference of the
boundary charge between the shifted and unshifted lattice and $\bar{\rho}={\nu\over Z}$ is the
average charge per site for the infinite system without a boundary. 
\begin{figure}
  \centering
  \includegraphics[width=\columnwidth]{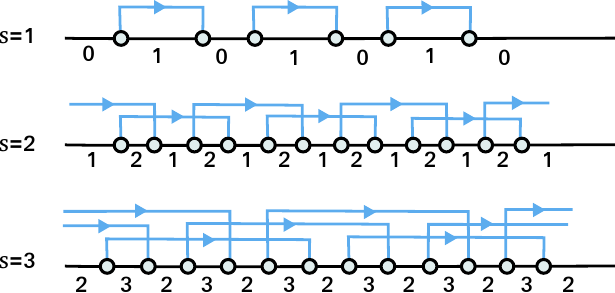}  
  \caption{The allowed configurations for  $s=1$, $s=2$, and $s=3$ 
    when all contractions have the same direction. The topological charge 
    $\Delta F\in\{s-1,s\}$ is indicated between all adjacent vertices. Changing the direction
    of all contractions leads to the configurations with $s'=-s+1=0,-1,-2$, 
    see Eq.~(\ref{eq:Delta_F_minus}). 
  }
  \label{fig:equal_contractions}
\end{figure}
\begin{figure}
  \centering
  \includegraphics[width=\columnwidth]{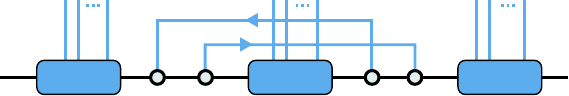}  
  \caption{The construction how a pair of two contractions with different directions is inserted
    into an old configuration, formed by the three boxes. The
    condition is that between the two right and the two left vertices of the new contractions no
    other vertex of the old configuration is allowed to appear. Depending on the old
    configuration it might be necessary to invert the two directions of the new contractions in 
    order to get an allowed configuration after inserting the two new contractions.
  }
  \label{fig:new_pair}
\end{figure}
\begin{figure*}
  \centering
  \includegraphics[width=1.3\columnwidth]{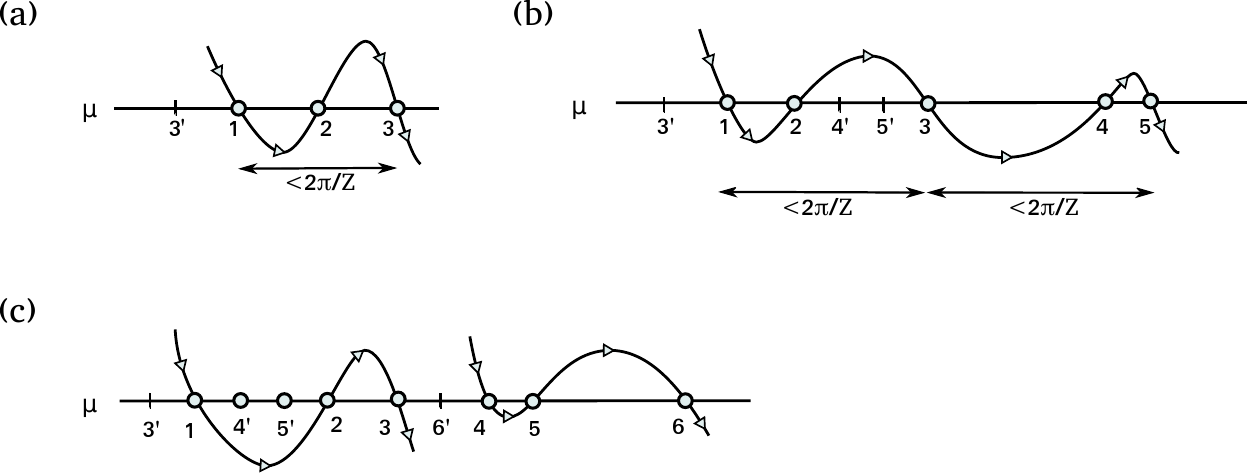}  
  \caption{Illustration of configurations which arise when one adds subsequently pairs of edge states 
    crossing $\mu$ from above and below. For simplicity we have taken a constant $\mu$ independent
    of $\varphi$. Here, the points $i$ indicate the phase points $\phi_i$ 
    and $i'$ the shifted ones $\phi_i^\prime=\phi_i-{2\pi\over Z}$. (a) Adding the pair $(2,3)$ 
    to an edge state crossing $\mu$ at $1$ from above leads to an additional oscillation around
    $\mu$ with length smaller than ${2\pi\over Z}$ such that the edge state remains to run from the
    upper to the lower band. (b) Adding another pair $(4,5)$ leads to a second oscillation of
    length smaller than ${2\pi\over Z}$. Note that $(4',5')$ lies here between $2$ and $3$ such
    that the topological constraint is fulfilled. (c) Two consecutive edge states with one
    additional oscillation each. Here the shifted points $4'$ and $5'$ of the right edge state
    are located between the crossing points $1$ and $2$ of the left edge state. Note that the
    topological constraint does not allow $4'$ and $5'$ to lie between $2$ and $3$. This gives the
    rule that if the crossing points $i$ and $j$ are connected by an edge state with a maximum
    (minimum) in between then the shifted points $i'$ and $j'$ must be located in an interval
    where the edge state has also a maximum (minimum). 
    The connection of the edge states between adjacent crossing points is just a guide for the eye, these
    connections can also be cut (e.g., when $\mu$ is identical to the band edge this is even 
    necessary). 
  }
  \label{fig:constraint_1}
\end{figure*}
\begin{figure*}
  \centering
  \includegraphics[width=1.3\columnwidth]{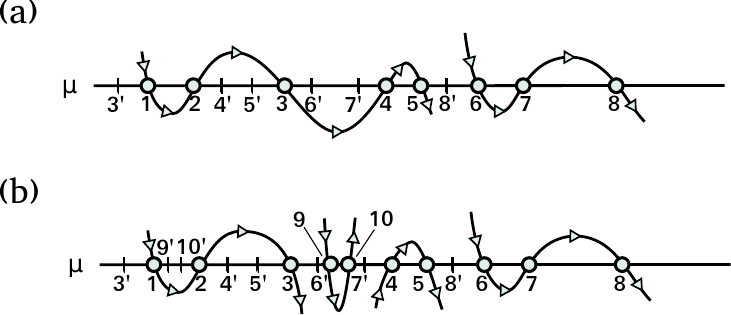}  
  \caption{(a) Two consecutive edge states with two (one) oscillation of the left (right) edge state
    around $\mu$, analog to the cases discussed in Fig.~\ref{fig:constraint_1}. (b) A more complicated
    configuration where we insert the pair $(9,10)$ between $6'$ and $7'$ in (a) and cut the edge 
    mode connecting $3$ and $4$. As a result we get four consecutive edge states, where the
    two outer ones connect the upper with the lower band whereas the two middle ones return to the 
    same band, the left (right) one to the upper (lower) band. 
    Note that the crossing points $1$ and $2$ are essential, otherwise there would be a
    mismatch between the consecutive points $3'$ and $9'$ which correspond both to outgoing vertices.
    The crossing points $4$ and $5$ are necessary to get a new configuration compared to 
    Fig.~\ref{fig:constraint_1}. Without $4$ and $5$ one can connect the edge state 
    between $10$ and $6$ and gets two consecutive edge states oscillating around $\mu$. Furthermore the 
    crossing points $6$ and $7$ are needed since without $6'$ and $7'$ there is a mismatch between 
    $3$ and $9$ and between $10$ and $4$ corresponding both to incoming or outgoing vertices, respectively. 
    Therefore, {\it all} crossing points are essential to construct such an involved configuration.    
  }
  \label{fig:constraint_2}
\end{figure*}
To get a feeling which configurations are possible for the edge states according to the
constraint (\ref{eq:constraint}), we proceed by providing an iterative scheme how more complicated
configurations can be obtained from simpler ones. As a starting point we consider the possible 
configurations when all contractions have the same direction. 
These configuration are shown in Fig.~\ref{fig:equal_contractions}(a,b,c) for different values of
$s=1,2,3$. The construction for larger values is obvious and the ones for $s=0,-1,-2,\dots$ are
obtained by reversing the directions of all contractions according to Eq.~(\ref{eq:Delta_F_minus}).
In terms of the edge states these configurations correspond to the cases where all edge states either 
enter or leave the band. For $s=1$ ($s=0$) the entering (leaving) points of adjacent edge states 
have a distance larger than ${2\pi\over Z}$ and increasing (decreasing) $s$ by one means that 
$Z$ additional edge states enter (leave) the band. For these configurations we
proof in Appendix \ref{app:diophantine} the Diophantine equation 
[\onlinecite{dana_jpc_85}-\onlinecite{hatsugai_prb_93}]
\begin{align}
\label{eq:diophantine_1}
M = M_-- M_+ &= \nu - sZ\\ 
\label{eq:diophantine_2}
\Leftrightarrow -M = M_+ - M_- &= \nu' - s'Z \,,
\end{align}
with $\nu'=Z-\nu$ and $1\le\nu,\nu'\le Z-1$.
That $\nu$ is precisely the index corresponding to the considered gap will be shown in 
Section \ref{sec:universal}. As shown below in this section the Diophantine equation holds not only 
for the case when all contractions have the same direction but for {\it all} allowed configurations. 

The way to obtain all configurations by mixing contractions with different directions is then
quite obvious. As shown in Appendix \ref{app:construction_edge} they are obtained iteratively by inserting
into a given configuration one pair of contractions with different directions such that no other 
vertex appears between the two right and the two left vertices of the two new contractions, see
Fig.~\ref{fig:new_pair} for illustration.
Obviously by choosing the direction of the two new contractions appropriately one obtains again
an allowed configuration and proceeding in this way every configuration can be obtained, see
the proof in Appendix \ref{app:construction_edge}. We note from this construction that neither
the number $M=M_--M_+$ nor the integer $s$ are changed by adding a new pair of contractions. 
Therefore, the Diophantine equations (\ref{eq:diophantine_1}) and (\ref{eq:diophantine_2})
remain valid.

To visualize what kinds of edge state configuration are generated by this construction we have
shown in Fig.~\ref{fig:constraint_1}(a,b) what happens if one adds pairs of edge states
crossing the chemical potential from different sides one after the other to a single crossing point.
This leads to more and more oscillations of the edge state around the chemical potential where each
additional oscillation must have a length smaller than ${2\pi\over Z}$. We note that there is no need 
to connect the edge state between two adjacent crossing points, all these lines can be optionally cut. 
The topological constraint only fixes the allowed configuration of the crossing points. 
Therefore, it is possible that not all edge states connect the two bands in the same direction, both
the numbers $M_+$ and $M_-$ can be unequal to zero. In particular,
when $\mu$ is chosen as the band edge (either the bottom of the upper band or the top of the lower
band) it is even necessary to leave out the connection lines above or below $\mu$. In
Fig.~\ref{fig:constraint_1}(c) and Fig.~\ref{fig:constraint_2}(a) we have shown how several edge states 
oscillating around the chemical potential can arise. If we connect all possible adjacent crossing 
points we find that all edge states run from one band to the other (either all downwards or all upwards). 
Furthermore we find the rule that if two crossing points $i$ and $j$ are connected by an edge state with a 
maximum (minimum) of the edge state between these two points, then the shifted points $i'$ and $j'$ must be located 
in an interval where the edge state has also a maximum (minimum), at least in the case when an edge
mode is present there. How more complicated configurations can appear where not all edge states connect the
bands in the same direction is shown in Fig.~\ref{fig:constraint_2}(b). Here, two consecutive edge states
return to the same band, one to the upper and the other to the lower band.
However, these are rather exotic configurations which in practice occur 
very unlikely, except for very special functions $F_v$ and $F_\gamma$ with many random Fourier components 
defining the model.

\section{Boundary charge and density}
\label{sec:boundary_charge}

In this section we will define the boundary charge via a macroscopic average over the microscopic
density, similar to Ref.~[\onlinecite{park_etal_prb_16}] and Chapter 4.5.1 in 
Ref.~[\onlinecite{vanderbilt_book_2018}]. As proposed in Ref.~[\onlinecite{paper_prl}] this definition allows for
a gauge invariant decomposition of the boundary charge in a Friedel, polarization, and edge part. 
We present all the necessary formulas for the Friedel and 
polarization part in terms of the Bloch states useful for many further investigations. We also 
review the particle-hole duality used in Ref.~[\onlinecite{paper_prl}]. 
In addition to Ref.~[\onlinecite{paper_prl}], we will also present the calculation of the boundary
charge for a half-infinite system with a right boundary and establish a unique relation between the
boundary charges of $H_R$ and $H_L$. In Section~\ref{sec:localization} we will present an analytical 
calculation of the density based on the analytic continuation of 
Bloch states, similar to previous treatments for continuum systems \cite{kallin_halperin_prb_84}.
We prove the exponential localization of the boundary charge, calculate the 
localization length, and show that a nontrivial pre-exponential function appears which falls off with
a generic exponent $1/\sqrt{n}$ for large distances. 

\subsection{Definition and splitting of boundary charge}
\label{sec:definition}

\begin{figure}
\centering
\includegraphics[width= 0.7\columnwidth]{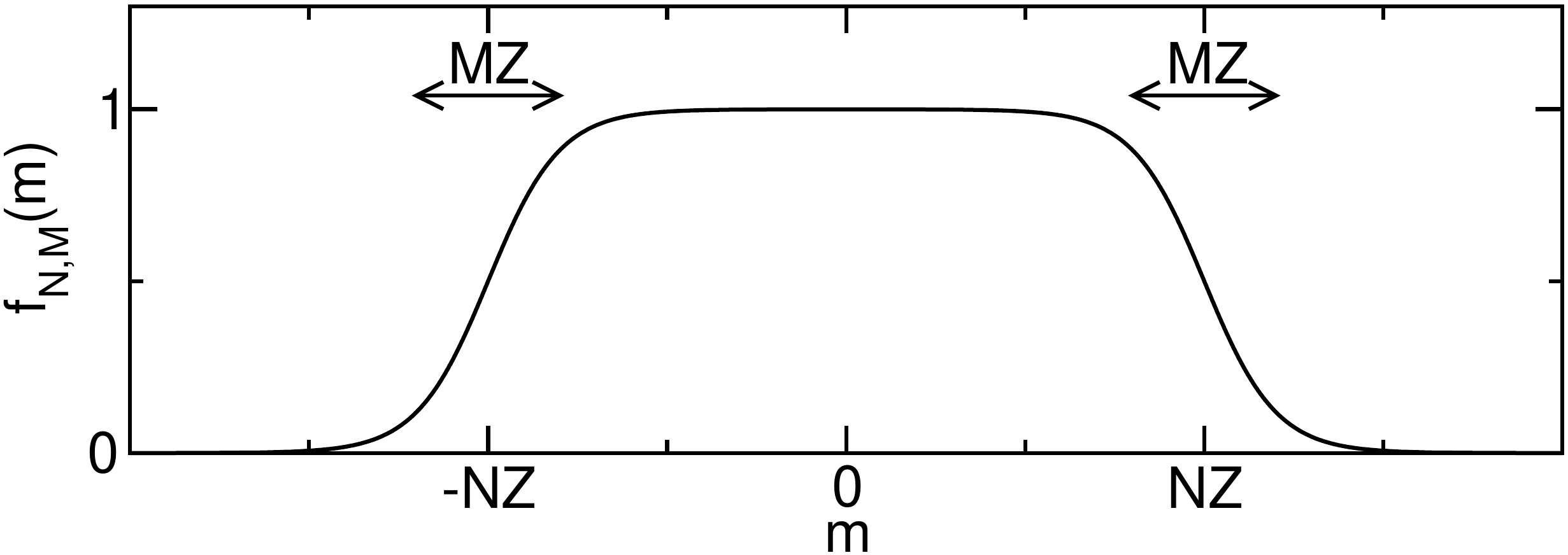} 
 \caption{Sketch of the envelope function $f_{N,M}(m)$ with $N\gg M\gg 1$.}
\label{fig:envelope}
\end{figure}

Following Ref.~[\onlinecite{park_etal_prb_16}], we define the boundary charges $Q_B^{R/L}$ 
corresponding to the half-infinite systems $H_{R/L}$ via the macroscopic average
\begin{align}
\label{eq:QB_def}
Q_B^{R/L} = \lim_{M\rightarrow\infty}\lim_{N\rightarrow\infty}
\sum_{m=\pm 1}^{\pm\infty} \left(\rho(m)-\bar{\rho}\right) f_{N,M}(m)\,,
\end{align}
where $\rho(m)={\big\langle |m\rangle\langle m| \big\rangle}_{H_{R/L}}$ is the
density at site $m$ for the ground state of $H_{R/L}$ and $f_{N,M}(m)$ is an 
envelope function of a charge 
measuring device which, at site $m\sim NZ$, varies smoothly from unity to zero 
on a scale $MZ$ much larger than the scale $Z$ of a unit cell, i.e., $N\gg M\gg 1$, 
see Fig.~\ref{fig:envelope} for a sketch of the envelope function. 
$\bar{\rho}$ is defined as the average density per site of the infinite system
\begin{align}
\label{eq:bar_rho_def}
\bar{\rho} = {1\over Z} \sum_{j=1}^Z \rho_{\text{bulk}}(j) \,,
\end{align}
where $\rho_{\text{bulk}}(j)= {\big\langle |nj\rangle\langle nj| \big\rangle}_{H_{\text{bulk}}}$
is the bulk density in the ground state of the infinite system described by
$H_{\text{bulk}}$. This bulk density is independent of the unit cell index $n$ but can 
depend on the site index $j$ within a unit cell. In the following we consider the 
insulating regime and assume that the chemical potential $\mu=\mu_\nu$ is 
located in gap $\nu$, i.e., the bands $\alpha=1,2,\dots,\nu$ are filled. 
In this case we get from (\ref{eq:bloch_form})
\begin{align}
\label{eq:rho_bulk_total}
\rho_{\text{bulk}}(j)&= \sum_{\alpha=1}^\nu\rho^{(\alpha)}_{\text{bulk}}(j)\,,\\
\nonumber
\rho^{(\alpha)}_{\text{bulk}}(j)&= \int_{-\pi}^\pi dk \,|\psi^{(\alpha)}_{k,\text{bulk}}(n,j)|^2 \\
\label{eq:rho_bulk_alpha}
&={1\over 2\pi}\int_{-\pi}^\pi dk\,|\chi^{(\alpha)}_k(j)|^2\,,
\end{align}
and, due to the normalization (\ref{eq:bulk_orthogonality}) of the Bloch states, 
\begin{align}
\label{eq:bar_rho_result}
\bar{\rho} = {\nu \over Z} \,.
\end{align}
To get a well-defined expression independent of the envelope function we use the
central property that the density of the half-infinite system approaches the bulk 
density very far away from the boundary
\begin{align}
\label{eq:far_from_boundary}
\lim_{n\rightarrow\infty} \rho(n,j) = \rho_{\text{bulk}}(j)\,.
\end{align}
As we will show in Section~\ref{sec:localization} this happens on an exponential
scale $\xi_B\sim 1/\kappa_{\text{e}}^{(\nu)},1/\kappa_{\text{bp}}^{(\nu)}$, see Eq.~(\ref{eq:xi_B}).  
The limit $N,M\rightarrow\infty$ in Eq.~(\ref{eq:QB_def}) has to be understood in
the sense $NZ\gg\xi$ and $MZ\gg Z$, i.e., the measuring device should be smooth on the
scale of the unit cell and should probe the charge on a scale much larger than the
localization length $\xi_B$ of the boundary charge. This guarantees that the deviation
of the experimental measurement from the mathematical limit in Eq.~(\ref{eq:QB_def}) is
exponentially small. Splitting the summand of (\ref{eq:QB_def})
in two terms via $\rho(m)-\bar{\rho}=\left(\rho(m)-\rho_{\text{bulk}}(j)\right)
+ \left(\rho_{\text{bulk}}(j)-\bar{\rho}\right)$, 
we can then set $f_{N,M}(m)\approx 1$ for the first term.  
For the second term we expand 
$f_{N,M}(m)=f_{N,M}(Z(n-1)+j)\approx f_{N,M}(Zn) + f^\prime_{N,M}(Zn)(-Z+j)$ up to linear
order which becomes exact in the limit $N\gg M\rightarrow\infty$. Using in addition
$\rho_{R/L}(m=0)=0$ and $\sum_{j=1}^Z(\rho_{\text{bulk}}(j)-\bar{\rho})=0$ together with
\begin{align}
\label{eq:integration_1}
\lim_{M\rightarrow\infty}\lim_{N\rightarrow\infty} \sum_{n=1}^\infty f^\prime(Zn) = -{1\over Z}\,,\\
\label{eq:integration_2}
\lim_{M\rightarrow\infty}\lim_{N\rightarrow\infty} \sum_{n=-\infty}^0 f^\prime(Zn) = {1\over Z}\,,
\end{align}
we get the decomposition
\begin{align}
\label{eq:QRL_zw1}
Q_B^{R/L} = \sum_{m=\pm 1}^{\pm\infty} \left(\rho(m)-\rho_{\text{bulk}}(j)\right) 
+ Q_P^{R/L} \,,
\end{align}
where
\begin{align}
\label{eq:QP_R}
Q_P^R &= -{1\over Z} \sum_{j=1}^Z j \left(\rho_{\text{bulk}}(j)-\bar{\rho}\right) \,,\\
\nonumber
Q_P^L &= {1\over Z} \sum_{j=1}^Z j \left(\rho_{\text{bulk}}(j)-\bar{\rho}\right) 
- (\rho_{\text{bulk}}(Z) - \bar{\rho})\\
\label{eq:QP_L}
&={1\over Z} \sum_{j=1}^{Z-1} j \left(\rho_{\text{bulk}}(j)-\bar{\rho}\right) 
\end{align}
is the contribution from the bulk polarization to the boundary charges, analog to the 
surface charge of a dielectric medium in classical electrodynamics. We note that this 
decomposition is analog to decompositions to define the boundary charge within the
MTP, see e.g. Eq.~(4.85) in Chapter 4.5 of Ref.~[\onlinecite{vanderbilt_book_2018}]. The
essential difference is that we have chosen a particular representation of the polarization
charge such that the remaining part can be expressed uniquely via the Friedel and edge state
charge, see below. 

Using (\ref{eq:bar_rho_def})
and (\ref{eq:rho_bulk_total}) we can split the polarization charge into the contributions of the 
individual bands
\begin{align}
\label{eq:QPRL_splitting_bands}
Q^{R/L}_P = \sum_{\alpha=1}^\nu Q_P^{R/L,(\alpha)} \,,
\end{align}
with 
\begin{align}
\label{eq:QPR_alpha}
Q_P^{R,(\alpha)} &\equiv Q_P^{(\alpha)} \,,\\
\label{eq:QPL_alpha}
Q_P^{L,(\alpha)} &= - Q_P^{(\alpha)} - \rho_{\text{bulk}}^{(\alpha)}(Z) + {1\over Z} \,,\\
\label{eq:QP_alpha}
Q_P^{(\alpha)}  &= -{1\over Z} \sum_{j=1}^Z j 
\left(\rho^{(\alpha)}_{\text{bulk}}(j)-{1\over Z}\right) \,.
\end{align}
This gives for the sum
\begin{align}
\label{eq:QP_RL_relation}
Q_P^{L,(\alpha)} + Q_P^{R,(\alpha)} = -\rho^{(\alpha)}_{\text{bulk}}(Z) + {1\over Z}  \,.
\end{align}

Finally, we split the total density 
\begin{align}
\label{eq:band_edge}
\rho(m)=\rho_{\text{band}}(m)+\rho_{\text{edge}}(m)
\end{align} 
into the contribution of the filled bands and the edge states and define the Friedel density via
\begin{align}
\label{eq:friedel_density}
\rho_F(m) = \rho_{\text{band}}(m)-\rho_{\text{bulk}}(j) 
= \sum_{\alpha=1}^\nu \rho^{(\alpha)}_F(m)\,,
\end{align}
where
\begin{align}
\label{eq:friedel_band}
\rho^{(\alpha)}_F(n,j) &= \rho_{\text{band}}^{(\alpha)}(n,j)-\rho^{(\alpha)}_{\text{bulk}}(j) 
\end{align}
is the contribution from a single band. Thus, together with $\rho_{\text{band}}(m=0)=0$,
we arrive at the final splitting
\begin{align}
\label{eq:QRL_splitting}
Q_B^{R/L} &= Q_F^{R/L} + Q_E^{R/L} + Q_P^{R/L} \,,
\end{align}
where
\begin{align}
\label{eq:QFE_RL}
Q_F^{R/L} = \sum_{m=\pm 1}^{\pm\infty} \rho_F(m)\quad,\quad
Q_E^{R/L} = \sum_{m=\pm 1}^{\pm\infty} \rho_{\text{edge}}(m)
\end{align}
define the contributions to the boundary charge from the Friedel and edge charge densities
of $H_{R/L}$. Using (\ref{eq:friedel_density}) we can split $Q_F^{R/L}$ into the contributions
of the individual bands
\begin{align}
\label{eq:QFRL_splitting_bands}
Q^{R/L}_F = \sum_{\alpha=1}^\nu Q_F^{R/L,(\alpha)} \quad,\quad
Q_F^{R/L,(\alpha)} = \sum_{m=\pm 1}^{\pm\infty} \rho^{(\alpha)}_F(m)\,.
\end{align}

Using (\ref{eq:rho_bulk_alpha}) together with the form (\ref{eq:psi_H_LR}) 
of the eigenfunctions $\psi_k^{(\alpha)}(n,j)$ of band $\alpha$ of $H_{\text{bulk}}$, we can
write the Friedel density of band $\alpha$ in the form 
\begin{align}
\nonumber
\rho_F^{(\alpha)}(n,j) &= \int_0^\pi dk\,|\psi_k^{(\alpha)}(n,j)|^2 - 
\int_{-\pi}^\pi dk \,|\psi^{(\alpha)}_{k,\text{bulk}}(n,j)|^2\\
\label{eq:rho_friedel_alpha}
&= -{1\over 2\pi} \int_{-\pi}^\pi dk\,{\left(\chi_k^{(\alpha)}(j)\right)}^2 e^{2ikn}\,.
\end{align}
This gives for $Q_F^{R,(\alpha)}$ the form
\begin{align}
\label{eq:QF_R_alpha}
Q_F^{R,(\alpha)} &\equiv Q_F^{(\alpha)}\,,\\
\label{eq:QF_alpha}
Q_F^{(\alpha)} &= -{1\over 2\pi} 
\sum_{n=1}^\infty \int_{-\pi}^\pi dk\,(\chi_k^{(\alpha)})^T \chi_k^{(\alpha)} e^{2ikn} \,.
\end{align}
Using (\ref{eq:chi_form}) and (\ref{eq:N_k}) we can write
\begin{align}
\nonumber 
(\chi_k^{(\alpha)})^T \chi_k^{(\alpha)} &= 
{(a_k^{(\alpha)})^T a_k^{(\alpha)} + s(\epsilon_k^{(\alpha)})^2 \over
(a_k^{(\alpha)})^T a_{-k}^{(\alpha)} + s(\epsilon_k^{(\alpha)})^2}\\
\label{eq:chiT_chi}
&= 1 + {1\over N_k^{(\alpha)}} (a_k^{(\alpha)})^T (a_k^{(\alpha)} - a_{-k}^{(\alpha)}) \,.
\end{align}
The unity on the r.h.s. does not contribute to (\ref{eq:QF_alpha}). The second term
can be evaluated with the form (\ref{eq:a_k_explicit}) for $a_k$ leading to
\begin{align}
\nonumber
(a_k^{(\alpha)})^T (a_k^{(\alpha)} - a_{-k}^{(\alpha)}) &= \\
\label{eq:akT_ak_m_a_mk}
&\hspace{-2cm}
= \left((f^Tf)(\epsilon_k^{(\alpha)}) + (f^Tg)(\epsilon_k^{(\alpha)})e^{ik}\right)(e^{-2ik}-1)\,.
\end{align}
Inserting (\ref{eq:chiT_chi}) and (\ref{eq:akT_ak_m_a_mk}) in (\ref{eq:QF_alpha}) and
using 
\begin{align}
\label{eq:sum_identity}
\sum_{n=1}^\infty e^{2ikn}(e^{-2ik}-1) = \sum_{n=0}^\infty e^{2ikn} - \sum_{n=1}^\infty e^{2ikn} = 1 \,,
\end{align}
we get the compact form
\begin{align}
\label{eq:QF_alpha_2}
Q_F^{(\alpha)} = - {1 \over 2\pi}\int_{-\pi}^\pi dk \, 
{(f^Tf)(\epsilon_k^{(\alpha)}) + (f^Tg)(\epsilon_k^{(\alpha)}) \cos(k) \over N^{(\alpha)}_k} \,,
\end{align}
which allows for a straightforward numerical evaluation. 

A similar analysis yields for the Friedel charge $Q_F^{L,(\alpha)}$
\begin{align}
\nonumber
Q_F^{L,(\alpha)}- \rho^{(\alpha)}_{\text{bulk}}(Z)  &= \sum_{m=-\infty}^0 \rho^{(\alpha)}_F(m)\\
\nonumber
&\hspace{-2cm}
=-{1\over 2\pi} \sum_{n=-\infty}^0 \int_{-\pi}^\pi dk\,(\chi_k^{(\alpha)})^T \chi_k^{(\alpha)} e^{2ikn} \\
\nonumber
&\hspace{-2cm}
= -{1\over 2\pi} \sum_{n=-\infty}^0 \int_{-\pi}^\pi dk\,e^{2ikn}\,\cdot\\
\label{eq:QF_L_zw}
& \hspace{-1.5cm}
\cdot \,\left\{1 + 
{(f^Tf)(\epsilon_k^{(\alpha)}) + (f^Tg)(\epsilon_k^{(\alpha)}) e^{ik} \over N^{(\alpha)}_k}
(e^{-2ik}-1)\right\} \,.
\end{align}
The unity in the brackets contributes only for $n=0$ and gives $-1$. For the second term we use
\begin{align}
\label{eq:sum_identity_2}
\sum_{n=-\infty}^0 e^{2ikn}(e^{-2ik}-1) 
= \sum_{n=-\infty}^{-1} e^{2ikn} - \sum_{n=-\infty}^0 e^{2ikn} = -1 
\end{align}
and, by using (\ref{eq:QF_alpha_2}), we get $-Q_F^{(\alpha)}$ for this term since the $\sin(k)$ part 
of $e^{ik}$ does not contribute due to $\epsilon_k^{(\alpha)}=\epsilon_{-k}^{(\alpha)}$ and 
$N_k^{(\alpha)}=N_{-k}^{(\alpha)}$. Thus, we get from (\ref{eq:QF_L_zw})
the relation
\begin{align}
\label{eq:QF_RL_relation}
Q_F^{L,(\alpha)} + Q_F^{R,(\alpha)} = -1 + \rho^{(\alpha)}_{\text{bulk}}(Z)  \,.
\end{align}

Defining the boundary charges of band $\alpha$ by
\begin{align}
\label{eq:QBRL_alpha}
Q_B^{R/L,(\alpha)} = Q_F^{R/L,(\alpha)} + Q_P^{R/L,(\alpha)} \,,
\end{align}
we get from (\ref{eq:QP_RL_relation}) and (\ref{eq:QF_RL_relation}) the universal result
\begin{align}
\label{eq:QB_RL_alpha_relation}
Q_B^{L,(\alpha)} + Q_B^{R,(\alpha)} = -{Z-1\over Z} \,.
\end{align}
This is an interesting relation between the boundary charges at the left and right boundary stated
in Eq.~(\ref{eq:left_right_relation}) in the introduction. In Section~\ref{sec:zak} we will see that
it leads to a corresponding relation between the Zak-Berry phases for $H_L$ and $H_R$, see

As shown in Section~\ref{sec:edge} each gap $\nu=1,\dots,Z-1$ contains either an edge state of
$H_L$ or one of $H_R$. Since the chemical potential $\mu_\nu$ is located somewhere in gap $\nu$
we get for all $1\le\nu\le Z-1$
\begin{align}
\label{eq:QE_RL_relation}
Q_E^R + Q_E^L = \nu-1 + \theta(\mu_\nu-\epsilon_{\text{e}}^{(\nu)}) \,.
\end{align}
Using (\ref{eq:QPRL_splitting_bands}), (\ref{eq:QRL_splitting}), (\ref{eq:QFRL_splitting_bands}), 
(\ref{eq:QBRL_alpha}), (\ref{eq:QB_RL_alpha_relation}) and (\ref{eq:QE_RL_relation}), 
we obtain finally the following useful relation between the boundary charges of $H_{R/L}$
\begin{align}
\label{eq:QB_RL_relation}
Q_B^L + Q_B^R = \bar{\rho} -1 + \theta(\mu_\nu-\epsilon_{\text{e}}^{(\nu)})\,.
\end{align}
For the case when the chemical potential is lying above the highest band
all states are filled and we get 
$\rho(m)=\bar{\rho}=1$ such that, using the definition (\ref{eq:QB_def}), we get zero
boundary charge
\begin{align}
\label{eq:QB_LR_full}
Q_B^L = Q_B^R = 0 \quad\text{for}\quad \nu=Z\,.
\end{align}

Since the two boundary charges are not independent, we will discuss in the following
only the boundary charge $Q_B^R$ of $H_R$ and use the simplified notation
$Q_B\equiv Q_B^R$, $Q_F\equiv Q_F^R$, $Q_P\equiv Q_P^R$ and $Q_E\equiv Q_E^R$ such that
\begin{align}
\label{eq:QB_splitting}
Q_B = Q_F + Q_P + Q_E = \sum_{\alpha=1}^\nu Q_B^{(\alpha)} + Q_E\,,
\end{align}
where 
\begin{align}
\label{eq:QB_band}
Q_B^{(\alpha)} = Q_F^{(\alpha)} + Q_P^{(\alpha)}
\end{align}
is the boundary charge of a single band and 
\begin{align}
\label{eq:QFP_band}
Q_F = \sum_{\alpha=1}^\nu Q_F^{(\alpha)} \quad,\quad
Q_P = \sum_{\alpha=1}^\nu Q_P^{(\alpha)} \,.
\end{align}
Here, $Q_F^{(\alpha)}$ and $Q_P^{(\alpha)}$ are defined in Eqs.~(\ref{eq:QF_alpha}) and (\ref{eq:QP_alpha}).
If all states are filled we get from (\ref{eq:QB_LR_full}) that $Q_B=0$ or
\begin{align}
\label{eq:QB_full}
\sum_{\alpha=1}^Z Q_B^{(\alpha)} = -Q_E^{\text{tot}} \,,
\end{align}
where $Q_E^{\text{tot}}$ denotes the total number of edge states of $H_R$ in all gaps $\nu=1,\dots,Z-1$. 
Furthermore, due to $\sum_{\alpha=1}^Z \rho^{(\alpha)}_{\text{bulk}}(j)=1$, we get from
(\ref{eq:QP_alpha}) that the sum of the polarization charges of all bands is zero 
\begin{align}
\label{eq:QP_full}
\sum_{\alpha=1}^Z Q_P^{(\alpha)} = 0\,.
\end{align}
Using (\ref{eq:QB_band}) and (\ref{eq:QB_full}) this implies that the
sum of the Friedel charges of all bands is quantized in integer units given by
the negative charge of all edge states
\begin{align}
\label{eq:QF_full}
\sum_{\alpha=1}^Z Q_F^{(\alpha)} = -Q_E^{\text{tot}} \,.
\end{align}

\subsection{Particle-hole duality}
\label{sec:ph_duality}

For the discussion of the universal properties of the boundary charge $Q_{B,p}\equiv Q_B$ 
of the particles it will turn out to be very important to look also at the boundary charge  
$Q_{B,h}$ of the holes. To define this quantity we note that the particle and hole charge densities
are given by
\begin{align}
\label{eq:rho_p}
\rho_p(m) &= \rho(m) \,,\\
\label{eq:rho_h}
\rho_h(m) &= -(1-\rho(m)) = \rho(m) - 1 \,.
\end{align}
In the same way the average particle and hole charge densities per site are given for the infinite
system by
\begin{align}
\label{eq:bar_rho_ph}
\bar{\rho}_p = \bar{\rho} \quad,\quad 
\bar{\rho}_h = \bar{\rho} - 1 \,.
\end{align}
The corresponding boundary charges for particles and holes are then defined via (\ref{eq:QB_def}) as
\begin{align}
\label{eq:QB_p_def}
Q_{B,p} &= \lim_{M\rightarrow\infty}\lim_{N\rightarrow\infty}
\sum_{m=\pm 1}^{\pm\infty} \left(\rho_p(m)-\bar{\rho}_p\right) f_{N,M}(m)\,,\\
\label{eq:QB_h_def}
Q_{B,h} &= \lim_{M\rightarrow\infty}\lim_{N\rightarrow\infty}
\sum_{m=\pm 1}^{\pm\infty} \left(\rho_h(m)-\bar{\rho}_h\right) f_{N,M}(m)\,.
\end{align}
Using Eqs.~(\ref{eq:rho_p}-\ref{eq:QB_h_def}) we find the important property that
the boundary charges of the particles and holes are the same
\begin{align}
\label{eq:QB_ph_relation}
Q_B = Q_{B,p} = Q_{B,h}\,.
\end{align}
We emphasize that looking at the same physics from the hole point of view is not just reproducing the
same in a different language, it involves in addition the Pauli principle and is related to the
fact that the boundary charge is zero when all states are filled, see Eq.~(\ref{eq:QB_LR_full}).
When inversing the occupation of all states by populating all states above $\mu_\nu$ 
instead of below $\mu_\nu$ the charge density changes to $\rho(m)\rightarrow 1-\rho(m)$ 
due to the Pauli principle. Thus, the hole charge density $\rho_h(m)=\rho(m)-1$ can also be viewed
as the negative charge density after population inversion which describes a new physical situation
providing new information about the properties of the system. This will be very essential for the
discussion in Section~\ref{sec:invariant_physics}. We note that the two different viewpoints in terms
of particles and holes have nothing to do with particle-hole symmetry but just involves the physics of the
Pauli principle.

\subsection{Density and localization of boundary charge}
\label{sec:localization}

In this section we discuss the localization of the boundary charge $Q_B$ of $H_R$ given 
by $Q_B=Q_F+Q_E+Q_P$ with
\begin{align}
\nonumber 
Q_F + Q_E &= \sum_{m=1}^\infty \left[\rho_F(m) + \rho_{\text{edge}}(m)\right] \\
\label{eq:QFE_density}
&=\sum_{m=1}^\infty [\rho(m)-\rho_{\text{bulk}}(j)] \,,
\end{align}
see Eqs.~(\ref{eq:band_edge}), (\ref{eq:friedel_density}), (\ref{eq:QRL_splitting}) and (\ref{eq:QFE_RL}).
The polarization part $Q_P\equiv Q_P^R$, given by (\ref{eq:QP_R}), is by definition a contribution
to the boundary charge localized on the scale $Z$ of the unit cell since it occurs for all scales
$N,M\gg 1$ of the envelope function $f_{N,M}(m)$ of the charge measurement probe. In contrast, the
Friedel density $\rho_F(m)$ and the edge state density $\rho_{\text{edge}}(m)$ can have 
localization lengths at the boundary much larger than the length scale of one unit cell 
$\xi_F,\xi_{\text{e}}\gg Z$. We expect that the two length scales of $\rho_F$ and $\rho_{\text{edge}}$ are of the same
order $\xi_F\sim\xi_{\text{e}}$ since, due to charge conservation, an edge state leaving/entering 
a certain band by adiabatically changing the phase variable $\varphi$ will decrease/increase the charge 
of this band by one and it is very unlikely that this happens nonlocally. The localization length 
$\xi_{\text{e}}$ can be arbitrarily large since the localization length 
$\xi_{\text{e}}^{(\nu)}\sim 1/\kappa_{\text{e}}^{(\nu)}$ of an edge state of $H_R$ located in gap $\nu$ 
will go to infinity when its energy $\epsilon_{\text{e}}^{(\nu)}$ approaches the band edges. 
Correspondingly, also the localization length of the Friedel density can be very large. 
However, the localization length $\xi_B$ of the boundary charge or of the difference of the 
{\it total} density between the half-infinite and infinite system 
$\rho(m)-\rho_{\text{bulk}}(j)=\rho_F(m) + \rho_{\text{edge}}(m)$ 
will turn out to be rather small since the
edge state density is cancelled by a corresponding contribution of the Friedel density.
A similar result has been obtained for the density of half-infinite continuum systems discussed in
Ref.~[\onlinecite{kallin_halperin_prb_84}] based on the analytic continuation introduced in 
Ref.~[\onlinecite{rehr_kohn_prb_74}]. We summarize the results here and refer to 
Appendix~\ref{app:density} for the technical details.

If the chemical potential $\mu_\nu\equiv\mu$ is placed in gap $\nu$, $\rho_F(m)+\rho_{\text{edge}}(m)$ 
results from summing up all Friedel densities of the filled bands $\alpha=1,\dots,\nu$, together
with all edge states from gaps $\nu'=1,\dots,\nu-1$ and the one from gap $\nu$ if it is
occupied, i.e., if $\epsilon_{\text{e}}^{(\nu)}<\mu_\nu$. This gives
\begin{align}
\label{eq:splitting_bands}
\rho(m)-\rho_{\text{bulk}}(j) = \sum_{\alpha=1}^\nu \rho_F^{(\alpha)}(m) + \rho_{\text{edge}}(m)\,,
\end{align}
with
\begin{align}
\label{eq:splitting_edge}
\rho_{\text{edge}}(m) = \sum_{\alpha=1}^{\nu-1}\rho_{\text{edge}}^{(\alpha)}(m)
+ \rho_{\text{edge}}^{(\nu)}(m) \theta(\mu_\nu-\epsilon_{\text{e}}^{(\nu)})\,,
\end{align}
where $\rho_{\text{edge}}^{(\nu')}(n,j)=[\psi_{k_{\text{e}}^{(\nu')}}^{\text{e}}(n,j)]^2$ denotes the density of
the edge state in gap $\nu'$. To calculate $\rho_F^{(\alpha)}(m)$ we close the integration of 
Eq.~(\ref{eq:QF_alpha}) in the
upper half. According to Fig.~\ref{fig:bc_edge} one can split the integration into branch cut contributions
starting at $k_{\text{bc}}^{\alpha,\alpha-1}$ and integrations around the positions $k_e^{\alpha,\alpha-1}$ of the 
edge state poles (if present in the upper half). As shown in Appendix~\ref{app:density} one obtains 
for the pole contribution 
\begin{align}
\nonumber
\rho_{F,P}^{(\alpha)}(m) &=\\
\nonumber
&\hspace{-1cm} 
- \rho_{\text{edge}}^{(\alpha)}(m)\,\theta(\epsilon_{\text{bc}}^{(\alpha)}-\epsilon_{\text{e}}^{(\alpha)})
\,\delta_{Z>\alpha}\\
\label{eq:friedel_pole_density}
&\hspace{-1cm} 
- \rho_{\text{edge}}^{(\alpha-1)}(m)\,\theta(\epsilon_{\text{e}}^{(\alpha-1)}-\epsilon_{\text{bc}}^{(\alpha-1)})
\,\delta_{\alpha>1}
\,,
\end{align}
As a result the pole contribution of the Friedel density of band $\alpha$ cancels exactly the
edge state density of those edge states which belong to band $\alpha$. For the total charge one
would have expected this due to charge conservation but that it happens locally for each lattice 
site even if the edge state energies are far away from the band edges is quite surprising. It
shows that edge states are not the only special effects happening at the boundary: if they appear, 
they always leave a corresponding fingerprint in the density of the scattering states at the boundary. 

Summing all branch cut contributions over the occupied bands it turns out (see Appendix~\ref{app:density})
that the common ones of adjacent bands $\alpha$ and $\alpha+1$ starting at 
$k_{\text{bp}}^{(\alpha)}$ cancel each other exactly since the values
of the integrand left and right to the branch cut are interchanged, see  
Fig.~\ref{fig:bc}(b,c). What remains is only the branch cut contribution 
from the valence band $\alpha=\nu$ in gap $\nu$, which can be written as
\begin{align}
\nonumber
\rho_{F,\text{bc}}(n,j) &= \sum_{\alpha=1}^\nu \rho^{(\alpha)}_{F,\text{bc}}(n,j) \\
\label{eq:rho_friedel_total_bc_im}
&\hspace{-1cm}
={1\over \pi} e^{-2\kappa_{\text{bp}}^{(\nu)} n}
\text{Im} \int_0^\infty d\kappa\,\left[\chi_{k_{\text{bp}}^{(\nu)}+i\kappa+0^+}^{(\nu)}(j)\right]^2 e^{-2\kappa n} \,.
\end{align}
As a result the localization length of the branch cut contribution of the total Friedel density
is given by 
\begin{align}
\label{eq:xi_F_total}
\xi_{F,\text{bc}}\sim 1/\kappa_{\text{bp}}^{(\nu)}\,.
\end{align} 
The pole contribution $\rho_{F,P}(m)=\sum_{\alpha=1}^\nu\rho_{F,P}^{(\alpha)}(m)$ of the total Friedel density
will cancel the density from all occupied edge states in gaps $\nu'=1,\dots,\nu-1$. However, the sum of
the pole contribution of the Friedel density from gap $\nu$ and the density of the edge state
in gap $\nu$ is only zero if both are present, i.e., for 
$\epsilon_{\text{e}}^{(\nu)} < \mu_\nu,\epsilon_{\text{bp}}^{(\nu)}$, or if both are absent,
i.e., for $\epsilon_{\text{e}}^{(\nu)} > \mu_\nu,\epsilon_{\text{bp}}^{(\nu)}$. Therefore we obtain
the following final result for the sum of the Friedel and edge state density
\begin{align}
\nonumber 
\rho(n,j) - \rho_{\text{bulk}}(j) &= \rho_F(n,j) + \rho_{\text{edge}}(n,j) \\
\nonumber
&\hspace{-2cm} 
= \rho_{F,\text{bc}}(n,j) + \\
\label{eq:total_density}
&\hspace{-2cm} + \,[\psi_{k_{\text{e}}^{(\nu)}}^{\text{e}}(n,j)]^2 \cdot
\begin{cases}
1 \quad\,\,\,\,\text{for}\,\,\epsilon_{\text{bp}}^{(\nu)} < \epsilon_{\text{e}}^{(\nu)} < \mu_\nu \\
-1\quad\text{for}\,\,\mu_\nu<\epsilon_{\text{e}}^{(\nu)} < \epsilon_{\text{bp}}^{(\nu)} \\
0 \hspace{1cm} \text{otherwise}
\end{cases}\,.
\end{align} 
Thereby, the second term on the r.h.s. can only occur if $\epsilon_{\text{e}}^{(\nu)}$ 
corresponds to the energy of an edge state of $H_R$, i.e., if $\text{Im}(k_{\text{e}}^{(\nu)})>0$.
Since the localization length of $\psi_{k_{\text{e}}^{(\nu)}}^{\text{e}}(n,j)$ is given by
$\xi_{\text{e}}^{(\nu)}\sim 1/\kappa_{\text{e}}^{(\nu)}$ we get together with (\ref{eq:xi_F_total})
the following compact form for the localization length $\xi_B$ of the boundary charge
\begin{align}
\label{eq:xi_B}
\xi_B = 
\begin{cases}
1/\kappa_{\text{e}}^{(\nu)} 
\quad\text{for}\,\,\mu_\nu<\epsilon_{\text{e}}^{(\nu)} < \epsilon_{\text{bp}}^{(\nu)}
\,\,\text{or}\,\,\epsilon_{\text{bp}}^{(\nu)} < \epsilon_{\text{e}}^{(\nu)} < \mu_\nu \\
1/\kappa_{\text{bp}}^{(\nu)} 
\quad\text{for}\,\,\epsilon_{\text{e}}^{(\nu)} < \mu_\nu,\epsilon_{\text{bp}}^{(\nu)}
\,\,\text{or}\,\,\epsilon_{\text{e}}^{(\nu)} > \mu_\nu,\epsilon_{\text{bp}}^{(\nu)}
\end{cases}.
\end{align} 
As a surprising result we find the important property that the density and the localization length 
of the boundary charge depend only on the properties of the valence band $\alpha=\nu$ and its 
analytic continuation into gap $\nu$ (which is controlled by the $k$-dependence of the Bloch states 
close to the top of the valence band) and on
the properties of the last edge state in gap $\nu$ between the valence and conduction band.
There is no dependence on the bands $\alpha=1,\dots,\nu-1$ and the properties of the edge states in gaps
$\nu'=1,\dots,\nu-1$. Since $\kappa_{\text{e}}^{(\nu)}<\kappa_{\text{bp}}^{(\nu)}$ we find that $\xi_B$
can only become very large if the chemical potential $\mu_\nu$ approaches the top (bottom) of band 
$\alpha=\nu$ ($\alpha=\nu+1$) and is slightly below (above) the edge state energy.

Our result is not changed when the edge state energy is identical to the energy at the branching
point $\epsilon_{\text{e}}^{(\nu')} =\epsilon_{\text{bp}}^{(\nu')}$ for some $\nu'=1,\dots,\nu$ such that
a branching pole arises for the analytic continuation of the bands $\alpha=\nu'$ and 
$\alpha=\nu'+1$. Due to (\ref{eq:N_k_expansion_k_e_bp}) the contribution from the branching pole of
each of these bands will cancel only half of the density of the edge state, such that the sum of both 
cancels the whole one. This means that for all gaps $\nu'=1,\dots,\nu-1$ the edge state density
is again cancelled by the pole contribution of all Friedel densities. If
$\epsilon_{\text{e}}^{(\nu)} =\epsilon_{\text{bp}}^{(\nu)}$ happens for the gap in which the chemical potential
is located, we have to add in the result (\ref{eq:total_density}) a factor ${1\over 2}$ 
in front of $[\psi_{k_{\text{e}}^{(\nu)}}^{\text{e}}(n,j)]^2$.

Finally, we note that the asymptotic behaviour of $\rho_{F,\text{bc}}(n,j)$ for $n\gg 1$ can
be evaluated from (\ref{eq:rho_friedel_total_bc_im}) by expanding $N_k^{(\nu)}$ around
$k=k_{\text{bp}}^{(\nu)}$ with the help of (\ref{eq:expansion_bp}) and (\ref{eq:sND_relation}).
After a straighforward calculation one obtains
\begin{align}
\nonumber
\rho_{F,\text{bc}}(n,j) &\,\stackrel{n\gg 1}{\sim}\, {1\over\sqrt{n}}\,e^{-2\kappa_{\text{bp}}^{(\nu)} n}\\
\label{eq:rho_F_bc_asymptotic}
&\hspace{-1cm}\times
\begin{cases}
a^{(\nu)}_{k_{\text{bp}}^{(\nu)}}(j)^2\quad\text{for}\,\,j=1,\dots,Z-1 \\
s(\epsilon_{k_{\text{bp}}^{(\nu)}})^2\quad\text{for}\,\,j=Z 
\end{cases}\,.
\end{align}
This gives rise to a universal asymptotic behaviour $\sim {1\over\sqrt{n}}$ of the pre-exponential
function of the density. We note that similar observations have been reported for matrix
elements of the one-particle density matrix 
$\rho(x,x')={a\over 2\pi}\int_{-\pi/a}^{\pi/a}dk\,\psi_{-k,\text{bulk}}(x)\psi_{k,\text{bulk}}(x')$ of
infinite continuum systems in the asymptotic regime $|x-x'|\rightarrow\infty$ \cite{he_vanderbilt_prl_01} but
it is so far not clear how the results are related to each other.

\section{The surface charge theorem} 
\label{sec:zak}

In this section we will present the central proof of the unique formulation 
of the surface charge theorem (\ref{eq:surface_charge_theorem}) for
a single band, relating the boundary charge to the bulk polarisation in terms of the Zak-Berry phase
evaluated in a particular gauge. In particular, we will discuss the difference of the Zak-Berry phases
defined with respect to the two different ways (\ref{eq:bloch_1}) and (\ref{eq:bloch_2}) to represent
the Bloch wave. Based on this theorem we will present an alternative proof for the 
universal relation (\ref{eq:left_right_relation}) between the boundary charges and the Zak-Berry phases
of the half-infinite systems $H_R$ and $H_L$ with a left or a right boundary, respectively. Furthermore, 
we will use this theorem to relate the change of the boundary charge under a shift of the lattice by one site
to the winding number of the phase difference of the Bloch wave function between adjacent sites,
analog to Ref.~[\onlinecite{paper_prl}] but generalizing it to a shift by an arbitrary 
number of sites.

The Zak-Berry phase $\gamma_\alpha$ for a certain band $\alpha$ is defined by 
\begin{align}
\label{eq:zak_phase_1}
\gamma_\alpha &= i \int_{-\pi}^\pi dk \,(\chi^{(\alpha)}_k)^\dagger {d\over dk} \chi^{(\alpha)}_k \\
\label{eq:zak_phase_2}
&= - \text{Im} \int_{-\pi}^\pi dk \, (\chi^{(\alpha)}_{-k})^T {d\over dk} \chi^{(\alpha)}_k \,,
\end{align} 
where we used (\ref{eq:bloch_vector_property}) and (\ref{eq:bloch_vector_periodicity}) together 
with partial integration to derive the second equation. Using the form (\ref{eq:chi_form}) of the
Bloch state we find with the help of $a^{(\alpha)}_{-k}=(a^{(\alpha)}_k)^*$ and 
$N^{(\alpha)}_k=N^{(\alpha)}_{-k}=(N^{(\alpha)}_k)^*$
\begin{align}
\label{eq:zak_phase_3}
\gamma_\alpha = - \int_{-\pi}^\pi dk \, {1\over N^{(\alpha)}_k} 
\text{Im}\left\{ (a^{(\alpha)}_{-k})^T {d\over dk} a^{(\alpha)}_k\right\} \,.
\end{align} 
Inserting the form (\ref{eq:a_k_explicit}) for $a_k$ and using the central property (\ref{eq:fg_relation})
we get
\begin{align}
\nonumber
- \text{Im} \left\{(a^{(\alpha)}_{-k})^T {d\over dk} a^{(\alpha)}_k\right\} &=\\
\nonumber
&\hspace{-3cm}
= -{1\over 2i} \left((a^{(\alpha)}_{-k})^T {d\over dk} a^{(\alpha)}_k + (k\rightarrow -k)\right)\\
\label{eq:Im_aT_ddk_a}
&\hspace{-3cm}
= (f^Tf)(\epsilon_k^{(\alpha)}) + (f^Tg)(\epsilon_k^{(\alpha)}) \cos(k) \,,
\end{align}
Therefore, the Zak-Berry phase can be written in the form
\begin{align}
\label{eq:zak_phase_4}
\gamma_\alpha = \int_{-\pi}^\pi dk \, 
{(f^Tf)(\epsilon_k^{(\alpha)}) + (f^Tg)(\epsilon_k^{(\alpha)}) \cos(k) \over N^{(\alpha)}_k} \,.
\end{align} 
Comparing with (\ref{eq:QF_alpha}) we get the result
\begin{align}
\label{eq:friedel_zak}
Q_F^{(\alpha)} = - {\gamma_\alpha \over 2\pi} \,.
\end{align}
Combining this result with (\ref{eq:QB_full}) we find that the total number of edge states 
of $H_R$ is related to the total Zak-Berry phase
\begin{align}
\label{eq:total_zak_edge}
Q_E^{\text{tot}} = {\gamma_{\text{tot}}\over 2\pi}\quad,\quad 
\gamma_{\text{tot}} = \sum_{\alpha=1}^Z \gamma_\alpha\,.
\end{align}
We note that the Zak-Berry phase of an individual band is neither gauge invariant nor quantized. However,
the total Zak-Berry phase 
\begin{align}
\nonumber
\gamma_{\text{tot}} &=  i \int_{-\pi}^\pi dk\,\text{Tr} U_k^\dagger {d\over dk} U_k \\
\label{eq:total_zak}
&= i \int_{-\pi}^\pi dk\,{d\over dk} \ln(\text{det} U_k) 
\end{align} 
can be written as a winding number of the determinant of the unitary matrix 
$U_k=(\chi_k^{(1)}\cdots\chi_k^{(Z)})$. Therefore, the total Zak-Berry phase is 
quantized in units of $2\pi$.
This is consistent with our result (\ref{eq:total_zak_edge}) since
the total number of edge states must be quantized in integer units. However, we note that
also the total Zak-Berry phase is not gauge-invariant and the precise relation (\ref{eq:total_zak_edge})
holds only in the particular gauge where $\chi_k^{(\alpha)}(Z)$ is real.  

Also the sum of the Friedel and polarization charge of a single band can be written in terms of
a Zak-Berry phase. To achieve this we use a different gauge of the Bloch state $\chi_k^{(\alpha)}$ by
writing the Bloch wave function (\ref{eq:bloch_form}) in the form
(\ref{eq:bloch_2}) which is the standard one within solid state physics
\begin{align}
\label{eq:bloch_form_2}
\psi^{(\alpha)}_{k,\text{bulk}}(m)={1\over\sqrt{2\pi}}\bar{\chi}_k^{(\alpha)}(j) e^{ikm/Z}\,,
\end{align}
with $m=Z(n-1)+j$ labelling the lattice site and 
\begin{align}
\label{eq:bar_chi}
\bar{\chi}_k^{(\alpha)}(j) = \chi_k^{(\alpha)}(j) e^{ik{Z-j\over Z}}\,.
\end{align}
This form can be written in the standard form 
\begin{align}
\label{eq:bloch_standard}
\psi^{(\alpha)}_{\tilde{k},\text{bulk}}(x) = {1\over \sqrt{2\pi}} u_{\tilde{k}}(x)e^{i\tilde{k}x}\,,
\end{align}
with $\tilde{k}\equiv{k\over L}$, $x\equiv ma$, and 
$u_{\tilde{k}}(x)=u_{\tilde{k}}(x+L)\equiv \bar{\chi}_k(j)$, 
where $L=Za$ denotes the length of the unit cell.
We note that $\bar{\chi}_k^{(\alpha)}\ne\bar{\chi}_{k+2\pi}^{(\alpha)}$ is no longer periodic in $k$, 
except for the component $\bar{\chi}_k^{(\alpha)}(Z)=\chi_k^{(\alpha)}(Z)$. It fulfils the condition
\begin{align}
\label{eq:periodicity_bar_chi}
\bar{\chi}_{k+2\pi}(j) = \bar{\chi}_k(j) e^{-i{2\pi\over Z}j}\,,
\end{align}
which is equivalent to $u_{\tilde{k}+{2\pi\over L}}(x) = u_{\tilde{k}}(x) e^{-i{2\pi\over L}x}$ such
that $\psi^{(\alpha)}_{\tilde{k},\text{bulk}}$ is periodic under
$\tilde{k}\rightarrow \tilde{k}+{2\pi\over L}$. This is the standard gauge
chosen for the definition of the Zak-Berry phase within the MTP
\begin{align}
\label{eq:zak_phase_bar}
\bar{\gamma}_\alpha = i \int_{-\pi}^\pi dk \,
(\bar{\chi}^{(\alpha)}_k)^\dagger {d\over dk} \bar{\chi}^{(\alpha)}_k \,.
\end{align} 
Inserting (\ref{eq:bar_chi}) and using 
${1\over 2\pi}\int_{-\pi}^\pi dk |\chi_k^{(\alpha)}(j)|^2=\rho_{\text{bulk}}^{(\alpha)}(j)$, we find after
some straightforward manipulations
\begin{align}
\nonumber
-{\bar{\gamma}_\alpha \over 2\pi} &= -{\gamma_\alpha \over 2\pi}
- {1\over Z}\sum_{j=1}^Z j \,\rho_{\text{bulk}}^{(\alpha)}(j) + \sum_{j=1}^Z\rho_{\text{bulk}}^{(\alpha)}(j)\\
\label{eq:zak_phase_bar_zw}
&= Q_F^{(\alpha)} + Q_P^{(\alpha)} - {1\over Z^2}\sum_{j=1}^Z j + 1\,,
\end{align}
leading to the central result of a unique formulation of the 
{\it \underline{surface charge theorem for a single band}}
\begin{align}
\label{eq:zak_QB}
\fbox{\parbox{4cm}{
\begin{center}$
Q_B^{(\alpha)} = -{\bar{\gamma}_\alpha \over 2\pi} - {Z-1\over 2Z}
$\end{center}}}
\,,
\end{align}
where the second term on the right hand side 
$- {Z-1\over 2Z}={1\over Z}\sum_{j=1}^Z (j-Z) {1\over Z}$
can be written as a certain representation for the polarization of the ions 
(note that $e=1$ denotes the charge of the electron). 
This result applies to $H_R$, i.e., for a left boundary where the Zak-Berry phase 
$\bar{\gamma}_\alpha^R\equiv\bar{\gamma}_\alpha$ is defined with respect to the unit cell
$j=1,\dots,Z$, starting with $j=1$ at the boundary. The result for $H_L$ with a right boundary can be obtained
in the same way by defining the Zak-Berry phase $\bar{\gamma}_\alpha^L$ with respect to the unit
cell $j=Z-1,Z-2,\dots,1,Z$, starting with $j=Z-1$ at the boundary, and reversing the direction of the
Bloch wave. Relabelling the sites of $H_L$ such that they look the same as for $H_R$ we obtain after
a straightforward consideration that the Bloch wave function for $H_L$ can be written as
$\psi^{L,(\alpha)}_{k,\text{bulk}}(n,j)=\chi^{L,(\alpha)}_k(j)e^{ikn}$ with
\begin{align}
\label{eq:chi_L}
\chi^{L,(\alpha)}_k(j) =
\begin{cases} 
\chi^{R,(\alpha)}_{-k}(Z-j)e^{-ik} \,\, & \text{for}\,\,j=1,\dots,Z-1 \\
\chi^{R,(\alpha)}_{-k}(Z) \,\, & \text{for}\,\,j=Z
\end{cases}\,,
\end{align}
where $\chi^{R,(\alpha)}_k\equiv\chi^{(\alpha)}_k$. This leads with (\ref{eq:bar_chi}) to
\begin{align}
\label{eq:bar_chi_L}
\bar{\chi}^{L,(\alpha)}_k(j) = \bar{\chi}^{R,(\alpha)}_{-k}(Z-j)\,,
\end{align}
where $\bar{\chi}_k^{R,(\alpha)}(0)\equiv\bar{\chi}_k^{R,(\alpha)}(Z)$. 
Using the definition (\ref{eq:zak_phase_bar}) of the Zak phases $\bar{\gamma}_\alpha^{R/L}$ with
respect to $\bar{\chi}^{R/L,(\alpha)}_k(j)$ we get the following
universal relationship between the Zak phases for left/right boundaries
\begin{align}
\label{eq:zak_RL}
\fbox{\parbox{4cm}{
\begin{center}$
\bar{\gamma}^R_\alpha + \bar{\gamma}^L_\alpha = 0
$\end{center}}}
\,.
\end{align}
Since the boundary charges $Q_B^{R,L(\alpha)}$ of $H_{R/L}$ are given by 
\begin{align}
\label{eq:zak_QB_RL}
Q_B^{R,L(\alpha)} = -{\bar{\gamma}^{R,L}_\alpha \over 2\pi} - {Z-1\over 2Z}\,,
\end{align}
we find that (\ref{eq:zak_RL}) is equivalent to the universal relationship between 
the boundary charges for left/right boundaries 
\begin{align}
\label{eq:QB_RL_alpha_relation_consistent}
\fbox{\parbox{4cm}{
\begin{center}$
Q_B^{L,(\alpha)} + Q_B^{R,(\alpha)} = -{Z-1\over Z} 
$\end{center}}}
\,,
\end{align}
which is consistent with (\ref{eq:QB_RL_alpha_relation}).

The surface charge theorem (\ref{eq:zak_QB}) is very helpful to understand how the boundary charge 
of a single band changes when we change the position of the boundary by $\Delta m$ lattice sites 
to the right, or, equivalently,
shift the lattice by $\Delta m$ sites to the left. This is achieved by changing the phase
$\varphi\rightarrow \varphi + {2\pi\over Z}\Delta m$. To fulfil the boundary condition for the
half-infinite system we need that $\bar{\chi}_k^{(\alpha)}(Z)=\chi_k^{(\alpha)}(Z)$ is real. Since,
for the shifted system, the site $m=\Delta m$ is the last site $j=Z$ of the unit cell, we have
to take the following Bloch wave function for the shifted system
\begin{align}
\label{eq:Bloch_shifted}
\tilde{\psi}_{k,\text{bulk}}(m) = e^{-i\theta_k^{(\alpha)}} \psi^{(\alpha)}_{k,\text{bulk}}(m+\Delta m) \,,
\end{align}
where $\theta_k^{(\alpha)}$ is the phase of $\psi_{k,\text{bulk}}^{(\alpha)}(\Delta m)$. 
Inserting (\ref{eq:bloch_form_2}) and using (\ref{eq:zak_phase_bar}) we find after a
straightforward algebra that this leads to a change of $\bar{\gamma}_\alpha$ by 
\begin{align}
\label{eq:zak_change}
\bar{\gamma}_\alpha \rightarrow \bar{\gamma}_\alpha 
- 2\pi \left({\Delta m\over Z} - w_{\alpha,\Delta m}\right)\,,
\end{align}
where $w_{\alpha,\Delta m}=w[\theta_k^{(\alpha)}]$ is the winding number of the phase 
factor $e^{i\theta_k^{(\alpha)}}$
\begin{align}
\label{eq:winding_number}
w[\theta^{(\alpha)}_k] = {1\over 2\pi i} \int_{-\pi}^\pi dk \,
e^{-i\theta^{(\alpha)}_k} {d\over dk} e^{i\theta^{(\alpha)}_k}\,.
\end{align}
Together with (\ref{eq:zak_QB}) this gives for the change of the boundary charge
\begin{align}
\label{eq:QB_change_Delta_m}
\fbox{\parbox{8cm}{
\begin{center}$
Q_B^{(\alpha)}(\varphi + {2\pi\over Z}\Delta m) - Q_B^{(\alpha)}(\varphi) =
{\Delta m\over Z} - w_{\alpha,\Delta m}
$\end{center}}}
\,.
\end{align}
Since $\psi_{k,\text{bulk}}^{(\alpha)}(m=0)$ has been chosen real we note that $\theta_k^{(\alpha)}$ is the
gauge invariant phase difference of the Bloch wave function between site $m=\Delta m$ and
$m=0$. In an analogous way we find the same result when the unshifted half-infinite system starts at 
$m=m'$ and the shifted one at $m=m'+\Delta m$. In this case, the phase difference of the Bloch
wave function between site $m=m'+\Delta m$ and $m=m'$ enters into the winding number. We note
that (\ref{eq:QB_change_Delta_m}) respects the periodicity $Q_B(\varphi+2\pi)=Q_B(\varphi)$
for $\Delta m=Z$ since $\theta_k^{(\alpha)}=k$ and $w[k]$=1 for this case.

The central result (\ref{eq:QB_change_Delta_m}) provides a very nice interpretation of
the winding numbers of phase differences of the Bloch wave function between two different
sites in terms of the change of the boundary charge when one shifts the boundary. It will
be discussed in all detail in Section~\ref{sec:universal} for $\Delta m=1$
\begin{align}
\label{eq:winding_number_1}
w_\alpha \equiv w_{\alpha,1}\,,
\end{align}
together with the possible values for $w_\alpha$. We note that 
this winding number can also be expressed by the winding number $\bar{w}_\alpha$ of the 
phase $\varphi^{(\alpha)}_k(1)$ of the first component $\chi^{(\alpha)}_k(1)$. Since
$\theta_k^{(\alpha)} = k + \varphi_k^{(\alpha)}(1)$ we get 
\begin{align}
\label{eq:relation_winding}
w_\alpha = 1 + \bar{w}_\alpha \quad,\quad \bar{w}_\alpha=w[\varphi^{(\alpha)}_k(1)]\,.
\end{align}

\section{Topological indices} 
\label{sec:winding_zak_chern}

Since the winding number $w_\alpha$ introduced in the preceding Section~\ref{sec:zak} plays 
a very central role for universal properties of the boundary charge and is directly related 
to the phase variable of the Bloch wave function, we present in this section
a comparison of this winding number to other topological indices used for the classification
of topological insulators, in particular, to the Zak-Berry phase $\gamma_\alpha$ and to the 
Chern number $C^{(\alpha)}$ of a single band.
This provides a more detailed analysis compared to Ref.~[\onlinecite{paper_prl}] and shows 
that the winding number $w_\alpha$ contains much more information compared to the
Zak-Berry phase and the Chern number. 

First, we note that $w_\alpha$ is a gauge invariant and quantized topological index irrespective 
of any symmetry constraints. This is in contrast to the Zak-Berry phase $\gamma_\alpha$ which is only quantized 
in case of a local inversion symmetry as was first shown in Ref.~[\onlinecite{zak}]. Local inversion
symmetry is defined by a unitary operator $\Pi$ acting on the local system with the property
\begin{align}
\label{eq:local_inversion_def}
\Pi h(\delta) \Pi^\dagger = h(-\delta) \quad &\Leftrightarrow\quad
\Pi h_k \Pi^\dagger = h_{-k}\,.
\end{align}
For our model a concrete realization is given by 
\begin{align}
\label{eq:local_inversion_specific}
\Pi |j\rangle = |Z-j+1\rangle \,,
\end{align}
which requires
\begin{align}
\label{eq:condition_local_inversion}
v_j = v_{Z-j+1}\quad,\quad t_j = t_{Z-j} \,,
\end{align}
with $t_0\equiv t_Z$. To show the quantization we note that (\ref{eq:local_inversion_def}) implies
that $\Pi\chi^{(\alpha)}_{-k}$ is also an eigenvector of $h_k$ with the same eigenvalue. 
As a consequence $\chi^{(\alpha)}_k$ and $\Pi\chi^{(\alpha)}_{-k}$ must agree up to a phase factor 
$e^{i\eta^{(\alpha)}_k}$
\begin{align}
\label{eq:relation_chi_Pi_chi}
\Pi\chi^{(\alpha)}_{-k} = e^{i\eta^{(\alpha)}_k}\chi^{(\alpha)}_k\,.
\end{align}
According to the definition (\ref{eq:winding_number}) we denote the quantized winding number of 
this phase factor by $w[\eta^{(\alpha)}_k]$, whereas $\gamma[\chi^{(\alpha)}_k]$ defines the Zak-Berry phase 
of the Bloch state $\chi^{(\alpha)}_k$. Using $\chi_{-k}^{(\alpha)}=(\chi_k^{(\alpha)})^*$ we get
\begin{align}
\nonumber
\gamma_\alpha&=\gamma[\chi_k^{(\alpha)}]=\gamma[e^{-i\eta^{(\alpha)}_k}\Pi(\chi_k^{(\alpha)})^*]\\
\nonumber
&= 2\pi w[\eta^{(\alpha)}_k] 
+\gamma[\Pi(\chi_k^{(\alpha)})^*]\\
\nonumber
&= 2\pi w[\eta^{(\alpha)}_k]+\gamma[(\chi_k^{(\alpha)})^*]\\
\label{eq:zak_inversion_symmetry}
&= 2\pi w[\eta^{(\alpha)}_k]-\gamma[\chi_k^{(\alpha)}]
= 2\pi w[\eta^{(\alpha)}_k]-\gamma_\alpha \,,
\end{align}
leading to the quantization in units of $\pi$ which was the central result of
Ref.~[\onlinecite{zak}]
\begin{align}
\label{eq:zak_quantization}
\gamma_\alpha = \pi w[\eta^{(\alpha)}_k] \,.
\end{align}
The relationship to the winding number $w_\alpha$ is obtained by combining (\ref{eq:relation_chi_Pi_chi})
with the concrete definition (\ref{eq:local_inversion_specific}) of the operator $\Pi$ leading to
\begin{align}
\nonumber
e^{i\eta^{(\alpha)}_k}\chi^{(\alpha)}_k(Z) &= (\Pi\chi^{(\alpha)}_{-k})(Z)  \\
\label{eq:zak_zw}
&= \chi^{(\alpha)}_{-k}(1) 
= (\chi^{(\alpha)}_k(1))^*\,.
\end{align}
Since $\chi^{(\alpha)}_k(Z)$ is real this implies $e^{i\eta^{(\alpha)}_k}=e^{-i\varphi^{(\alpha)}_k(1)}$
and $w[\eta_k^{(\alpha)}] = - w[\varphi_k^{(\alpha)}(1)]$, where 
$\varphi^{(\alpha)}_k(1)$ is the phase of the first component $\chi^{(\alpha)}_k(1)$. 
Together with (\ref{eq:relation_winding}) this leads to the central result
\begin{align}
\nonumber
w_\alpha &= 1 + w[\varphi_k^{(\alpha)}(1)] \\
\label{eq:w_zak_relation}
&= 1 - w[\eta_k^{(\alpha)}] = 1 - {\gamma_\alpha\over \pi}\,,
\end{align}
where we used (\ref{eq:zak_quantization}) in the last step. As a consequence we find in the
special case of local inversion symmetry that the quantization of the Zak-Berry phase in units of 
$\pi$ is directly related to the quantization of the winding number $w_\alpha$ which, up to 
our knowledge, is an interesting fact not noticed previously. It shows that
the winding number $w_\alpha$ is a much more general topological index which can be also used in the
{\it absence} of inversion symmetry, in contrast to the standard class of topological crystalline insulators
[\onlinecite{hughes_etal_prb_83}-\onlinecite{lau_etal_prb_16}] which rely on the quantization
of the Zak-Berry phase. The Zak-Berry phase is not a gauge invariant quantity and, in the absence of inversion 
symmetry, it is even not quantized.

In summary we conclude that the Zak-Berry phase $\gamma_\alpha$ itself is only a useful physical 
concept to characterize topological properties in case of local inversion symmetry. 
In this case its quantization in units of $\pi$
leads via (\ref{eq:friedel_zak}) to a quantization of the Friedel charge $Q_F^{(\alpha)}$ 
in half-integer units. Since the polarization charge (\ref{eq:QP_alpha}) is zero in the 
presence of local inversion symmetry, due to the property 
$\rho^{(\alpha)}_{\text{bulk}}(j)=\rho^{(\alpha)}_{\text{bulk}}(Z-j+1)$, this means that also the
boundary charge $Q_B^{(\alpha)}$ is quantized in half-integer units which can be measured
experimentally. In the absence of local inversion symmetry the winding number $w_\alpha$
is a more general index to characterize topological properties and is 
related to the universal properties of the {\it change} of the boundary charge when the 
boundary is shifted to a different position. 

\begin{figure}
  \centering
    \includegraphics[width= \columnwidth]{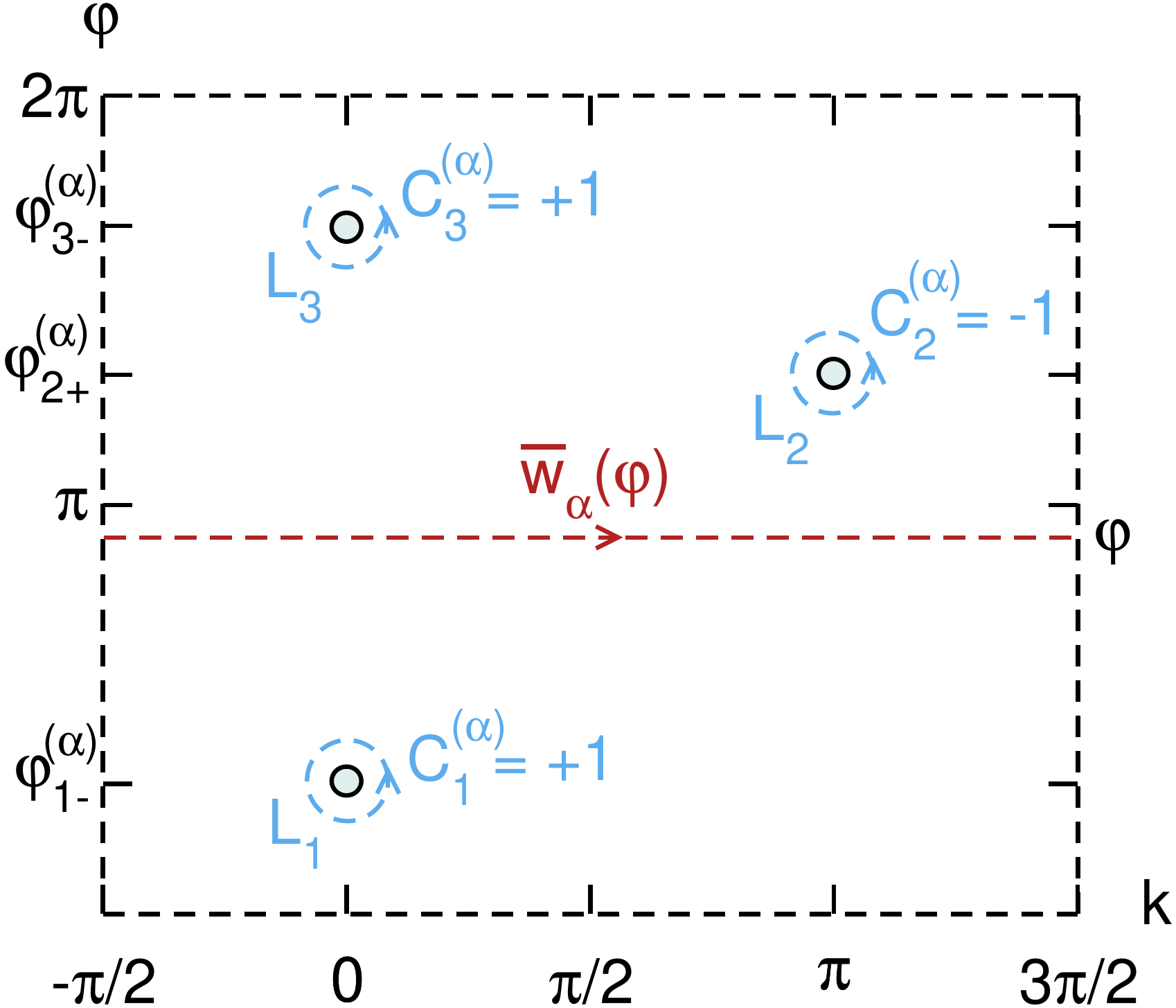} 
  \caption{Singularites of the Berry connection in $(k,\varphi)$-space at the points where
  edge states enter or leave band $\alpha$. We have shown one edge state entering the band at 
  $\varphi_{2+}^{(\alpha)}$ for $k=\pi$ and two edge states leaving the band at $\varphi_{1-}^{(\alpha)}$ 
  and $\varphi_{3-}^{(\alpha)}$ for $k=0$. The Chern number $C^{(\alpha)}=\sum_i C^{(\alpha)}_i$ is 
  given by the negative sum over all closed integrals over the paths $L_i$ of the Berry connection 
  around the singularities at $(k_0,\varphi_{i\pm}^{(\alpha)}$), with $k_0=0$ or $k_0=\pi$. 
  According to Eq.~(\ref{eq:chern_sum_singularities}), this is identical
  to the sum over the windings $C_i^{(\alpha)}$ of the phase of $a_k^{(\alpha)}(1)$ around the singularities.
  The dashed line indicates the path along which the winding number $\bar{w}_\alpha(\varphi)$ is defined 
  which determines the invariant $I_\alpha(\varphi)$ via Eq.~(\ref{eq:invariant_alpha}). When 
  $\varphi$ crosses $\varphi_{i\pm}^{(\alpha)}$ from below the winding number $\bar{w}_\alpha(\varphi)$ jumps
  by $\pm 1$ according to Eq.~(\ref{eq:w_alpha_QB}) leading to $C_i^{(\alpha)}=\mp 1$, 
  see Eq.~(\ref{eq:Cw_relation}). Note that $\bar{w}_\alpha(\varphi)$ jumps also by $\mp 1$ at
  the phase values $\varphi_{i\pm}^{(\alpha)}-{2\pi\over Z}$ according to Eq.~(\ref{eq:w_alpha_QB})
  such that $\bar{w}_\alpha(\varphi)=\bar{w}_\alpha(\varphi+2\pi)$ is fulfilled.}
  \label{fig:singularities}
\end{figure}

In contrast to the Zak-Berry phase $\gamma_\alpha$ the Chern number $C^{(\alpha)}$ is a topological index 
which, similar to $w_\alpha$, is also quantized and gauge invariant in the absence of any symmetry 
constraints (cf. Refs.~[\onlinecite{hatsugai_prl_93,kohmoto_annals}]). It is defined via an integral 
over the Berry curvature $F^{(\alpha)}$ as
\begin{align}
\label{eq:chern}
C^{(\alpha)} = {1\over 2\pi}\int_{-\pi/2}^{3\pi/2}dk \int_0^{2\pi}d\varphi \,F^{(\alpha)}(k,\varphi)\,,
\end{align}
with 
\begin{align}
\label{eq:berry_curvature}
F^{(\alpha)} = \partial_k A^{(\alpha)}_\varphi - \partial_\varphi A^{(\alpha)}_k\,.
\end{align}
Here, the vector $\vec{A}^{(\alpha)}=(A^{(\alpha)}_k,A^{(\alpha)}_\varphi)$ is the Berry connection defined by
\begin{align}
\label{eq:berry_connection}
A^{(\alpha)}_k = i (\chi^{(\alpha)}_k)^\dagger \partial_k \chi^{(\alpha)}_k \quad,\quad
A^{(\alpha)}_\varphi = i (\chi^{(\alpha)}_k)^\dagger \partial_\varphi \chi^{(\alpha)}_k \,.
\end{align}
We note that the Chern number is gauge invariant and does not change when one
replaces $\chi_k^{(\alpha)}$ by the different gauge $\bar{\chi}_k^{(\alpha)}$ defined in 
(\ref{eq:bar_chi}). This leads to an additional term
$\partial_\varphi\sum_j |\chi_k^{(\alpha)}(j)|^2{Z-j\over Z}$ for the Berry curvature which gives
zero when inserted in (\ref{eq:chern}).

We have written the integral over $k$ in (\ref{eq:chern}) from $k=-\pi/2$ to $k=3\pi/2$ since
the singularities of the Berry connection appear at the entering/leaving points $\varphi^{(\alpha)}_{i\pm}$ of
the edge states which occur for $k=k_0$, with $k_0=0$ or $k_0=\pi$, 
see Fig.~\ref{fig:singularities}. This follows from the fact that all analytic quantities 
$a_{k_0}^{(\alpha)}(j)$, $s(\epsilon^{(\alpha)}_{k_0})$ and $N_{k_0}^{(\alpha)}$ are zero for these 
particular phase values but the Bloch state remains finite. In addition, at this point
the phases of all components of the Bloch state become the same
and identical to the phase of the first component $\chi^{(\alpha)}_k(1)=a_k^{(\alpha)}(1)/\sqrt{N_k}$.
This is very important to prove the quantization of the Chern number and its relation to $w_\alpha$. 
To show this we first note that close to one of the singularities $(k_0,\varphi_{i\sigma_i}^{(\alpha)})$  
we get up to linear order in $k-k^{(\nu)}_{\text{e}}$ and $\varphi-\varphi_{i\sigma_i}^{(\alpha)}$
(compare with (\ref{eq:edge_band_connection}))
\begin{align}
\label{eq:dispersion_touching_point}
\epsilon_k^{(\alpha)}(\varphi) \approx \epsilon_{\text{e}}^{(\nu)}(\varphi) \,,
\end{align}
where $\nu$ is the index of the edge state corresponding to gap $\nu=\alpha$ ($\nu=\alpha-1$) if
the edge state enters/leaves at the top (bottom) of the band. This follows 
since the edge state leaves/enters the band in a smooth way as function of 
$\varphi$ together with $\partial_k\epsilon_k^{(\alpha)}=0$ at $k=k_0$, see (\ref{eq:eps_der_1}).
Inserting (\ref{eq:dispersion_touching_point}) in (\ref{eq:aj_a1}) and using 
$s(\epsilon_{\text{e}}^{(\nu)})=0$ we find
\begin{align}
\label{eq:aj_a1_touching_point}
t_1\cdots t_{j-1} a_k^{(\alpha)}(j) \approx d_{1,j-1}(\epsilon_{\text{e}}^{(\nu)}) a_k^{(\alpha)}(1)
\end{align}
up to linear order in $k-k^{(\nu)}_{\text{e}}$ and $\varphi-\varphi_{i\sigma_i}^{(\alpha)}$,
showing that close to all singularities the phases of all components are identical to the 
phase $\varphi_k^{(\alpha)}(1)$ of the first component of the Bloch state. Using the normalization
$\sum_{j=1}^Z|\chi_k^{(\alpha)}(j)|^2=1$, this means, that for $(k,\varphi)$ close to 
the singularity $(k_0,\varphi_{i\sigma_i}^{(\alpha)})$, we get for the Berry connection vector
\begin{align}
\label{eq:berry_connection_singularity}
\vec{A}^{(\alpha)}\approx -(\partial_k,\partial_\varphi)\varphi_k^{(\alpha)}(1)\,.
\end{align}
Taking this result together with Stokes theorem, relating the Chern number (\ref{eq:chern}) 
to the sum over the closed integrals of the Berry connection around the singularities, we get
\begin{align}
\label{eq:chern_sum_singularities}
C^{(\alpha)}=-\sum_i \oint_{L_i}\vec{A}^{(\alpha)}d\vec{l}
= \sum_i \oint_{L_i}d\varphi_k^{(\alpha)}(1)\,,
\end{align}
where $L_i$ denotes the counterclockwise closed curve around the corresponding singularity 
(see Fig.~\ref{fig:singularities}). As a result the Chern number is given by the
sum $C^{(\alpha)} = \sum_i C^{(\alpha)}_i$ over the winding numbers $C^{(\alpha)}_i$ of $a_k^{(\alpha)}(1)$
around the singularity at $(k_0,\varphi_{i\sigma_i}^{(\alpha)})$. We note that this winding number
has to be distinguished from the winding number $\bar{w}_\alpha(\varphi)$ defined in 
(\ref{eq:relation_winding}) which describes the winding of $a_k^{(\alpha)}(1)$
along the path $k=0\rightarrow k=2\pi$ at fixed phase $\varphi$, see Fig.~\ref{fig:singularities}. 
Obviously the two are related by
\begin{align}
\label{eq:Cw_relation}
C^{(\alpha)}_i = \bar{w}_\alpha(\varphi_{i\sigma_i}^{(\alpha)}-0^+) 
- \bar{w}_\alpha(\varphi_{i\sigma_i}^{(\alpha)}+0^+) \,,
\end{align}
i.e., $C^{(\alpha)}_i$ is identical to the {\it negative jump} of $\bar{w}_\alpha(\varphi)$ when $\varphi$ crosses 
a point $\varphi_{i\sigma_i}^{(\alpha)}$ where an edge state enters/leaves the band. 

Since $\bar{w}_\alpha(\varphi)$ can be related via (\ref{eq:QB_change_Delta_m}) and
(\ref{eq:relation_winding}) to the change of the boundary charge as
\begin{align}
\label{eq:w_alpha_QB}
\bar{w}_\alpha(\varphi) = -1 - {1\over Z} - Q_B^{(\alpha)}(\varphi+{2\pi\over Z}) + Q_B^{(\alpha)}(\varphi)\,,
\end{align}
we find that the jump of $\bar{w}_\alpha(\varphi)$ is given by the jump of $Q_B^{(\alpha)}(\varphi)$ 
at $\varphi=\varphi_{i\sigma_i}^{(\alpha)}$ which is identical to $\sigma_i$. We note that 
$Q_B^{(\alpha)}(\varphi+{2\pi\over Z})$ does not jump at $\varphi=\varphi_{i\sigma_i}^{(\alpha)}$ since
no edge state can appear at $\varphi=\varphi_{i\sigma_i}^{(\alpha)}+{2\pi\over Z}$, 
see Fig.~\ref{fig:contraction} and the discussion in Section~\ref{sec:constraints_edge}. 
Denoting by $M^{(\alpha)}_\pm$ the total number of entering/leaving points of edge states, we find
\begin{align}
\nonumber
C^{(\alpha)} &= \sum_i C^{(\alpha)}_i = - \sum_i \sigma_i \\
\label{eq:chern_M}
&= M^{(\alpha)}_- - M^{(\alpha)}_+ \equiv M^{(\alpha)} \,.
\end{align}
Due to (\ref{eq:w_alpha_QB}) we note that $\bar{w}_\alpha(\varphi)$ jumps also at all phase 
values $\varphi_{i\sigma}-{2\pi\over Z}$ such that 
$\bar{w}_\alpha(\varphi)=\bar{w}_\alpha(\varphi+2\pi)$ is a periodic function and the total Chern number
can not be calculated by the overall jump of $\bar{w}_\alpha$ across a certain phase interval.  Note that
the relation (\ref{eq:chern_M}) was originally established in Ref.~[\onlinecite{hatsugai_prl_93}], and 
it is often referred to as the bulk-boundary correspondence.

In summary, we find that the winding number $w_\alpha(\varphi)$ contains much more information
compared to the Chern number $C^{(\alpha)}$. The Chern number is a phase-integrated quantity which
measures the sum of the jumps of $w_\alpha(\varphi)$ at the phase values $\varphi=\varphi_{i,\pm}$
where edge states enter or leave the band. Neither the value of $w_\alpha$ itself nor the precise
positions of the jumps enter into the Chern number. If $\varphi$ is an effective phase resulting
from the quasimomentum $k_y$ perpendicular to the boundary of a two-dimensional system, it can
be shown that the total Chern number of all filled bands $C_\nu=\sum_{\alpha=1}^\nu C^{(\alpha)}$ 
is related to the plateau values of the transverse conductance \cite{thouless_etal_prl_82}. 
In contrast, $w_\alpha(\varphi)$ measures via (\ref{eq:w_alpha_QB}) the whole phase-dependence 
of the change $\Delta Q^{(\alpha)}_B(\varphi)$ of the boundary charge when the phase
is changed by ${2\pi\over Z}$.    

We note that the winding numbers indicated in Fig.~\ref{fig:singularities} along the various
paths defining $C_i^{(\alpha)}$ and $\bar{w}_\alpha$ are the windings of the phase of the first component
$a_k^{(\alpha)}(1)$ of the Bloch state. The corresponding line integrals of the Berry connection 
are only the same (up to a sign) for the curves $L_i$ (defining $C_i^{(\alpha)}$) but {\it not}
for the horizontal lines. The latter integral would result in the Zak-Berry phase
(\ref{eq:zak_phase_1}) and one can also define the Chern number via the sum of the 
jumps of the Zak-Berry phase at the phase values $\varphi=\varphi_{i,\pm}$ where edge states 
enter or leave the band. This can also be interpreted as a winding number \cite{qi_wu_zhang_prb_06} but we 
emphasize that, in contrast to $w_\alpha(\varphi)$, the Zak-Berry phase $\gamma_\alpha(\varphi)$ itself
is {\it not} a winding number and is neither quantized nor gauge-invariant
as outlined above.

Finally, we note that when the Chern number is summed over all filled bands, 
we find that many terms cancel since
the number of entering (leaving) modes at the bottom of band $\alpha$ is identical
to the number of leaving (entering) modes at the top of band $\alpha-1$. Therefore, in 
consistency with many previous works 
\cite{thouless_etal_prl_82,hatsugai_prl_93,kohmoto_prb_89_jpsj_92,hatsugai_prb_93,hatsugai_jphys,kohmoto_annals},
only those edge states entering/leaving the top of band $\alpha=\nu$ remain and the following result is 
obtained for the total Chern number
\begin{align}
\nonumber
C_\nu &= \sum_{\alpha=1}^\nu C^{(\alpha)} = M^{(\nu)}_- - M^{(\nu)}_+ \\
\label{eq:chern_total}
&= M_-(\mu_\nu) - M_+(\mu_\nu) = M_\nu \,,
\end{align}
where $M_\pm(\mu_\nu)$ denotes the total number of edge states crossing the chemical potential
$\mu_\nu$ from above/below, already introduced in (\ref{eq:topological_charge}). According
to the discussion in Section~\ref{sec:constraints_edge} we note that the difference
$M_\nu = M_-(\mu_\nu) - M_+(\mu_\nu)$ does not depend on the precise position of the 
chemical potential inside the gap but only on the index $\nu$ of the gap. 

\begin{figure*}
\centering
\includegraphics[width= 1.\columnwidth]{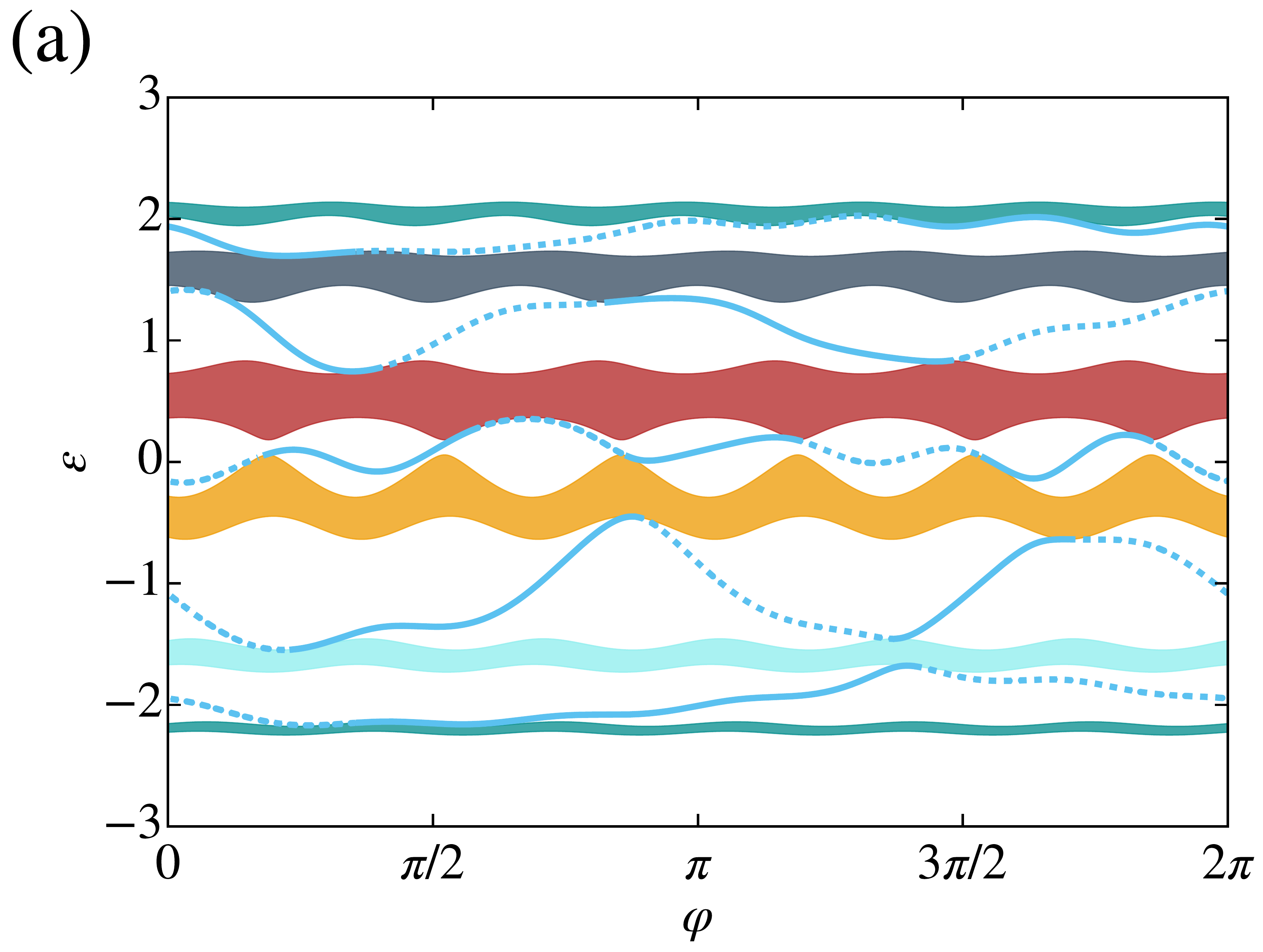} 
\includegraphics[width= 1.\columnwidth]{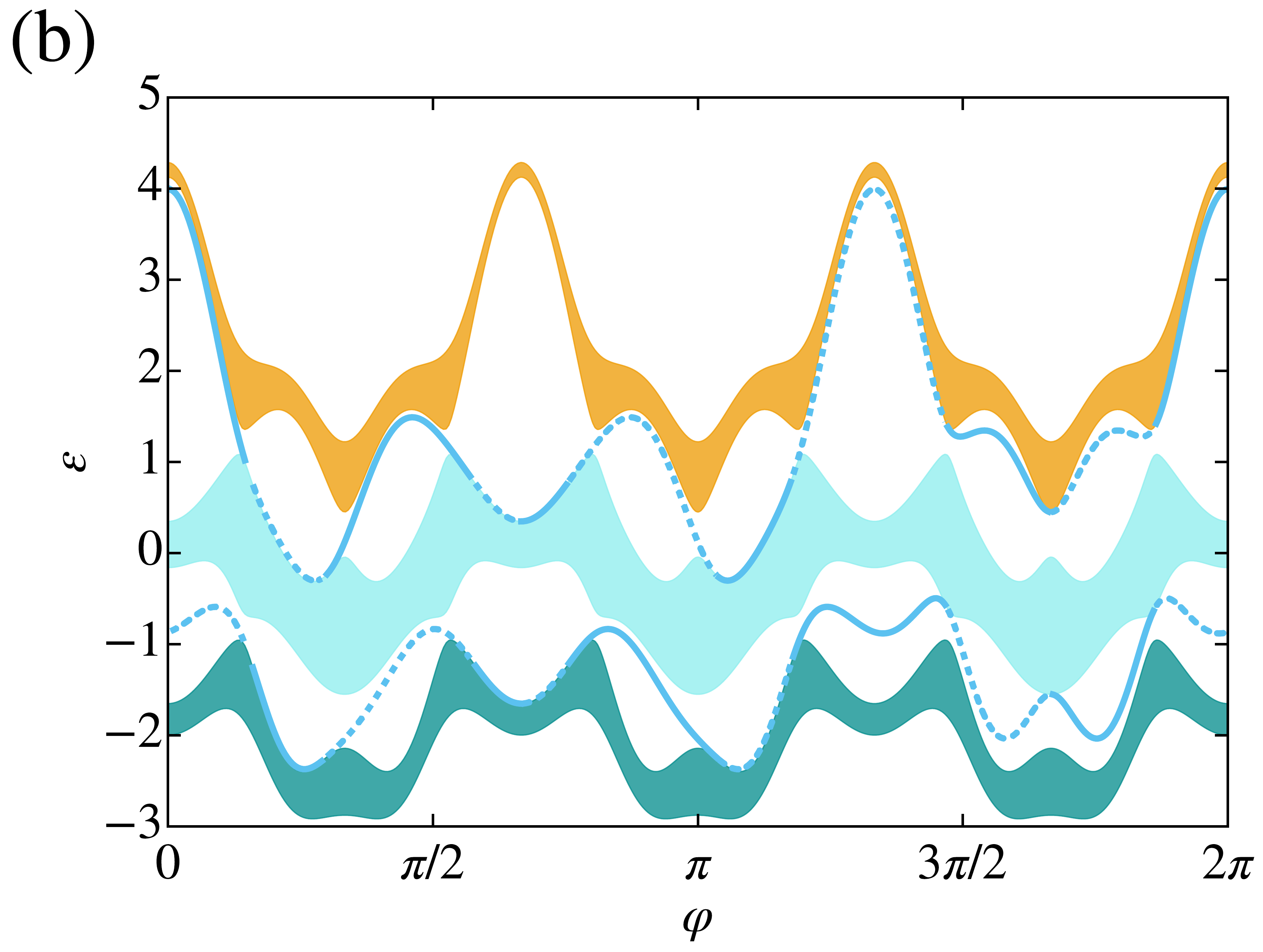} 
 \caption{(a) The band structure and edge states analog to Fig.~\ref{fig:edge}(b) for 
   $Z=6$, $V=0.5$, $t=1$, $\delta t=0.1$ and three random Fourier coefficients for the real functions  
   $F_v$ and $F_t$ in Eqs.~(\ref{eq:v_form}) and (\ref{eq:t_form}) according to the form (\ref{eq:F_random1}), 
   see Supplemental Material for the precise parameters \cite{SM}. For the gaps $\nu=1,2,3$ there are $\nu$ 
   edge states of $H_R$ (blue solid lines) moving upwards whereas, for the gaps $\nu=4,5$, there are
   $Z-\nu$ edge states of $H_R$ moving downwards. For band $\alpha=4$ all edge states of $H_R$ are entering
   the band, $Z-\alpha=2$ from above and $\alpha-1=3$ from below, i.e., in total $Z-1=5$ edge states.
   (b) The band structure and edge states analog to Fig.~\ref{fig:edge}(b) for 
   $Z=3$, $V=0.5$, $t=1$, $\delta t=0$ and five random Fourier coefficients for the real function 
   $F_v$ in Eq.~(\ref{eq:v_form}) according to the form (\ref{eq:F_random1}), 
   see Supplemental Material for the precise parameters \cite{SM}. Several edge states of $H_R$ return
   to the same band in both gaps $\nu=1,2$. For band $\alpha=2$ all edge
   states of $H_R$ are entering the band (disregarding the ones which return to the same band), 
   $Z-\alpha=1$ from above and $\alpha-1=1$ from below, i.e., in total $Z-1=2$ edge states.
}
\label{fig:edge_2}
\end{figure*}

\section{Universal properties}
\label{sec:universal}

The topic of this section are the universal properties of the boundary charge as proposed in
Ref.~[\onlinecite{paper_prl}] that will be proven rigorously here based on the topological constraints
for edge states derived in Section~\ref{sec:constraints_edge}. We review the physical picture based
on charge conservation and particle-hole duality as in Ref.~[\onlinecite{paper_prl}] and provide
a proof of the essential ingredient that the phase-dependence of the model parameters can always be
chosen such that no edge states cross the chemical potential in a certain gap within a phase interval
of size ${2\pi\over Z}$. We provide a rigorous proof of the two central results of 
Ref.~[\onlinecite{paper_prl}] why the invariants defined for a single band or for a 
given chemical potential are quantized and which values are allowed. In addition, we treat the case
where the wavelength of the modulations is any rational number and discuss the expectations for
multi-channel systems.

\subsection{Physical picture}
\label{sec:invariant_physics}

We start with physical arguments what we expect for the change of $Q_B$ if the lattice is shifted
by one site towards the boundary. This means that the lattice of $H_R$ starts not with site $(n,j)=(1,1)$
(as in Fig.~\ref{fig:model_RL}) but with $(n,j)=(1,2)$. Formally, we achieve this by changing
the phase variable $\varphi$ by ${2\pi\over Z}$ such that $v_j\rightarrow v_{j+1}$ and
$t_j\rightarrow t_{j+1}$. This is a very fundamental question of how observables defined at 
the boundary depend on the way one cuts off an infinite system to define the boundary. Since 
the boundary charge contains also the charge $Q_E$ of the edge states this is also 
related to the question of how the appearance and energy of edge states depend on the way 
one defines the boundary. As can be seen already from Fig.~\ref{fig:edge}(b) the appearance 
and energy of edge states of $H_R$ (shown as solid blue lines) depend crucially on the phase variable 
$\varphi$, a fact very well known from the integer QHE in 2D systems (where the quasimomentum
$k_y$ in $y$-direction plays the role of $\varphi$) \cite{thouless_etal_prl_82,hatsugai_prb_93}. 
An illustrative but very fundamental example in this respect is the SSH model for $Z=2$ and $v_j=0$. 
This model is parametrized only by two hopping parameters $t_1$ and $t_2$
and it is obviously not important for the bulk properties whether $t_1>t_2$ or $t_1<t_2$, one can
just interchange the two hoppings to get one or the other case. If the
system starts with site $(n,j)=(1,1)$ it is obvious that $t_1 < t_2$ leads to the appearance
of an edge state of $H_R$ (since an electron can not leave the first site of the system in the
limiting case $t_1=0$). Similarly, if the system starts with site $(n,j)=(1,2)$ one needs the
condition $t_1 > t_2$ to get an edge state of $H_R$. This property is generic for any $Z$, edge
states of $H_R$ appear only in certain regions of $\varphi$. The underlying physics for the appearance of
edge states as function of $\varphi$ is very obvious and is related to charge pumping and charge
conservation \cite{thouless_prb_83,hatsugai_fukui_prb_16}. If the phase variable is changed 
adiabatically in time by $2\pi$, the charge $\nu$ of a whole unit cell (corresponding to 
the number $\nu$ of filled bands when the chemical potential $\mu_\nu$ is located in gap $\nu$) has been moved
into the boundary. Since the boundary charge does not change for a phase change by $2\pi$ (the
Hamiltonian $H_R$ is exactly the same) the charge $\nu$ must
be taken away by exactly $\nu$ edge states which move above the chemical potential during this process 
and move the charge to higher bands, see Fig.~\ref{fig:edge}(b). As can be seen for the gaps $\nu=4,5$ 
in Fig.~\ref{fig:edge_2}(a) it can also happen that $Z-\nu$ edge states of $H_R$ run downwards from 
band $\alpha=\nu+1$ to band $\alpha=\nu$. This can be understood by an adiabatic pumping process 
in terms of the hole picture described in Section~\ref{sec:ph_duality}. Due to Eq.~(\ref{eq:bar_rho_ph})
on average the hole charge $\nu-Z$ is shifted into the boundary when the phase changes by $2\pi$. 
Since the boundary charges of the holes and particles are the same this means that $Z-\nu$ edge states
have to move below the chemical potential to compensate this charge. Furthermore, it can happen that
edge states return to the same band, see Fig.~\ref{fig:edge_2}(b). Which case appears, depends
on the model parameters and how the phase-dependence is chosen via the functions 
$F_v$ and $F_t$ in Eqs.~(\ref{eq:v_form}) and (\ref{eq:t_form}).
\begin{figure*}
\centering
\begin{minipage}{0.44\textwidth}
  \includegraphics[width= 0.8\columnwidth]{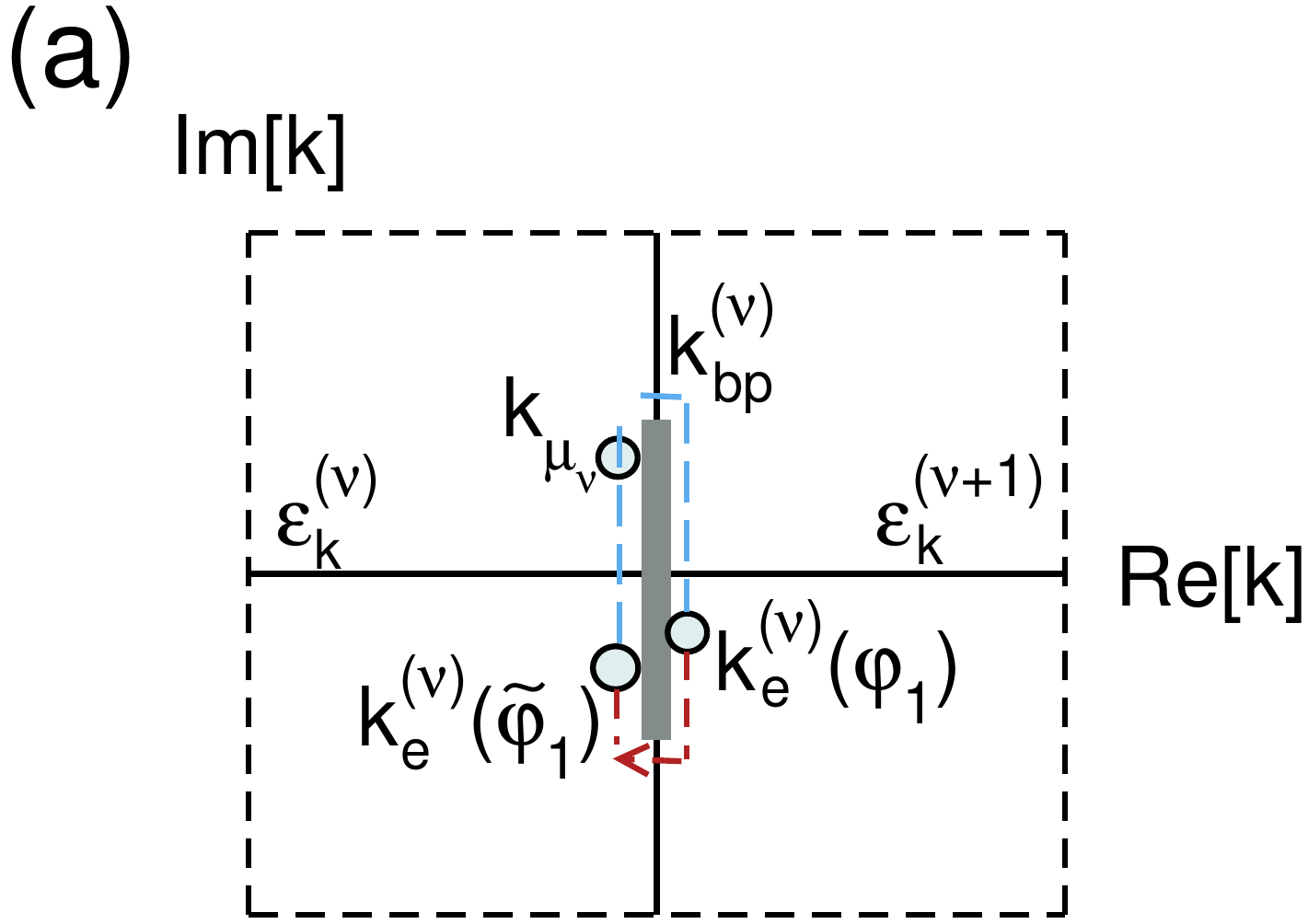} 
\end{minipage}
\begin{minipage}{0.44\textwidth}
  \includegraphics[width= 0.8\columnwidth]{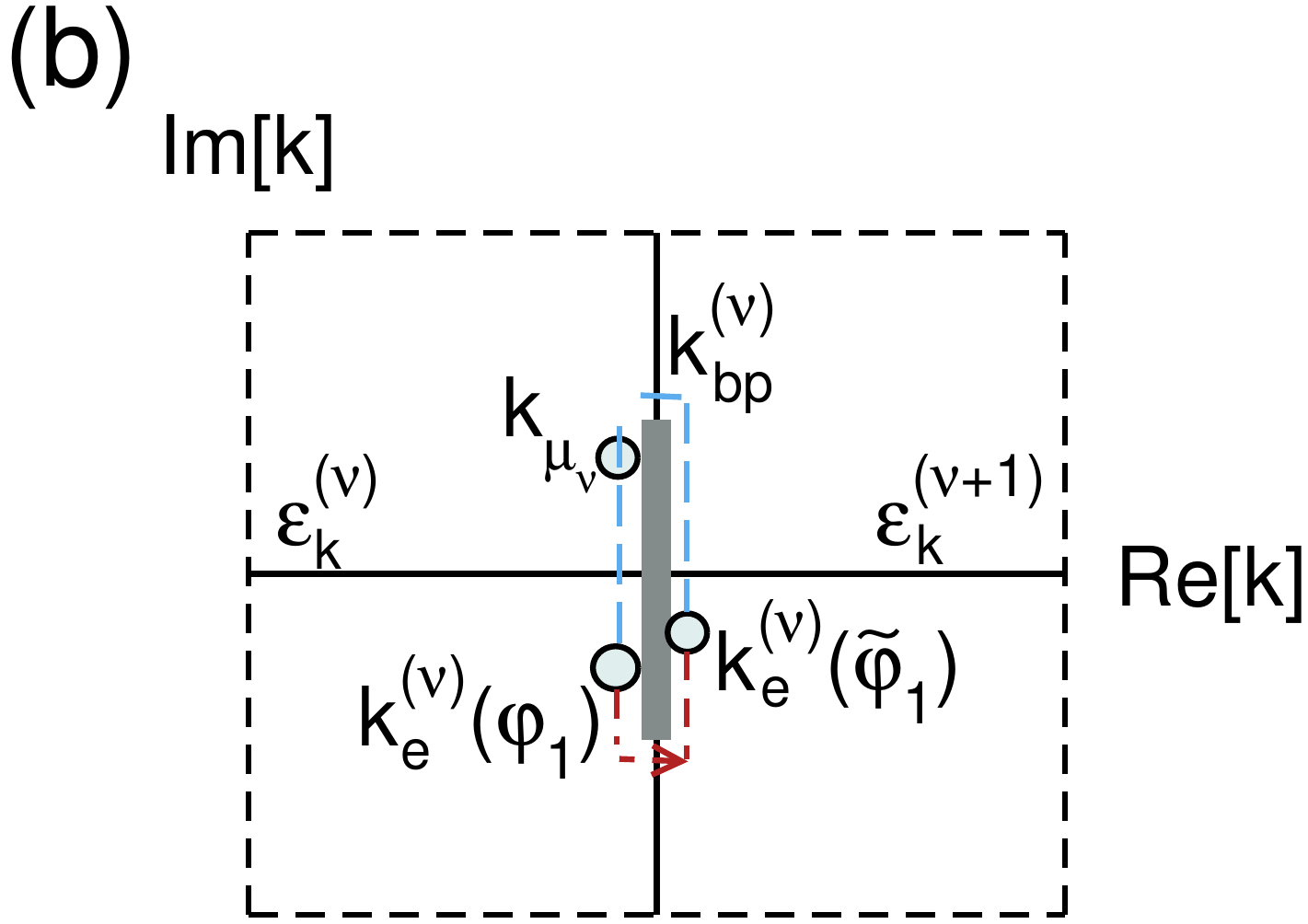} 
\end{minipage}
\caption{Sketch of the branch cut separating band $\nu$ from $\nu+1$, corresponding to 
  the analytic continuation of Fig.~\ref{fig:bc_edge}(a). The position of $k_{\mu_\nu}$ corresponds to 
  $\mu_\nu$ and is defined via $\epsilon_{k_{\mu_\nu}}=\mu_\nu$ and $\text{Im}(k_{\mu_\nu})>0$.
  The positions of $k_e^{(\nu)}(\varphi_1)$ and $k_e^{(\nu)}(\tilde{\varphi_1})$ correspond to the 
  edge state poles at the phases $\varphi_1$ and $\tilde{\varphi}_1=\varphi_1+{2\pi\over Z}$, respectively. 
  Two possibilities are shown in (a) and (b) for their position. Obviously, for both cases one can
  choose the orientation of the path $k_e^{(\nu)}(\varphi)$ of the edge state pole for all phases 
  $\varphi_1<\varphi<\tilde{\varphi}_1$ in such a way that no crossing through $k_{\mu_\nu}$ 
  occurs (red part of the contour). We have not indicated in the figure the weak dependence of
  $k_{\mu_\nu}$ and $k_{\text{bp}}^{(\nu)}$ on $\varphi$.
}
\label{fig:no_edge}
\end{figure*}

\begin{figure*}
\centering
\begin{minipage}{0.44\textwidth}
  \includegraphics[width= 1.\columnwidth]{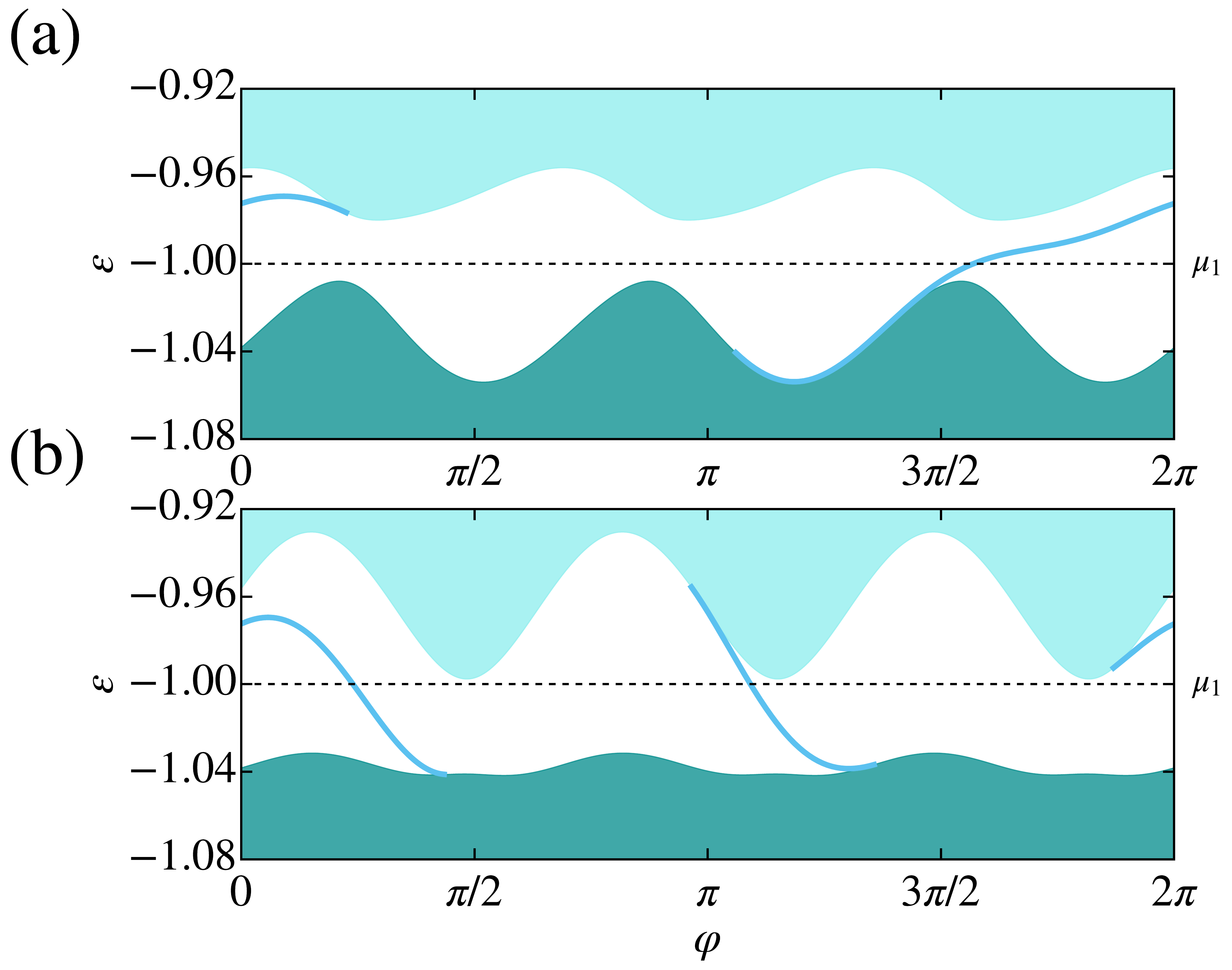} 
\end{minipage}
\begin{minipage}{0.53\textwidth}
 \includegraphics[width= 0.8\columnwidth]{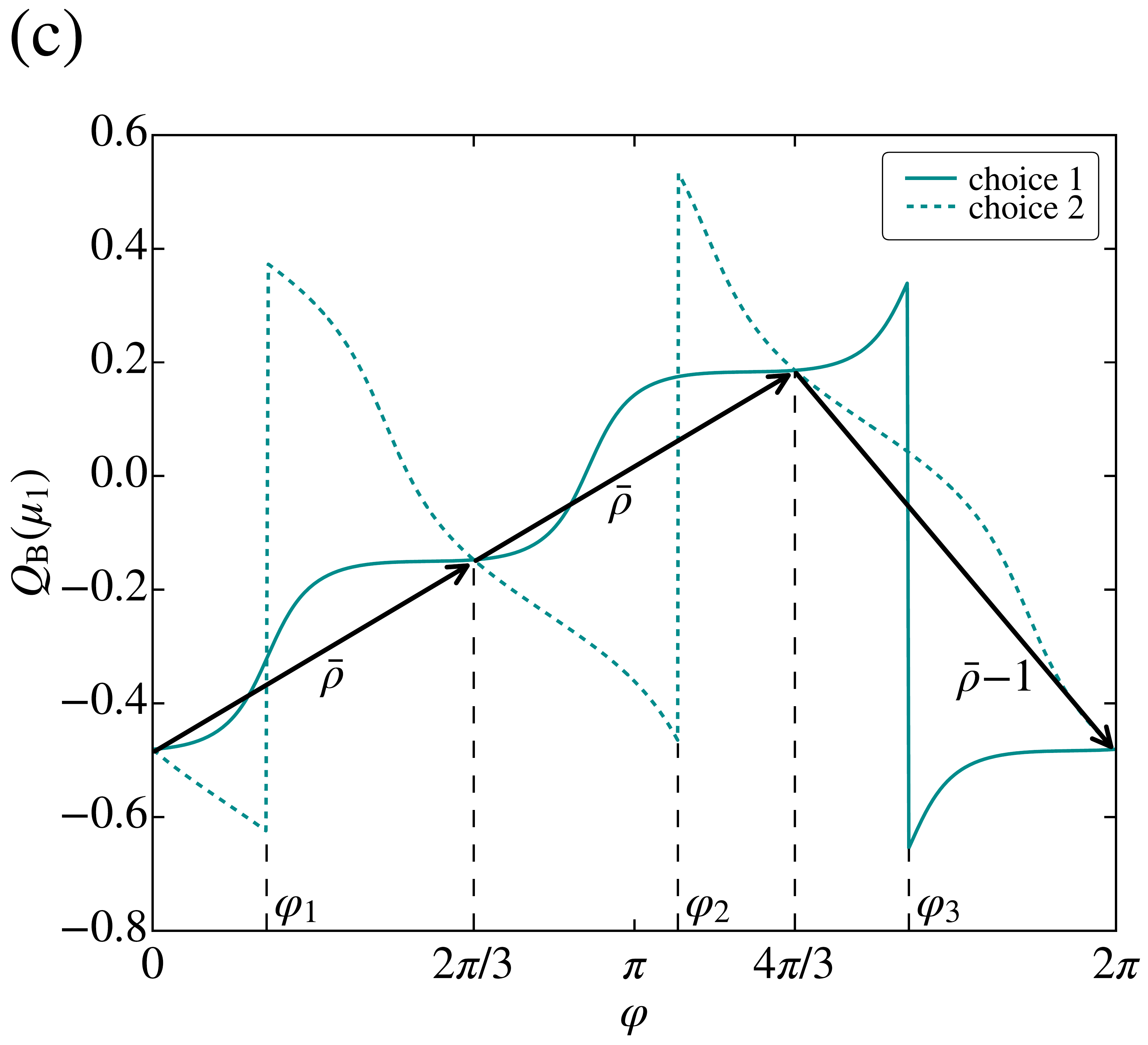}
\end{minipage}
\hfill
 \caption{In (a) and (b) we show the phase-dependence of the band structure of the first two bands and the
   edge states of $H_R$ in the first gap (blue lines) for $Z=3$, $V=0.1$, $t=1$, $\delta t=0.1$ for
   two functions $F_v$ and $F_t$ in Eqs.~(\ref{eq:v_form}) and (\ref{eq:t_form}) taken
   from (\ref{eq:F_random2}) with fixed and random parameters for $v_j(0)$ and $t_j(0)-t$
   at phase $\varphi=0$ but different choices for the phase-dependence in between [via the random parameters 
   $v_j^{(1)}$ and $t_j^{(1)}$ used in (\ref{eq:g_n})], see the Supplemental Material \cite{SM}
   for the concrete parameters. In (a) one edge state appears running from the lower to the
   upper band whereas in (b) two edge states run from the upper to the lower band. In (c) the phase
   dependence of $Q_B$ is shown for the two choices of (a) and (b) with $\mu_1$ located in the 
   first gap [see dashed line in (a) and (b)]. At $\varphi=0$ and $\varphi=2\pi/3$, we get the change 
   $\Delta Q_B(0)=\Delta Q_B({2\pi\over 3})=\bar{\rho}={1\over 3}$ (see the two left arrow). 
   For the first choice $Q_B$ increases on average by ${1\over Z}={1\over 3}$ on both intervals
   whereas for the second choice $Q_B$ decreases on average by 
   ${1\over Z}-1=-{2\over 3}$ and obtains a jump by $+1$ at $\varphi=\varphi_{1,2}$ where 
   an edge state moves below $\mu_1$. For $\varphi=4\pi/3$, we get 
   $\Delta Q_B({4\pi\over 3})=\bar{\rho}-1=-{2\over 3}$ (see right arrow). Here, for the first choice, an edge 
   state moves above $\mu_1$ at $\varphi=\varphi_3$ leading to a jump of $Q_B$ by $-1$ whereas for
   the second choice no edge state is involved.
}
\label{fig:edge_unique}
\end{figure*}

The two ways of how our Gedankenexperiment of charge pumping can be interpreted in terms of particles and holes 
is very fundamental for an intuitive understanding of the physics. That each configuration of the edge states
in a certain gap can be explained either by the particle or hole version of our Gedankenexperiment 
but not by the other relies on the Pauli principle which has no classical analog. Whereas in the
particle picture the boundary charge {\it increases} by $\nu$ during a phase change of $2\pi$ and has
to be compensated by $\nu$ edge states moving {\it above} $\mu_\nu$, in the hole picture the boundary
charge {\it decreases} by $Z-\nu$ and has to be compensated by $Z-\nu$ edge states moving {\it below} 
$\mu_\nu$. This will give rise to two completely different line shapes of the boundary charge as function of
the phase variable $\varphi$, see Fig.~\ref{fig:edge_unique}(c) for an example which will be discussed
below in all detail.

One can also think of occupying only one single band $\alpha$ such that one charge is moved into the 
boundary after a phase change of $2\pi$. Therefore, one expects that during this process in total 
one edge state has to leave this band. In fact, in Figs.~\ref{fig:edge}(b) and \ref{fig:edge_2}(a,b) 
it is the case for most of the bands that either $\alpha-1$ edge states enter into the band bottom 
and $\alpha$ edge states leave from the band top or $Z-\alpha$ edge states enter into the band top and
$Z-\alpha+1$ edge states leave from the band bottom, such that in both cases one edge state leaves in total.
However, in each of the figures one band is special in the sense that $Z-1$ edge states enter the band, 
$Z-\alpha$ ones from above and $\alpha-1$ ones from below. This case can be understood in the hole
picture since the hole charge $1-Z$ has been moved into the boundary after a phase change of $2\pi$
which has to be compensated by $Z-1$ edge states entering the band. 

We now consider the more fundamental issue of how the boundary charge changes when we move 
the system only by {\it one} site towards the boundary. We call this change
\begin{align}
\nonumber 
\Delta Q_B &\equiv \Delta Q_B(\varphi,\mu_\nu) \\
\label{eq:QB_change}
&= Q_B(\varphi+{2\pi\over Z},\mu_\nu) - Q_B(\varphi,\mu_\nu)\,,
\end{align}
which depends on the phase $\varphi$ and the chemical potential $\mu_\nu$ in gap $\nu$ (note that 
edge states of $H_R$ can be present in the gap such that $Q_B$ depends on the precise value of $\mu_\nu$
in gap $\nu$). In this case we expect that {\it on average} the charge $\bar{\rho}={\nu\over Z}$ is moved 
into the boundary. Since $Q_B$ is defined via a {\it macroscopic average} on length scales much larger
than the size of a unit cell, we expect the same for $\Delta Q_B=\bar{\rho}$ if, in addition, no edge state is 
moving above/below $\mu_\nu$ during the shift. So far we have only involved the physics of 
classical charge conservation to get this result. If we take in addition the Pauli principle into account 
and look at the same process
from the hole point of view on average the hole charge $\Delta Q_{B,h} = \bar{\rho}_h=\bar{\rho}-1$ 
is moved into the boundary during the shift, which gives $\Delta Q_B = \Delta Q_{B,h} = \bar{\rho}-1$.
Therefore, we expect that the following two universal values are possible for the change of $\Delta Q_B$
\begin{align}
\label{eq:QB_change_universal}
\Delta Q_B(\varphi,\mu_\nu) \in \{\bar{\rho},\bar{\rho}-1\} \,.
\end{align}
It is very important to realize that this result does {\it not} involve any edge state physics but relies only on 
classical charge conservation together with the Pauli principle. If additional edge states move 
above/below $\mu_\nu$ during the shift one expects Eq.~(\ref{eq:QB_change_universal}) to change $\text{mod}(1)$.
However, for a given set of parameters $\{v_j(\varphi_1),t_j(\varphi_1)\}$ at phase $\varphi=\varphi_1$ and,
correspondingly, via Eq.~(\ref{eq:shift}) also at all phases shifted by multiples of ${2\pi\over Z}$,
the value of $\Delta Q_B(\varphi_1,\mu_\nu)$ does {\it not} depend on the particular choice of the phase
dependence of $v_j(\varphi)$ and $t_j(\varphi)$ for all phases $\varphi_1<\varphi<\varphi_1+{2\pi\over Z}$.
For our $1$-channel model with a non-degenerate spectrum it is shown in Appendix~\ref{app:edge_tuning} 
that, for any given phase interval of size ${2\pi\over Z}$, the phase-dependence can always be chosen such that
no edge state of $H_R$ crosses the chemical potential $\mu_\nu$ in gap $\nu$ as function of $\varphi$ in
this interval. This is related to the fact that the complex quasimomentum $k_e^{(\nu)}(\varphi)$ of the edge 
state pole moves around the branch cut defined between band $\nu$ and $\nu+1$ as function of $\varphi$, see 
Fig.~\ref{fig:bc_edge}(a) and movies provided in the Supplemental Material \cite{SM}. This movement can
have both orientations depending on the choice of the phase-dependence of the parameters. 
Defining the complex quasimomentum $k_{\mu_\nu}$ corresponding
to the chemical potential uniquely via $\epsilon_{k_{\mu_\nu}}=\mu_\nu$ and $\text{Im}(k_{\mu_\nu})>0$, the condition
that an edge state of $H_R$ does not cross $\mu_\nu$ for all $\varphi_1<\varphi<\varphi_1+{2\pi\over Z}$
is equivalent to the condition that $k_e^{(\nu)}(\varphi)$ does not cross $k_{\mu_\nu}(\varphi)$ in this interval.
This can always be achieved by choosing the phase-dependence of the parameters such that the orientation 
of the movement of $k_e^{(\nu)}(\varphi)$ has the appropriate sign, see Fig.~\ref{fig:no_edge}(a,b).
As a result we find that the two values predicted in Eq.~(\ref{eq:QB_change_universal}) are the
only allowed values for the single-channel case and are {\it not} related to any edge state physics.

As an example we demonstrate in Fig.~\ref{fig:edge_unique} that a different parametrization for 
the phase-dependence gives rise to very different interpretations of the change of the boundary charge.
In Fig.~\ref{fig:edge_unique}(a,b) we show the edge states of $H_R$ in the first gap $\nu=1$ 
for $Z=3$ and for given parameter sets of $t_j(\varphi)$ and $v_j(\varphi)$ at all phases 
$\varphi=2\pi j/3$, with $j=1,\dots,3$, but with two different choices of how the phase-dependence
is defined in between. In Fig.~\ref{fig:edge_unique}(a) one edge state is moving from the lower to 
the upper band whereas in Fig.~\ref{fig:edge_unique}(b) two edge states are moving from the upper to the
lower band. Fig.~\ref{fig:edge_unique}(c) shows the corresponding phase-dependence of the boundary charge 
$Q_B$ for the two choices when the chemical potential $\mu_1$ is located in the first gap [dashed line
in Figs.~\ref{fig:edge_unique}(a,b)]. Obviously, at $\varphi=2\pi j/3$, with $j=1,2,3$, the value of 
$Q_B$ must be the same for the two choices but the line shape in between is completely different. 
Comparing the two points $\varphi=0, {2\pi\over 3}$ or $\varphi={2\pi\over 3}, {4\pi\over 3}$, we get 
$\Delta Q_B = \bar{\rho} = {1\over 3}$ in both cases [see the two left arrows in Fig.~\ref{fig:edge_unique}(c)]. 
For the first choice $Q_B$ is increasing monotonously by ${1\over 3}$ without any edge state involved. In contrast, 
for the second choice, $Q_B$ decreases monotonously by $\bar{\rho}-1=-{2\over 3}$ but at
$\varphi=\varphi_{1,2}$ an edge state moves below $\mu_1$ such that $Q_B$ gets a discontinuous jump
by $+1$ and the same result is obtained for $\Delta Q_B={1\over 3}$. 
Similarly, comparing $Q_B$ between
$\varphi={4\pi\over 3}$ and $\varphi=2\pi$ we get $\Delta Q_B = \bar{\rho}-1=-{2\over 3}$ for both
choices. Here, the situation is the other way around, for the first choice an edge state moves 
above $\mu_1$ at $\varphi=\varphi_3$ whereas for the second choice no edge state is involved, leading
again to the same net result for $\Delta Q_B$. 
Therefore, we conclude that just by looking at the change of $Q_B$ when the system is cut off
at a different site at the boundary, it is not unambiguous to interpret the value of $\Delta Q_B$
in terms of edge state physics. Instead, the correct interpretation is in terms of charge conservation
and the Pauli principle as explained above, the edge states just play the role of ``followers'' and can
appear in one or the other form depending on the concrete choice of the phase-dependence. 

The result (\ref{eq:QB_change_universal}) holds if the chemical potential $\mu_\nu$ is located in
gap $\nu$. We note that it is not essential that the chemical potential is a constant, it can as well have a phase
dependence provided it is ${2\pi\over Z}$-periodic: $\mu_\nu(\varphi)=\mu_\nu(\varphi+{2\pi\over Z})$.
This is necessary for band structures like in Fig.~\ref{fig:edge_2}(b) where two bands can not
be separated by a fixed energy. If we consider the boundary charge 
$Q_B(\varphi,\mu_\nu,\mu_{\nu'})=Q_B(\varphi,\mu_\nu)-Q_B(\varphi,\mu_{\nu'})$
of all states lying between two energies $\mu_{\nu'}$ and $\mu_{\nu}$ located in gaps $\nu'<\nu$ and $\nu$, 
respectively, the corresponding
change is obtained by taking the difference of (\ref{eq:QB_change_universal}) for $\nu$ and $\nu'$
\begin{align}
\nonumber
\Delta Q_B(\varphi,\mu_\nu,\mu_{\nu'}) &= \Delta Q_B(\varphi,\mu_\nu) - \Delta Q_B(\varphi,\mu_{\nu'}) \\
\label{eq:QB_change_universal_nu_nu'}
&\in \{\bar{\rho}-\bar{\rho}',\bar{\rho}-\bar{\rho}'\pm 1\}\, 
\end{align}
with $\bar{\rho}={\nu\over Z}$ and $\bar{\rho}'={\nu'\over Z}$, giving rise to three possible values. 
For example, for a single band $\alpha$ [with $\mu_\nu(\varphi)$ chosen as the band top and 
$\mu_{\nu'}(\varphi)=\mu_{\nu-1}(\varphi)$ as the band bottom] we obtain
\begin{align}
\label{eq:QB_alpha_change_universal}
\Delta Q^{(\alpha)}_B(\varphi) \in \{{1\over Z},{1\over Z}\pm 1\}\,.
\end{align}
Here, we see that not only the values ${1\over Z}$ and ${1\over Z}-1$ are possible, 
corresponding to the particle and hole picture, respectively, when no edge state enters/leaves the band
during the shift. For a single band the edge states can enter and leave at the bottom or the top of the
band. Therefore, not only the edge pole encircling the branch cut between band $\alpha$ and $\alpha+1$
is relevant but also the one between band $\alpha-1$ and $\alpha$. It is obvious that the phase
dependence can not always be chosen such that {\it both} edge poles avoid crossing the energy of the
band edges. As a result the edge states have to be taken into account and another value becomes 
possible for $\Delta Q_B^{(\alpha)}$. Comparing (\ref{eq:QB_alpha_change_universal}) with
(\ref{eq:QB_change_Delta_m}) for $\Delta m=1$ we see that the winding number
$w_\alpha$ of the phase difference of the Bloch wave function between site
$m=1$ and $m=0$ can only take three possible values
\begin{align}
\label{eq:winding_bloch_values}
w_\alpha\in\{0,\pm 1\}\,.
\end{align}
This result will be proven analytically in the next Section~\ref{sec:invariant_single_band}.

Finally, we note that the result (\ref{eq:QB_change_universal}) holds only for $1$-channel 
tight-binding models where the spectrum is non-degenerate. For multi-channel systems 
with $N_c$ weakly coupled channels several edge states encircle each branch cut of 
Fig.~\ref{fig:no_edge} and it is no longer possible to choose the phase-dependence
of $v_j(\varphi)$ and $t_j(\varphi)$ in such a way that no edge state crosses $\mu_\nu$ 
on a phase interval of size ${2\pi\over Z}$. \cite{multi-channel} For $N_c$ weakly 
coupled channels (such that we still have $Z-1$ gaps) $N_c\nu$ bands are filled when
the chemical potential is located in gap $\nu$. Therefore, instead of 
Eqs.~(\ref{eq:rho_p}) and (\ref{eq:rho_h}), we get the following result for the 
particle and hole charge densities 
\begin{align}
\label{eq:rho_ph_Nc}
\rho_p(m) = \rho(m) \quad&,\quad \rho_h(m) = \rho(m) - N_c \,,\\ 
\label{eq:rho_ph_bar_Nc}
\bar{\rho}_p = \bar{\rho} \quad&,\quad \bar{\rho}_h = \bar{\rho} - N_c\,, 
\end{align}
with $\bar{\rho}=N_c\nu/Z$.
Since the particle and hole boundary charges are again the same we find the following
two possibilities for the change of the boundary charge when no edge states move above/below
$\mu_\nu$ during the shift by one lattice site
\begin{align}
\label{eq:QB_Nc_change_no_edge_1}
\Delta Q_B &= \Delta Q_{B,p} = \bar{\rho}_p = \bar{\rho} \,,\\
\label{eq:QB_Nc_change_no_edge_2}
\Delta Q_B &= \Delta Q_{B,h} = \bar{\rho}_h = \bar{\rho} - N_c \,.
\end{align}
Since this result can only be changed $\text{mod}(1)$ when edge states move above or below
$\mu_\nu$ during the shift it is reasonable that the allowed values are given by 
\begin{align}
\label{eq:QB_Nc_change_universal}
\Delta Q_B(\varphi,\mu_\nu)\in \{\bar{\rho},\bar{\rho}-1,\dots,\bar{\rho}-N_c\}\,,
\end{align}
since this is obviously the limiting result for vanishing coupling between the channels 
(where we can just add up (\ref{eq:QB_change_universal}) independently). The case of several
channels will be discussed in more detail in a future work \cite{multi-channel}.

\subsection{Invariant and boundary charge for a single band}
\label{sec:invariant_single_band}

In this section we consider the case of a single band. We define the invariant
\begin{align}
\label{eq:invariant_alpha}
I_\alpha(\varphi) = \Delta Q_B^{(\alpha)} - {1\over Z} = -w_\alpha(\varphi)\,,
\end{align}
which is identical to the negative winding number and is an integer for all $\varphi$. 
We start with the proof of (\ref{eq:winding_bloch_values}), 
which states that the invariant can only take the values
\begin{align}
\label{eq:I_w_alpha_values}
I_\alpha \in \{0,\pm 1\}\,.
\end{align}
We first use (\ref{eq:relation_winding}) to relate $w_\alpha=1+\bar{w}_\alpha$ to the winding number 
$\bar{w}_\alpha$ of the phase $\varphi_k^{(\alpha)}(1)$ of the first component $\chi_k^{(\alpha)}(1)$ 
of the Bloch state. (\ref{eq:I_w_alpha_values}) then requires the proof of 
\begin{align}
\label{eq:winding_w_alpha}
\bar{w}_\alpha\in\{0,-1,-2\}\,.
\end{align}
To show that only these particular values are allowed 
we use the form (\ref{eq:a_1_edge_tilde_d}) for $a^{(\alpha)}_k(1)=|a^{(\alpha)}_k(1)|e^{i\varphi^{(\alpha)}_k(1)}$,
which shows that $\bar{w}_\alpha$ is the winding of the complex number 
$q^{(\alpha)}_k=(t_Z/\bar{t}^Z)a^{(\alpha)}_k(1)$ (when $k$ changes from zero to $2\pi$) with
\begin{align}
\label{eq:q_k}
q^{(\alpha)}_k &= \tilde{d}^{(\alpha)}_k e^{-ik} + 1 \,,\\
\label{eq:d_k}
\tilde{d}^{(\alpha)}_k&=\tilde{d}^{(\alpha)}_{-k}=\tilde{d}_{2,Z-1}(\epsilon_k^{(\alpha)})\,.
\end{align}
The winding is determined from the number of crossings of $q^{(\alpha)}_k$ 
through the positive and negative real axis. This happens for 
\begin{align}
\label{eq:qk_real_1}
& (1) \,\, k=0:\quad  q^{(\alpha)}_0 = d^{(\alpha)}_0 + 1 \\
\label{eq:qk_real_2}
& (2) \,\, k=\pi:\quad  q^{(\alpha)}_\pi = -d^{(\alpha)}_\pi + 1 \\
\label{eq:qk_real_3}
& (3) \,\, \tilde{d}^{(\alpha)}_k=0:\quad  q^{(\alpha)}_k = q^{(\alpha)}_{-k} = 1  \,.
\end{align}
It can not happen that no crossing appears on the positive real axis since this would
require $d^{(\alpha)}_0 < -1$, $d^{(\alpha)}_\pi > 1$ and $d^{(\alpha)}_k\ne 0$ for all $k$. This
is obviously not possible.
The winding number then follows from the number of crossings through the negative real axis
(note that the number of crossings through $q^{(\alpha)}_k = 1$ for $k\ne 0,\pi$ is an even number due to
$q^{(\alpha)}_k = q^{(\alpha)}_{-k}$)
\begin{align}
\label{eq:w_alpha_values}
\bar{w}_\alpha = \begin{cases}
0 \quad \,\,\,\,\text{for} \quad \tilde{d}^{(\alpha)}_0 > -1\,,\,\tilde{d}^{(\alpha)}_\pi < 1 \\
-1 \quad \text{for} \quad \tilde{d}^{(\alpha)}_0 < -1\,,\,\tilde{d}^{(\alpha)}_\pi < 1 \\
-1 \quad \text{for} \quad \tilde{d}^{(\alpha)}_0 > -1\,,\,\tilde{d}^{(\alpha)}_\pi > 1 \\
-2 \quad \text{for} \quad \tilde{d}^{(\alpha)}_0 < -1\,,\,\tilde{d}^{(\alpha)}_\pi > 1 
\end{cases}.
\end{align}
This proves Eq.~(\ref{eq:I_w_alpha_values}). 

\begin{figure}
  \centering
  \includegraphics[width= 0.9\columnwidth]{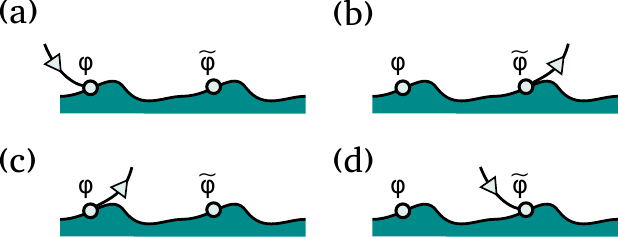} 
  \caption{Four scenarios how edge states can enter/leave a band for phase $\varphi$ and
    $\tilde{\varphi}=\varphi+{2\pi\over Z}$. We have shown a band top but the
    same behaviour occurs for a band bottom. The quasimomentum at the band edge can be either $k_0=0$
    or $k_0=\pi$. For (a) and (d) an edge state enters at $\varphi$ and $\tilde{\varphi}$, 
    respectively, such that the sign of ${d \over d\varphi}d_1^{(\alpha)}$ is given by $(-1)^{k_0/\pi}$
    at the entering point according to (\ref{eq:leaving_entering}). Analog, for (b) and (c) the edge 
    state leaves such that the sign is given by $-(-1)^{k_0/\pi}$ at the leaving point.
    Together with (\ref{eq:w_alpha_values}) and (\ref{eq:d_12_property3}) this gives the correct 
    correlation to the jump of $I_\alpha$ by $-1$ for (a) and (b) and by $+1$ for (c) and (d), see
    explanation in the main text.
  }
  \label{fig:edge_cases}
\end{figure}
We now determine the cases where the invariant jumps as function of $\varphi$. This are the points 
where either $\tilde{d}^{(\alpha)}_0(\varphi)=-1$ or $\tilde{d}^{(\alpha)}_\pi(\varphi)=1$.
In both cases we get $d_1^{(\alpha)}(\varphi)=\tilde{d}^{(\alpha)}_{k_0}(\varphi)=-e^{-ik_0}$, where
$k_0=0,\pi$ correspond to values at the band edges and $d_1^{(\alpha)}(\varphi)$ has been defined
in (\ref{eq:d_1}). According to (\ref{eq:property_3}) this means that an edge state appears
at $\varphi$ or at $\tilde{\varphi}=\varphi + {2\pi\over Z}$. Thus, we conclude that the invariant
$I_\alpha$ can only jump by $\pm 1$ at phase $\varphi$ when an edge state enters/leaves the band at phase 
$\varphi$ or at $\tilde{\varphi}$. This is quite obvious and consistent with the definition
(\ref{eq:invariant_alpha}) of the invariant which can only jump when the boundary charge jumps at
$\varphi$ or $\tilde{\varphi}$. Due to charge conservation, the latter happens precisely when an edge
state leaves/enters the band. 

\begin{figure*}
\centering
\begin{minipage}{0.32\textwidth}
  \includegraphics[width= 1.\columnwidth]{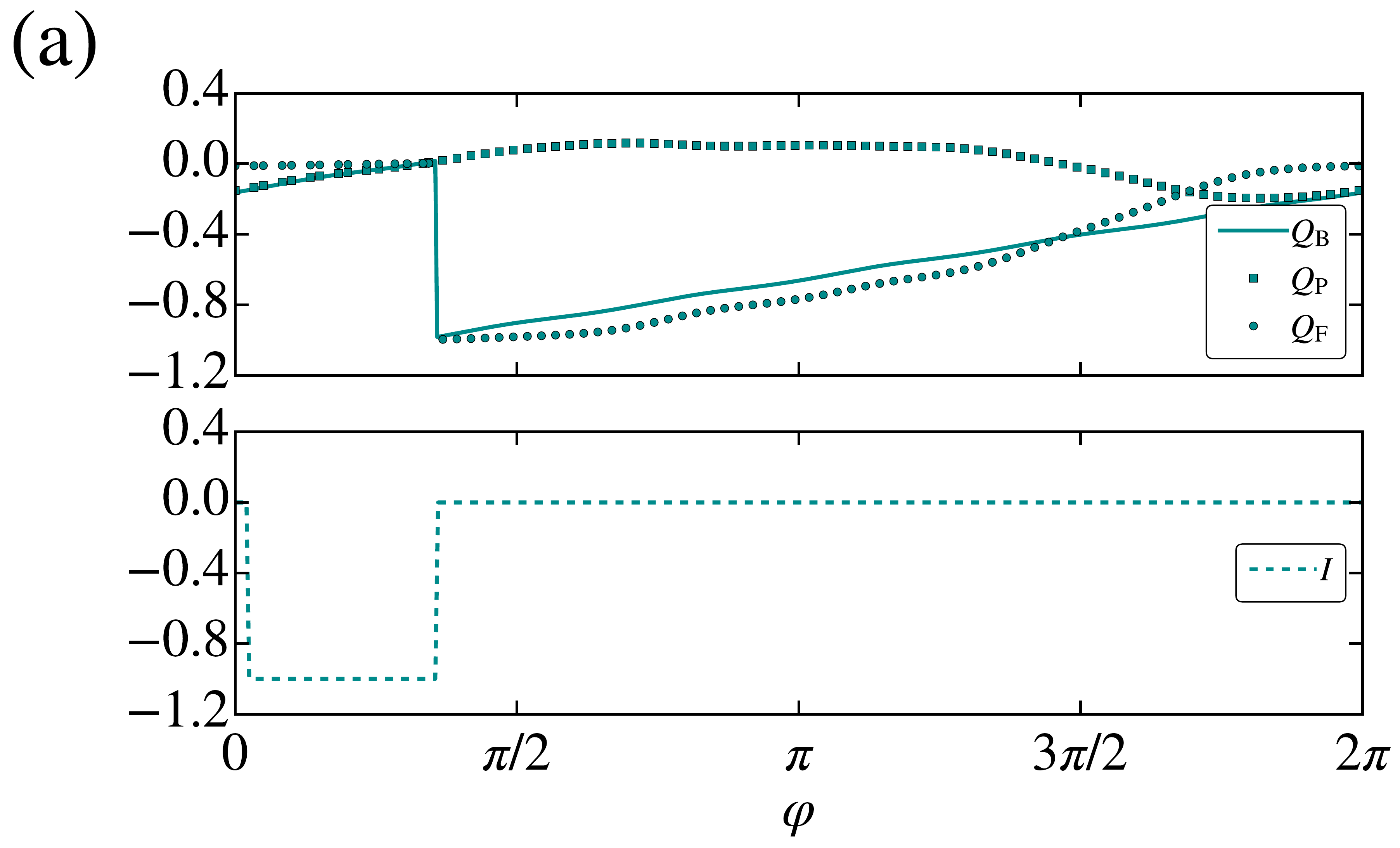}
\end{minipage}
\begin{minipage}{0.32\textwidth}
  \includegraphics[width= 1.\columnwidth]{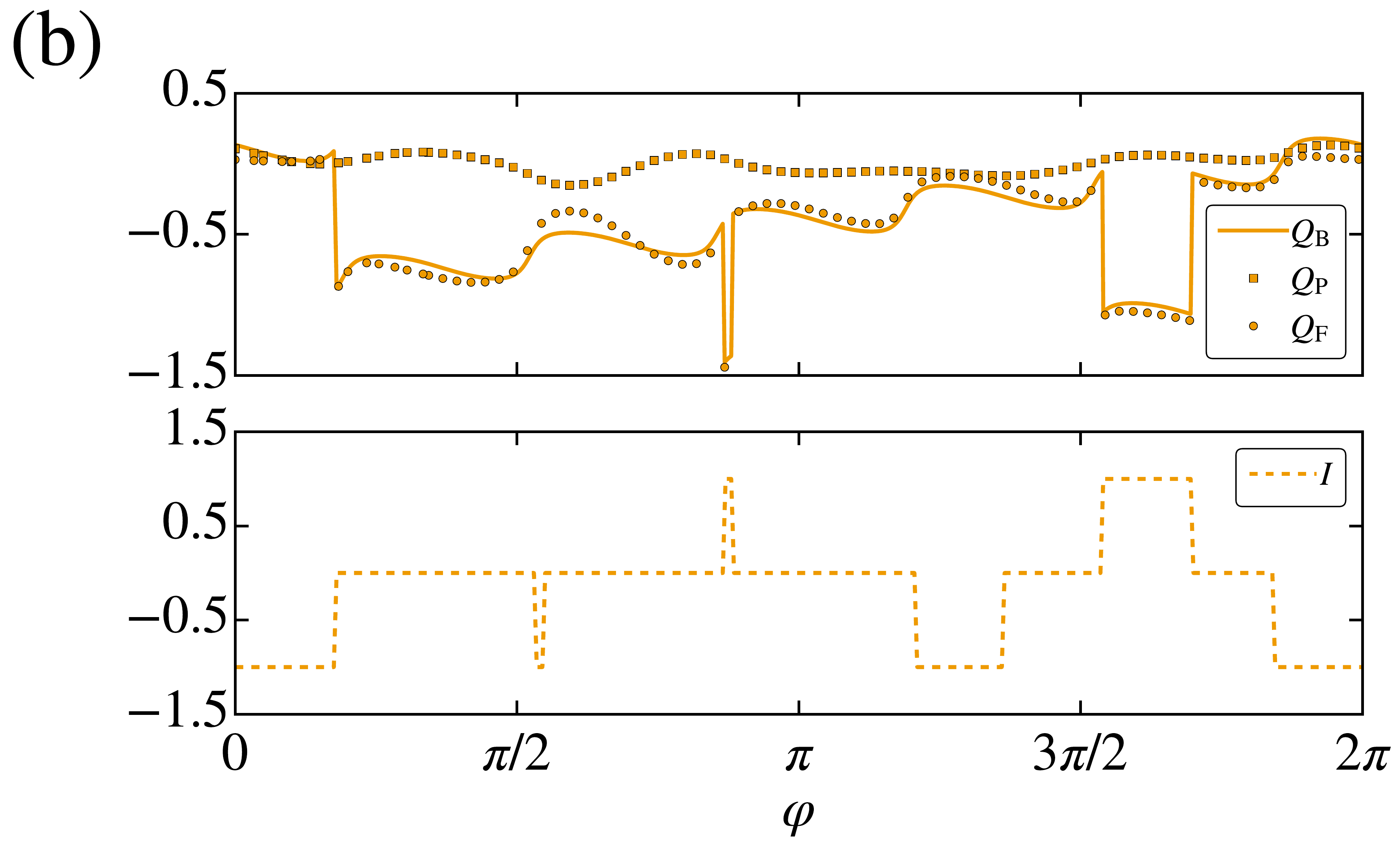}
\end{minipage}
\begin{minipage}{0.32\textwidth}
  \includegraphics[width= 1.\columnwidth]{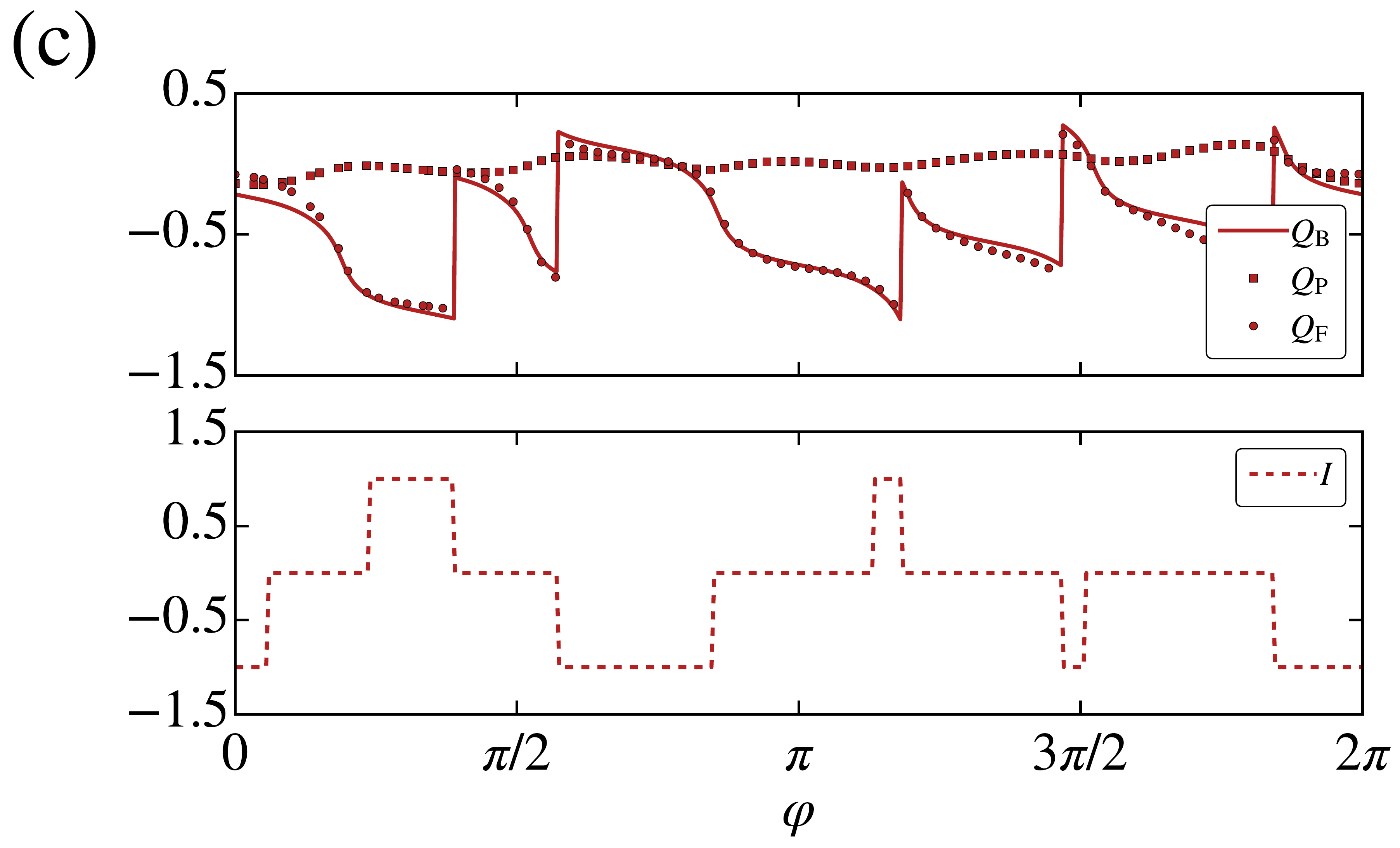}
\end{minipage}
\hfill
 \caption{phase-dependence of $Q_B^{\alpha}$, $Q_F^{(\alpha)}$, $Q_P^{(\alpha)}$ and $I_\alpha$ for
   $\alpha=1,3,4$ in (a), (b) and (c), respectively, using the parameters of Fig.~\ref{fig:edge_2}(a).  
}
\label{fig:QB_single_bands}
\end{figure*}
One can also understand that the jump of $I_\alpha(\varphi)$ by $\mp 1$ corresponds to the cases that
either $Q_B(\varphi)$ jumps by $\pm 1$ (i.e., an edge state enters/leaves band $\alpha$ at $\varphi$) or 
$Q_B(\tilde{\varphi})$ jumps by $\mp 1$ (i.e., an edge state leaves/enters band $\alpha$ at $\tilde{\varphi}$), 
see Fig.~\ref{fig:edge_cases}(a-d). We have depicted
the four possibilities of an edge state entering/leaving at phase $\varphi$ or at 
$\tilde{\varphi}=\varphi + {2\pi\over Z}$ in Figs.~\ref{fig:edge_cases}(a-d). For the cases shown in 
Figs.~\ref{fig:edge_cases}(a,b) and Figs.~\ref{fig:edge_cases}(c,d) the invariant changes 
by $-1$ and $+1$, respectively. Using (\ref{eq:leaving_entering}) we see that 
$\text{sign}\left\{{d \over d\varphi}d_1^{(\alpha)}(\varphi)\right\} = \pm (-1)^{k_0/\pi}$ for 
Fig.~\ref{fig:edge_cases}(a) and Fig.~\ref{fig:edge_cases}(c), respectively, 
where an edge state enters/leaves at $\varphi$, and 
$\text{sign}\left\{{d \over d\varphi}d_1^{(\alpha)}(\tilde{\varphi})\right\} = \mp(-1)^{k_0/\pi}$
for Fig.~\ref{fig:edge_cases}(b) and Fig.~\ref{fig:edge_cases}(d), respectively, 
where an edge state leaves/enters at $\tilde{\varphi}$. Using (\ref{eq:d_12_property3}), this gives 
$\text{sign}\left\{{d \over d\varphi}d_1^{(\alpha)}(\varphi)\right\} = \pm(-1)^{k_0/\pi}$ for
Figs.~\ref{fig:edge_cases}(a,b) and Figs.~\ref{fig:edge_cases}(c,d), respectively. From
(\ref{eq:w_alpha_values}) we conclude that the winding number $\bar{w}_\alpha$
jumps by $\pm 1$ for these two cases, respectively, which corresponds to the invariant jumping by $\mp 1$ since
$I_\alpha=-1-\bar{w}_\alpha$, see Eqs.~(\ref{eq:invariant_alpha}) and (\ref{eq:relation_winding}).

The fact that the boundary charge $Q_B^{(\alpha)}$ can only have discontinuous jumps by $\pm 1$ 
when edge states enter/leave the band at the phase values $\varphi_{i\pm}^{(\alpha)}$, with
$i=1,\dots,M_\pm^{(\alpha)}$, leads together with (\ref{eq:QB_alpha_change_universal}) to the 
following shape of the phase-dependence of $Q_B^{(\alpha)}$
\begin{align}
\label{eq:QB_alpha_phase}
Q_B^{(\alpha)}(\varphi) = f_\alpha(\varphi) + {M^{(\alpha)}\over 2\pi}\varphi 
+ F_\alpha(\varphi) \,,
\end{align}
where 
\begin{align}
\label{eq:f_alpha}
f_\alpha(\varphi) = f_\alpha(\varphi+{2\pi\over Z})
\end{align}
is an unknown non-universal ${2\pi\over Z}$-periodic function and 
\begin{align}
\label{eq:F_alpha}
F_\alpha(\varphi) = \sum_{\sigma=\pm} \sum_{i=1}^{M_\sigma^{(\alpha)}} \sigma \theta(\varphi-\varphi_{i\sigma}^{(\alpha)})
\end{align}
is the part describing the discontinuous jumps from edge states entering/leaving the band. 
Here, 
\begin{align}
\label{eq:M_alpha}
M^{(\alpha)} = M_-^{(\alpha)} - M_+^{(\alpha)}
\end{align}
is the net number of edge states leaving the band for a phase change
by $2\pi$ which is given by the Chern number $C^{(\alpha)}$ of band $\alpha$, see (\ref{eq:chern_M}).
Using the form (\ref{eq:QB_alpha_phase}) 
we get $\Delta Q_B^{(\alpha)}(\varphi)=M^{(\alpha)}/Z + \Delta F_\alpha(\varphi)$ which,
together with $\Delta Q_B^{(\alpha)}(\varphi)=I_\alpha(\varphi)+{1\over Z}$ leads to
\begin{align}
\label{eq:diophantine_alpha}
M^{(\alpha)} = 1 - s^{(\alpha)} Z \,,
\end{align}
where 
\begin{align}
\label{eq:s_alpha} 
s^{(\alpha)} = \Delta F_\alpha(\varphi) - I_\alpha(\varphi) 
\end{align}
is a characteristic and phase independent integer for band $\alpha$. Eq.~(\ref{eq:diophantine_alpha}) 
just describes charge conservation. When the phase changes by $2\pi$ the charge $1\,\text{mod}(Z)$
is pumped into the boundary which has to be taken away by a corresponding number $M^{(\alpha)}$ of edge 
states. This equation is also called the Diophantine equation discussed within the integer QHE 
[\onlinecite{dana_jpc_85}-\onlinecite{hatsugai_prb_93}], here derived for a single band. 
To get (\ref{eq:QB_alpha_phase}), we have only used that the invariant $I_\alpha$ is an integer.
An additional topological constraint how the edge states can enter/leave a band follow from the allowed 
values $I_\alpha\in\{0,\pm 1\}$. For given $s^{(\alpha)}$ we get
\begin{align}
\label{eq:selection_rule_alpha}
\Delta F_\alpha(\varphi)\in \{s^{(\alpha)},s^{(\alpha)}\pm 1\}\,.
\end{align}

It is important to notice that $M^{(\alpha)}$ and, consequently, also the integer 
$s^{(\alpha)}$ depend crucially on the choice of the functions $F_v$ and $F_t$ 
defining the model parameters via Eqs.~(\ref{eq:v_form}) and (\ref{eq:t_form}). As we have discussed in
Section~\ref{sec:invariant_physics} for given parameters at phase $\varphi=0$ the phase-dependence can
always be chosen such that, for a given band $\alpha$, one of the following two cases can be realized 
\begin{align}
\label{eq:delta_M_case1}
M^{(\alpha)} = 1 \quad &\Leftrightarrow \quad s^{(\alpha)} = 0 \,,\\
\label{eq:delta_M_case2}
M^{(\alpha)} = 1-Z \quad &\Leftrightarrow \quad s^{(\alpha)} = 1 \,.
\end{align}
This are also the most frequently obtained values for rather smooth functions $F_v(\varphi)$ and $F_t(\varphi)$
where either of the two cases occurs for {\it all} bands.  
Additional multiples of $Z$ are obtained for $M^{(\alpha)}$ if the phase-dependence contains higher Fourier
components which is rather exotic. Therefore, in the following we will only consider a phase-dependence
where one of the two cases of Eqs.~(\ref{eq:delta_M_case1}) and (\ref{eq:delta_M_case2}) occurs for each band.

We note that for a rational wave length $\lambda={Z\over p}$ of the modulation where the phase 
parametrization is chosen according to 
Eqs.~(\ref{eq:v_form_p}) and (\ref{eq:t_form_p}) we get analog results for the invariant but it 
has to be defined differently since a shift by one lattice site corresponds to a phase change 
${2\pi p\over Z}$
\begin{align}
\label{eq:invariant_alpha_p}
I_\alpha(\varphi) &= \Delta Q^{(\alpha)}_B(\varphi) - {1\over Z} \in \{0,\pm 1\}\,,\\
\label{eq:Delta_QB_alpha_p}
\Delta Q^{(\alpha)}_B(\varphi) &= Q^{(\alpha)}_B(\varphi+{2\pi p\over Z}) - Q^{(\alpha)}_B(\varphi)\,.
\end{align}
The form (\ref{eq:QB_alpha_phase}) of the phase-dependence of 
$Q^{(\alpha)}_B(\varphi)=Q^{(\alpha)}_B(\varphi+2\pi)$ 
does not change, in particular the function $f_\alpha(\varphi)$ has precisely the same periodicity 
(\ref{eq:f_alpha}) with period ${2\pi\over Z}$. The latter is proven as follows. First, shifting
the lattice by one site via a phase change of ${2\pi p\over Z}$ requires the periodicity
$f_\alpha(\varphi)=f_\alpha(\varphi + {2\pi p\over Z})$ due to (\ref{eq:invariant_alpha_p}) and 
the smoothness of $f_\alpha(\varphi)$. Furthermore the overall system does not change when $\varphi$
changes by $2\pi$ leading to the second condition $f_\alpha(\varphi)=f_\alpha(\varphi+2\pi)$. Since
$p$ and $Z$ are incommensurate one can always find integers $n$ and $m$ such that $mp=1+nZ$ or
${2\pi\over Z}=2\pi n+{2\pi p\over Z}m$. As a consequence, the function $f_\alpha(\varphi)$ must have the 
period ${2\pi\over Z}$. The only equation which changes is the Diophantine equation 
(\ref{eq:diophantine_alpha}) which gets the different form
\begin{align}
\label{eq:diophantine_alpha_p}
p M^{(\alpha)} = 1 - s^{(\alpha)} Z \,,
\end{align}
where $p$ and $s^{(\alpha)}$ must be such that $M^{(\alpha)}$ is an integer. Again we note that this can always be
solved if $Z$ and $p$ are incommensurate. The topological constraint formulated via Eqs.~(\ref{eq:s_alpha})
and (\ref{eq:selection_rule_alpha}) remains the same.

In Fig.~\ref{fig:QB_single_bands}(a)-(c) we show the phase-dependence of the boundary charge
$Q_B^{(\alpha)}$, the Friedel charge $Q_F^{(\alpha)}$, the polarization charge $Q_P^{(\alpha)}$, and
the invariant $I_\alpha$ for the bands $\alpha=1,3,4$ of Fig.~\ref{fig:edge_2}(a). In
Fig.~\ref{fig:edge_2}(a) we find $M_-^{(\alpha)}-M_+^{(\alpha)}=1$ for $\alpha\ne 4$ and 
$M_-^{(4)}-M_+^{(4)}=1-Z$, corresponding to the particle and hole picture, respectively. 
Therefore, except for $\alpha=4$, the linear term in Eq.~(\ref{eq:QB_alpha_phase})
leads to an average increase of $Q_B{(\alpha)}$ by ${1\over Z}$ on the phase interval ${2\pi\over Z}$. 
On top of this linear function edge states can enter or leave the band 
leading to discontinuous jumps of $Q_B^{(\alpha)}$ by $\pm 1$ and the function $f_\alpha(\varphi)$ 
can lead to further oscillations with period ${2\pi\over Z}$. As can be seen in 
Fig.~\ref{fig:QB_single_bands}(a) for $\alpha=1$ the 
invariant $I_1$ takes only the values $I_1\in\{0,-1\}$ since no edge states can enter/leave at the
band bottom. Up to a jump by $-1$ when an edge state leaves the band $Q_B^{(1)}$ is almost a 
linear function with average slope ${1\over 2\pi}$. However, this holds only for the sum
of the Friedel and polarization charge, neither $Q_F^{(1)}$ nor $Q_P^{(1)}$ alone show any linear behaviour 
on average. In Fig.~\ref{fig:QB_single_bands}(b) we show the phase-dependence for band $\alpha=3$.  
The average slope of $Q_B^{(3)}$ is the same but the function $f_3(\varphi)$ is rather large and influences
the result strongly, leading even to a negative slope of $Q_B^{(3)}$ between the jumps. The invariant  
can take all values $I_3\in\{0,\pm 1\}$ since edge states can enter/leave from both sides of the band. The
same happens for band $\alpha=4$ shown in Fig.~\ref{fig:QB_single_bands}(c). However, in this case, the
average slope of $Q_B^{(4)}$ is negative corresponding to the hole picture. Here, all edge states enter the
band leading to $Z-1$ discontinuous jumps by $+1$ which have to be compensated by the large negative slope 
${1-Z\over 2\pi}$ of the linear term.

\begin{figure*}
  \centering
  \includegraphics[width= 0.8\columnwidth]{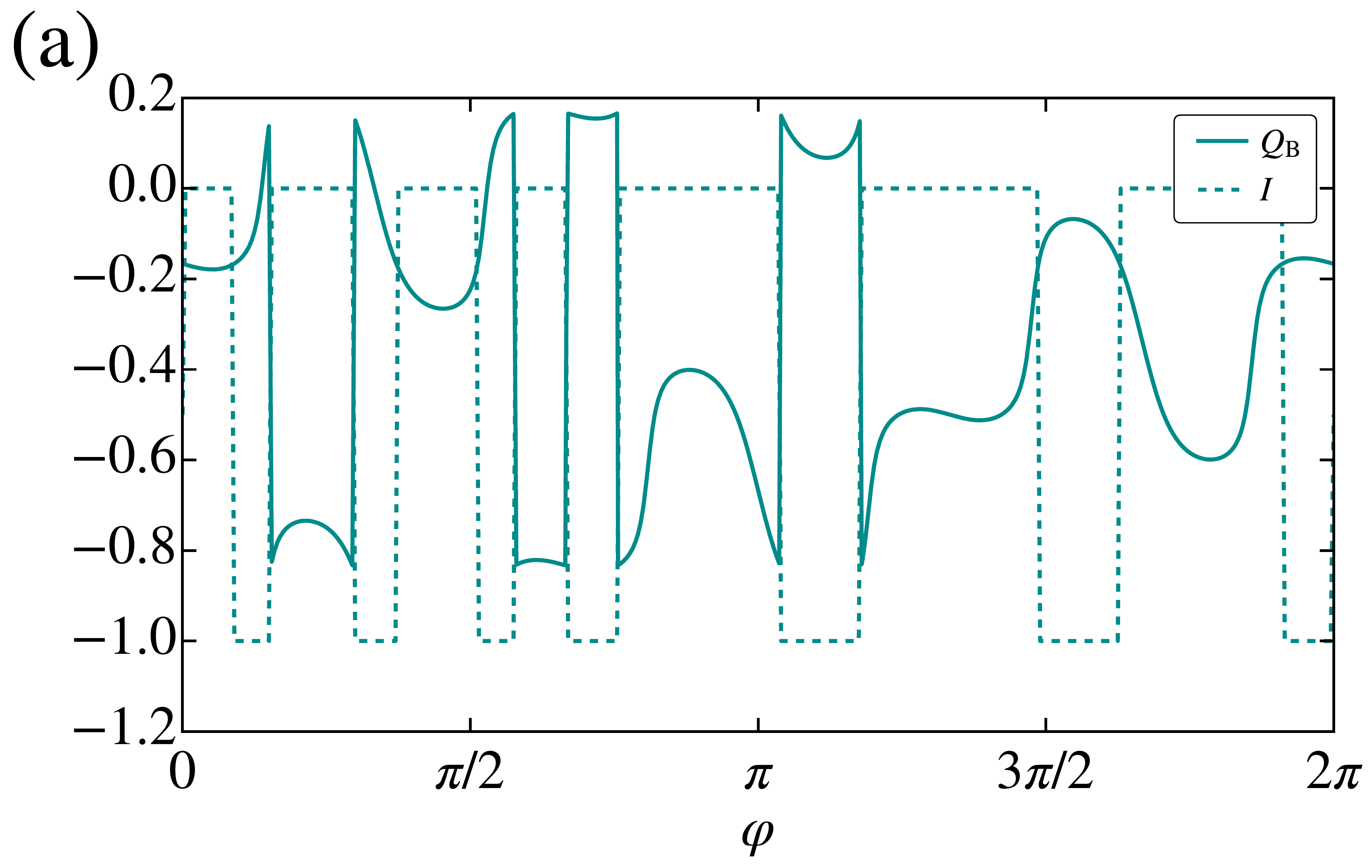}
  \includegraphics[width= 0.8\columnwidth]{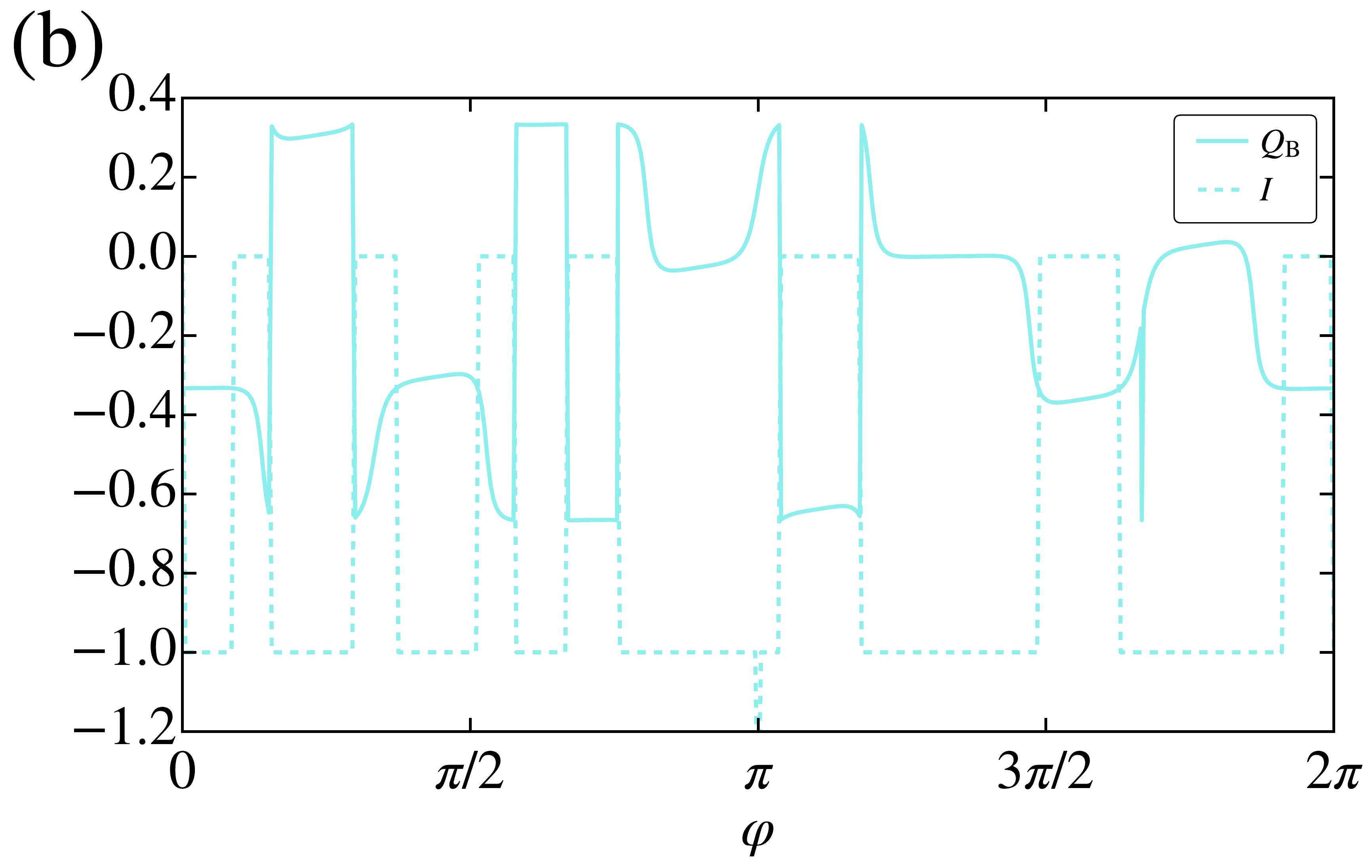}
  \caption{phase-dependence of boundary charge $Q_B(\varphi,\mu_\nu)$ and invariant $I(\varphi,\mu_\nu)$
    for $Z=3$ and (a) $\nu=1$, (b) $\nu=2$ for the parameters of Fig.~\ref{fig:edge_2}(b).}
  \label{fig:QB_invariant_total_Z3}
\end{figure*}
\begin{figure*}
  \centering
  \includegraphics[width= 0.65\columnwidth]{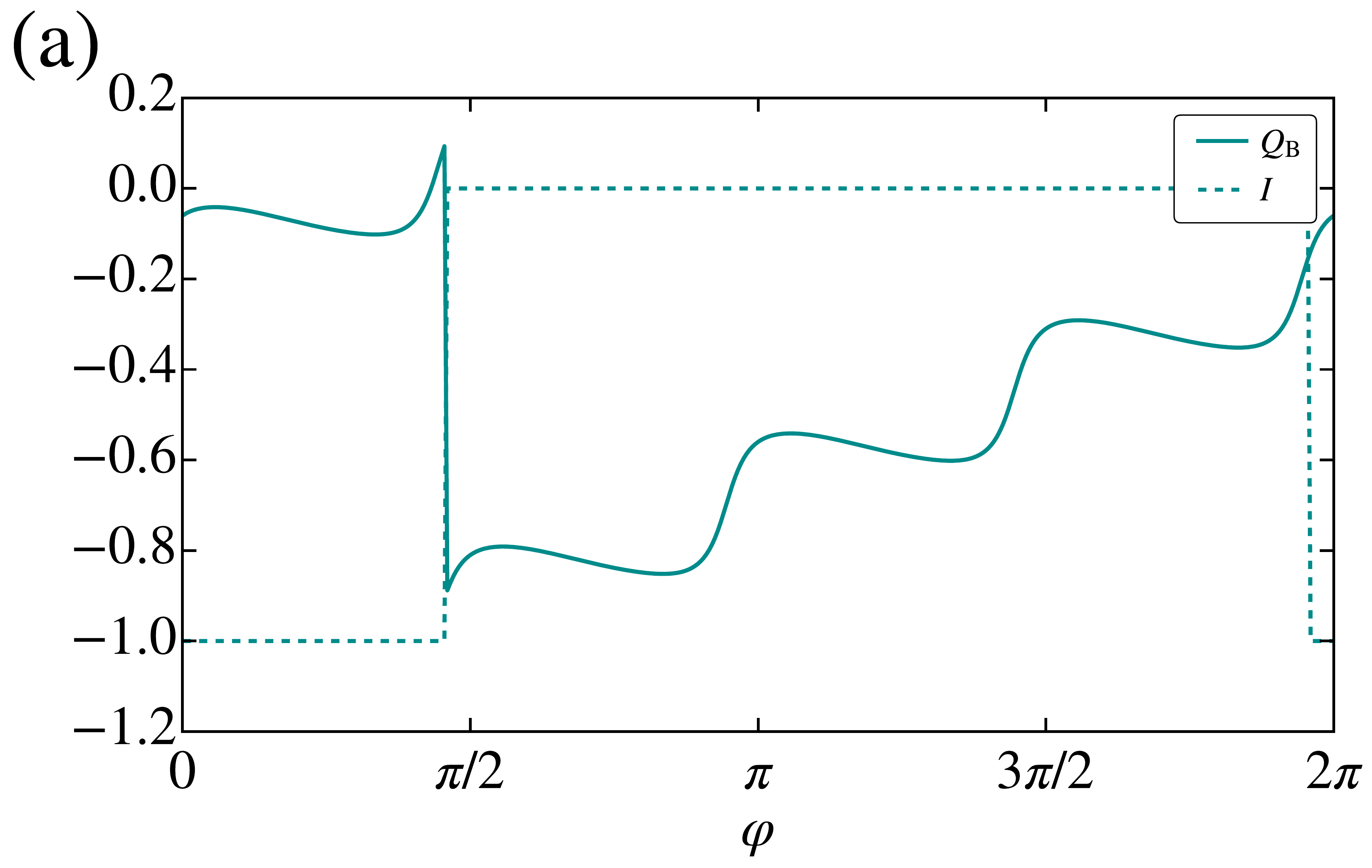}
  \includegraphics[width= 0.65\columnwidth]{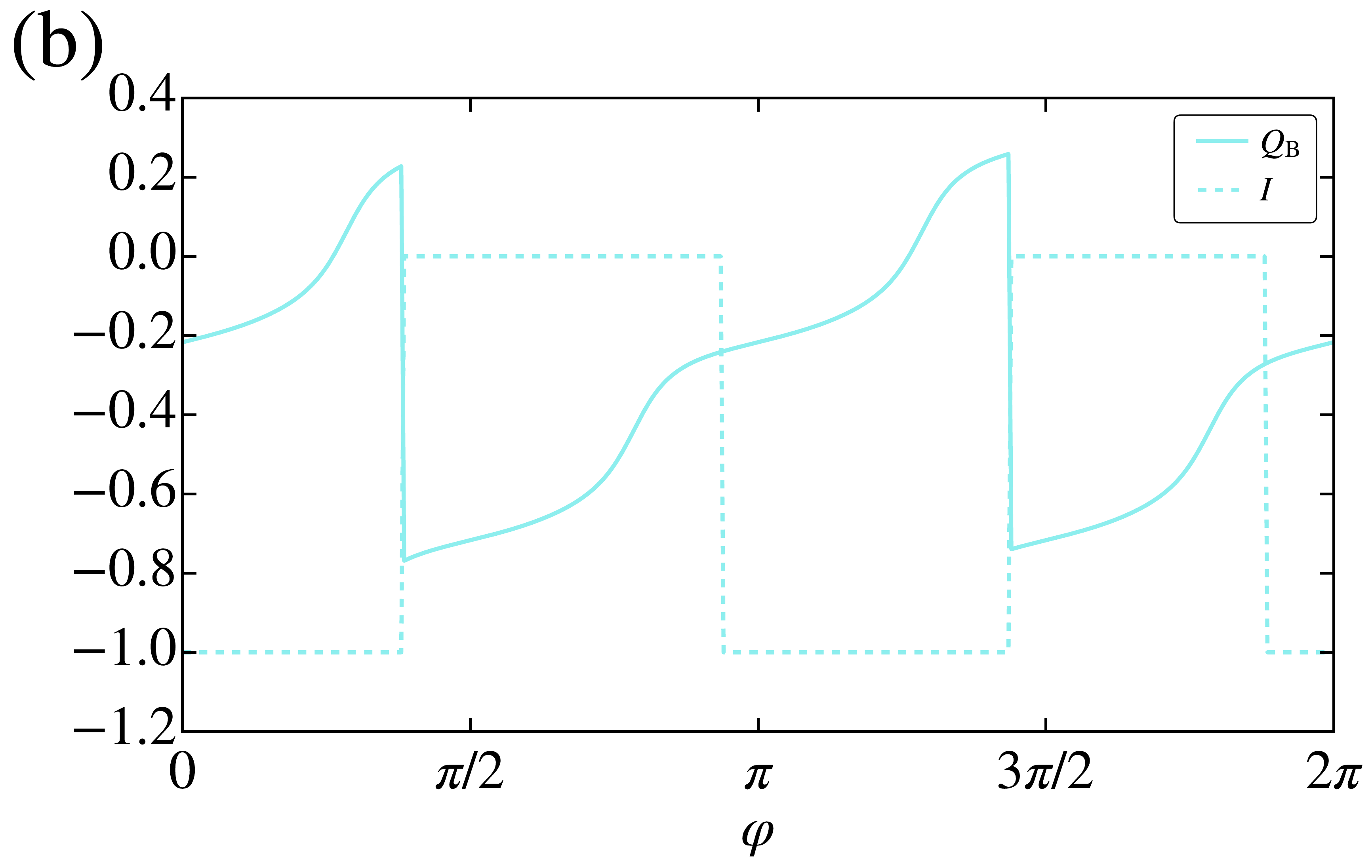}
  \includegraphics[width= 0.65\columnwidth]{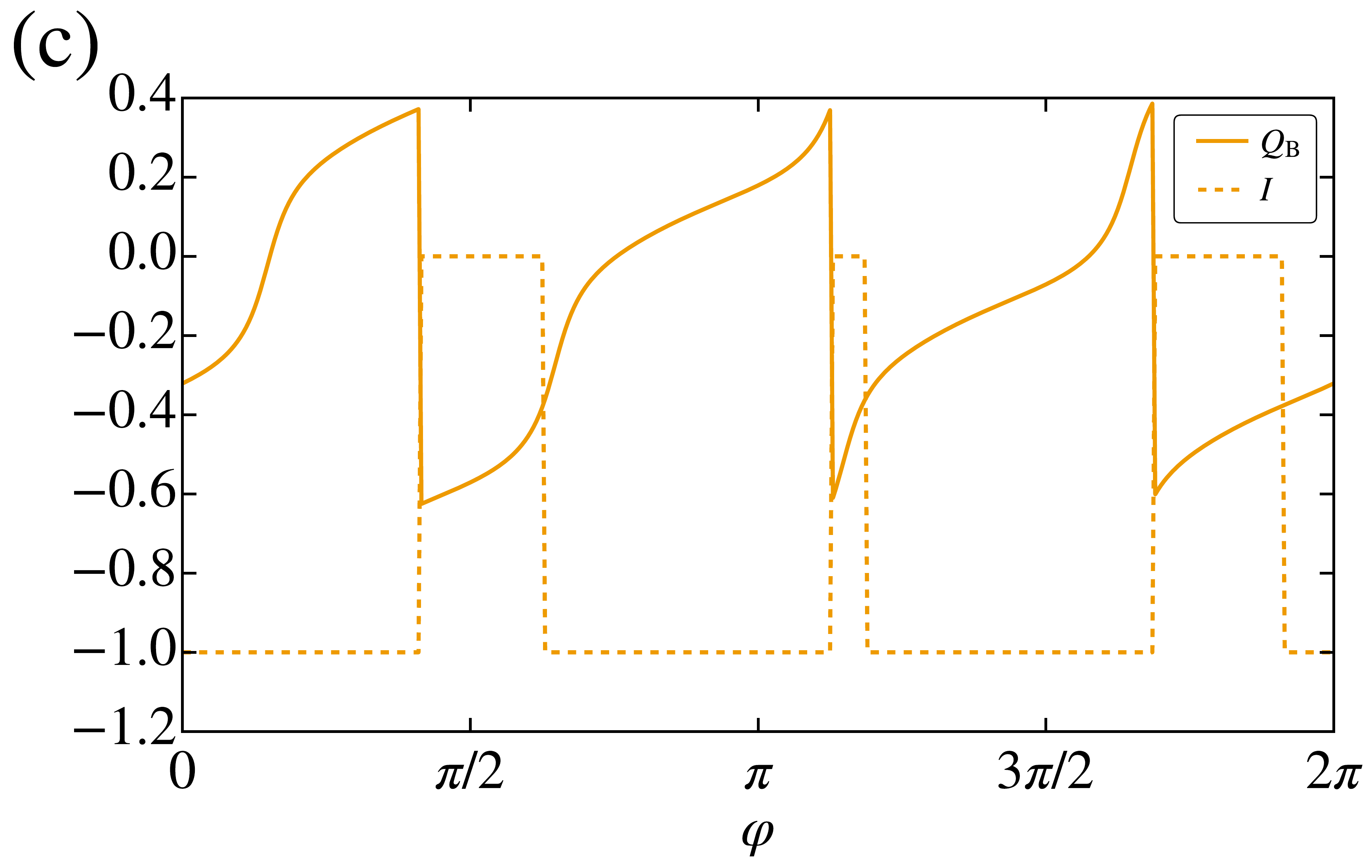}
  \caption{phase-dependence of boundary charge $Q_B(\varphi,\mu_\nu)$ and invariant $I(\varphi,\mu_\nu)$
    for $Z=4$ and (a) $\nu=1$, (b) $\nu=2$, (c) $\nu=3$ for the parameters of Fig.~\ref{fig:edge}(b).}
  \label{fig:QB_invariant_total_Z4}
\end{figure*}
\begin{figure*}
\centering
 \includegraphics[width= 0.65\columnwidth]{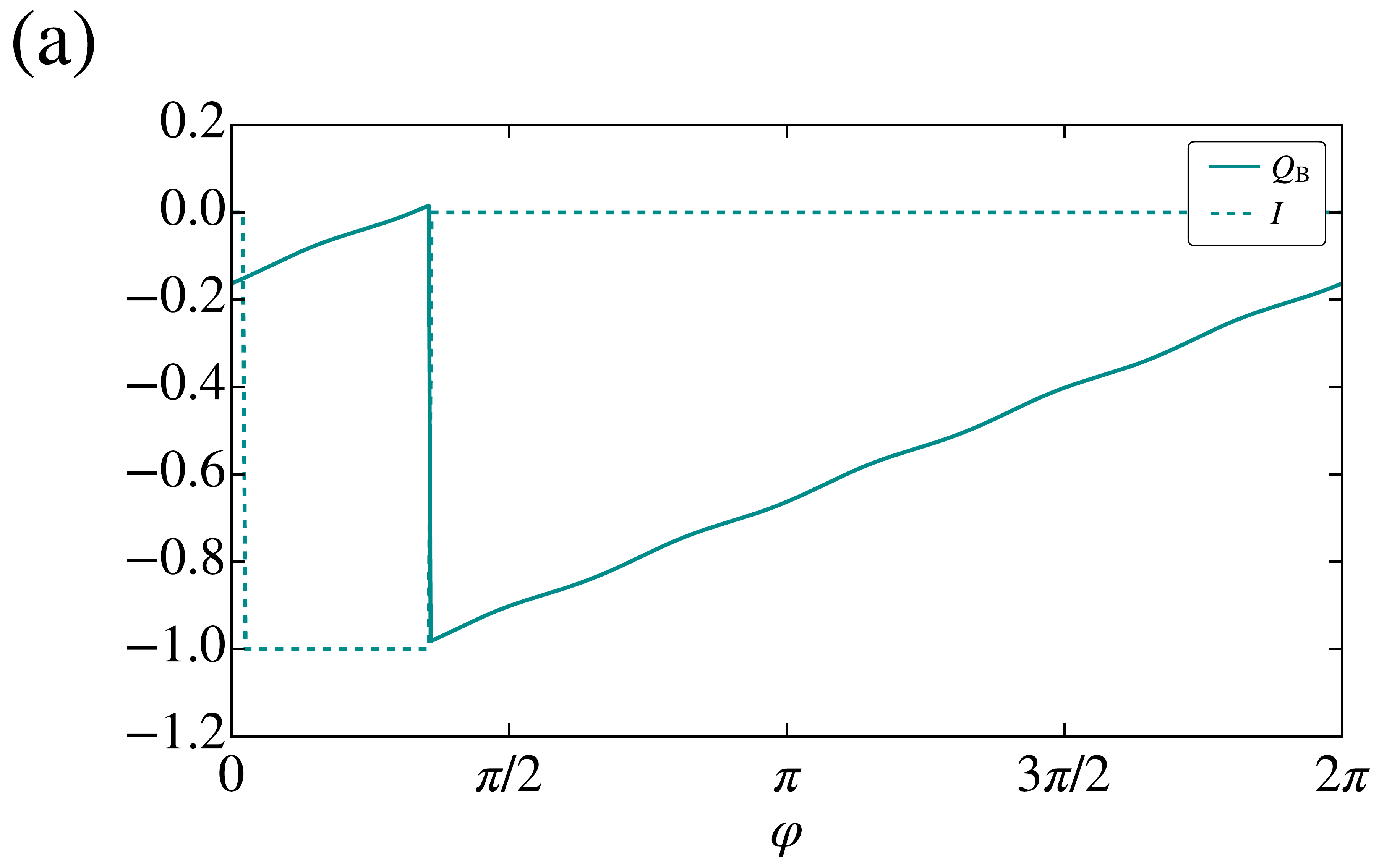}
 \includegraphics[width= 0.65\columnwidth]{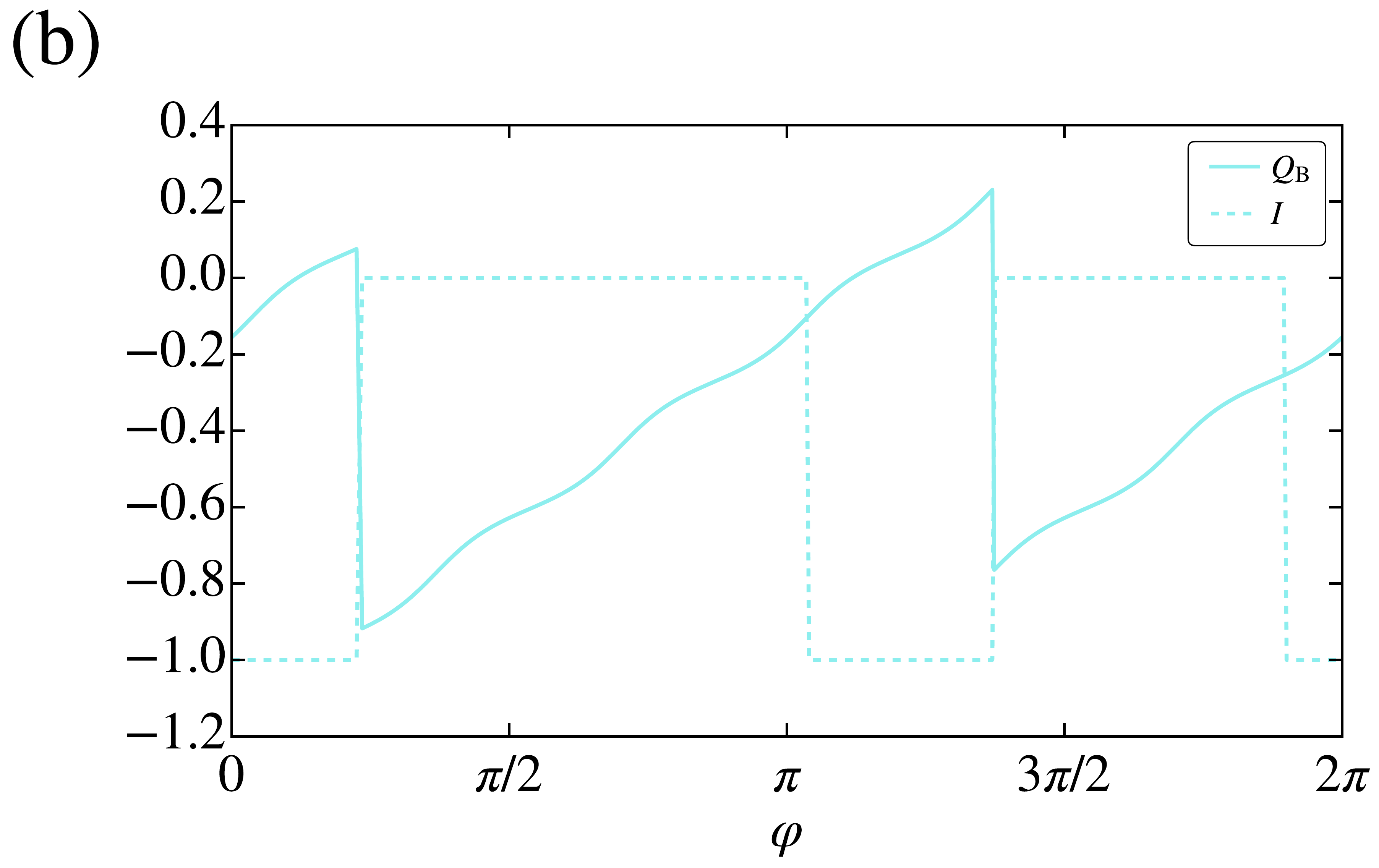}
 \includegraphics[width= 0.65\columnwidth]{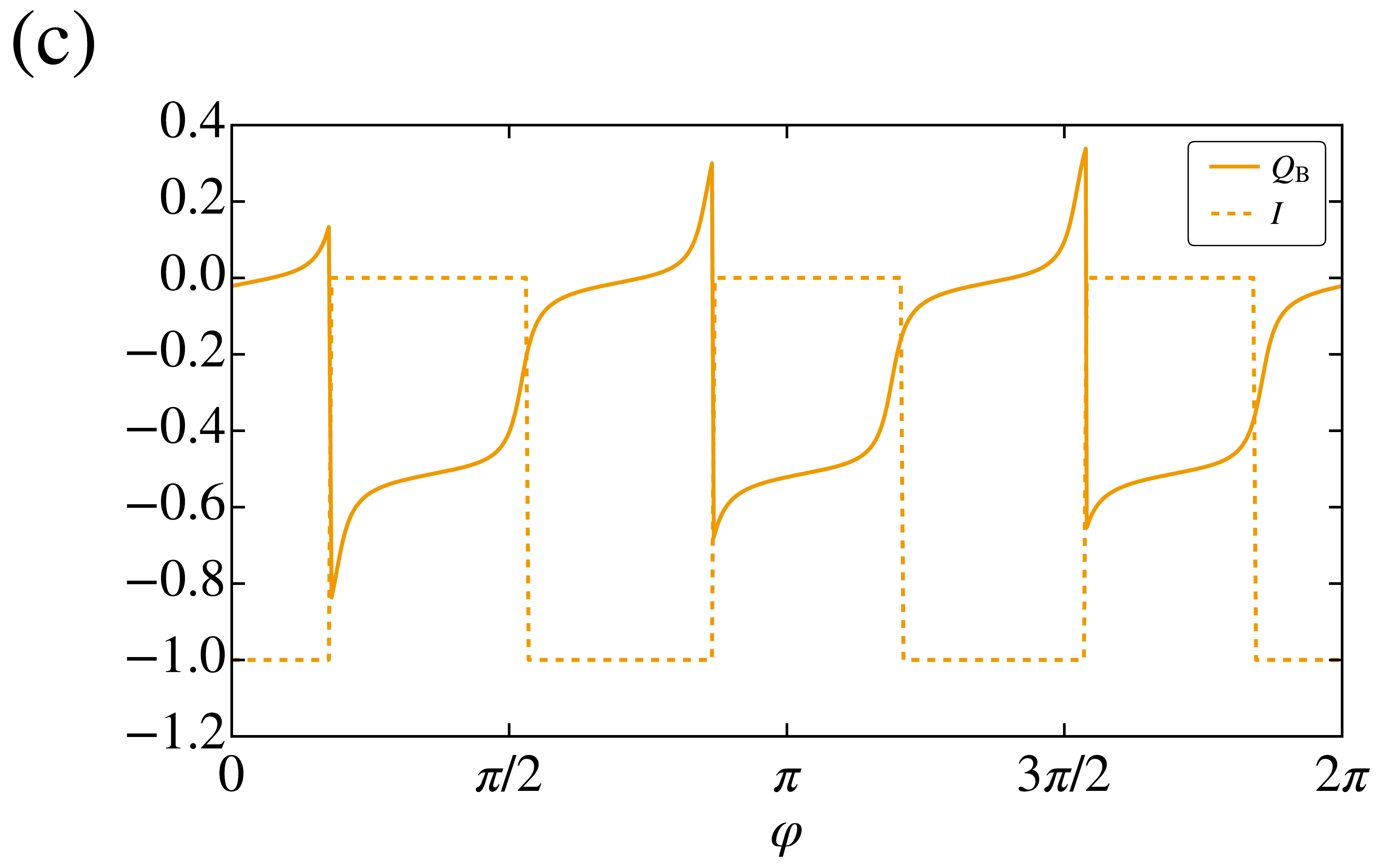}
 \includegraphics[width= 0.65\columnwidth]{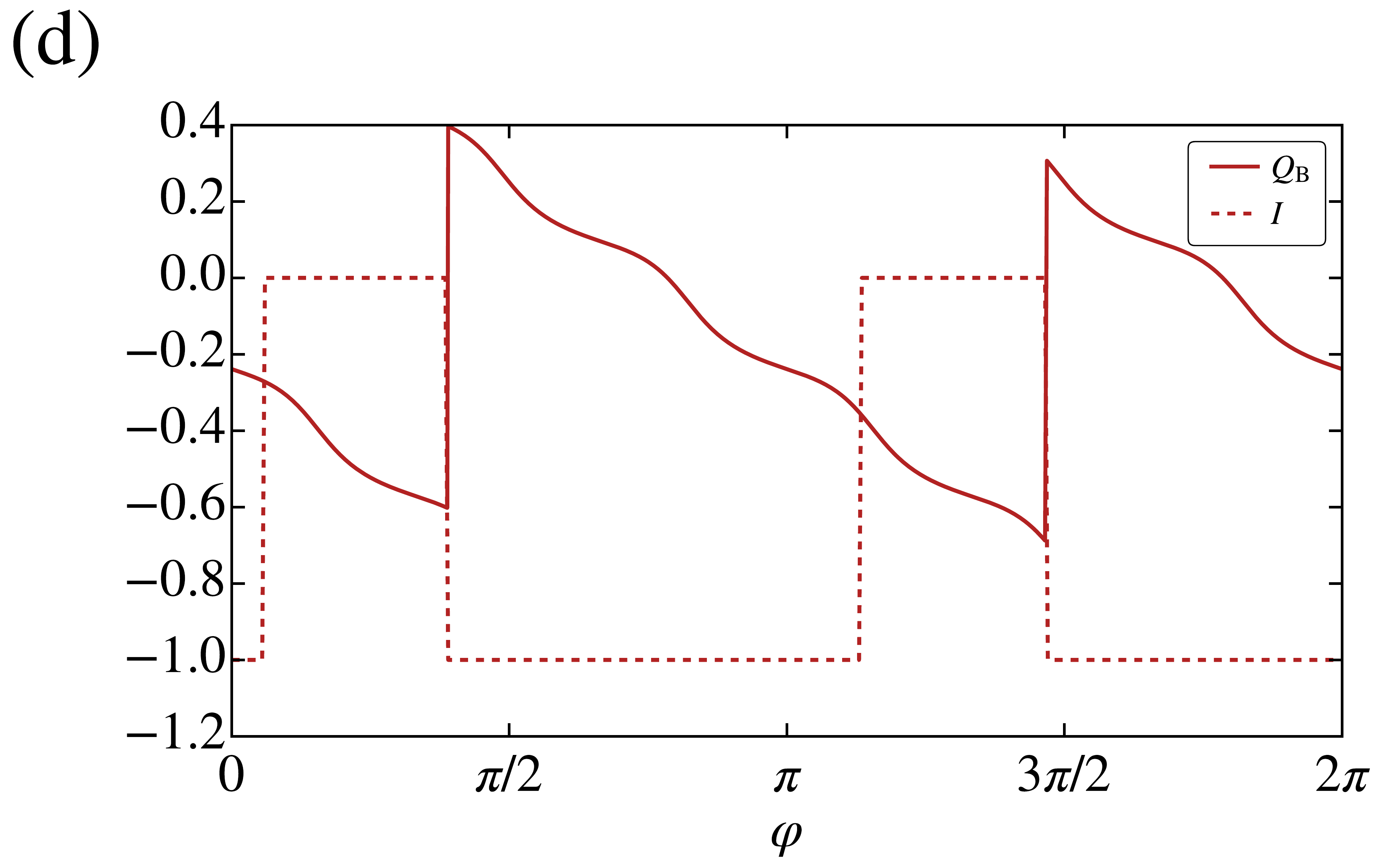}
 \includegraphics[width= 0.65\columnwidth]{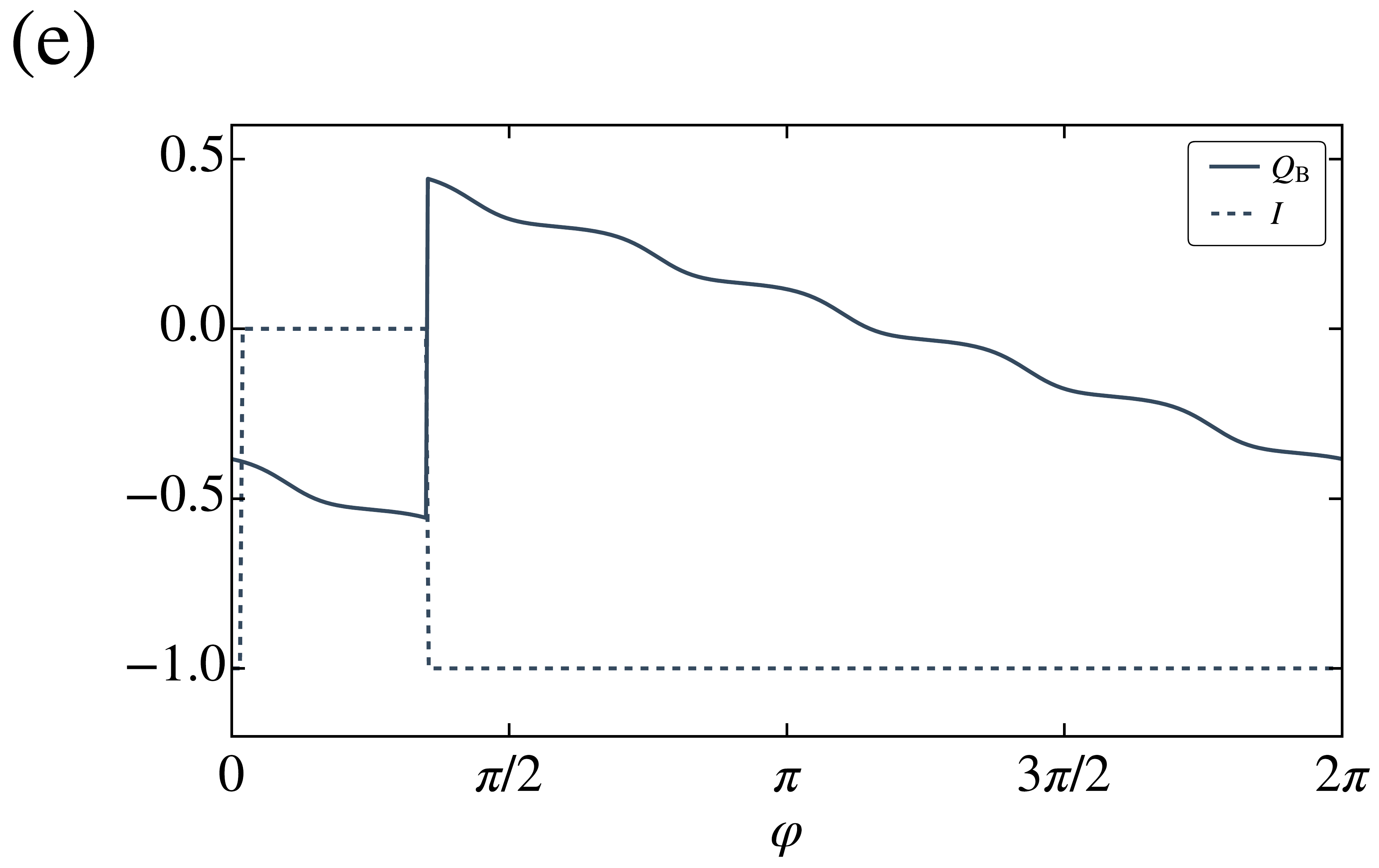}
 \caption{phase-dependence of boundary charge $Q_B(\varphi,\mu_\nu)$ and invariant $I(\varphi,\mu_\nu)$
   for $Z=6$ and (a) $\nu=1$, (b) $\nu=2$, (c) $\nu=3$, (d) $\nu=4$, (e) $\nu=5$ 
   for the parameters of Fig.~\ref{fig:edge_2}(a).}
 \label{fig:QB_invariant_total_Z6}
\end{figure*}
\subsection{Total invariant and boundary charge}
\label{sec:invariant_total}
 
We now consider the case where the chemical potential $\mu_\nu$ is placed somewhere in gap $\nu$
and calculate the total boundary charge $Q_B(\varphi,\mu_\nu)$. 
All our results are also valid if the chemical potential is chosen phase-dependent as long as
it is ${2\pi\over Z}$-periodic
\begin{align}
\label{eq:mu_phase}
\mu_\nu(\varphi) = \mu_\nu\left(\varphi+{2\pi\over Z}\right)\,.
\end{align}
Therefore, we can take for $\mu_\nu$ also the top of band $\alpha=\nu$ or the bottom of band $\alpha=\nu+1$
since $\epsilon_k^{(\alpha)}(\varphi)=\epsilon_k^{(\alpha)}(\varphi+{2\pi\over Z})$ is 
${2\pi\over Z}$-periodic in the phase. The boundary charge $Q_B$ is given via (\ref{eq:QB_splitting}) by summing over 
the boundary charges $Q_B^{(\alpha)}$ of the occupied bands together with the charge of the 
occupied edge states. Therefore, we get for the change under a phase shift by ${2\pi\over Z}$
\begin{align}
\nonumber
\Delta Q_B(\varphi,\mu_\nu) &= \sum_{\alpha=1}^\nu \Delta Q_B^{(\alpha)}(\varphi) + \Delta Q_E(\varphi,\mu_\nu) \\
\label{eq:QB_nu}
&= \sum_{\alpha=1}^\nu I_\alpha(\varphi) + {\nu\over Z} + \Delta Q_E(\varphi,\mu_\nu) \,.
\end{align}
Using (\ref{eq:invariant_alpha}) this can be expressed by a total invariant $I(\varphi,\mu_\nu)$ via
\begin{align}
\label{eq:invariant_nu_def}
I(\varphi,\mu_\nu) &\equiv \Delta Q_B(\varphi,\mu_\nu) - {\nu \over Z} \\
\label{eq:invariant_nu_1}
&= \sum_{\alpha=1}^\nu I_\alpha(\varphi) + \Delta Q_E(\varphi,\mu_\nu) \\
\label{eq:invariant_nu_2}
&= - \sum_{\alpha=1}^\nu w_\alpha(\varphi) + \Delta Q_E(\varphi,\mu_\nu)\,.
\end{align}
As a result the total invariant is an integer or, with $\bar{\rho}={\nu\over Z}$, that 
$\Delta Q_B - \bar{\rho}$ is an integer. To proof (\ref{eq:QB_change_universal}) we have to
show that the invariant can only take two values
\begin{align}
\label{eq:invariant_nu_values}
I(\varphi,\mu_\nu) \in \{0,-1\}\,.
\end{align}
To show this we first state the generic form of the phase-dependence of $Q_B$. As explained in 
detail in Section~\ref{sec:localization} the boundary charge can only jump if an edge state
moves below or above $\mu_\nu$. We denote all these phase values by 
$\varphi_{i\pm}(\mu_\nu)$, with $i=1,\dots,M_\pm(\mu_\nu)$, which depend on the choice of
the chemical potential (for simplicity we do not indicate a possible phase-dependence of $\mu_\nu$). 
However, we note that the difference 
\begin{align}
\label{eq:M_nu}
M_\nu = M_-(\mu_\nu) - M_+(\mu_\nu) 
\end{align}
depends only on $\nu$ since $M_\nu$ describes the number of those edge states 
connecting the bands $\alpha=\nu$ and $\alpha=\nu+1$ when the phase changes by $2\pi$ 
counted positive (negative) when all these edge states move upwards (downwards).
As shown in (\ref{eq:chern_total}) $M_\nu=C_\nu$ is identical to the sum over the 
Chern numbers of all filled bands $\alpha=1,\dots,\nu$. Since 
$\Delta Q_B(\varphi,\mu_\nu) - {\nu\over Z}$ is an integer for all phases $\varphi$, 
the phase-dependence is similar to (\ref{eq:QB_alpha_phase}) of a single band
\begin{align}
\label{eq:QB_nu_phase}
Q_B(\varphi,\mu_\nu) = f(\varphi,\mu_\nu) + {M_\nu\over 2\pi}\varphi + F(\varphi,\mu_\nu) \,,
\end{align}
where 
\begin{align}
\label{eq:f_nu}
f(\varphi,\mu_\nu) = f(\varphi+{2\pi\over Z},\mu_\nu)
\end{align}
is an unknown non-universal ${2\pi\over Z}$-periodic function and 
\begin{align}
\label{eq:F_nu}
F(\varphi,\mu_\nu) = \sum_{\sigma=\pm} \sum_{i=1}^{M_\sigma(\mu_\nu)} 
\sigma \theta[\varphi-\varphi_{i\sigma}(\mu_\nu)]
\end{align}
is the part describing the discontinuous jumps from edge states moving below/above $\mu_\nu$. 
This is precisely the definition we introduced in (\ref{eq:topological_charge}) where we used this
quantity to define the topological charge $\Delta F(\varphi,\mu_\nu)$. 
Analog to (\ref{eq:diophantine_alpha}) we get from (\ref{eq:QB_nu_phase}) and (\ref{eq:invariant_nu_def}) 
the Diophantine equation
\begin{align}
\label{eq:diophantine_nu}
M_\nu = \nu - s_\nu Z \,,
\end{align}
where 
\begin{align}
\label{eq:s_nu} 
s_\nu = \Delta F(\varphi,\mu_\nu) - I(\varphi,\mu_\nu) 
\end{align}
is a characteristic and phase independent integer for gap $\nu$. Similar to $M_-^{(\alpha)}-M_+^{(\alpha)}$
and $s^{(\alpha)}$ we note that $M_\nu$ and $s_\nu$ depend crucially on the choice of the functions 
$F_v$ and $F_t$ to define the parameters of the model via Eqs.~(\ref{eq:v_form}) and (\ref{eq:t_form}).
From (\ref{eq:s_nu}) we conclude that the property for the invariant to take only the values
$I \in\{0,-1\}$ is equivalent to the following topological constraint of how the edge states 
can move below/above $\mu_\nu$ 
\begin{align}
\label{eq:selection_rule_nu}
\Delta F(\varphi,\mu_\nu)\in \{s_\nu,s_\nu- 1\}\,,
\end{align}
such that the integer $s_\nu$ is related via (\ref{eq:diophantine_nu}) to $M_\nu$. This is precisely
the topological constraint which we have proven in Section~\ref{sec:constraints_edge} via the
explicit conditions how edge states can appear in the gaps, see (\ref{eq:Delta_F_constraint}),
from which we have also obtained the Diophantine equation (\ref{eq:diophantine_1}). Here, we have obtained
the Diophantine equation (\ref{eq:diophantine_nu}) in a different way from the quantization of the 
invariant but, in addition, we have proven that the parameter $\nu$ is indeed the index of the
gap under consideration. We conclude that the approach described in Section~\ref{sec:invariant_physics}
to derive the topological constraint, following along the lines of Ref.~[\onlinecite{paper_prl}], 
is indeed correct and can be rigorously proven. This means that charge conservation and particle-hole 
duality together with the fact that the phase-dependence of the parameters can always be chosen such that 
no edge states cross $\mu$ in a certain gap $\nu$ within some phase interval of size ${2\pi\over Z}$ 
describe precisely the right physical picture. This is a very surprising and remarkable fact since
the two ways to derive the same result are based on a complete different approach. Whereas in
Section~\ref{sec:constraints_edge} the proof is based on the precise way the edge states can run 
between the bands, in Ref.~[\onlinecite{paper_prl}] the proof
is based on the {\it absence} of edge states crossing $\mu$. This shows very clearly that it is 
{\it not} the physics of edge states which drives the topological constraint but rather the edge
modes are followers which have to adjust to a certain choice of the phase-dependence (or the way
the boundary is shifted continuously through the lattice) in order to respect the topological
constraint. 

Similar to (\ref{eq:delta_M_case1}) 
we note that for rather smooth functions $F_v(\varphi)$ and $F_t(\varphi)$ the integer $s_\nu$ 
is typically given by one of the following cases for each gap $\nu$  
\begin{align}
\label{eq:M_value_1} 
M_\nu = \nu \quad &\Leftrightarrow \quad s_\nu = 0 \,,\\
\label{eq:M_value_2}
M_\nu = \nu-Z \quad &\Leftrightarrow \quad s_\nu = 1 \,.
\end{align}
$M_\nu=\nu$ means that $\nu$ edge states move upwards in gap $\nu$, corresponding to the particle picture
where the charge $\bar{\rho}$ is moved into the boundary under a phase change of $2\pi$ which has
to be compensated by $\nu$ edge states moving above $\mu_\nu$.
$M_\nu=\nu-Z$ is the case where $Z-\nu$ edge states move downwards, describing the case where
the hole charge $\bar{\rho}-1$ is moved into the boundary which has to be compensated by
$Z-\nu$ edge states moving below $\mu_\nu$. 

Analog to (\ref{eq:Delta_QB_alpha_p}) and (\ref{eq:diophantine_alpha_p}) we note that 
for rational wave lengths with phase parametrization according to Eqs.~(\ref{eq:v_form_p}) and 
(\ref{eq:t_form_p}) all equations remain the same only the invariant is redefined as
\begin{align}
\label{eq:invariant_total_p}
I(\varphi,\mu_\nu) &= \Delta Q_B(\varphi,\mu_\nu) - {\nu\over Z} \in \{0,-1\}\,,\\
\label{eq:Delta_QB_total_p}
\Delta Q_B(\varphi,\mu_\nu) &= Q_B(\varphi+{2\pi p\over Z},\mu_\nu) - Q_B(\varphi,\mu_\nu)\,,
\end{align}
and the Diophantine equation changes to
\begin{align}
\label{eq:diophantine_total_p}
p M_\nu = \nu - s_\nu Z \,.
\end{align}

In Figs.~\ref{fig:QB_invariant_total_Z3}(a,b), Figs.~\ref{fig:QB_invariant_total_Z4}(a-c) and
Figs.~\ref{fig:QB_invariant_total_Z6}(a-e) we show the phase-dependence of the boundary charge and the
invariant for $Z=3,4,6$ corresponding to the band structures of Fig.~\ref{fig:edge_2}(b), 
Fig.~\ref{fig:edge}(b) and Fig.~\ref{fig:edge_2}(a), respectively. For each case we consider all 
possibilities for the gap $\nu=1,\dots,Z-1$ and take the chemical potential $\mu_\nu$ at the top
of band $\alpha=\nu$ (if $\mu_\nu$ is placed somewhere in the gap only the position of the jumps
from edge states are shifted). The three terms on the r.h.s. of Eq.~(\ref{eq:QB_nu_phase}) can
be clearly identified. The part $F(\varphi,\mu_\nu)$ shows up in the discontinuous jumps of
$Q_B(\varphi,\mu_\nu)$ by $\pm 1$ at the phase values $\varphi=\varphi_{i\pm}$ where edge states 
enter or leave the top of band $\alpha=\nu$. Correspondingly, the invariant $I(\varphi,\mu_\nu)$ 
jumps at $\varphi=\varphi_{i\pm}$ by $\mp 1$ and at $\varphi=\varphi_{i\pm}-{2\pi\over Z}$ by $\pm 1$.
The invariant is always given by the two values $I\in\{0,-1\}$, even for the rather chaotic 
case $Z=3$ of Figs.~\ref{fig:QB_invariant_total_Z3}(a,b), where many edge states return to the same band.
The linear term ${M_\nu\over 2\pi}\varphi$ on the r.h.s. of (\ref{eq:QB_nu_phase}) is visible
by an overall positive or negative average slope of $Q_B$ in Figs.~\ref{fig:QB_invariant_total_Z4}(a-c)
for $Z=4$ and in Figs.~\ref{fig:QB_invariant_total_Z6}(a-e) for $Z=6$. A positive slope occurs on
average for $M_\nu=\nu=1,2,3$ in Figs.~\ref{fig:QB_invariant_total_Z4}(a-c) and
Figs.~\ref{fig:QB_invariant_total_Z6}(a-c) whereas a negative slope is observed on average for
$M_\nu=\nu-Z$ and $\nu=4,5$ in Figs.~\ref{fig:QB_invariant_total_Z6}(d,e). The non-universal 
${2\pi\over Z}$-periodic function $f(\varphi,\mu_\nu)$ turns out to be large for small $Z$ or for
gaps $\nu$ close to $Z/2$. When $Z$ is large the boundary charge is almost a linear function
for the gaps $\nu$ close to $1$ or $Z$, see $\nu=1,5$ in Figs.~\ref{fig:QB_invariant_total_Z6}(a,e).

\section{Summary and outlook}
\label{sec:summary}

In this work we have presented a rigorous basis for a unique relationship
between the boundary charge and the bulk Zak-Berry phase together with the analytical understanding of universal
properties of the boundary charge for a wide class of half-infinite nearest-neighbor tight-binding models 
with one channel per site in 1D beyond symmetry constraints. The proposed representation of the 
exact eigenstates and their analytic continuation, underlying essentially our
analytical treatment, might be of interest for a wider class of models including multi-channel
cases. We addressed the very fundamental issue of the topological constraints for the edge states 
when the boundary of a half-infinite system is shifted to a different position and the relation to
the universal change of the boundary charge.
We introduced a measurable topological invariant to characterize this dependence and revealed a link to the
winding number of a fundamental phase, namely the phase difference of the Bloch wave for the 
infinite system between the sites left and right to the boundary. We analysed this winding number
in comparison to other topological indices classifying topological systems and found that the
winding number contains more information and probes this phase directly. Another important insight of
this work is the proof that the derived universal properties can indeed be described by using
charge conservation of particles and holes alone, as it was proposed in Ref.~[\onlinecite{paper_prl}].
The edge states were shown to play the role of followers obeying certain topological constraints
such that particles and holes fulfil charge conservation at the same time. This reflects a very
simple physical picture analog to charge pumping that piling up particle or hole charge at the
boundary by shifting the lattice towards a boundary leads inevitably to a linear increase of the
boundary charge until edge states entering or leaving the band guarantee that the charge does not
change when a whole unit cell has been shifted into the boundary. As a consequence, edge states
are driven by this mechanism, leaving a fingerprint in the density of the scattering states as we 
have demonstrated, in agreement with previous works on continuum systems, by calculating the 
pole contributions of the Friedel density. Besides this ``edge'' part of the density of the 
scattering states we have also analysed the part from branch cuts leading to a different 
localization length and a nontrivial pre-exponential power-law that deserves further 
investigations, in particular in the presence of interactions. 

So far our results refer to all single-channel models falling into the wide class of 
commensurate Aubry-Andr\'e-Harper models but with generic modulation functions for the potentials
and hoppings. However, since our proposed representation and analytic continuation of Bloch states is quite
general and since the principle of charge conservation of particle and holes is always 
fulfilled, we expect that our results can be generalized to multi-channel systems as well 
\cite{multi-channel}. As we already outlined in this work, one expects a weakening of the topological 
constraint in the sense
that, if $N_c$ channels are present, the topological invariant can take $N_c+1$ different values.
An interesting question for future research will be the development of a non-Abelian version
of the winding number and the determination of the precise gauge of the Bloch states such that
an unambiguous link can be set up between the boundary charge and the Zak-Berry phase. 

Although the scattering states of the half-infinite system consist of a linear combination of an incoming and 
outgoing plane wave for the single-channel model under consideration here, we have found that the 
scattering states have a nontrivial influence on the density and the boundary charge. This is even 
expected to be more 
dramatic for multi-channel systems since, in this case, the scattering states have to fulfil more boundary
conditions which can only be fulfilled if they contain, in addition to the purely oscillating waves,
also exponentially decaying contributions. This has been shown in a recent article on STM setups
for probing the spectral density at the boundary of a Floquet topological insulator
\cite{floquet_stm}. There it was shown, that the exponentially decaying contributions of the 
scattering states lead to a dramatic effect for the STM signal at bifurcation points which are even more
pronounced and stable than the ones from topological states in the gap. It is quite intuitive that the 
exponential localization of the boundary charge will certainly hold always for an insulator (since a 
typical velocity devided by the inverse gap is the only relevant length scale besides the localization 
length of the edge states) but it is not obvious how the pre-exponential function will look like in the 
presence of several channels. 

Since the principle of charge conservation of particle and holes is not violated by interactions
we expect that our results are quite robust against disorder and Coulomb interactions. The stability 
against disorder has already been demonstrated in Ref.~[\onlinecite{paper_prl}] provided that disorder is
so weak that the gaps are not closed. Weak Coulomb interaction can be treated very effectively
by functional renormalization group methods \cite{lin_etal_preprint} or bosonization techniques
\cite{gangadharaiah_etal_prl_12,piasotski_etal_preprint}. An interesting issue concerns the properties
of the boundary charge for the case of strong Coulomb interaction, where bosonization methods have suggested 
the generation of charge and spin density wave instabilities, possibly relevant for the occurrence of
fractional charges as they appear, e.g., for the fractional quantum Hall effect \cite{FQHE}. 

An issue touched only slightly in this work is a precise discussion of the unknown and non-universal
function $f(\varphi,\mu_\nu)$ in (\ref{eq:QB_nu_phase}). For rather smooth choices of the 
phase-dependence of the model parameters this function is observed to be rather small for not too small
$Z$ and for sufficiently large gaps located at rather low or high energies. This 
behaviour is expected to drastically change when the system is close to special symmetry points, where
the boundary charge is quantized for all phases at particular filling factors, leading to a completely
flat curve for $Q_B(\varphi)$, up to discrete jumps from edge states. This happens typically when the
gap closes at particular values of the phase corresponding to Weyl semimetal physics \cite{weyl_preprint}.
Breaking the symmetry slightly leads only to a small deviation from a flat curve although the Chern
number $C_\nu=M_\nu$ determining the slope of the linear term in (\ref{eq:QB_nu_phase}) might be nonzero.  
In this case the function $f(\varphi,\mu_\nu)$ is very strong and plays a very important role. As a result
this function seems to have two tendencies driven by two different physical mechanism. When it is small
the boundary charge adjusts to charge conservation of particles and holes for all values of the phase leading to a 
nearly perfect linear form of $Q_B(\varphi)$ between the jumps. On the other hand, it can be driven by
symmetry constraints, which have the tendency to adjust the phase-dependence to certain quantized values 
of the boundary charge. Which mechanism wins depends on how strong certain symmetries are broken
as will be discussed in a future work \cite{weyl_preprint}.

\section*{Acknowledgments}
We thank P. W. Brouwer, C. Bruder, F. Hassler, V. Meden, M. Thakurathi and S. Wessel for fruitful discussions. 
This work was supported by the Deutsche Forschungsgemeinschaft via RTG 1995, the 
Swiss National Science Foundation (SNSF) and NCCR QSIT. Simulations were performed with computing resources 
granted by RWTH Aachen University under project prep0010.
Funding was received from the European Union's Horizon 2020 research, innovation 
program (ERC Starting Grant, grant agreement No 757725) as well as from the independence grant 
from the CRC 183 network.

M.P. and D.M.K. contributed equally to this work.

\begin{appendix}

\section{Complex hoppings}
\label{app:hoppings}

Here we show that the phases of the hoppings of any nearest-neighbor tight-binding model with one
orbital per site defined on a half-infinite system can be gauged away by a unitary transformation.
Starting from the generic Hamiltonian
\begin{align}
\label{eq:h_generic}
H = \sum_{m=1}^\infty \left\{\hat{v}_m|m\rangle\langle m| - (\hat{t}_m e^{i\theta_m}|m+1\rangle\langle m|+\text{h.c.})\right\},
\end{align}
where $\hat{v}_m={\hat{v}}_m^*$, $\hat{t}_m={\hat{t}}_m^*$, and $\theta_m=\theta_m^*$ are real, 
we define a unitary transformation $U|m\rangle = e^{i\phi_m}|m\rangle$ via the phases $\phi_1=0$, 
$\phi_2=\theta_1$, $\phi_3=\theta_2+\phi_2$, etc., such that $\theta_m=\phi_{m+1}-\phi_m$. It follows that
\begin{align}
\label{eq:h_transformed}
U^\dagger H U = \sum_{m=1}^\infty \left\{\hat{v}_m|m\rangle\langle m| - (\hat{t}_m |m+1\rangle\langle m|+\text{h.c.})\right\}
\end{align}
contains only real hoppings. The unitary transformation gives each site only a phase factor, i.e., the
density on each site remains invariant. Furthermore, we find that the hoppings can be chosen 
positive $\hat{t}_m>0$. The case $\hat{t}_m=0$ is excluded since it would correspond to a finite system.

\section{Choice of phase-dependence}
\label{app:phase_dependence}

Here, we present two different ways how we parametrize the two real and periodic functions
$F_\gamma(\varphi)=F_\gamma(\varphi+2\pi)$, with $\gamma=v,t$, in Eqs.\,(\ref{eq:v_form}) and (\ref{eq:t_form})
in case we take a random choice in the figures. 

The first choice is to take a random periodic and real function of the form
\begin{align}
\label{eq:F_random1}
F_\gamma(\varphi) 
= 2 \sum_{n=1}^{N_\gamma} r_\gamma^n \cos(n\varphi + \theta_\gamma^n)\,,
\end{align}
where $0<r_\gamma^n<1, \theta_\gamma^n$ are $2 N_\gamma$ random and real parameters. Due to (\ref{eq:vt_average})
there is no Fourier component with $n=0$.

The second choice consists in fixing the values of $\gamma_j(0)\equiv v_j(0),t_j(0)$ for all
$j=1,\dots,Z$ at phase $\varphi=0$ via 
\begin{align}
\label{eq:vt_zero}
v_j(0) = V v_j^{(0)} \quad,\quad t_j(0) = t + \delta t \,t_j^{(0)} 
\end{align}
by the real parameters $\gamma_j^{(0)}=F_\gamma(2\pi j/Z)$ (for $\gamma=v,t$), with zero averave
\begin{align}
\label{eq:gamma_average}
{1\over Z}\sum_{j=1}^Z\gamma_j^{(0)} = 0
\end{align}
due to (\ref{eq:vt_average}). The phase-dependence is then chosen in a random way via the function
\begin{align}
\label{eq:F_random2}
F_\gamma(\varphi)=\text{Re}\left\{\sum_{n=1}^Z F_\gamma^n e^{in\varphi}\right\} + G_\gamma(\varphi)\,,
\end{align}
where
\begin{align}
\label{eq:F_n}
F_\gamma^n = {1\over Z} \sum_{j=1}^Z \gamma_j^{(0)} e^{-in2\pi j/Z}\,,
\end{align}
is the discrete Fourier transform of $\gamma_j^{(0)}$, with $n=1,\dots,Z$, and $G_\gamma(\varphi)$
is some random periodic function with $G_\gamma(2\pi j/Z)=0$. Therefore, by construction the
condition $\gamma_j^{(0)}=F_\gamma(2\pi j/Z)$ is fulfilled. The second term involving the function
$G_\gamma$ is introduced since the first term is zero for $\varphi={\pi\over Z}(2j+1)$
\begin{align}
\label{eq:n_zero}
F_\gamma\left({\pi \over Z}(2j+1)\right) = G_\gamma\left({\pi \over Z}(2j+1)\right)\,.
\end{align}
This follows by inserting (\ref{eq:F_n}) in (\ref{eq:F_random2}) and performing the sum over $n$ 
with the result
\begin{align}
\nonumber
F_\gamma(\varphi) - G_\gamma(\varphi) &=
{1\over Z}\sum_{j'=1}^Z \gamma_{j'}^{(0)}\sin({\varphi Z\over 2})\\
\label{eq:F_random2_explicit}
&\hspace{-1cm}
\times\left\{-\sin\left({\varphi Z\over 2}\right) + 
\cos\left({\varphi Z \over 2}\right)\cot\left({\varphi\over 2}-{\pi j'\over Z}\right)\right\} \,.
\end{align}
Inserting $\varphi={\pi\over Z}(2j+1)$ and using (\ref{eq:gamma_average}) we find (\ref{eq:n_zero}).
For $G_\gamma=0$ this means that all potentials and hopping are the same at these phase values
\begin{align}
\nonumber
& G_\gamma = 0 \quad\Rightarrow \\
\label{eq:G_zero}
& \quad v_j\left({\pi\over Z}(2j+1)\right)=0 \quad,\quad t_j\left({\pi\over Z}(2j+1)\right)=t\,,
\end{align}
leading to the special case of gap closings. Therefore, to cover the generic case the second term 
$G_\gamma(\varphi)$ is chosen randomly via  
\begin{align}
\label{eq:G}
G_\gamma(\varphi) = \text{Im} \sum_{n=1}^Z G_\gamma^n e^{in\varphi}\,,
\end{align}
where
\begin{align}
\label{eq:g_n}
G_\gamma^n ={1\over Z} \sum_{j=1}^Z \gamma_j^{(1)} e^{-in2\pi j/Z}
\end{align}
is defined analog to (\ref{eq:F_n}) but with some different real and random numbers for $\gamma_j^{(1)}$. 
Explicitly one obtains analog to (\ref{eq:F_random2_explicit}) 
\begin{align}
\nonumber
G_\gamma(\varphi) &=
{1\over Z}\sum_{j'=1}^Z \gamma_{j'}^{(1)}\sin\left({\varphi Z\over 2}\right)\cdot\\
\label{eq:G_explicit}
&\hspace{0cm}
\cdot\left\{\cos\left({\varphi Z\over 2}\right) - 
\sin\left({\varphi Z\over 2}\right)\cot\left({\varphi\over 2}-{\pi j'\over Z}\right)\right\} \,.
\end{align}

\section{Useful identities}
\label{app:identities}

We first proof (\ref{eq:aj_a1}) which in terms of the components $\chi_k^{(\alpha)}(j)$ of the
Bloch state reads for $j=2,\dots,Z-1$
\begin{align}
\nonumber
t_1\cdots t_{j-1} \chi_k^{(\alpha)}(j) &=\\
\label{eq:chi_j_chi_1}
&\hspace{-2cm}
= d_{1,j-1} \,\chi_k^{(\alpha)}(1) - d_{2,j-1}\,\chi_k^{(\alpha)}(Z)\,  t_Z e^{-ik}\,,
\end{align}
where we omitted for simplicity the dependence of the determinants on $\epsilon_k^{(\alpha)}$. 
First we note that the eigenvalue equation (\ref{eq:bloch_eigenstate}) together with the
form (\ref{eq:h_k}) of the Bloch Hamiltonian implies the recurrence relation
\begin{align}
\label{eq:chi_recurrence}
- t_j \chi_k^{(\alpha)}(j+1) + \bar{v}_j \chi_k^{(\alpha)}(j) -t_{j-1} \chi_k^{(\alpha)}(j-1)  = 0\,,
\end{align}
for $j=2,\dots,Z-1$, together with
\begin{align}
\label{eq:chi_recurrence_Z}
-t_{1} \chi_k^{(\alpha)}(2) + \bar{v}_1 \chi_k^{(\alpha)}(1) - t_Z e^{-ik}\chi_k^{(\alpha)}(Z) = 0\,,
\end{align}
where $\bar{v}_j = v_j-\epsilon_k^{(\alpha)}$ has been defined in (\ref{eq:bar_v}).
(\ref{eq:chi_recurrence_Z}) gives (\ref{eq:chi_j_chi_1}) for $j=2$. The other values are obtained
by induction. Assuming that (\ref{eq:chi_j_chi_1}) holds for all $j=2,\dots,l$, we find for $j=l+1$
\begin{align}
\nonumber
t_1\cdots t_l \chi_{k}^{(\alpha)}(l+1) &= \\
\nonumber
&\hspace{-1.5cm}
=t_1\cdots t_{l-1} [\bar{v}_l \chi_k^{(\alpha)}(l) - t_{l-1} \chi_k^{(\alpha)}(l-1)] \\
\nonumber
&\hspace{-1.5cm}
= \bar{v}_l [d_{1,l-1} \,\chi_k^{(\alpha)}(1) - d_{2,l-1} \,\chi_k^{(\alpha)}(Z) t_Z e^{-ik}] -\\
\nonumber
&\hspace{-1cm}
- t_{l-1}^2[d_{1,l-2}\, \chi_k^{(\alpha)}(1) - d_{2,l-2}\,  \chi_k^{(\alpha)}(Z) t_Z e^{-ik}] \\
\label{eq:induction}
&\hspace{-1.5cm}
= d_{1l} \,\chi_k^{(\alpha)}(1) - d_{2l}\,  \chi_k^{(\alpha)}(Z) t_Z e^{-ik}\,,
\end{align}
where we used (\ref{eq:chi_recurrence}) in the first step, (\ref{eq:chi_j_chi_1}) in the second step,
and the recurrence relation (\ref{eq:det_recursion_2}) for the determinants in the last step. This
is identical to (\ref{eq:chi_j_chi_1}) for $j=l+1$. 

Next we proof
\begin{align}
\label{eq:det_property}
(t_1\cdots t_{j-1})^2 d_{j+1,Z-1} = d_{1,j-1}d_{2,Z-1}-d_{2,j-1}d_{1,Z-1}\,,
\end{align}
with $2\le j \le Z-1$. This relation follows directly from inserting (\ref{eq:a_k_explicit}) together
with (\ref{eq:f}) and (\ref{eq:g}) in (\ref{eq:aj_a1}).

To proof the relation
\begin{align}
\label{eq:det_derivative}
{d\over d\epsilon} d_{ij}(\epsilon) = -\sum_{j'=i}^j d_{i,j'-1}(\epsilon) d_{j'+1,j}(\epsilon)\,,
\end{align}
for $i\le j$, we use the property
\begin{align}
\label{eq:det_der}
{d\over d\epsilon} \text{det}R(\epsilon) = \text{Tr} {\text{det}R(\epsilon)\over R(\epsilon)}
{d\over d\epsilon} R(\epsilon)\,,
\end{align}
valid for any matrix $R(\epsilon)$. Taking $R(\epsilon)=A^{(ij)}-\epsilon$ with $A^{(ij)}$ defined
in (\ref{eq:Aij}), we get with (\ref{eq:d_ij})
\begin{align}
\nonumber
{d\over d\epsilon} d_{ij}(\epsilon) &= - \text{Tr} {\text{det}(A^{(ij)}-\epsilon)\over A^{(ij)}-\epsilon}\\
\label{eq:det_derivative_zw1}
&= -\sum_{j'=i}^j \left({\text{det}(A^{(ij)}-\epsilon)\over A^{(ij)}-\epsilon}\right)_{j'j'}\,,
\end{align}
where we labelled the matrix elements of the $j-i+1$-dimensional matrix $A^{(ij)}$ with the index 
$j'=i,\dots,j$. Using in analogy to (\ref{eq:B_jj'}) (with $1\rightarrow i$ and $Z-1\rightarrow j$)
that
\begin{align}
\label{eq:B_ij_matrix_element}
\left({\text{det}(A^{(ij)}-\epsilon)\over A^{(ij)}-\epsilon}\right)_{j'j'} = 
d_{i,j'-1}(\epsilon) d_{j'+1,j}(\epsilon)\,,
\end{align}
we arrive at (\ref{eq:det_derivative}).

Using (\ref{eq:det_derivative}) we now proof the relation (\ref{eq:fg_relation}) for 
the vectors $f(\epsilon)$ and $g(\epsilon)$. Using the definitions (\ref{eq:f}) and (\ref{eq:g})
we get 
\begin{align}
\nonumber
f^T {d\over d\epsilon} g &= \bar{t}^Z t_Z^2 \sum_{j=1}^{Z-1} d_{j+1,Z-1} {d\over d\epsilon} d_{1,j-1} \\
\nonumber
&= \bar{t}^Z t_Z^2 \sum_{j=1}^{Z-1} d_{j+1,Z-1} \sum_{k=1}^{j-1} d_{1,k-1} d_{k+1,j-1} \\
\label{eq:fg_1}
&= \bar{t}^Z t_Z^2 \sum_{1\le k<j \le Z-1} d_{1,k-1} d_{k+1,j-1} d_{j+1,Z-1}  \\
\nonumber
g^T {d\over d\epsilon} f &= \bar{t}^Z t_Z^2 \sum_{j=1}^{Z-1} d_{1,j-1} {d\over d\epsilon} d_{j+1,Z-1} \\
\nonumber
&= \bar{t}^Z t_Z^2 \sum_{j=1}^{Z-1} d_{1,j-1} \sum_{k=j+1}^{Z-1} d_{j+1,k-1} d_{k+1,Z-1} \\
\label{eq:fg_2}
&= \bar{t}^Z t_Z^2 \sum_{1\le j<k \le Z-1} d_{1,j-1} d_{j+1,k-1} d_{k+1,Z-1} \,.
\end{align}
Interchanging $j\leftrightarrow k$ in (\ref{eq:fg_2}) we find that it agrees with (\ref{eq:fg_1}),
which proves (\ref{eq:fg_relation}).

We now derive useful relations for the derivative of the $B$-matrix defined by (\ref{eq:a_k_solution})
and (\ref{eq:s}). Taking successively the derivatives ${d^n\over d\epsilon^n}$ of
\begin{align}
\label{eq:AB}
(A-\epsilon)B(\epsilon) = B(\epsilon)(A-\epsilon)=s(\epsilon) \,,
\end{align}
we find for $n=1,2,\dots$ with $B^{(n)}\equiv B^{(n)}(\epsilon)\equiv {d^n\over d\epsilon^n}B(\epsilon)$ 
and $s^{(n)}\equiv s^{(n)}(\epsilon)\equiv {d^n\over d\epsilon^n}s(\epsilon)$
\begin{align}
\nonumber
-n B^{(n-1)} + (A-\epsilon)B^{(n)} & = \\
\label{eq:AB_n}
&\hspace{-2cm} 
= -n B^{(n-1)} + B^{(n)}(A-\epsilon) = s^{(n)} \,.
\end{align}
Multiplying this equation from the left or right with $B$ and using (\ref{eq:AB}) we find in addition
\begin{align}
\nonumber
-n B B^{(n-1)} + s B^{(n)} &= \\
\label{eq:B_n}
&\hspace{-2cm} 
= -n B^{(n-1)} B + s B^{(n)} = s^{(n)}B \,.
\end{align}
These relations imply the following useful identities 
\begin{align}
\label{eq:AB_com}
& (A-\epsilon)B^{(n)} = B^{(n)}(A-\epsilon)\,,\\
\label{eq:BB_der_com}
& B B^{(n-1)} = B^{(n-1)} B \,,\\  
\label{eq:AB_prop}
& -n B^{(n-1)} + (A-\epsilon)B^{(n)} = s^{(n)} \,,\\
\label{eq:BB__der_prop}
& -n B B^{(n-1)} + s B^{(n)} = s^{(n)}B \,.
\end{align}

Next we try to set up useful identities of the function $D(\epsilon_k)=\cos(k)$ defined in
(\ref{eq:dispersion}) and (\ref{eq:D_function}), together with its derivatives. Using the
form (\ref{eq:b_k}) of the vector $b_k$ we find
\begin{align}
\nonumber
b_k^T B b_{-k} &= t_Z^2 B_{11} + t_{Z-1}^2 B_{Z-1,Z-1} + 2 t_Z t_{Z-1} D B_{1,Z-1}\\
\label{eq:bBb}
&= t_Z^2 B_{11} + t_{Z-1}^2 B_{Z-1,Z-1} + 2 \bar{t}^Z D \,,
\end{align}
where we used (\ref{eq:B_jj'}) to get $B_{1,Z-1}=t_1\cdots t_{Z-1}$ in the last step.
Taking the derivatives of this equation w.r.t. $\epsilon_k$ we get
with $B^{(n)}_{1,Z-1}=0$ for $n=1,2,\dots$
\begin{align}
\label{eq:bBb_der}
{d^n\over d\epsilon_k^n}\left\{b_k^T B b_{-k}\right\} = 
b_k^T B^{(n)} b_{-k} + 2 \bar{t}^Z D^{(n)}  \,.
\end{align}
From (\ref{eq:b_k_equation}) and (\ref{eq:a_k_solution}) we get
\begin{align}
\label{eq:bBb_vs}
s(v_Z-\epsilon_k) = b_k^T B b_{-k} \,,
\end{align}
Using (\ref{eq:bBb_der}) and taking derivatives of this equation w.r.t. $\epsilon$ we get
for $n=1,2,\dots$
\begin{align}
\nonumber
-2 s^{(n-1)} + s^{(n)} (v_Z-\epsilon_k) &= \\
\label{eq:D_n_identity}
&\hspace{-1cm}
= b_k^T B^{(n)} b_{-k} + 2 \bar{t}^Z D^{(n)} \,. 
\end{align}

\section{Expansion of $N_k$ around poles}
\label{app:N_k_expansion}

To prove Eqs.~(\ref{eq:N_k_expansion}) and (\ref{eq:N_k_expansion_k_e_bp}) 
we expand $N_k$ around the quasimomentum $k=k_{\text{e}}\equiv k^{(\nu)}_{\text{e}}$ 
of the edge pole up to linear order where, 
according to Section~\ref{sec:edge}, we have the conditions
\begin{align}
\label{eq:edge_condition_1}
s(\epsilon_{k_{\text{e}}})=0\quad,\quad a_{-k_{\text{e}}}=0\,,
\end{align}
and
\begin{align}
\label{eq:edge_condition_2}
(A-\epsilon_{k_{\text{e}}}) a_{k_{\text{e}}} =  0 \quad,\quad
b_{-k_{\text{e}}}^T a_{k_{\text{e}}} &= 0 \,.
\end{align}
In the following we use the subindex $\text{e}$ to indicate that quantities are evaluated at
$\epsilon_{k_{\text{e}}}$, e.g., $B_{\text{e}}\equiv B(\epsilon_{k_{\text{e}}})$. 

We first assume that this is an isolated pole and does not agree with the branching point 
$k^{(\nu)}_{\text{e}}\ne k^{(\nu)}_{\text{bp}}$. This means that 
${d\epsilon_k\over dk}|_{k=k_{\text{e}}}$ is
finite, i.e., $\epsilon_k-\epsilon_{k_{\text{e}}}\sim O(k-k_{\text{e}})$. Therefore, the term 
$\left[s(\epsilon_k)\right]^2$ in $N_k$ is of $O(k-k_{\text{e}})^2$ and can be neglected. We get
\begin{align}
\nonumber
N_k &= a_k^T a_{-k} + \left[s(\epsilon_k)\right]^2 \\
\nonumber
&\approx a_{k_{\text{e}}}^T {d\over dk}(a_{-k})|_{k=k_{\text{e}}} (k-k_{\text{e}}) \\
\nonumber
&= - a_{k_{\text{e}}}^T {d\over dk}(B(\epsilon_k) b_{-k})|_{k=k_{\text{e}}} (k-k_{\text{e}})  \\
\nonumber
&= - {d\epsilon_k\over dk}|_{k=k_{\text{e}}} a_{k_{\text{e}}}^T B^{(1)}_{\text{e}} b_{-k_{\text{e}}}(k-k_{\text{e}})  \\
\label{eq:Nk_expansion}
& \hspace{1cm} - a_{k_{\text{e}}}^T B_{\text{e}} {d\over dk}(b_{-k})|_{k=k_{\text{e}}} (k-k_{\text{e}}) \,.
\end{align}
The first term of the last equation is zero since, due to the Faddeev-LeVerrier algorithm, the 
matrix $B(\epsilon)$ can be written as a polynomial of degree $Z-2$ in the matrix $A-\epsilon$. 
This holds also for its derivatives since ${d\over d\epsilon}(A-\epsilon)=-1$.
Therefore, by using (\ref{eq:edge_condition_2}), we get no contribution of this term.
For the second term we use 
\begin{align}
\label{eq:b_der}
{d\over dk}(b_{-k}) = {i\over e^{-2ik}-1} (b_k - b_{-k})\,,
\end{align} 
and get with $B_{\text{e}} b_{k_{\text{e}}}=-a_{k_{\text{e}}}$ and $B_{\text{e}} b_{-k_{\text{e}}}=-a_{-k_{\text{e}}}=0$ the
result
\begin{align}
\label{eq:Nk_expansion_result}
N_k = {i\over e^{-2ik_{\text{e}}} - 1} a_{k_{\text{e}}}^T a_{k_{\text{e}}} (k-k_{\text{e}}) + O(k-k_{\text{e}})^2 \quad,
\end{align}
which proves Eq.~(\ref{eq:N_k_expansion}).

Next we consider the case when an edge pole agrees with a branching point
$k_{\text{e}}\equiv k^{(\nu)}_{\text{e}}= k^{(\nu)}_{\text{bp}}$. In this case we have in addition
to (\ref{eq:edge_condition_1}) the condition
\begin{align}
\label{eq:edge_condition_3}
D^{(1)}_{\text{e}}\equiv {d D\over d\epsilon}(\epsilon_{k_{\text{e}}})=0 \,.
\end{align}
Therefore, we have to be more careful since expanding the equation $D(\epsilon_k)=\cos(k)$ around
$k=k_{\text{e}}$ we get
\begin{align}
\label{eq:k_eps}
{1\over 2} D^{(2)}_{\text{e}} (\epsilon_k-\epsilon_{k_{\text{e}}})^2 \approx -\sin(k_{\text{e}})(k-k_{\text{e}})\,,
\end{align}
i.e., we have to expand all terms up to $(\epsilon_k-\epsilon_{k_{\text{e}}})^2\sim (k-k_{\text{e}})$. We get
\begin{align}
\nonumber
N_k &= a_k^T a_{-k} + s(\epsilon_k)^2 \\
\nonumber
&= -a_k^T B(\epsilon_k) b_{-k} + s(\epsilon_k)^2 \\ 
\nonumber
&\approx -a_k^T  B(\epsilon_k) b_{-k_{\text{e}}} 
+ (s^{(1)}_{\text{e}})^2 (\epsilon-\epsilon_{k_{\text{e}}})^2  \\
\label{eq:Nk_expansion_2}
&\hspace{1cm}
- a_{k_{\text{e}}}^T B_{\text{e}} {d\over dk} (b_{-k})|_{k=k_{\text{e}}} (k-k_{\text{e}})  \,.
\end{align}
The last term has already been evaluated above and leads to (\ref{eq:Nk_expansion_result}). For the
first term we use
\begin{align}
\nonumber
& -a_k^T  B(\epsilon_k) b_{-k_{\text{e}}} = 
b_k^T  B(\epsilon_k)B(\epsilon_k) b_{-k_{\text{e}}}\\
\nonumber
&=
\left(-a_{k_{\text{e}}}^T  + b_{k_{\text{e}}}^T  B^{(1)}_{\text{e}} (\epsilon_k-\epsilon_{k_{\text{e}}})
+ O(\epsilon_k-\epsilon_{k_{\text{e}}})^2\right)B(\epsilon_k) b_{-k_{\text{e}}}\,,
\end{align}
where we used $k-k_{\text{e}}\sim (\epsilon_k-\epsilon_{k_{\text{e}}})^2$ in the last step, see (\ref{eq:k_eps}).
The first term in the bracket gives zero analog to the first term on the r.h.s. of 
Eq.~(\ref{eq:Nk_expansion}) since $B(\epsilon_k)$ 
can be written as a power series in $A-\epsilon_k$ or in $A-\epsilon_{k_{\text{e}}}$. 
In addition, since $B_{k_{\text{e}}}b_{-k_{\text{e}}}=-a_{-k_{\text{e}}}=0$ we get
$B(\epsilon_k) b_{-k_{\text{e}}}\sim (\epsilon_k-\epsilon_{k_{\text{e}}})$ and obtain
\begin{align}
\label{eq:zw_1}
& -a_k^T  B(\epsilon_k) b_{-k_{\text{e}}} \approx 
b_{k_{\text{e}}}^T  (B^{(1)}_{\text{e}})^2 b_{-k_{\text{e}}}(\epsilon-\epsilon_{k_{\text{e}}})^2 \,.
\end{align}
Inserting this result in (\ref{eq:Nk_expansion_2}) we get
\begin{align}
\nonumber
N_k &= \left(b_{k_{\text{e}}}^T (B^{(1)}_{\text{e}})^2 b_{-k_{\text{e}}} + (s^{(1)}_{\text{e}})^2\right)
(\epsilon_k-\epsilon_{k_{\text{e}}})^2 + \\
\label{eq:Nk_expansion_3}
&\hspace{0cm}
+ {i\over e^{-2ik_{\text{e}}} - 1} a_{k_{\text{e}}}^T a_{k_{\text{e}}} (k-k_{\text{e}}) + 
O(\epsilon_k-\epsilon_{k_{\text{e}}})^3 \,.
\end{align}
Finally, we prove that the first term on the r.h.s. is identical to the second one. 
We rewrite the second term by inserting
\begin{align}
\nonumber
a_{k_{\text{e}}} &= - B_{\text{e}} b_{k_{\text{e}}} \\
\nonumber
&= a_{-k_{\text{e}}} - B_{\text{e}} (b_{k_{\text{e}}} - b_{-k_{\text{e}}}) \\
\nonumber
& = - 2i t_Z \sin(k_{\text{e}}) B_{\text{e}}e_1 \,,
\end{align}
where we used (\ref{eq:edge_condition_1}) and (\ref{eq:b_k}) in the last equality, and defined 
the $Z-1$-dimensional unit vector $e_1=(10\cdots 0)^T$. This gives 
$a_{k_{\text{e}}}^T a_{k_{\text{e}}}=-4 t_Z^2 \sin^2(k_{\text{e}}) \left((B_{\text{e}})^2\right)_{11}$ and the
second term on the r.h.s. of Eq.~(\ref{eq:Nk_expansion_3}) can be written with the help of (\ref{eq:k_eps}) as
\begin{align}
\nonumber
{i\over e^{-2ik_{\text{e}}} - 1} a_{k_{\text{e}}}^T a_{k_{\text{e}}} (k-k_{\text{e}}) &= \\
\label{eq:second_term}
&\hspace{-2cm}
= - t_Z^2 e^{ik_{\text{e}}} \left((B_{\text{e}})^2\right)_{11} D^{(2)}_{\text{e}} (\epsilon_k-\epsilon_{k_{\text{e}}})^2\,.
\end{align}
The matrix element $\left((B_{\text{e}})^2\right)_{11}$ can be rewritten with the help of 
(\ref{eq:BB__der_prop}) for $n=1$ and $k=k_{\text{e}}$, and using (\ref{eq:B_jj'}), (\ref{eq:tilde_d_2_Z-1})
and (\ref{eq:tilde_d_2_Z-1_edge})
\begin{align}
\nonumber
\left((B_{\text{e}})^2\right)_{11} &= - s_{\text{e}}^{(1)} (B_{\text{e}})_{11} 
= - s_{\text{e}}^{(1)} d_{2,Z-1}(\epsilon_{k_{\text{e}}})  \\
\label{eq:matrix_element}
&= - s_{\text{e}}^{(1)} {\bar{t}^Z\over t_Z^2} \tilde{d}_{2,Z-1}(\epsilon_{k_{\text{e}}})  
= s_{\text{e}}^{(1)} {\bar{t}^Z\over t_Z^2} e^{-i k_{\text{e}}}\,.
\end{align}
With the help of (\ref{eq:D_n_identity}) for $n=1,2$ and $k=k_{\text{e}}$ we get 
\begin{align}
\label{eq:zw_a}
s_{\text{e}}^{(1)} (v_Z-\epsilon_{k_{\text{e}}}) &= b_{k_{\text{e}}}^T B_{\text{e}}^{(1)} b_{-k_{\text{e}}} \,,\\ 
\label{eq:zw_b}
-2 s_{\text{e}}^{(1)} + s_{\text{e}}^{(2)} (v_Z-\epsilon_{k_{\text{e}}}) &= b_{k_{\text{e}}}^T B^{(2)} b_{-k_{\text{e}}} 
+ 2 \bar{t}^Z D_{\text{e}}^{(2)} \,. 
\end{align}
Using (\ref{eq:matrix_element}), (\ref{eq:zw_a}) and (\ref{eq:zw_b}), we can calculate the factor 
in front of $(\epsilon_k-\epsilon_{k_{\text{e}}})^2$ on the r.h.s. of Eq.~(\ref{eq:second_term})
\begin{widetext}
\begin{align}
\nonumber
- t_Z^2 e^{ik_{\text{e}}} \left((B_{\text{e}})^2\right)_{11} D^{(2)}_{\text{e}} 
&= - s_{\text{e}}^{(1)} \bar{t}^Z D^{(2)}_{\text{e}}  
= - s_{\text{e}}^{(1)} \left(-s_{\text{e}}^{(1)} + {1\over 2} s_{\text{e}}^{(2)} (v_Z-\epsilon_{\text{e}}) 
- {1\over 2} b_{k_{\text{e}}}^T B_{\text{e}}^{(2)} b_{-k_{\text{e}}}\right) \\
\label{eq:zw_3}
&=  (s_{\text{e}}^{(1)})^2 + {1\over 2} b_{k_{\text{e}}}^T 
\left(-s_{\text{e}}^{(2)} B_{\text{e}}^{(1)} + s_{\text{e}}^{(1)} B_{\text{e}}^{(2)}\right) b_{-k_{\text{e}}} 
\end{align}
\end{widetext}
Finally, we use (\ref{eq:AB_prop}) for $n=1,2$ and $k=k_{\text{e}}$
\begin{align} 
\label{eq:zw_4}
& s_{\text{e}}^{(1)} = - B_{\text{e}} + (A-\epsilon_{k_{\text{e}}}) B_{\text{e}}^{(1)}  \,,\\
\label{eq:zw_5}
& s_{\text{e}}^{(2)} = - 2 B_{\text{e}}^{(1)} + (A-\epsilon_{k_{\text{e}}}) B_{\text{e}}^{(2)} \,, 
\end{align} 
which leads to
\begin{align}
\nonumber
-s_{\text{e}}^{(2)} B_{\text{e}}^{(1)} + s_{\text{e}}^{(1)} B_{\text{e}}^{(2)} 
&= 2 (B_{\text{e}}^{(1)})^2 - B_{\text{e}}B_{\text{e}}^{(2)}\\
\label{eq:zw_6}
&= 2 (B_{\text{e}}^{(1)})^2 - B_{\text{e}}^{(2)} B_{\text{e}} \,,
\end{align}
where we used (\ref{eq:BB_der_com}) in the last step.
Inserting this result in (\ref{eq:zw_3}) and using $B_{\text{e}}b_{-k_{\text{e}}} = - a_{-k_{\text{e}}}=0$
we arrive at
\begin{align}
\nonumber
- t_Z^2 e^{ik_{\text{e}}} \left((B_{\text{e}})^2\right)_{11} D^{(2)}_{\text{e}} &=\\
\label{eq:zw_final}
&\hspace{-1cm}
= (s_{\text{e}}^{(1)})^2 + b_{k_{\text{e}}}^T (B_{\text{e}}^{(1)})^2 b_{-k_{\text{e}}}\,. 
\end{align}
This proves that the first and second term on the r.h.s. of Eq.~(\ref{eq:Nk_expansion_3}) are the
same and we get the final result of Eq.~(\ref{eq:N_k_expansion_k_e_bp}) 
\begin{align}
\label{eq:Nk_expansion_e_bp}
N_k = {2i\over e^{-2ik_{\text{e}}} - 1} a_{k_{\text{e}}}^T a_{k_{\text{e}}} (k-k_{\text{e}}) + 
O(\epsilon_k-\epsilon_{k_{\text{e}}})^3 \,.
\end{align}

\section{Topological constraints for edge states}
\label{app:constraints_edge}

In this Appendix we prove the central Eq.~(\ref{eq:constraint}) for the topological constraint of
edge state that outgoing and incoming vertices must alternate. In a first step we discuss in more 
detail the phase-dependence of following determinants evaluated with the energy at the top of the band
(i.e., $k_0=0$ for $\nu$ even and $k_0=\pi$ for $\nu$ odd)
\begin{align}
\label{eq:d_1}
d_1(\varphi) &\equiv \tilde{d}_{2,Z-1}(\epsilon_{k_0}(\varphi),\varphi)\,,\\
\label{eq:d_2}
d_2(\varphi) &\equiv \tilde{d}_{1,Z-2}(\epsilon_{k_0}(\varphi),\varphi)\,,
\end{align}
where we have left out the band index for simplicity. Using (\ref{eq:det_relation_edge})
and the periodicity of the band dispersion $\epsilon_k(\varphi)=\epsilon_k(\varphi+{2\pi\over Z})$,
we note the relation
\begin{align}
\label{eq:d_12_relation}
d_1(\varphi) = d_2\left(\varphi+{2\pi\over Z}\right) \,.
\end{align}

When an edge state of $H_R$ enters or leaves at $\varphi=\varphi_0$ the phase-dependence of the 
edge state energy will connect smoothly to the band dispersion with the same first derivative
\begin{align}
\label{eq:edge_band_connection}
\epsilon^\prime_{\text{e}}(\varphi_0) = 
\epsilon^\prime_{k_0}(\varphi_0) \quad.
\end{align}
This means that up to linear order in $\varphi-\varphi_0$ the two dispersion 
are the same $\epsilon_{\text{e}}=\epsilon_{k_0}+O(\varphi-\varphi_0)^2$. Thus, up to this
order we can replace $\epsilon_{k_0}(\varphi)$ by $\epsilon_{\text{e}}(\varphi)$ in
(\ref{eq:d_1}) and (\ref{eq:d_2}) and get with the help of (\ref{eq:tilde_d_2_Z-1_edge})
and (\ref{eq:tilde_d_1_Z-2_edge})
\begin{align}
\label{eq:d_1_edge}
d_1(\varphi) &= - e^{-ik_{\text{e}}(\varphi)} + O(\varphi-\varphi_0)^2 \,,\\
\label{eq:d_2_edge}
d_2(\varphi) &= - e^{ik_{\text{e}}(\varphi)} + O(\varphi-\varphi_0)^2\,.
\end{align}
Taking the derivate of these relations at $\varphi=\varphi_0$ and using 
$k_{\text{e}}(\varphi)=k_0 + i \kappa_{\text{e}}(\varphi)$ together with 
$\kappa_{\text{e}}(\varphi_0)=0$ and (\ref{eq:d_12_relation}) we get
\begin{align}
\label{eq:d_12_property1}
d_1(\varphi_0) &= d_2(\varphi_0) = d_1(\varphi_0-{2\pi\over Z})
= -e^{-ik_0} \,,\\
\label{eq:d_12_property2}
d_1^\prime(\varphi_0) &=-\kappa^\prime_{\text{e}}(\varphi_0)\,e^{-ik_0}\,,\\
\label{eq:d_12_property3}
d_1^\prime(\varphi_0)  &= -d_2^\prime(\varphi_0)
= - d_1^\prime(\varphi_0-{2\pi\over Z})\,.
\end{align}
Since $\kappa^\prime_{\text{e}}(\varphi_0)\gtrless 0$ corresponds to an 
edge state of $H_R$ leaving/entering the band, we get with $e^{-ik_0}=(-1)^{k_0/\pi}$ from (\ref{eq:d_12_property2})
the central property
\begin{align}
\nonumber
&\text{Entering/leaving edge state of $H_R$}\quad \Rightarrow \\
\label{eq:leaving_entering}
&\hspace{1cm}
\text{sign}\left\{d_1^\prime(\varphi_0)\right\} = \pm(-1)^{k_0/\pi} \,.
\end{align}
In summary, we conclude that the determinant $d_1(\varphi)$ has always the same
value $-e^{-ik_0}$ at the points $\varphi=\varphi_0$ where edge states of $H_R$ enter or leave the band,
the derivative $d^\prime_1(\varphi)$ of the determinant has different sign for
entering and leaving edge states at $\varphi=\varphi_0$, and $d^\prime_1(\varphi)$ has a different sign for 
$\varphi=\varphi_0$ and $\varphi=\varphi_0^\prime=\varphi_0-{2\pi\over Z}$. In terms of the
contractions this means that 
\begin{align}
\nonumber
\text{(I)} & \,\,d_1(\varphi) \,\,\text{has the same value $-e^{-ik_0}$ at all}\\
\label{eq:property_1}
& \,\, \text{vertices,}\\
\nonumber
\text{(II)} & \,\,\text{sign}\{d^\prime_1(\varphi)\}=\pm(-1)^{k_0/\pi}\,\,\text{for} \\
\label{eq:property_2}
& \,\,\text{outgoing/incoming vertices,}
\end{align}
and these two properties hold irrespective of whether the vertices are the right or left vertices 
of a contraction.

In addition, we get the property that if the determinant fulfils the condition 
$d_1(\varphi)=-e^{-ik_0}$, either an edge state of $H_R$ must enter/leave at $\varphi$ or 
at $\varphi+{2\pi\over Z}$, i.e., in terms of the contraction
\begin{align}
\label{eq:property_3}
\text{(III) $d_1(\varphi)=-e^{-ik_0} \Rightarrow$ a vertex appears at $\varphi$.}
\end{align} 
To prove this we first note that $e^{-ik_0}=e^{ik_0}$ due to $k_0=0,\pi$.
With (\ref{eq:a_1_edge_tilde_d}) this implies $a_{k_0}(1)=0$ and (\ref{eq:aj_a1}) leads to
\begin{align}
\label{eq:aj_jump}
t_1\cdots t_{j-1} a_{k_0}(j) = 
- d_{2,j-1}(\epsilon_{k_0}) s(\epsilon_{k_0}) t_Z e^{-ik_0}\,,
\end{align}
for $j=2,\dots,Z-1$.
We consider two cases: (a) $s(\epsilon_{k_0})=0$ and (b) $s(\epsilon_{k_0})\ne 0$.
For case (a) an edge state of $H_R$ enters/leaves the band and $a_{k_0}=0$ which is 
consistent with (\ref{eq:a_minus_k_edge}) since the two quasimomenta $\pm k_0$ are equivalent. 
However, note that the Bloch vector $\chi_{k_0}$ is still well-defined since $N_{k_0}$
is also zero. For case (b) we get $a_{k_0}\ne 0$ and $N_{k_0}\ne 0$ such that 
$\chi_{k_0}(1)=a_{k_0}(1)/\sqrt{N_{k_0}}=0$. This means that we have found an edge state when the
boundary of $H_R$ is defined between the sites $m=1$ and $m=2$ which corresponds to the shifted
system at $\varphi+{2\pi\over Z}$. Therefore, in this case an edge state must enter/leave the band at 
phase $\varphi+{2\pi\over Z}$. 

Since the derivative of the determinant must alternate between two consecutive points where it
takes the same value, we get from the three properties (\ref{eq:property_1}), (\ref{eq:property_2}) 
and (\ref{eq:property_3}) the central condition that outgoing and incoming vertices must alternate,
which proves Eq.~(\ref{eq:constraint}).

\section{Diophantine equation}
\label{app:diophantine}

Here we prove the Diophantine equation (\ref{eq:diophantine_1}) for the case when
all contraction lines go to the right, see Fig.~\ref{fig:equal_contractions}. We split the 
phase interval of size $2\pi$ in $Z$ subintervals of size ${2\pi\over Z}$. Starting at $\varphi=0$
with an outgoing vertex, we find in the first subinterval $s$ incoming vertices including the
one at $\varphi={2\pi\over Z}$. At phase $\varphi={2\pi\over Z}+0^+$ the topological charge is
$s-1$, corresponding to $s-1$ right-going contractions. Therefore, in the second subinterval 
we obtain $s-1$ incoming vertices. At all
following phase values $\varphi={2\pi n\over Z}+0^+$, with $n=2,\dots,Z-1$, the topological charge
is either $s-1$ or $s$. This gives either $s$ or $s-1$ incoming vertices for the remaining
subintervals. As a consequence, we obtain for the maximal and minimal value of the number
$M_+$ of incoming vertices the result
\begin{align}
\label{eq:Mp_max_result}
M_+^{\text{max}} &= 2s-1 + s(Z-2) = - 1 + sZ \,,\\
\label{eq:Mp_min_result}
M_+^{\text{min}} &= 2s-1 + (s-1)(Z-2) = 1- Z + sZ \,,
\end{align}
which gives the following result for $M=M_--M_+=-M_+$
\begin{align}
\label{eq:M_min_result}
M^{\text{min}} &= - M_+^{\text{max}} = 1 - sZ \,,\\
\label{eq:M_max_result}
M^{\text{max}} &= - M_+^{\text{min}} = Z-1 - sZ \,.
\end{align}
The proof when all contractions go to the left is analog and gives the same result for $M=M_-$. 
This proves (\ref{eq:diophantine_1}).

\section{Construction of all edge state configurations}
\label{app:construction_edge}

Here we show that all configurations of contractions can be constructed by starting from 
the ones where all contractions have the same direction and iteratively adding pairs of 
contractions with different directions according to Fig.~\ref{fig:new_pair}.
To prove this statement we go the other way around and show that we can eliminate all such pairs 
iteratively starting from any given configuration until we end in a configuration where all
contractions have the same direction. 

\begin{figure}
  \centering
    \includegraphics[width= 0.9\columnwidth]{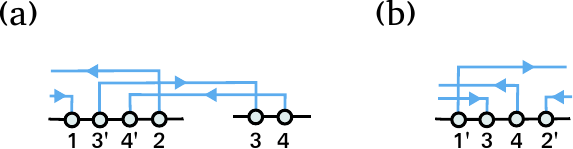} 
  \caption{(a) The two vertices $1$ and $2$ are assumed to be a pair of left-going vertices with
    shortest distance where $1$ is an incoming and $2$ an outgoing vertex. As shown it is not
    possible that two other right-going vertices $3'$ and $4'$ occur between $1$ and $2$ since 
    the two vertices $3$ and $4$ would form a pair with shorter distance compared to the pair $(1,2)$.
    (b) Here we show the two right-going vertices $1'$ and $2'$ corresponding to $1$ and $2$ of (a)
    which have the same distance.
    As shown it is not possible that two left-going vertices $3$ and $4$ can appear in between
    since they would have a shorter distance compared to the pair $(1,2)$. 
}
  \label{fig:elimination}
\end{figure}

Taking any configuration we first identify that pair $(1,2)$ of left-going vertices, one of them
outgoing and the other incoming, which have shortest distance such that no other left-going 
vertex is allowed to occur in between, see the two vertices $1$ and $2$ in Fig.~\ref{fig:elimination}(a).
W.l.o.g. we assume that vertex $1$ is incoming and $2$ is outgoing (the other case can be
proven analog). We first prove that it is also not possible that right-going vertices can occur
between $1$ and $2$. In Fig.~\ref{fig:elimination}(a) we have shown a pair $(3',4')$ of two 
right-going vertices between $1$ and $2$. These two vertices are connected to $(3,4)$. Thereby, 
it is not possible that $4=2$ since otherwise the contraction between $4'$ and $4$ would be
shorter than the one between $3'$ and $3$ (note that $3$ can not occur between $1$ and $2$ since
we assumed that no other left-going vertex can be in this interval). As a consequence we find
that the two vertices $3$ and $4$ have a shorter distance compared to $1$ and $2$ which leads
to a contradiction since the pair $(1,2)$ was assumed to be the one with the shortest distance.
Similarly, we can prove that no other vertex can occur between $1'$ and $2'$. First, a
right-going vertex $3'$ is not allowed between $1'$ and $2'$ since this would lead to a left-going 
vertex $3$ between $1$ and $2$ which is not allowed. As shown in Fig.~\ref{fig:elimination}(b) it is
also not allowed that two left-going vertices $3$ and $4$ can occur between $1'$ and $2'$
since their distance would be smaller than the one between $1$ and $2$, again leading to a 
contradiction. As a consequence, we find that no other vertex can appear between $1$ and $2$
and between $1'$ and $2'$ such that this pair can be taken out without violating the 
topological constraint. Proceeding in this way one finds that all pairs of contractions with
different directions can be eliminated until one arrives at a configuration where all
left-going vertices are either incoming or outgoing vertices. This are the configurations
where all contractions have the same direction, see Fig.~\ref{fig:equal_contractions}.

\section{Density}
\label{app:density}

\begin{figure}
\centering
\includegraphics[width= 1.\columnwidth]{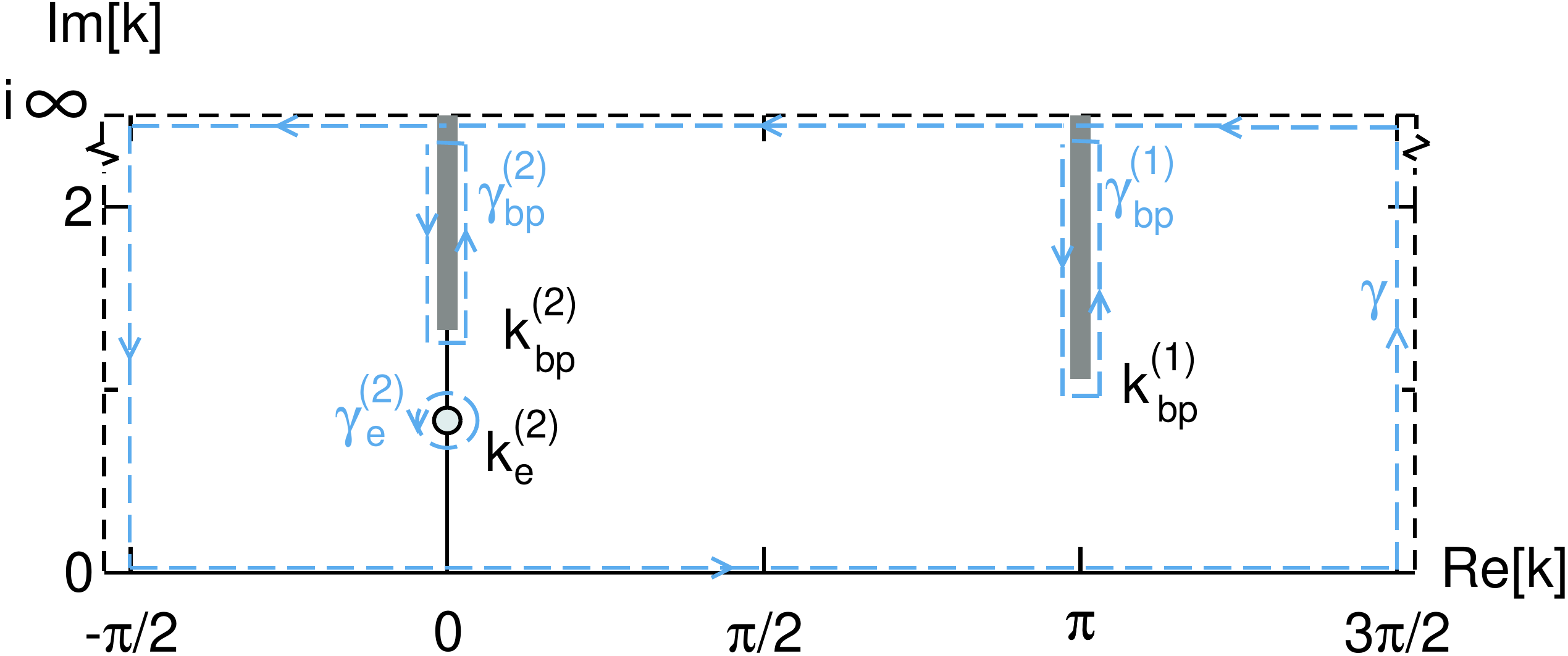} 
 \caption{Closing of the integration contour in the upper half via the path $\gamma$ 
   to calculate the Friedel density $\rho_F^{(2)}(n,j)$ of band $\alpha=2$, corresponding to the
   analytic continuation shown in Fig.~\ref{fig:bc_edge}(b).
   The two contributions from $3\pi/2\rightarrow 3\pi/2+i\infty$ and $-\pi/2+i\infty\rightarrow -\pi/2$ 
   cancel each other due to the periodicity of $\chi_k^{(2)}=\chi_{k+2\pi}^{(2)}$. The asymptotic part from
   $3\pi/2+i\infty\rightarrow -\pi/2+i\infty$ is zero. Using the analytic properties from 
   Fig.~\ref{fig:bc_edge}(b) the integration along $\gamma$ is the same as the sum of the two 
   integrals along $\gamma_{\text{bc}}^{(1)}$ and $\gamma_{\text{bc}}^{(2)}$ surrounding the two branch cuts,
   together with the integral along $\gamma_{\text{e}}^{(2)}$ encircling the edge pole at $k_{\text{e}}^{(2)}$.}
\label{fig:integration_contour}
\end{figure}

Here we present the technical details of Section~\ref{sec:localization} to calculate the density
of a half-infinite system, analog to Ref.~[\onlinecite{kallin_halperin_prb_84}] for a system in 
continuum. First we analyse the Friedel density $\rho_F^{(\alpha)}(n,j)$ of a certain band $\alpha$
starting from (\ref{eq:QF_alpha}) and, by using the periodicity $\chi_k^{(\alpha)}=\chi_{k+2\pi}^{(\alpha)}$,
shift the integration region to $\int_{-\pi/2}^{3\pi/2} dk$. We close the integration
contour in the upper half via the closed path $\gamma$ shown in Fig.~\ref{fig:integration_contour}
and obtain
\begin{align}
\label{eq:rho_friedel_alpha_gamma}
\rho_F^{(\alpha)}(n,j) = 
-{1\over 2\pi} \oint_{\gamma} dk\,{\left[\chi_k^{(\alpha)}(j)\right]}^2 e^{2ikn}\,.
\end{align}
To obtain this we have used that the two contributions from $3\pi/2\rightarrow 3\pi/2+i\infty$ 
and $-\pi/2+i\infty\rightarrow -\pi/2$ cancel each other due to the periodicity of 
$\chi_k^{(\alpha)}=\chi_{k+2\pi}^{(\alpha)}$. To show that the asymptotic part from
$3\pi/2+i\infty\rightarrow -\pi/2+i\infty$ is zero we need to analyse the asymptotic form
of the integrand for $k=x+iy$ with $x,y$ real and $y\rightarrow\infty$. 
Using the results (\ref{eq:eps_asym}) and (\ref{eq:chi_2_j_asym}) derived in 
Appendix~\ref{app:asymptotic}, we get for $n\ge 1$
\begin{align}
\nonumber
[\chi_k^{(\alpha)}(j)]^2 e^{2ikn} &\rightarrow 
{1\over Z} (t_1\dots t_{j-1})^2 t_Z^2 (-\epsilon_k^{(\alpha)})^{-2j} e^{2ik(n-1)}\\
\label{eq:friedel_asym}
&\sim e^{2ik(n-1+{j\over Z})} \rightarrow 0\,.
\end{align}
Using the analytic continuation of $\left[\chi_k^{(\alpha)}(j)\right]^2$ 
according to Fig.~\ref{fig:bc_edge}(b), we can
deform the contour $\gamma$ to a sum over the contours $\gamma_{\text{bc}}^{(\nu)}$ around the
branch cuts starting at $k_{\text{bp}}^{(\nu)}$ together with the contours $\gamma_{\text{e}}^{(\nu)}$ 
encircling edge poles, with $\nu\in\{\alpha-1,\alpha\}$ and $1\le\nu\le Z-1$. Thereby, the pole
lying below $k_{\text{bp}}^{(\nu)}$ occurs only under the conditions 
\begin{align}
\label{eq:pole_occurrence_1}
\text{Im}(\kappa_{\text{e}}^{(\nu)}) &> 0 \,,\\
\label{eq:pole_occurrence_2}
\epsilon_{\text{bp}}^{(\nu)} \lessgtr \epsilon_{\text{e}}^{(\nu)} \quad&\text{for}\quad
\nu=\begin{cases} \alpha-1 \\ \alpha \end{cases}\,.
\end{align}
The first condition requires the edge state to be an eigenstate of $H_R$ and the second condition
is necessary since edge poles arising from the analytic continuation of band $\alpha$ must either
lie in gap $\alpha-1$ with energy above $\epsilon^{(\alpha-1)}_{\text{bp}}$ or in gap $\alpha$ with
energy below $\epsilon^{(\alpha)}_{\text{bp}}$, see Fig.~\ref{fig:edge}(b). 
Eq.~(\ref{eq:rho_friedel_alpha_gamma}) for the Friedel density can therefore be split into
two parts from branch cut integrations and pole contributions  
\begin{align}
\label{eq:rho_friedel_alpha_splitting}
\rho_F^{(\alpha)}(n,j) = \rho_{F,\text{bc}}^{(\alpha)}(n,j) + \rho_{F,\text{p}}^{(\alpha)}(n,j) \, ,
\end{align}
with
\begin{align}
\label{eq:rho_friedel_alpha_bc}
\rho_{F,\text{bc}}^{(\alpha)}(n,j) &= - {1\over 2\pi} \sum_{\nu=\alpha,\alpha-1 \atop 1\le \nu \le Z-1}
\oint_{\gamma_{\text{bc}}^{(\nu)}} dk\,{\left[\chi_k^{(\alpha)}(j)\right]}^2 e^{2ikn}\,,\\
\label{eq:rho_friedel_alpha_p}
\rho_{F,\text{p}}^{(\alpha)}(n,j) &= - {1\over 2\pi} \sideset{}{'}\sum_{\nu=\alpha,\alpha-1}
\oint_{\gamma_{\text{p}}^{(\nu)}} dk\,{\left[\chi_k^{(\alpha)}(j)\right]}^2 e^{2ikn}\,,
\end{align}
where the prime at the last $\sum^\prime$ refers to the conditions stated in (\ref{eq:pole_occurrence_1})
and (\ref{eq:pole_occurrence_2}). To evaluate the branch cut contribution we write
$k=k_{\text{bp}}^{(\nu)}+i\kappa\pm 0^+$, with $\kappa>0$, and use
$\chi_{k_{\text{bp}}^{(\nu)}+i\kappa-0^+}^{(\alpha)}(j)=[\chi_{k_{\text{bp}}^{(\nu)}+i\kappa+0^+}^{(\alpha)}(j)]^*$ which
follows from (\ref{eq:bloch_vector_property}), (\ref{eq:bloch_vector_periodicity}) 
and (\ref{eq:k_bp}). This gives 
\begin{align}
\nonumber
\rho_{F,\text{bc}}^{(\alpha)}(n,j) &= {1\over \pi}\sum_{\nu=\alpha,\alpha-1 \atop 1\le \nu \le Z-1}
e^{-2\kappa_{\text{bp}}^{(\nu)} n} \\
\label{eq:rho_friedel_alpha_bc_im}
&\hspace{0cm}\times
\text{Im} \int_0^\infty d\kappa\,\left[\chi_{k_{\text{bp}}^{(\nu)}+i\kappa+0^+}^{(\alpha)}(j)\right]^2 e^{-2\kappa n} \,.
\end{align}
Summing up all branch cut contributions from the occupied bands $\alpha=1,\dots,\nu$, we find
that the contributions from adjacent bands $\alpha$ and $\alpha+1$ for the common branch cut starting at 
$k_{\text{bp}}^{(\alpha)}$ will exactly cancel each other since the values
of the integrand left and right due to branch cut are interchanged, see  
Fig.~\ref{fig:bc}(b,c). What remains is only the branch cut from band $\alpha=\nu$ in gap $\nu$,
leading to Eq.~(\ref{eq:rho_friedel_total_bc_im}).

To calculate the edge pole contribution we use (\ref{eq:chi_edge}) and (\ref{eq:N_k_expansion})
to get for $k\rightarrow k_{\text{e}}^{(\nu)}$
\begin{align}
\nonumber
\chi_k^{(\alpha)}(j)^2 &\rightarrow 
{N_{k_{\text{e}}^{(\nu)}}^{\text{e}}\over N_k^{(\alpha)}} \left[\chi_{k_{\text{e}}^{(\nu)}}^{\text{e}}(j)\right]^2 \\
\label{eq:chi_chi_edge}
&\rightarrow 
{-i\over k-k_{\text{e}}^{(\nu)}} \left[\chi_{k_{\text{e}}^{(\nu)}}^{\text{e}}(j)\right]^2 \,.
\end{align}
This gives for (\ref{eq:rho_friedel_alpha_p}) with (\ref{eq:edge_state}) 
\begin{align}
\label{eq:rho_friedel_alpha_p_explicit}
\rho_{F,\text{p}}^{(\alpha)}(n,j) = - \sideset{}{'}\sum_{\nu=\alpha,\alpha-1}
[\psi_{k_{\text{e}}^{(\nu)}}^{\text{e}}(n,j)]^2 \,.
\end{align}
As a result the pole contribution of the Friedel density of band $\alpha$ cancels exactly the
edge state density of those edge states which belong to band $\alpha$, see Eq.~(\ref{eq:friedel_pole_density}).

\section{Asymptotic values}
\label{app:asymptotic}

In this Appendix we determine the asymptotic form of various quantities 
for $k=x+iy$ with $x,y$ real and $y\rightarrow\infty$. Since 
$e^{-ik}=e^{-ix}e^{y}\rightarrow\infty$ and $e^{ik}=e^{ix}e^{-y}\rightarrow 0$ we find from 
(\ref{eq:dispersion}) that $e^{-ik}\rightarrow {1\over\bar{t}^Z}(-\epsilon_k^{(\alpha)})^Z$ or
\begin{align}
\label{eq:eps_asym}
(-\epsilon_k^{(\alpha)})^Z \rightarrow \bar{t}^Z e^{-ik}\,.
\end{align}
We note that this asymptotic condition has no unique solution for the individual bands $\epsilon_k^{(\alpha)}$.
A detailed analysis (not important for the following) gives
\begin{align}
\label{eq:eps_alpha_asym}
\epsilon_k^{(\alpha)} \rightarrow - \bar{t} e^{-ik/Z} e^{i 2\pi \alpha/Z}\,.
\end{align}
\begin{figure*}
  \centering
  \begin{minipage}{0.32\textwidth}
    \includegraphics[width= \columnwidth]{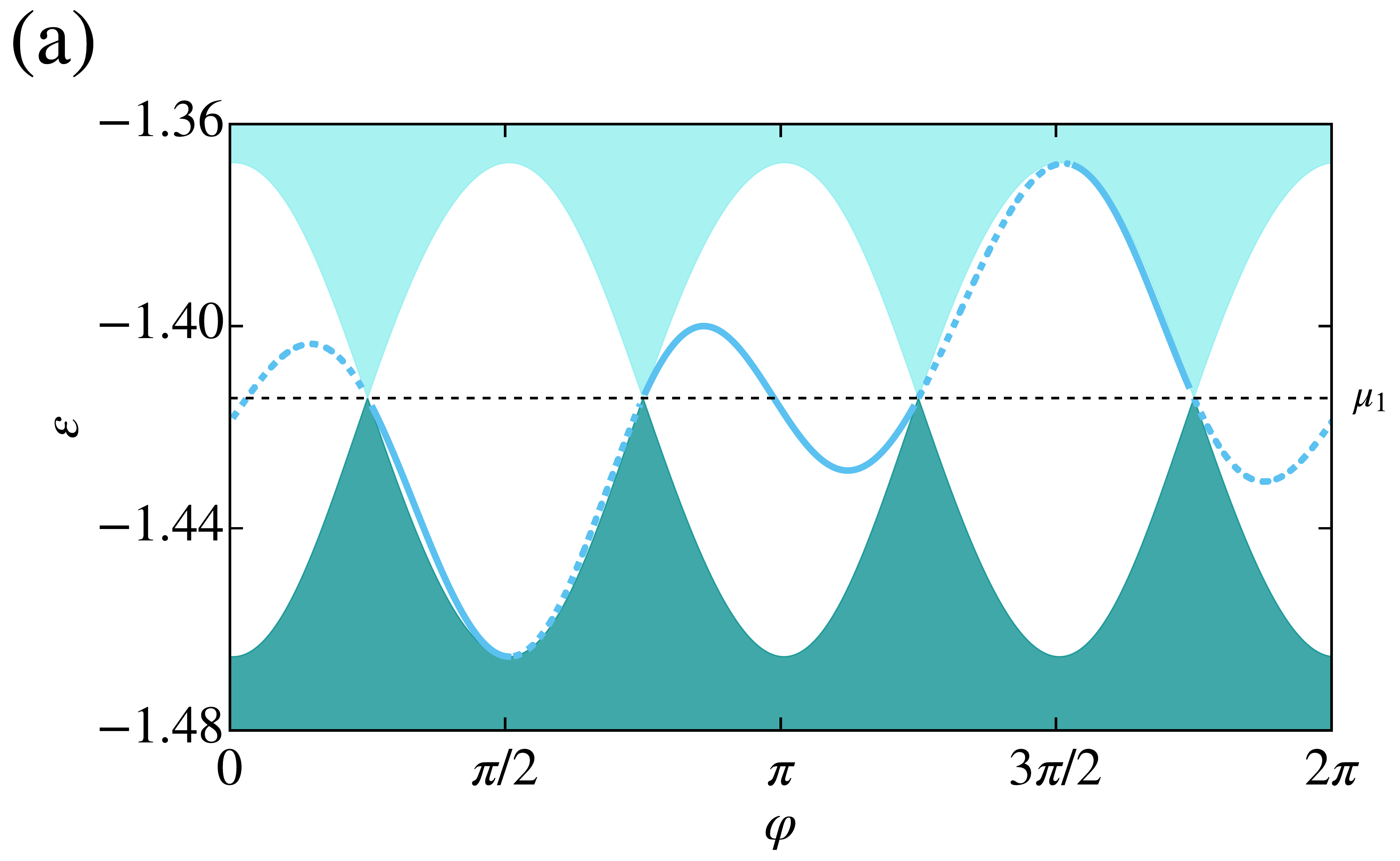} 
  \end{minipage}
  \begin{minipage}{0.32\textwidth}
    \includegraphics[width= \columnwidth]{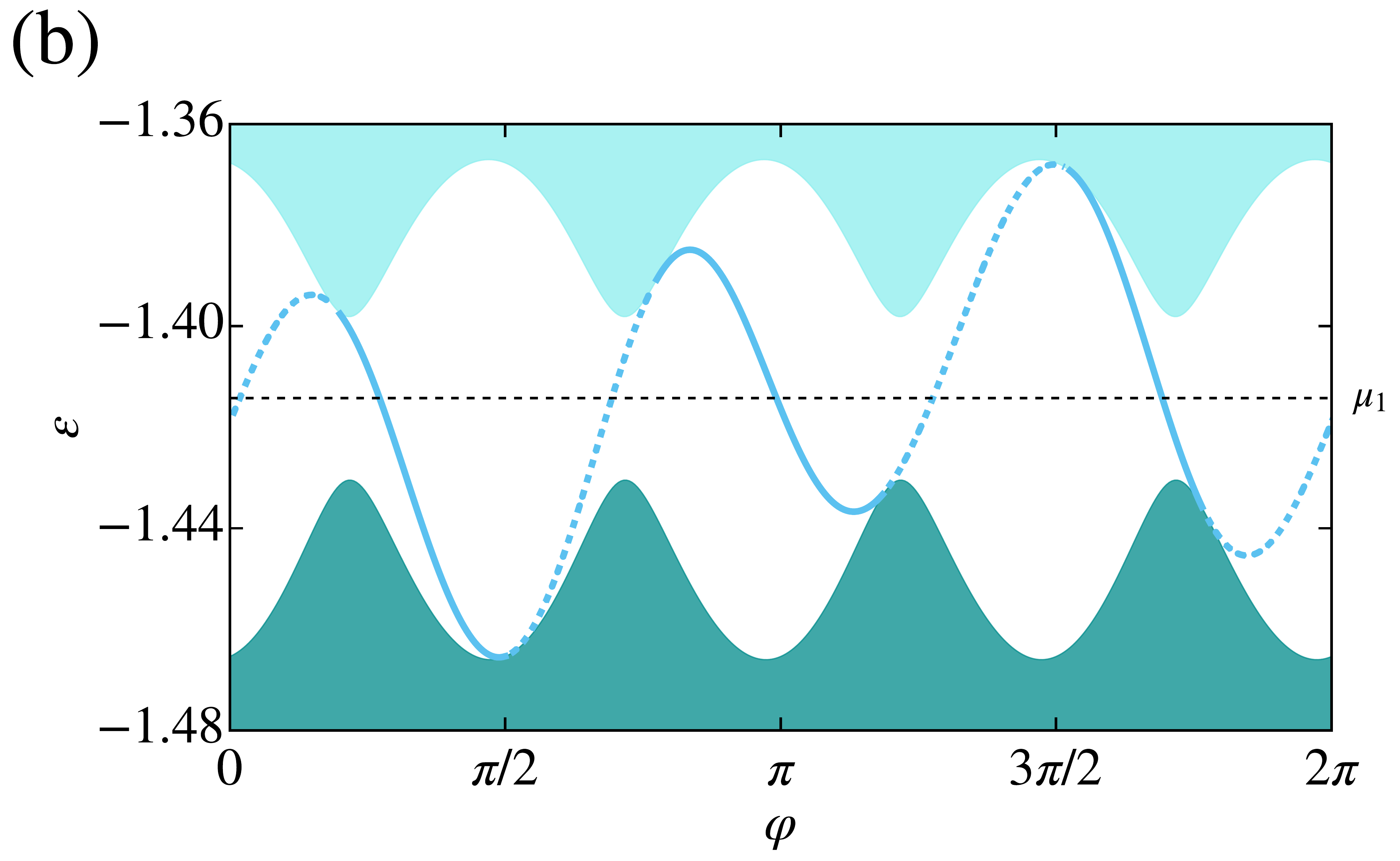} 
    \includegraphics[width= \columnwidth]{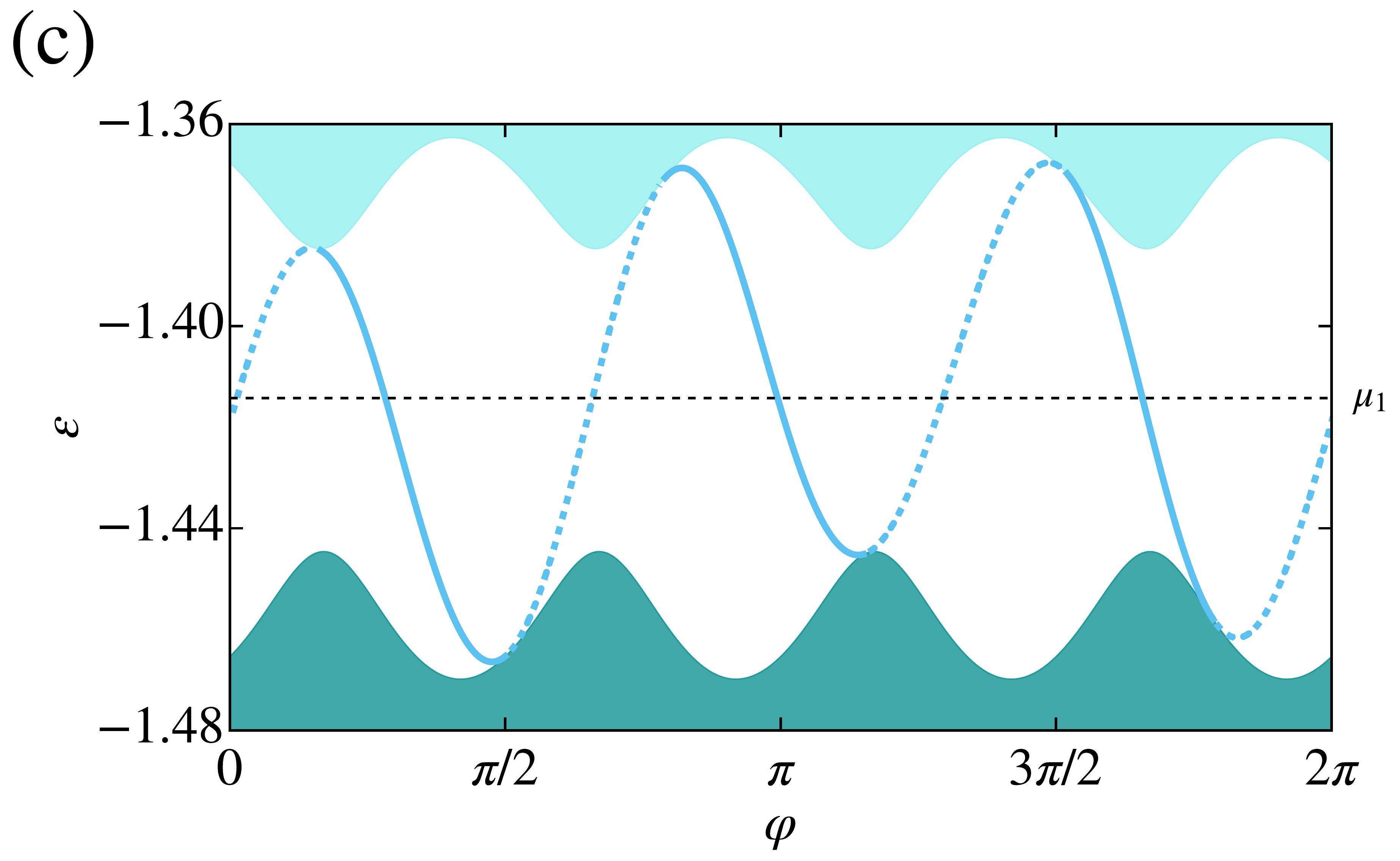} 
  \end{minipage}
  \begin{minipage}{0.32\textwidth}
    \includegraphics[width= \columnwidth]{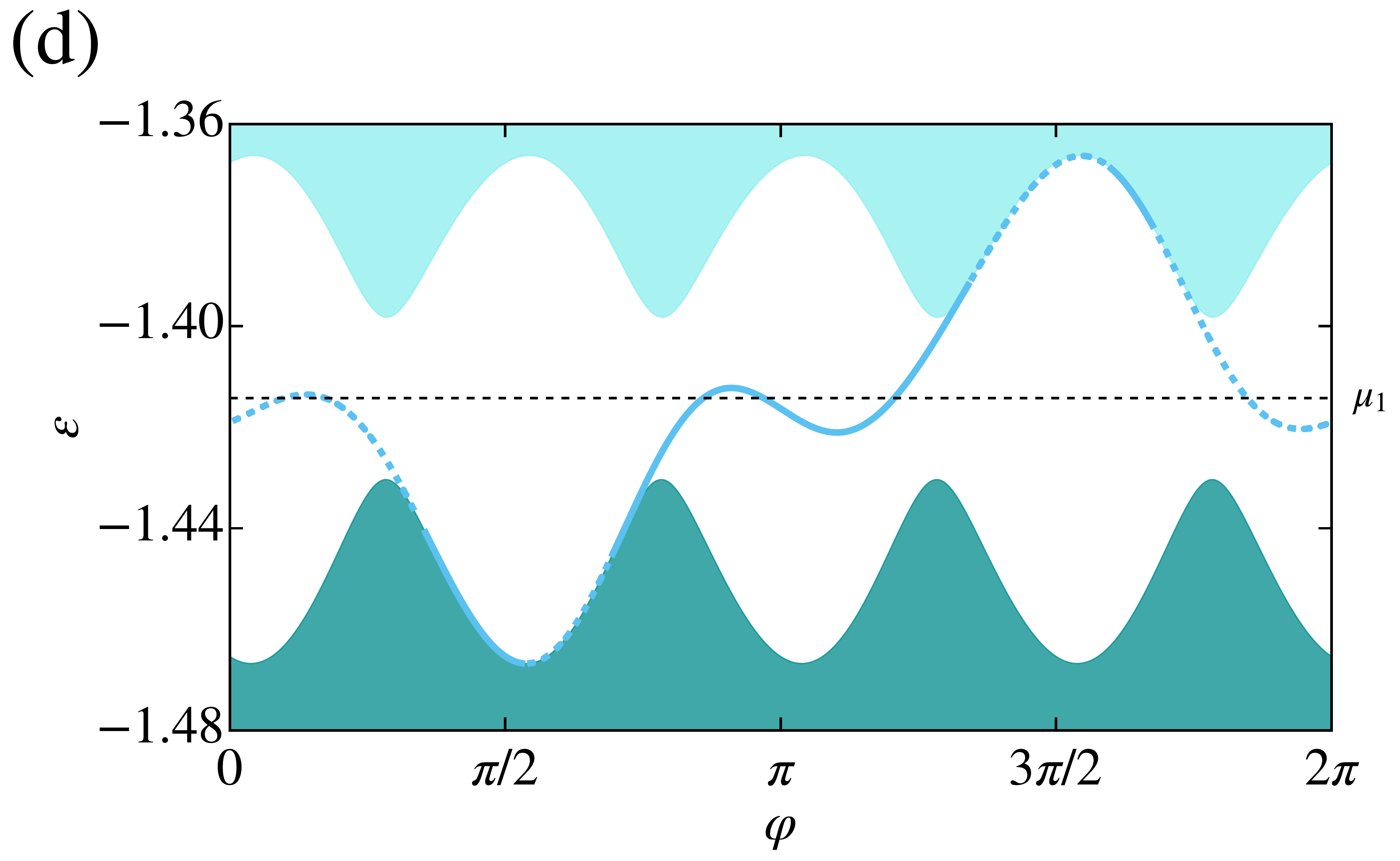} 
    \includegraphics[width= \columnwidth]{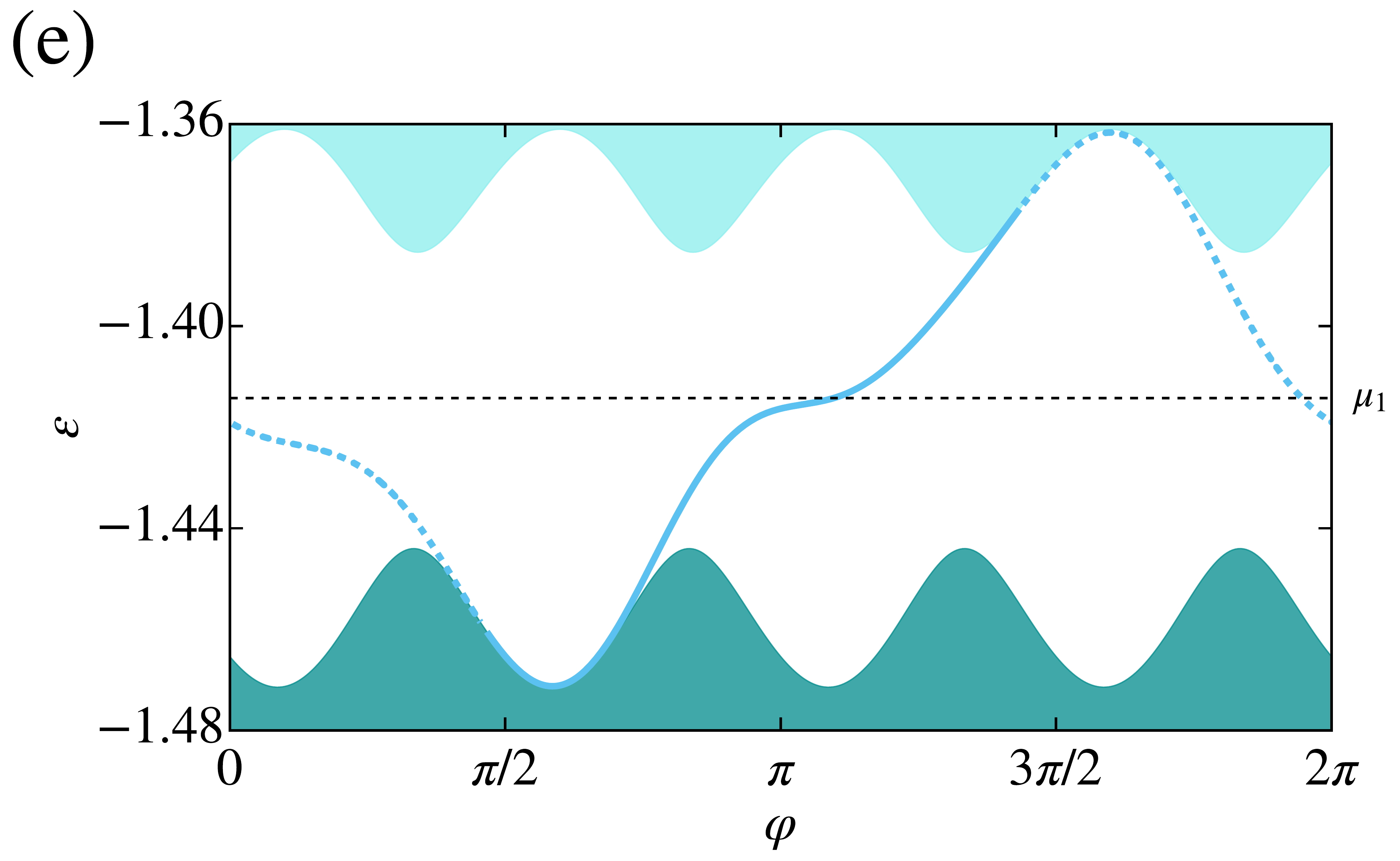} 
  \end{minipage}
  \caption{The phase-dependence of the edge state energy (solid blue line for $H_R$ and dashed blue line for $H_L$) 
    in the first gap $\nu=1$ for $Z=4$ and $V=0.3$, $t=1$, $\delta t=0.2$, for various functions 
    $F_v$ and $F_t$ in Eqs.~(\ref{eq:v_form}) and (\ref{eq:t_form}) taken from (\ref{eq:F_random2}) with 
    fixed and random parameters for $v_j(0)$ and $t_j(0)-t$ at phase $\varphi=0$ but different choices 
    for the phase-dependence in between (via the random parameters $\gamma_j^{(1)}$ for $\gamma=v,t$ 
    used in (\ref{eq:g_n})), see the Supplemental Material \cite{SM} for the concrete parameters. 
    The chemical potential $\mu_1$ is indicated by a dashed black line. 
    In (a) we have chosen $\gamma_j^{(1)}=0$ which leads to a gap
    closing at $\varphi={\pi\over 4}$. In (c) the $\gamma_j^{(1)}$ are by a
    factor of $2$ larger compared to (b). (d) and (e) correspond to
    (b) and (c), respectively, but the sign of all $\gamma_j^{(1)}$ has been
    reversed. As can be seen for the choices in (d) and (e) no edge state of $H_R$ crosses $\mu_1$ in
    the phase interval $[0,{2\pi\over Z}]=[0,{\pi\over 2}]$.
  }
  \label{fig:G_parametrization}
\end{figure*}
As a result of (\ref{eq:eps_asym}) we get $\epsilon_k^{(\alpha)}\rightarrow\infty$ such that all 
determinants have the asymptotic condition 
$d_{nm}(\epsilon_k^{(\alpha)})\rightarrow (-\epsilon_k^{(\alpha)})^{m-n+1}$ for $m-n+1\ge 0$ (note that
$d_{nm}\equiv 1$ for $n>m$). Using (\ref{eq:a_k_explicit}), (\ref{eq:f}), and (\ref{eq:g}) this
gives for $j=1,\dots,Z-1$
\begin{align}
\label{eq:ak_asym}
a^{(\alpha)}_k(j) &\rightarrow t_1\cdots t_{j-1} t_z (-\epsilon_k^{(\alpha)})^{Z-1-j} e^{-ik} \,,\\
\label{eq:a_mk_asym}
a^{(\alpha)}_{-k}(j) &\rightarrow {\bar{t}^Z\over t_1\cdots t_{j-1}}{1\over t_z} (-\epsilon_k^{(\alpha)})^{j-1} \,.
\end{align}
Furthermore, from (\ref{eq:s}), we get $s(\epsilon_k^{(\alpha)})\rightarrow (-\epsilon_k^{(\alpha)})^{Z-1}$ 
or with (\ref{eq:eps_asym})  
\begin{align}
\label{eq:s2_asym}
\left[s(\epsilon_k^{(\alpha)})\right]^2 \rightarrow \bar{t}^Z (-\epsilon_k^{(\alpha)})^{Z-2} e^{-ik}\,.
\end{align} 
Since $a_k^{(\alpha)}(1)\rightarrow \bar{t}^Z (-\epsilon_k^{(\alpha)})^{Z-2}e^{-ik}$ is largest and
$a_k^{(\alpha)}(j)a_{-k}^{(\alpha)}(j)\rightarrow \bar{t}^Z(-\epsilon_k^{(\alpha)})^{Z-2}e^{-ik}$, we obtain
\begin{align}
\nonumber
(a_k^{(\alpha)})^Ta_k^{(\alpha)} + \left[s(\epsilon_k^{(\alpha)})\right]^2 &\rightarrow [a_k^{(\alpha)}(1)]^2 \\
\label{eq:ak_ak_asym}
&\rightarrow t_Z^2(-\epsilon_k^{(\alpha)})^{2(Z-2)} e^{-ik}\,,\\ 
\nonumber
N^{(\alpha)}_k&=(a_k^{(\alpha)})^Ta_{-k}^{(\alpha)} + \left[s(\epsilon_k^{(\alpha)})\right]^2\\
\label{eq:Nk_asym}
&\rightarrow Z \bar{t}^Z (-\epsilon_k^{(\alpha)})^{Z-2} e^{-ik}\,.
\end{align}
Using (\ref{eq:chiT_chi}) this gives for 
$\sum_{j=1}^Z[\chi_k^{(\alpha)}(j)]^2=(\chi_k^{(\alpha)})^T \chi_k^{(\alpha)}$ the asymptotic form
\begin{align}
\label{eq:chi_T_chi_asym}
\sum_{j=1}^Z[\chi_k^{(\alpha)}(j)]^2 \rightarrow {1\over Z} t_Z^2 (-\epsilon_k^{(\alpha)})^{-2} e^{-2ik}\,.
\end{align}
For $[\chi_k^{(\alpha)}(j)]^2={1\over N^{(\alpha)}_k}[a_k^{(\alpha)}(j)]^2$ for $j=1,\dots,Z-1$
and $[\chi_k^{(\alpha)}(Z)]^2={1\over N^{(\alpha)}_k}\left[s(\epsilon_k^{(\alpha)})\right]^2$ we get from
(\ref{eq:ak_asym}), (\ref{eq:s2_asym}), and (\ref{eq:Nk_asym}) 
\begin{align}
\label{eq:chi_2_j_asym}
[\chi_k^{(\alpha)}(j)]^2 \rightarrow {1\over Z} (t_1\dots t_{j-1})^2 t_Z^2 (-\epsilon_k^{(\alpha)})^{-2j} e^{-2ik}\,.
\end{align}

\section{Tuning of edge states via choice of the phase-dependence}
\label{app:edge_tuning}

Here we show that the phase-dependence of $v_j(\varphi)$ and $t_j(\varphi)$ in the phase interval
$0<\varphi<{2\pi\over Z}$ for given parameters $v_j(0), t_j(0)$ and 
$v_j({2\pi\over Z})=v_{j+1}(0), t_j({2\pi\over Z})=t_{j+1}(0)$ at the boundaries $\varphi=0,{2\pi\over Z}$ 
can always be chosen such that no edge state crosses $\mu_\nu$ in this interval. 
This can be achieved by using the parametrization (\ref{eq:F_random2}) for the phase-dependence.
For $G_\gamma=0$ a gap closing occurs in the middle of the interval at $\varphi={\pi\over Z}$ where
all $v_j({\pi\over Z})=0$ and $t_j({\pi\over Z})=t$, see (\ref{eq:G_zero}). The spectrum of the 
edge state between the first and second band is shown in Fig.~\ref{fig:G_parametrization}(a) for $Z=4$.
At the gap closing at $\varphi={\pi\over 4}$ the edge state changes from an edge state of $H_L$ to an
edge state of $H_R$. In Fig.~\ref{fig:G_parametrization}(b,c) we show the same for a slightly 
nonzero function $G_\gamma$ by choosing a random set of parameters $\gamma_j^{(1)}$ in (\ref{eq:g_n})
on two scales $\sim 0.05$ and $\sim 0.1$, respectively. In Fig.~\ref{fig:G_parametrization}(d,e) 
we have changed the sign of all $\gamma_j^{(1)}$. As can be seen in 
Figs.~\ref{fig:G_parametrization}(b,c) close to $\varphi\sim{\pi\over 4}$ the edge state of $H_R$ 
is taken upwards together with the upper band whereas in 
Figs.~\ref{fig:G_parametrization}(d,e) it is taken downwards together with the lower band. This has
the effect that the edge state of $H_R$ does no longer cross the chemical potential in
Figs.~\ref{fig:G_parametrization}(d,e). This behaviour is generic and can be observed for any $Z$ and
for any initial parameters for $\gamma_j^{(0)}$ and for all kinds of random parameters for $\gamma_j^{(1)}$.
Changing the sign of $G_\gamma$ corresponds precisely to the change of the orientation of the edge pole
encircling the branch cut in order to avoid the crossing of $k_{\text{e}}^{(\nu)}(\varphi)$ through 
the value of $k_{\mu_\nu}$ corresponding to the chemical potential, see Fig.~\ref{fig:no_edge}. 
For arbitrary $\nu$ we find that this point describes the crossover from the edge state encircling
the branch cut $\nu$ times in clockwise direction to the case of $Z-\nu$ windings with
counter-clockwise orientation. This corresponds to the two cases of $\nu$ edge states moving
upwards compared to $Z-\nu$ edge states moving downwards, see Eqs.~(\ref{eq:M_value_1}) and
(\ref{eq:M_value_2}). We note that additional windings by multiples of $Z$ do not occur
[corresponding to values $s\ne 0,1$ in the Diophantine equation (\ref{eq:diophantine_nu})] since
with our choice (\ref{eq:F_random2}) of the phase parametrization we have only chosen at most
$Z$ Fourier modes for $F_\gamma(\varphi)$.

\end{appendix}


\begin{thebibliography}{99}


\bibitem{klitzing_dorda_pepper_prl_80}
K. von Klitzing, G. Dorda, and M. Pepper, Phys. Rev. Lett. {\bf 45}, 494 (1980).

\bibitem{thouless_etal_prl_82}
D. J. Thouless, M. Kohmoto, M. P. Nightingale, and M. den Nijs, 
Phys. Rev. Lett. {\bf 49}, 405 (1982).



\bibitem{volkov_pankratov_JETP_85}
B. A. Volkov and O. A. Pankratov, Pis’ma Zh. Eksp. Teor. Fiz. {\bf 42}, 145
(1985) [JETP Lett. {\bf 42}, 178 (1985)].

\bibitem{pankratov_etal_SSC_87}
O. A. Pankratov, S. V. Pakhomov, and B. A. Volkov, Solid State Commun. {\bf 61}, 93 (1987).

\bibitem{kane_mele_prl_95}
C. L. Kane and E. J. Mele, Phys. Rev. Lett. {\bf 95}, 146802 (2005).

\bibitem{bernevig_etal_science_06}
B. A. Bernevig, T. L. Hughes, and S.-C. Zhang, Science {\bf 314}, 1757 (2006).

\bibitem{fu_kane_mele_prl_07}
L. Fu, C. L. Kane, and E. J. Mele, Phys. Rev. Lett. {\bf 98}, 106803 (2007).



\bibitem{koenig_etal_science_07}
M. K\"onig, S. Wiedmann, C. Brune, A. Roth, H. Buhmann, L. W. Molenkamp,
X.-L. Qi, and S.-C. Zhang, Science {\bf 318}, 766 (2007).

\bibitem{hsieh_etal_nature_08}
D. Hsieh, D. Qian, L. Wray, Y. Xia, Y. S. Hor, R. J. Cava, and M. Z. Hasan, 
Nature (London) {\bf 452}, 970 (2008).




\bibitem{hasan_kane_RMP_10}
M. Z. Hasan and C. L. Kane, Rev. Mod. Phys. {\bf 82}, 3045 (2010).

\bibitem{qi_zhang_RMP_11}
X.-L. Qi and S.-C. Zhang, Rev. Mod. Phys. {\bf 83}, 1057 (2011).

\bibitem{bernevig_book_13}
B. A. Bernevig, {\it Topological Insulators and Topological Superconductors}, 
Princeton University Press (2013).

\bibitem{tkachov_book_15}
G. Tkachov, {\it Topological Insulators: The Physics of Spin Helicity in Quantum Transport},
(Pan Stanford, 2015).

\bibitem{asboth_book_16}
J. K. Asb\'oth, L. Oroszl\'any, and A. P\'alyi, {\it A Short Course on Topological Insulators},
Lecture Notes in Physics, Springer 2016. 



\bibitem{charge_fractionalization}
R. Jackiw and J. Schrieffer, Nucl. Phys. B {\bf 190}, 253 (1981); 
W. P. Su and J. R. Schrieffer, Phys. Rev. Lett. {\bf 46}, 738 (1981); 
M. J. Rice and E. J. Mele, Phys. Rev. Lett. {\bf 49}, 1455 (1982); 
R. Jackiw and G. Semenoff, Phys. Rev. Lett. {\bf 50}, 439 (1983); 
S. Ryu, C. Mudry, C.-Y. Hou, and C. Chamon, Phys. Rev. B {\bf 80}, 205319 (2009); 
J. Klinovaja and D. Loss, Phys. Rev. Lett. {\bf 110}, 126402 (2013); 
R. Wakatsuki, M. Ezawa, Y. Tanaka, and N. Nagaosa, Phys. Rev. B {\bf 90}, 014505 (2014).

\bibitem{com_1}
Here, {\it local} means that the symmetry operation can be defined within a unit cell.



\bibitem{hughes_etal_prb_83}
T. L. Hughes, E. Prodan, and B. A. Bernevig,
Phys. Rev. B {\bf 83}, 245132 (2011).

\bibitem{chiu_etal_prb_88}
C.-K. Chiu, H. Yao, and S. Ryu, Phys. Rev. B {\bf 88}, 075142 (2013).

\bibitem{shiozaki_sato_prb_90}
K. Shiozaki and M. Sato, Phys. Rev. B {\bf 90}, 165114 (2014).

\bibitem{alexandradinata_etal_prb_16}
A. Alexandradinata, Zhijun Wang, and B. A. Bernevig, Phys. Rev. X {\bf 6}, 021008 (2016).

\bibitem{trifunovic_brouwer_prb_17}
L. Trifunovic and P. Brouwer, Phys. Rev. B {\bf 96}, 195109 (2017). 

\bibitem{lau_etal_prb_16}
A. Lau, C. Ortix, and J. van den Brink, Phys. Rev. Lett. {\bf 115}, 216805 (2015);
A. Lau, J. van den Brink, and C. Ortix, Phys. Rev. B {\bf 94}, 165164 (2016);
A. Lau and C. Ortix, Eur. Phys. J. Spec. Top. 227, 1309 (2018).



\bibitem{schnyder_etal_prb_08}
A. P. Schnyder, S. Ryu, A. Furusaki, and A. W. W. Ludwig,
Phys. Rev. B {\bf 78}, 195125 (2008).

\bibitem{schnyder_etal_njp_10}
S. Ryu, A. P. Schnyder, A. Furusaki, and A. W. W. Ludwig,
New J. Phys. {\bf 12}, 065010 (2010).

\bibitem{kitaev_advphys_09}
A. Kitaev, in Advances in Theoretical Physics, edited by V. Lebedev, and M. Feigel’man, 
AIP Conf. Proc. No. 1134 (AIP, New York, 2009), p. 22.

\bibitem{slager_etal_natphys_12}
R.-J. Slager, A. Mesaros, V. Juri\v{c}i\'c, and J. Zaanen, Nat. Phys. {\bf 9}, 98 (2012).

\bibitem{jadaun_etal_prb_13}
P. Jadaun, D. Xiao, Q. Niu, and S. K. Banerjee, Phys. Rev. B {\bf 88}, 085110 (2013).

\bibitem{chiu_etal_prb_13}
C.-K. Chiu, H. Yao, and S. Ryu, Phys. Rev. B {\bf 88}, 075142 (2013).

\bibitem{zhang_kane_mele_prl_13}
F. Zhang, C. L. Kane, and E. J. Mele, Phys. Rev. Lett. {\bf 111}, 056403 (2013).

\bibitem{benalcazar_etal_prb_14}
W. A. Benalcazar, J.C.Y. Teo, and T. L. Hughes, Phys. Rev. B {\bf 89}, 224503 (2014).

\bibitem{morimoto_furusaki_prb_13}
T. Morimoto and A. Furusaki, Phys. Rev. B {\bf 88}, 125129 (2013).

\bibitem{diez_etal_njp_15}
M. Diez, D. I. Pikulin, I.C. Fulga, and J. Tworzydlo, New J. Phys. {\bf 17}, 043014 (2015).





\bibitem{kingsmith_vanderbilt_prb_93}
R.D. King-Smith and D. Vanderbilt, Phys. Rev. B(R) {\bf 47}, 1651 (1993).

\bibitem{vanderbilt_kingsmith_prb_93}
D. Vanderbilt and R.D. King-Smith, Phys. Rev. B {\bf 48}, 4442 (1993).

\bibitem{resta_revmodphys_94}
R. Resta, Rev. Mod. Phys. {\bf 66}, 899 (1994).

\bibitem{kudin_etal_chemphys_07}
K.N. Kudin and R. Car, J. of Chem. Phys. {\bf 126}, 234101 (2007).

\bibitem{marzari_etal_revmodphys_12}
N. Marzari, A.A. Mostofi, J.R. Yates, I. Souza, and D. Vanderbilt, 
Rev. Mod. Phys. {\bf 84}, 1419 (2012).

\bibitem{spaldin_solidstatechem_12}
N.A. Spaldin, J. of Solid State Chem. {\bf 195}, 2 (2012).

\bibitem{rhim_etal_prb_17}
J.-W. Rhim, J. Behrends and J.H. Bardarson, Phys. Rev. B {\bf 95}, 035421 (2017).

\bibitem{miert_ortix_prb_17}
G. van Miert and Carmine Ortix, Phys. Rev. B {\bf 96}, 235130 (2017).

\bibitem{vanderbilt_book_2018}
D. Vanderbilt, {\it Berry Phases in Electronic Structure Theory: 
Electric Polarization, Orbital Magnetization and Topological Insulators},
(Cambridge University Press, 2018).

\bibitem{zak}
J. Zak, Phys. Rev. Lett. {\bf 48}, 359 (1982);
J. Zak, Phys. Rev. Lett. {\bf 62}, 2747 (1989).



\bibitem{hatsugai_prl_93}
Y. Hatsugai, Phys. Rev. Lett. {\bf 71}, 3697 (1993).

\bibitem{qi_wu_zhang_prb_06}
X.-L. Qi, Y.-S. Wu, and S.-C. Zhang, Phys. Rev. B {\bf 74}, 045125 (2006).

\bibitem{delplace_ullmo_montambaux_prb_11}
P. Delplace, D. Ullmo, and G. Montambaux, Phys. Rev. B {\bf 84}, 195452 (2011).

\bibitem{bulk_boundary_correspondence}
L. Fidkowski, T.S. Jackson, and I. Klich, Phys. Rev. Lett. {\bf 107}, 036601 (2011);
R.S.K. Mong and V. Shivamoggi, Phys. Rev. B {\bf 83}, 125109 (2011);
V. Gurarie, Phys. Rev. B {\bf 83}, 085426 (2011);
A.M. Essin and V. Gurarie, Phys. Rev. B {\bf 84}, 125132 (2011);
T. Fukui, K. Shiozaki, T. Fujiwara, and S. Fujimoto, J. Phys. Soc. Jpn. {\bf 81}, 114602 (2012); 
Y. Yu, Y.-S. Wu, and X. Xie, Nucl. Phys. B {\bf 916}, 550 (2017);
J.-W. Rhim, J.H. Bardarson, and R.-J. Slager, Phys. Rev. B {\bf 97}, 115143 (2018); 
M. Silveirinha, Phys. Rev. X {\bf 9}, 011037 (2019).






\bibitem{SSH}
W.P. Su, J.R. Schrieffer, and A.J. Heeger, 
Phys. Rev. Lett. {\bf 42}, 1698 (1979);



\bibitem{park_etal_prb_16}
J.-H. Park, G. Yang, J. Klinovaja, P. Stano, and D. Loss, Phys. Rev. B {\bf 94}, 075416 (2016). 

\bibitem{thakurathi_etal_prb_18}
M. Thakurathi, J. Klinovaja, and D. Loss, Phys. Rev. B {\bf 98}, 245404 (2018).



\bibitem{thouless_prb_83}
D. J. Thouless, Phys. Rev. B {\bf 27}, 6083 (1982).

\bibitem{hatsugai_fukui_prb_16}
Y. Hatsugai and T. Fukui, Phys. Rev. B {\bf 94}, 041102(R) (2016).





\bibitem{paper_prl}
M. Pletyukhov, D.M. Kennes, J. Klinovaja, D. Loss, and H. Schoeller, 
arXiv:1911.06890, submitted to Phys. Rev. Lett.



 \bibitem{AAH}
 Y. Lahini, R. Pugatch, F. Pozzi, M. Sorel, R. Morandotti, N. Davidson, and Y. Silberberg, 
 Phys. Rev. Lett. {\bf 103}, 013901 (2009); 
 M. Schreiber, S. S. Hodgman, P. Bordia, H. P. L{\"u}schen, M. H. Fischer,
 R. Vosk, E. Altman, U. Schneider, and I. Bloch, Science  {\bf 349}, 842 (2015);
 S. Ganeshan, K. Sun, and S. Das Sarma, Phys. Rev. Lett. {\bf 110}, 180403 (2013).











\bibitem{dana_jpc_85}
I. Dana, Y. Avron, and J. Zak, J. Phys. C {\bf 18}, L679 (1985).

\bibitem{kohmoto_prb_89_jpsj_92}
M. Kohmoto, Phys. Rev. B {\bf 39}, 11943 (1989);
{\it ibid.}, J. Phys. Soc. Jpn. {\bf 61}, 2645 (1992).

\bibitem{hatsugai_prb_93}
Y. Hatsugai, Phys. Rev. B {\bf 48}, 11851 (1993).








\bibitem{floquet_paper}
D. M. Kennes, N. M\"uller, M. Pletyukhov, C. Weber, C. Bruder, F. Hassler, J. Klinovaja, 
D. Loss, and H. Schoeller, Phys. Rev. B {\bf 100}, 041103 (2019). 



\bibitem{diagonalization_finite_tight_binding}
A. Alase, E. Cobanera, G. Ortiz, and L. Viola, Phys. Rev. B {\bf 96}, 195133 (2017);
F. K. Kunst, G. van Miert, and E. J. Bergholtz, Phys. Rev. B {\bf 99}, 085427 (2019).



\bibitem{rehr_kohn_prb_74}
J.J. Rehr and W. Kohn, Phys. Rev. B {\bf 10}, 448 (1974).

\bibitem{kohn_pr_59}
W. Kohn, Phys. Rev. {\bf 115}, 809 (1959).


\bibitem{hatsugai_jphys}
Y. Hatsugai, J. Phys.: Condens. Matter {\bf 9}, 2507 (1997).



\bibitem{kallin_halperin_prb_84}
C. Kallin and B.I. Halperin, Phys. Rev. B {\bf 29}, 2175 (1984).

\bibitem{friedel_58}
J Friedel, Del Nuovo Cimento {\bf 2}, 287 (1958).

\bibitem{he_vanderbilt_prl_01}
L. He and D. Vanderbilt, Phys. Rev. Lett. {\bf 86}, 5341 (2001).




\bibitem{SM}
see Supplemental Material, where the parameters used in the Figures are listed.




\bibitem{kohmoto_annals}
M. Kohmoto, Ann. Phys. {\bf 160}, 343 (1985).






\bibitem{multi-channel}
N. M\"uller, D.M. Kennes, M. Pletyukhov, J. Klinovaja, D. Loss, and H. Schoeller, in preparation.

\bibitem{floquet_stm}
N. M\"uller, D.M. Kennes, J. Klinovaja, D. Loss, and H. Schoeller, 
arXiv:1911.02295, submitted to Phys. Rev. B. 

\bibitem{lin_etal_preprint}
Y.-T. Lin, D.M. Kennes, V. Meden, and H. Schoeller, in preparation. 

\bibitem{gangadharaiah_etal_prl_12}
S. Gangadharaiah, L. Trifunovic, and D. Loss, Phys. Rev. Lett. {\bf 108}, 136803 (2012).

\bibitem{piasotski_etal_preprint}
K. Piasotski, D.M. Kennes, M. Pletyukhov, J. Klinovaja, D. Loss, V. Meden, and H. Schoeller, in preparation. 

\bibitem{FQHE}
J. Klinovaja, Y. Tserkovnyak, and D. Loss, Phys. Rev. B {\bf 91}, 085426 (2015).

\bibitem{weyl_preprint}
M. Pletyukhov, D.M. Kennes, K. Piasotski, J. Klinovaja, D. Loss, and H. Schoeller, in preparation. 






\end{thebibliography}
\end{document}